# Enumeration and representation theory of spin space groups

Xiaobing Chen[1,†], Jun Ren[1,†], Yanzhou Zhu[2,†], Yutong Yu[1,†], Ao Zhang[1], Pengfei Liu[1], Jiayu Li[1], Yuntian Liu[1], Caiheng Li[2] and Qihang Liu[1,3,4,*]

[1] *Department of Physics and Shenzhen Institute for Quantum Science and Engineering, Southern University of Science and Technology, Shenzhen 518055, China*

[2] *National Center for Applied Mathematics Shenzhen, and Department of Mathematics, Southern University of Science and Technology, Shenzhen 518055, China*

[3] *Guangdong Provincial Key Laboratory of Computational Science and Material Design, Southern University of Science and Technology, Shenzhen 518055, China*

[4] *Shenzhen Key Laboratory of Advanced Quantum Functional Materials and Devices, Southern University of Science and Technology, Shenzhen 518055, China*

[†] These authors contributed equally to this work.

[*] Email: liuqh@sustech.edu.cn

## Abstract

Those fundamental physical properties, such as phase transitions, Weyl fermions, and spin excitation, in all magnetic ordered materials, were ultimately believed to rely on the symmetry theory of magnetic space groups. Recently, it has come to light that a more comprehensive group, known as the spin space group (SSG), which combines separate spin and spatial operations, is necessary to fully characterize the geometry and underlying properties of magnetic ordered materials. However, the basic theory of SSG has seldom been developed. In this work, we present a systematic study of the enumeration and the representation theory of SSG. Starting from the 230 crystallographic space groups and finite translation groups with a maximum order of 8, we establish an extensive collection of over 100000 SSGs under a four-index nomenclature as well as the International notation. We then identify inequivalent SSGs specifically applicable to collinear, coplanar, and noncoplanar magnetic configurations. To facilitate the identification of SSG, we develop an online program (findspingroup.com) that can determine the SSG symmetries of any magnetic ordered crystals. Moreover, we derive the irreducible co-representations of the little group in momentum space within the SSG framework. Finally, we illustrate the SSG symmetries and physical effects beyond the framework of magnetic space groups through several representative material examples, including a well-known altermagnet $RuO_2$, spiral spin polarization in the coplanar antiferromagnet $CeAuAl_3$, and geometric Hall effect in the noncoplanar antiferromagnet $CoNb_3S_6$. Our work advances the field of group theory in describing magnetic ordered materials, opening up avenues for deeper comprehension and further exploration of emergent phenomena in magnetic materials.

## I. Introduction

Symmetry has always been one of the core aspects of physics and material science. The crystallographic group theory, which Fedorov and Schönflies refined at the end of the 19th century, provides a fundamental framework for understanding and predicting the properties and behavior of crystal solids, facilitating advancements in fields ranging from solid-state physics and chemistry to materials engineering [1]. The complete set of symmetry elements exhibited by three-dimensional (3D) nonmagnetic or paramagnetic solids are depicted by 32 point groups (PGs) and 230 space groups (SGs), the latter of which include rotations, reflections, inversion ($P$), translations, and their combinations. The wavefunction properties of such crystals, like phase transition, selection rules, and band degeneracy, are successfully described by the representations (reps) of 230 SGs.

The crystallographic group theory was further explored in the early 20th century when Shubnikov and others realized that magnetic materials exhibit additional symmetries beyond the purely geometric symmetries found in nonmagnetic crystals. These symmetries arise from the magnetic moments associated with the atoms or ions within the crystal lattice. By introducing antiunitary time-reversal operation $T$ that flips the direction of spin, the 32 PGs and 230 SGs are extended to 122 magnetic point groups (MPGs) and 1651 magnetic space groups (MSGs), respectively [2]. With the development of X-ray diffraction and neutron diffraction techniques, the structure of a given magnetic crystal and the orientation of its spins can be determined and assigned a specific MSG. Since then, the theory of MSGs has been believed to be the ultimate theory in understanding the exotic properties such as ordering phenomena, phase transitions, spin excitation, and band topology in all magnetic ordered materials [3-9].

However, as an axial vector, spin symmetry contains not only $T$ that reverses its direction (like an inversion operator in spin space), but also spin rotation operations. The underlying concept of MPG and MSG is based on the assumption that a rotation operator acts simultaneously on spatial and spin coordinates. Exemplified by a simple collinear antiferromagnetic order shown in Fig. 1, the system does not exhibit any four-fold rotations within the regime of magnetic groups. However, if we consider the spatial coordinate and spin coordinate separately, the system thus has a spatial four-fold operation followed by a two-fold spin rotation. Therefore, from a geometric perspective, separating rotational operations from real space and spin space is necessary to fully describe the symmetry of a system with a vector field [10]. Such an enhanced magnetic group, first proposed in the 1960s, is referred to as a spin group [11-18], which includes both spin point groups (SPGs) and spin space groups (SSGs).

From a physical perspective, the spin group also constitutes the symmetry-operator group of magnetic Hamiltonians, capturing the behavior of corresponding quasiparticles. Examples include single-particle Hamiltonians for electrons and the Heisenberg Hamiltonian for spin waves when spin-orbit coupling (SOC) is not considered [15, 19, 20]. Even in the presence of SOC with certain forms, the system can manifest some hidden symmetries associated with a particular spin group [21-23]. With the recent prosperity of antiferromagnetic spintronics [24-26], it has been recognized that antiferromagnetic order can generate various spintronics effects that typically emerge in systems with strong SOC, such as spin splitting, spin current, and spin torque [27-44]. Consequently, spin group symmetries naturally serve as a starting point for investigating these Hamiltonians and related effects. For example, the recently discovered “altermagnetic phase” in collinear antiferromagnets is defined in the SOC-free limit, and thus is delimited by spin groups [45-48]. Besides, the spin group also helps to understand the band structures, topology, transport phenomena in systems with negligible SOC [49-52], and provides descriptions of complex magnetic orderings (e.g., helical, spiral, quasi-one-dimensional helimagnets) [53-55] and fruitful applications to magnetically ordered quasicrystals [56-58].

The development of crystallographic group theory includes the enumeration and classification of groups, as well as the rep theory. While the theories of crystallographic groups and magnetic groups have been firmly established in textbooks [1, 2], the theory of spin groups is still in its nascent stage, with only the enumeration of 598 SPGs completed by far [18]. The enumeration of SSGs and the related rep theory describing band degeneracies and wavefunction properties are still lacking. This is because including degrees of freedom in both real space and spin space poses significant challenges in classifying SSGs. For example, the infiniteness of the translation group implies an infinite number of SSGs.

In this work, we present a systematic study of SSGs' enumeration and rep theory:

(i) The entire enumeration procedure starts from the 230 SGs, combining a translational supercell having a maximum order of 8. We traverse all inequivalent normal subgroups and establish their mapping onto a PG in spin space through inequivalent isomorphism relationships. Consequently, using group extension, we establish a comprehensive collection of more than 100000 SSGs, each labeled with a four-index and an International notation (Sec. II). The number of SSGs exceeds that of MSGs by two orders of magnitude because SSGs involve not only spin-flip operation $T$ but also partially decoupled spin rotation and lattice rotation. In addition, SSGs contain combined symmetry operations of spin rotation and lattice translation, corresponding to the propagation vector. Therefore, SSGs offer a more comprehensive symmetry description for magnetic structures in comparison with MSGs. Interesting examples include collinear magnetic structures in cubic crystal systems, spiral magnetic structures with arbitrary angles between nonmagnetic unit cells, cubic magnetic configurations within trigonal or hexagonal lattice systems, etc.

(ii) We then distinguish inequivalent SSGs for collinear, coplanar, and noncoplanar magnetic configurations, for which the nontrivial operations in spin space play a crucial role. We find that collinear SSGs manifest one-by-one correspondence with the 1421 MSGs, in which the magnetic sublattices with opposite directions of local moments can be regarded as black and white. In addition, the number of inequivalent coplanar SSGs is significantly reduced to 16383 due to the mirror symmetry in spin space. Moreover, in collinear and coplanar magnetic structures, the mirror symmetry in spin space serves as an effective time-reversal operation, thereby preventing the occurrence of SOC-free anomalous Hall effect.

(iii) In addition, we derive the irreducible co-representations (co-irreps) of the little group at high-symmetry momenta within the framework of SSG (Sec. III). The band representations of SSGs supporting noncollinear (coplanar and noncoplanar) magnetic configurations are constructed using the projective rep theory and the modified

Dimmock and Wheeler's sum rule. On the other hand, band representations of collinear SSGs are considered separately due to the presence of infinite symmetry operations in the spin-only group. Particularly, the $SO(2)$ spin rotation symmetry provides a simpler perspective in understanding spin degeneracy, where the co-irreps can be obtained from the single-valued co-irreps of the type-II MSG, augmented by some additional degeneracy rules in spin space. From the perspective of SSG, the complete criteria for magnetic order induced spin degeneracy/spin splitting can be straightforwardly summarized.

(iv) We then introduce our online program of SSG identification for any given magnetic structures [59], where the SSGs of all the ~1970 experimentally reported magnetic structures in MAGNDATA database [60] are also identified (Sec. IV). We also provide several representative material examples to demonstrate their SSG symmetries and physical properties beyond the framework of MSG. The first example is altermagnet $RuO_2$, in which extra band degeneracies are observed along Z-R-A high symmetry lines while MSG only provides one-dimensional co-irreps. These band degeneracies can be attributed to the sticking effect of two conjugated 1D irreps of $SO(2)$ in SSG. The second example is coplanar antiferromagnet $CeAuAl_3$ with $U\tau$ symmetry, the combination of a two-fold spin rotation $U$ and a fractional lattice translation $\tau$. A spiral order induced antiferromagnetic spin splitting is found, of which the symmetry requirement differs that of collinear antiferromagnets, i.e., breaking $PT$ and $U\tau$ [46]. The third example is noncoplanar antiferromagnet $CoNb_3S_6$. While $PT$ symmetry is broken, it manifests a spin-degenerate band structure within the whole Brillouin zone. Such spin degeneracy originates from the $D_2$ double group in spin space at any arbitrary wavevector, providing two-dimensional co-irreps and thus prohibiting any components of spin polarization. Despite its spin-degenerate band structure, $CoNb_3S_6$ exhibits a geometric hall effect, i.e., SOC-free anomalous Hall effect originated from magnetic configuration, arising from the nonzero $z$-direction orbital magnetization. Finally, we provide a summary of the work in Sec. V.

## II. Enumeration of spin space groups

In this section, we first introduce the construction methodology of SSGs, which is based on the group extension approach of SPGs [17, 18]. This approach has also been adopted in constructing the full set of crystallographic PGs, MPGs, and MSGs. However, the complexity of SSGs lies in the possible cell expansions to accommodate complicated magnetic orders, leading to an infinite number of SSGs. We thus consider the magnetic supercells with reduced translation symmetry to have a maximum order of 8, obtaining a comprehensive collection of 100612 SSGs. We then develop the four-index nomenclature for each SSG according to the way of group construction, and the International notation according to the types of Bravais lattices and symmetry operations. Furthermore, we discuss the features of SSGs when considering pure spin operations in collinear and coplanar configurations.

### A. Basics and methodology

Compared with MSGs, the critical characteristic of SSGs is that the spatial and spin operations are considered separately (see Appendix A for the types of symmetry operations) [19]. Therefore, the symmetry elements of spin groups can be written as $\{g_s \| g_l\}$, where $g_l$ denotes the spatial operation $\{C_{\boldsymbol{n}}(\theta), PC_{\boldsymbol{n}}(\theta)|\tau\}$ of the lattice ($C_{\boldsymbol{n}}(\theta)$, $P$, and $\tau$ are the rotation of $\theta$ angle along $\boldsymbol{n}$ axes, inversion and translation, respectively). To the left of the double bar, $g_s = \{U_{\boldsymbol{m}}(\phi), TU_{\boldsymbol{m}}(\phi)\}$ consists of spin rotations $U_{\boldsymbol{m}}(\phi) \in SU(2)$ and the anti-unitary time-reversal operator $T$ that reverses spin and momentum simultaneously. Similar to that of SPGs, the general form of SSGs, named $G_{SS}$, could be expressed as the direct product of a *nontrivial SSG* $G_{NS}$ and a *spin-only group* $G_{SO}$. Nontrivial SSG is the SSG that has every spin operation combined with a spatial operation (except for the identity element $\{E||E|0\}$); spin-only group consists of pure spin operations [17]:

$$G_{SS} = G_{NS} \times G_{SO}, \tag{1}$$

where $G_{SO}$ only contains pure spin operations $\{g_s||E|0\}$; $G_{NS}$ contains no pure spin operations [17]. While for noncoplanar magnetic order $G_{SO}^{n} = \{E||E|0\}$ (identity group), for coplanar magnetic order $G_{SO}^{p} = \{E, TU_{\boldsymbol{n}}(\pi)\} = Z_2^K$, implying the mirror symmetry in spin space ($\boldsymbol{n}$ is perpendicular to the coplanar order). For collinear magnetic order, $G_{SO}^{l} = Z_2^K \ltimes SO(2)$, where $SO(2) = \{U_z(\phi), \phi \in [0,2\pi]\}$ contains full spin rotations along a specific axis $z$ [19]. Next, we first focus on the construction of nontrivial SSGs, while the full SSGs directly constructed using Eq. (1) are discussed later.

In analogy to the construction of 598 SPGs, the central idea of constructing nontrivial SSGs is based on the decomposition of normal subgroups of crystallographic SGs and group extension. First, a SG $G_0$ can be decomposed into cosets with respect to one of its normal subgroups $L_0$,

$$G_0 = L_0 \cup g_1 L_0 \cup \ldots \cup g_{n-1} L_0. \quad (2)$$

When the resulting quotient group $G_0/L_0$ is isomorphic to a PG, one can use such a PG as the operations for spin space, $G^s = \{E, g_{s_1}, \ldots, g_{s_{n-1}}\}$ ($g_{s_i} \in SO(3) \times \{E, T\} \cong O(3)$), and map $G^s$ to $G_0/L_0$ through isomorphism relationship, forming a SSG written as

$$G_{NS} = \{E||L_0\} \cup \{g_{s_1}||g_1 L_0\} \cup \ldots \cup \{g_{s_{n-1}}||g_{n-1} L_0\}. \quad (3)$$

The enumeration of all the possible SSGs includes exhaustively enumerating all $(G_0, L_0)$ pairs and inequivalent choices of coset representative elements, finding all $G^s$ isomorphic to $G_0/L_0$ in spin space, and finding all inequivalent mappings between $G^s$ and $G_0/L_0$. However, compared with that of SPGs, the enumeration of SSGs faces more challenges due to the existence of crystallographic translation groups of SGs. To distinguish different cases, three types of subgroups $L_0$ are categorized:

1) A subgroup $L_0$ of a SG $G_0$ is referred to as a *translationengleiche* subgroup [61] or a *t*-subgroup of $G_0$ if their translation subgroup $T(G_0)$ is retained, i.e., $T(L_0) = T(G_0)$, while the PG part $P(L_0)$ is a subgroup, i.e., $P(L_0) \leq P(G_0)$. A SSG formed in this way is called a ***t-type*** SSG, in which the t-index $i_t$ is used to present the reduction of the PG symmetry $i_t = |P(G_0)|/|P(L_0)|$. It is straightforward that the quotient groups $G_0/L_0$ of t-type SSGs must be isomorphic to one of the 32 crystallographic PGs. Therefore, the derivation of 598 nontrivial SPGs can be directly extended to obtain *t*-type SSGs.

2) A subgroup $L_0$ of a SG $G_0$ is referred to as a *klassengleiche* subgroup [61] or a *k*-subgroup of $G_0$ if the PG part $P(G_0)$ is retained, i.e., $P(L_0) = P(G_0)$, while the translation subgroup $T(L_0)$ is reduced to $T(L_0) < T(G_0)$. A SSG formed in this way is called a ***k-type*** SSG, in which the *k*-index $i_k$ is used to present the reduction of the translation symmetry $i_k = |T(G_0)|/|T(L_0)|$. In general, the *k*-subgroup implies cell multiplication to accommodate the magnetic configuration. Therefore, unlike $i_t$ which can be solely determined by the PG part of $G_0$ and $L_0$, $i_k$ is an independent condition to label an SSG. Importantly, since $G_0$ and $L_0$ have the same PG part, the quotient group $G_0/L_0$ is isomorphic to the quotient group of their translation part $T(G_0)/T(L_0)$, which forms a 3D lattice translation group $Z_{n_1} \times Z_{n_2} \times Z_{n_3}$, where $n_i$ are natural numbers. As a result, the group structures of $G_0/L_0$ are not limited to 32 crystallographic PGs (could even be a non-PG) and are countless, leading to infinite numbers of SSGs. In this work, we set a cutoff of $i_k = n_1 n_2 n_3 \leq 8$ to enumerate the SSGs that require cell expansion. For example, a SSG, whose translation quotient group $G_0/L_0$ has the structure of $Z_5$, allows for a five-fold rotation of spins propagating along a specific direction. For the commensurate magnets with $i_k > 8$, their SSGs can still be diagnosed case by case by our procedure.

3) Those $L_0$ that have lost PG operations as well as translation operations are called *general* subgroups of $G_0$ [$P(L_0) < P(G_0)$ and $T(L_0) < T(G_0)$]. A SSG formed in this way is called a ***g-type*** SSG, for which the quotient group $G_0/L_0$ is not necessarily Abelian. For example, $G_0/L_0$ could be isomorphic to a dihedral group $D_n$ with $n$ being any positive integers. In this case, $T(G_0)/T(L_0)$ is isomorphic to $Z_n$ which contributes a generator $r$ of order $n$, while $P(G_0)/P(L_0)$ is isomorphic to $Z_2$ that contributes a generator $h$ of order 2. When the coset representative $h$ is chosen as 2-fold rotation along an axis perpendicular to the translation vector of $r$, the two generators satisfy $hrh^{-1} = r^{-1}$, leading to $G_0/L_0 \cong D_n$. Through our construction procedure with a specific $i_k$ cutoff, we find that $g$-type subgroups constitute most of the total (over 85%).

To mathematically enumerate all group-normal subgroup pairs $(G_0, L_0)$ with a specific $i_k$ index, we adopt an inverse procedure, starting from $L_0$ and finding all supergroups $G_0$ whose subgroup contains $L_0$. The full list of group-normal subgroup pairs is obtained with the assistance of SUPERGROUPS program [62] implemented in Bilbao Crystallographic Server [63], leading to 10660 combinations of $(G_0, L_0, i_k)$. For each combination, there may be multiple possibilities with different coset representative elements, which relates to the symmetry operation between different magnetic sublattices. Furthermore, for a given $i_k$ there are also multiple possibilities fulfilling $i_k = n_1 n_2 n_3$, which correlates the cell expansion along different directions.

After coset decomposition, we next consider the coupled PG in spin space, $G^S \cong G_0/L_0$. The isomorphism relationship implies that given a specific combination of $(G_0, L_0, i_k)$, $G^S$ could have different choices. For example, if the quotient group $G_0/L_0$ is isomorphic to a $D_8$, $G^S$ could be $D_8$, $C_{8v}$, and $D_{4d}$. Moreover, for a given $G^S$, the isomorphic mapping between $G^S$ and $G_0/L_0$ also have multiple possibilities, leading to different nontrivial SSGs according to Eq. (3). Therefore, by enumerating inequivalent coset decompositions of $G_0/L_0$, different $G^S$ and inequivalent mappings between $G^S$ and $G_0/L_0$, we are able to identify a SSG by a four-index $(L_0, G_0, i_k, m)$ label. The mathematical procedure of finding inequivalent coset decompositions and inequivalent mappings is provided in Appendix A.

**B. Nontrivial SSGs**

Through the abovementioned methodology, we can mathematically obtain 122 types of $G^S$, and 100612 nontrivial SSGs ($G_{NS}$) with the cell expansion limited to $i_k \leq 8$. The complete list, provided in Supplementary Materials, includes 8505 $t$-type, 6738 $k$-type and 85369 $g$-type SSGs. In Table I, we show the statistics of SSGs classified by the crystal systems of $G_0$, which is the SG of the nonmagnetic symmetry of the magnetic primitive cell. It is shown that most SSGs concentrate in crystal systems with relatively lower symmetry, i.e., monoclinic, orthorhombic, and tetragonal (89% in total). In Table II, we show the 122 types of $G^S$ in spin space for all the SSGs with $i_k \leq 8$, 90 of which are not crystallographic PGs.

We next take a series of $t$-type nontrivial SSGs to illustrate the correspondence of the group construction and the realistic magnetic structure. For clarity, we use the International symbol to present group operations hereafter, where *P* and *T* are denoted by $\bar{1}$ in real space and spin space, respectively. Considering two SGs $L_0$ = *P2/c* (No. 13) and $G_0$ = *Pcca* (No. 54) fulfilling normal subgroup relationship, $G_0 = L_0 \cup \{2_{010}|0\ 0\ 1/2\}L_0$. One can define two different sublattices [marked by different colors in Fig. 2(a)] with the group elements of $L_0$ keeping the sublattice invariant. Thus, $L_0$ can be defined as the *sublattice group*. On the other hand, the coset elements $\{2_{010}|0\ 0\ 1/2\}L_0$ transform one sublattice to the other, as shown in Fig. 2(b). Therefore, there are three possibilities of $G^S$ being isomorphic to $Z_2$ ($G_0/L_0$), i.e., $\bar{1}$, 2, and $m$, leading to three inequivalent mappings for the magnetic moments, as shown in Figs. 2(c-e). Taking (13.54.1.1) as an example, the nontrivial SSG is constructed by mapping the $\bar{1}$ operator in spin space (i.e., *T*) to the representative element $\{2_{010}|0\ 0\ 1/2\}$, leading to $G_{NS} = \{E||L_0\} \cup \{\bar{1}||(2_{010}|0\ 0\ 1/2)L_0\}$.

While there is no multiplicity of different mappings for a specific order-2 $G^S$, there is another inequivalent coset decomposition $G_0 = L_0 \cup \{2_{001}|1/2\ 0\ 0\}L_0$ for $(L_0, G_0)$, owing to the inequivalence between *a* and *c* axis for SG *P2/c*. Consequently, there are another three nontrivial SSGs with a different choice of sublattices, as shown in Figs. 2(f-h). Overall, there are 6 *t*-type nontrivial SSGs for the given $(L_0, G_0)$ pair, with two supporting collinear magnetic structures ($G^S = \bar{1}$), and four of them supporting coplanar magnetic structures ($G^S = 2$ and *m*). However, we will show later that when considering spin-only group, the SSGs generated by $G^S = 2$ and $m$ are indeed equivalent. The group constructions of *k*-type and *g*-type nontrivial SSGs follow similar procedures, as exemplified by the SSGs of realistic materials $CeAuAl_3$ and $CoNb_3S_6$, respectively (see Sec. IVC and IVD).

### C. International notation

International notation, also known as Hermann-Mauguin notation, is a widely used nomenclature for PGs, SGs, MPGs, and MSGs. It is employed in the standard structural and symmetry reference, the International Tables for Crystallography [64], and is extensively utilized in various crystallography textbooks. Compared with Schoenflies notation, International notation is more favorable for presenting the directions of the symmetry axes and the translation symmetry elements in SGs.

We next develop the International notation for all the SSGs. The International notation of a SG is denoted as $Bg_1g_2g_3$, where the first uppercase letter $B$ describes the Bravais lattice, including primitive (P), based-centered (A, B, C), body-centered (I), face-centered (F) and rhombohedral (R) lattice. The following three letters describe the representative symmetry operations, as defined in Appendix Table B2. Since the translation operations in real space could also couple a PG operation in spin space in SSGs, we first need to expand the $Bg_1g_2g_3$ notation to a more comprehensive one $Bg_1g_2g_3t_at_bt_cb_1b_2b_3$. Here $t_{a\sim c}$ present the integral translation of cell expansion, while for SGs we simply have $t_a = (100), t_b = (010), t_c = (001)$. $b_{1\sim 3}$ stand for the fractional translations for the specific Bravais centering-type. The specific $b_1$, $b_2$, and $b_3$ for each Bravais lattice can be found in Appendix Table B5 (see more details on the International notation and operation symbols for SSG in Appendix B).

To construct a SSG, a SG $G_0 = Bg_1g_2g_3t_at_bt_cb_1b_2b_3$ is decomposed into cosets with respect to one of its normal subgroups $L_0$ by Eq. (2), and then map a PG $G^s$ to the quotient group $G_0/L_0$ through isomorphism relationship by Eq. (3). Consequently, each of the representative symmetry operation $g_{1\sim 3}$, Bravais centering-type fractional translation $b_{1\sim 3}$, and integral translation $t_{a\sim c}$ could combine with a spin PG operation under the basis of $G_0$. Such nomenclature is done for *t*-type and *g*-type SSGs. For *k*-type SSGs, we use a multi-color extension of the BNS convention of MSGs [65, 66], yielding more convenience and simpler notations (see Appendix C2).

Now we turn to the specific nomenclature of three types of SSGs. For *t*-type SSGs, $T(L_0) = T(G_0), P(L_0) \leq P(G_0)$. This implies that the group element of $G^s$ (excluding the identity) in a *t*-type SSG can only be combined with nontrivial SG operations $\{R|0\}$ or $\{R|\tau\}$ with $R \neq E$, rather than pure translations $\{E|\tau\}$. As a result, the Bravais fractional translation $b_{1\sim 3}$ and integral translation $t_{a\sim c}$ always combine with identity in spin space in *t*-type SSG, thus they can be omitted. In other words, *t*-type SSG can be described following Litvin's notation on SPGs [18], i.e., $B^{g_{s_1}}g_1{}^{g_{s_2}}g_2{}^{g_{s_3}}g_3$ in $G_0$ basis. The corresponding symmetry operations can be directly constructed by $\{g_{s_1}||g_1\}$, $\{g_{s_2} \| g_2\}$, and $\{g_{s_3} \| g_3\}$. We still take SSG (13.54.1.1) as an example, where the symmetry operations and the resultant general positions are listed in Table III. Note that Wyckoff positions and the corresponding magnetic configurations are constructed by using site-symmetry groups of the SSG, with the details provided in Appendix D. It can be found that for $G_0$ = $Pcca$ the $\bar{1}$ operator in spin space is mapped to the glide reflections $\{m_{100}|1/2\ 0\ 1/2\}$ and $\{m_{010}|0\ 0\ 1/2\}$, leading to the International notation $P^{-1}c^{-1}c^{1}a$. Similarly, the International notations for (13.54.1.2) to (13.54.1.6) shown in Fig. 2 are $P^{2_{001}}c^{2_{001}}c^{1}a$, $P^{m_{001}}c^{m_{001}}c^{1}a$, $P^{-1}c^{1}c^{-1}a$, $P^{2_{001}}c^{1}c^{2_{001}}a$ and $P^{m_{001}}c^{1}c^{m_{001}}a$, respectively (see Appendix C for details).

For *k*-type SSGs, $T(L_0) < T(G_0), P(L_0) = P(G_0)$. This indicates that a *k*-type SSG can be constructed directly using the sublattice SG $\{E \| L_0\}$ and an additional spin translation group $G_T^S = \left\{\{1\|1|0\}, \{g_{s_1}\|1|\tau_1\}, \ldots, \{g_{s_{n-1}}\|1|\tau_{n-1}\}\right\}$ generated by $\{g_{s_1}\|1|\tau_1\}$, $\{g_{s_2}\|1|\tau_2\}$ and $\{g_{s_3}\|1|\tau_3\}$. Among them, $\tau_i = (a_i\ b_i\ c_i)$ denote the fractional translations after cell expansion along the three basis vectors of $L_0$, and the product of the multiplicities of $\tau_{1-3}$ fulfills the imposed cutoff $i_k \leq 8$. Note that in *k*-type SSGs, $L_0$ and $G_0$ always belong to the same crystal systems, but not necessarily the same Bravais lattice. As a result, it is more convenient to denote *k*-type SSG as $B^{1}g_1{}^{1}g_2{}^{1}g_3{}^{g_{s_1}}\tau_1{}^{g_{s_2}}\tau_2{}^{g_{s_3}}\tau_3$ in $L_0$ basis, where $\tau_i = (a_i\ b_i\ c_i)$. In this sense, the symmetry operations of a *k*-type SSG can be directly constructed from the direct product of sublattice SG $\{E \| B g_1 g_2 g_3\}$ and $G_T^S$. Indeed, such nomenclature is a natural multi-color extension of the conventional BNS setting for type-IV MSGs [65, 66]. Specifically, Type-IV MSGs have the form $G + T\tau G$ ($i_k = 2$) and can thus be denoted by $B_X g_1 g_2 g_3$, where $X$ labels the black and white Bravais lattices. In comparison, the complexity of multi-color Bravais lattices is incorporated by the notation of ${}^{g_{s_1}}\tau_1{}^{g_{s_2}}\tau_2{}^{g_{s_3}}\tau_3$. We provide an example of a *k*-type SSG (99.107.4.1), whose International notation is written as $P^{1}4^{1}m^{1}m^{4^{1}_{001}}(1/2\ 1/2\ 1/4)$, of a realistic materials $CeAuAl_3$. A detailed explanation of this material example can be found in Sec. IVC and Appendix C.

Now we turn to the $g$-type SSGs with $T(L_0) < T(G_0)$ and $P(L_0) < P(G_0)$. The group element of $G^s$ in spin space will combine with the representative symmetry operation $g_{1\sim3}$, integral translation $t_{a\sim c}$ and Bravais fractional translation $b_{1\sim3}$, no matter in $G_0$ or $L_0$ basis. In $g$-type SSGs, $L_0$ and $G_0$ could belong to different crystal systems. Consequently, it is advantageous to write the International notation of the $g$-type SSGs in $G_0$ basis, as the first letter $B$ will reflect the information of the Bravais lattice of the magnetic cell (see Appendix C). Since each of the spatial operations is allowed to connect an independent rotation in spin space, the $g$-type SSG are thus denoted as $B^{g_{s_1}}g_1{}^{g_{s_2}}g_2{}^{g_{s_3}}g_3|(g_{s_4}, g_{s_5}, g_{s_6}; g_{s_7}, g_{s_8}, g_{s_9})$ in $G_0$ basis. Note that $g_{s_4}, g_{s_5}, g_{s_6}$ denote the spin rotation associated with $t_a, t_b, t_c$, while $g_{s_7}, g_{s_8}, g_{s_9}$ denote the spin rotation associated with $b_1, b_2, b_3$. Particularly, for $P$ Bravais lattice, $g_{s_7}, g_{s_8}, g_{s_9}$ can be simply omitted because all centering-type fractional translations $b_1, b_2, b_3$ are zero (Appendix Table B5). We provide an example of a $g$-type SSG (4.182.4.2), whose International notation is written as $P^{3^2_{-11-1}}6_3{}^{m_{110}}2^{m_{011}}2|(2_{001}, 2_{100}, 1)$, of a realistic material $CoNb_3S_6$. A detailed explanation of this material example can be found in Sec. IVD and Appendix C.

Besides the examples mentioned above, we present more examples of various complicated cases of the three types of SSGs in Appendix C. In addition, the International notations of all the enumerated nontrivial SSGs are provided in Supplementary Materials as well as our online program [59].

### D. SSGs for different magnetic configurations

In Table I we enumerate the nontrivial SSGs supporting collinear-only, noncoplanar-only, and coplanar magnetic configurations. We next implement spin-only group $G_{SO}$ into $G_{NS}$ to count all inequivalent SSGs for different magnetic configurations, including collinear, coplanar, and noncoplanar orders. We will elucidate that $G_{SO}$ plays a crucial role in finding equivalent collinear and coplanar SSGs. Without the loss of generality, we assume that each sublattice only contains one type of magnetic ions. The enumeration process is based on different $G^s$, and is divided into the following steps:

**1) Collinear SSGs**. In collinear magnets, all local moments point towards the same or opposite direction (e.g., the $z$ axis), and the corresponding SSG does not depend on the direction of the spins. The full SSG of a collinear magnet is written as $G_{SS} = G_{NS} \times Z_2^K \ltimes SO(2)$. The spin-only group $G_{SO}^l = Z_2^K \ltimes SO(2)$ (the International notation is ${}^{\infty m}1$) of collinear SSG ensures that only time-reversal $T$ needs to be considered for the spin space of $G_{NS}$, rendering only two possibilities, $G^s = 1$ and $\bar{1}$, corresponding to ferromagnets (ferrimagnets) and antiferromagnets, respectively. Therefore, collinear SSGs manifest one-by-one correspondence with the structure of MSGs, in which the magnetic sublattices with opposite directions of local moments can be regarded as black and white.

Specifically, there are 230 SSGs for collinear ferromagnets or ferrimagnets with their nontrivial SSG $G_{NS} = \{E||G_0\}$, where $G_0$ is one of the 230 SGs. These nontrivial $t$-type SSGs also have $G_0 = L_0$ and $G^s = 1$, corresponding to 230 type-I MSGs. For antiferromagnets, $G^s = \bar{1}$, the nontrivial SSG can thus be written as $\{E||L_0\} \cup \{T||AL_0\}$, where $A$ denotes the symmetry operation connecting the sublattices with opposite spins. If $A$ is a pure translation, i.e., $P(L_0) = P(G_0)$ and $i_k = 2$, the resulting SSGs correspond to 517 type-IV MSGs. On the other hand, if $A$ is inversion or rotation (proper or improper), the resulting SSGs with $T(L_0) = T(G_0)$ and $i_t = 2$ correspond to 674 type-III MSGs. Overall, there are 1421 inequivalent collinear SSGs in total, as marked by double stars in Supplementary Materials.

**2) Coplanar SSGs**. For those SSGs describing coplanar magnetic configurations, the spin-only group is $G_{SO}^p = \{E, TU_{\boldsymbol{n}}(\pi)\}$ (the International notation is ${}^{m}1$). Thus, all the spin rotation axes of $G^s$ should be either perpendicular or parallel to a specific axis $\boldsymbol{n}$, implying that polyhedral PGs are excluded for $G^s$. In addition, due to the existence of the spin-only group $G_{SO}^p$, we can further limit $G^s$ to unitary PGs, which do not contain $T$. To prove this, we consider an antiunitary nontrivial SSG $G_{NS}^{AU}$ with its maximal unitary subgroup $L_{NS}^{MU}$:

$$G_{NS}^{AU} \times \{E, TU_n(\pi)\} = [L_{NS}^{MU} \cup (G_{NS}^{AU} \backslash L_{NS}^{MU})] \times \{E, TU_n(\pi)\}$$
$$= [L_{NS}^{MU} \cup TU_n(\pi)(G_{NS}^{AU} \backslash L_{NS}^{MU})] \times \{E, TU_n(\pi)\} = G_{NS}^{U} \times \{E, TU_n(\pi)\}. \quad (4)$$

Eq. (4) suggests that for any antiunitary nontrivial SSG $G_{NS}^{AU}$, we can construct an equivalent unitary nontrivial SSG $G_{NS}^{U} = L_{NS}^{MU} \cup TU_n(\pi)(G_{NS}^{AU} \backslash L_{NS}^{MU})$ when taking $G_{SO}^{p}$ into account. Therefore, to obtain all inequivalent coplanar SSGs from the full $(G_0, L_0, i_k, m)$ SSG list, it is adequate to add the condition $G^s \cong C_n$ or $D_n$. Note that for $k$-subgroups and $g$-subgroups, $C_n$ and $D_n$ groups are not limited to crystallographic PGs (e.g., $Z_5$), resulting in 40 possibilities out of 122 $G^s$'s. Within considered cell multiplicity $i_k \leq 8$, we find 16383 inequivalent coplanar SSGs, which is a significantly reduced number compared with the number of $G_{NS}$ that support coplanar magnetic configurations (98834).

We note two special portions of $G_{NS}$ for coplanar configurations. The first one is $G^s = 2$ and $m$, for which the $G_{NS}$ support both collinear and coplanar (but not noncoplanar) configurations. However, when they describe collinear configurations, they are exactly equivalent to the SSGs with $G^s = \bar{1}$ due to the existence of $G_{SO}^{l}$, so they do not contribute new entries to 1421 inequivalent collinear SSGs. On the other hand, when they describe coplanar configurations, they are equivalent to each other because of Eq. (4), contributing 1191 inequivalent coplanar SSGs. Therefore, there are 1191 entries with $G^s = m$ rendering equivalent SSGs when considering $G_{SO}$ part. Specifically, as for our example, (13.54.1.2) and (13.54.1.3) [Figs. 2(d) and 2(e)] are equivalent coplanar SSGs.

The second case is $G^s = 2/m$, for which the $G_{NS}$ supports coplanar configurations only. However, considering $G_{SO}^{p}$, the SSGs generated by antiunitary $2/m$ is equivalent to those generated from $G^s = 222$ according to Eq. (4). Therefore, there are 9501 entries with $G^s = 2/m$ rendering equivalent SSGs when considering $G_{SO}$ part.

**3) Noncoplanar SSGs**. The spin-only group for noncoplanar SSGs is simply an identity group. When $G^s$ is a polyhedral PG ($T$, $T_d$, $T_h$, $O$, and $O_h$), there are 357 SSGs in total supporting only noncoplanar magnetic configurations. On the other hand, when taking other $G^s$ except 1, $\bar{1}$, 2, $m$, 2/$m$, and polyhedral PGs, the corresponding $G_{NS}$ (86951 entries) support both coplanar and noncoplanar configurations. Since $G_{SS} = G_{NS}$, all these SSGs for noncoplanar configurations are inequivalent. Finally, there are 87308 inequivalent noncoplanar SSGs in total.

We show in Fig. 3 the summary of the inequivalent SSGs for different magnetic configurations, indicating that the spin-only groups serve as another factor for equivalent SSGs. For example, the nontrivial SSGs supporting coplanar magnetic configurations indeed form a small subset of inequivalent coplanar SSGs (98834).

## III. Representation of spin space groups

Representation theory is the key to encoding the information of symmetry to quantum-mechanics wavefunctions, which determine the elementary excitations, geometric phases, selection rules, etc. In this section, we explore the general rep theory of SSGs, i.e., obtaining the projective co-irreps of the little group $G^k$, which consists of all symmetry operations that leave *k* invariant. In the following, we separately consider the SSGs supporting noncollinear (coplanar and noncoplanar) and collinear magnetic configurations, because the latter possesses infinite symmetry operations owing to the spin-only group $Z_2^K \ltimes SO(2)$.

For a noncollinear unitary SSG, we examine its regular projective reps, which are reducible reps containing all the possible irreducible representations (irreps). We then apply an approach that utilizes the eigenvalues and eigenvectors of a complete set of commuting operators (CSCO) to decompose the regular projective rep to obtain all irreps of the unitary SSG [67, 68]. Finally, we apply a modified Dimmock and Wheeler's character sum rule to derive the co-irreps of anti-unitary SSGs. To illustrate the procedure, we derive the co-irreps of the little groups for two materials, i.e., $CeAuAl_3$ and $CoNb_3S_6$. The comparison of co-irreps under SSG and MSG shows that the co-irreps in SSG may have higher dimensions.

The 1421 collinear SSGs have one-by-one correspondence to Type-I, III and IV MSGs. However, we show that the co-irreps of collinear SSGs can be obtained by considering the co-irreps of single-valued type-II MSGs plus the extra degeneracies that originated from some critical spin symmetries. To illustrate this, we compare the dimensions of the co-irreps under SSG and MSG for a collinear antiferromagnet $RuO_2$.

## A. Basics of representation theory for SGs and MSGs

We first briefly review the traditional approach to obtain the irreps of little group $G^k = \{R_1|\tau_1\}\mathbb{T} + \{R_2|\tau_2\}\mathbb{T} + \cdots + \{R_n|\tau_n\}\mathbb{T}$ ( $\mathbb{T}$ denotes the crystallographic translation group; $R_i$ and $\tau_i$ denote the real-space PG operation and fractional translation, respectively) in SGs and MSGs, with more details provided in Appendix A. In SGs, the complexity of reps comes from the fractional translations. Specifically, the group of rep matrices cover the corresponding little co-group $\breve{G}^k = \{R_1, R_2, \ldots, R_n\}$ multiple times because one $R_k$ could combine with multiple translations. To simplify the problem, the theory of projective rep is used to "mod" the influence of fractional translations and get all irreps of $\tilde{G}^k = G^k/\mathbb{T} = \{\{R_1|\tau_1\}, \{R_2|\tau_2\}, \ldots, \{R_n|\tau_n\}\}$. In SGs, a projective irrep $M_k^l(\{R_i|\tau_i\})$ of a little group $G^k$ leaves the complexity of fractional translation to the factor system, i.e., $d_k^l(\{R_i|\tau_i\}) = exp(-ik \cdot \tau_i) M_k^l(\{R_i|\tau_i\})$, where $d_k^l(\{R_i|\tau_i\})$ denote the *l*-th irrep of $\tilde{G}^k$. On the other hand, $M_k^l(\{R_i|\tau_i\})$ can be obtained by looking for the reps of an extended group constructed by $\breve{G}^k$ and a cyclic group formed by the factor system $exp(-iK_i \cdot \tau_i)$, which is also known as central extension ($K_i = R_i^{-1}k - k$). This means that to obtain all the irreps of a SG it is only necessary to find the reps of a few cyclic groups in addition to those of PGs ($\breve{G}^k$) that are already known [2].

For MSGs, the incorporation of antiunitary operations gives rise to co-irreps, which can be obtained in two steps. Since the little group $G_{MS}^k$ for a given MSG could contain antiunitary operations, the first step is to get the irreps of the maximal unitary subgroup of $G_{MS}^k$; they are calculated following the procedure of a SG discussed above. The second step is to take into account the antiunitary elements of $G_{MS}^k$, leading to three types of co-irreps according to the famous Dimmock and Wheeler's sum rule. The theory of co-irreps is useful for analyzing extra degeneracies caused by antiunitary operations and the corresponding degenerate states [2]. The details of projective rep, central extension, and Dimmock and Wheeler's sum rule are introduced in Appendix A.

**B. Representation of coplanar and noncoplanar (nontrivial) SSGs**

The rep theory of SSG developed here combined the theory of projective reps in SG and a modified version of Dimmock and Wheeler's sum rule in MSG. In the projective reps, the effect of unitary spin rotation is absorbed into the factor system. However, the little group of a SSG may contain some "spin nonsymmorphic symmetries" $\{g_s||E|\tau\}$, i.e., combined operations of spin operation $g_s$ and fractional translation $\tau$, rendering central extension method rather complicated. Instead, we focus on the regular projective reps that contain all the irreps. These regular projective reps are decomposed into projective irreps using CSCO method.

Unlike SGs, a single lattice operation $R_i$ in SSGs may correspond to multiple nonsymmorphic translations $\tau_i^{(a)}$ for the case of $i_k > 1$, as well as spin operations $g_{s_i}^{(a)}$, where $a$ labels different translations and spin operations accompanied with one $R_i$. In addition, the factor system can also absorb an additional phase of spin rotation $g_{s_i}^{(a)}$, where the rotation of $2\pi$ picks up a phase of -1. Considering $d_k^l\left(\left\{g_{s_i}^{(a)}\middle\|R_i\middle|\tau_i^{(a)}\right\}\right)$ an irrep of a unitary little SSG $G_{SS}^k$, we have:

$$d_k^l\left(\left\{g_{s_i}^{(a)}\middle\|R_i\middle|\tau_i^{(a)}\right\}\right) = exp\left(-ik\cdot\tau_i^{(a)}\right)M_k^l\left(\left\{g_{s_i}^{(a)}\middle\|R_i\middle|\tau_i^{(a)}\right\}\right). \quad (5)$$

Here $0 \le \phi\left(g_{s_i}^{(a)}\right) < 2\pi$ , in which $\phi\left(g_{s_i}^{(a)}\right)$ is the rotation angle of $g_{s_i}^{(a)}$. We note that projective reps $M_k^l$ do not distinguish translation operations that differ by integer multiples of lattice vectors, i.e., $M_k^l\left(\left\{g_{s_i}^{(a)}\middle\|R_i\middle|\tau_i^{(a)}+t_i\right\}\right) = M_k^l\left(\left\{g_{s_i}^{(a)}\middle\|R_i\middle|\tau_i^{(a)}+t_j\right\}\right)$ for $t_i, t_j \in T(L_0)$. That is to say, $M_k^l\left(\left\{g_{s_i}^{(a)}\middle\|R_i\middle|\tau_i^{(a)}\right\}\right)$ is a matrix-valued function on the elements $\left\{g_{s_i}^{(a)}\middle\|R_i\middle|\tau_i^{(a)}\right\}$ of the finite quotient group $\tilde{G}_{SS}^k = G_{SS}^k/T(L_0)$. According to the multiplication of reps $d_k^l$, for the projective reps of elements $\left\{g_{s_i}^{(a)}\middle\|R_i\middle|\tau_i^{(a)}\right\}$, $\left\{g_{s_j}^{(a)}\middle\|\ R_j\middle|\tau_j^{(a)}\right\} \in \tilde{G}_{SS}^k$ $[\,0 \le \phi\left(g_{s_i}^{(a)}\right), \phi\left(g_{s_j}^{(a)}\right) < 2\pi\,]$ we have:

$$M_k^l\left(\left\{g_{s_i}^{(a)}\middle\|R_i\middle|\tau_i^{(a)}\right\}\right)M_k^l\left(\left\{g_{s_j}^{(a)}\middle\|\ R_j\middle|\tau_j^{(a)}\right\}\right)$$
$$= (-1)^{\xi\left(g_{s_i}^{(a)},g_{s_j}^{(a)}\right)} exp(-iK_i\cdot\tau_j^{(a)})M_k^l\left(\left\{g_{s_l}^{(a)}\middle\|R_l\middle|\tau_l^{(a)}\right\}\right). \quad (6)$$

Here $\left\{g_{s_l}^{(a)}\middle\|R_l\middle|\tau_l^{(a)}\right\} = \left\{g_{s_i}^{(a)}g_{s_j}^{(a)}\middle\|\ R_iR_j\middle|\tau_i^{(a)} + R_i\tau_j^{(a)}\ mod\ T(L_0)\right\}$, $K_i = R_i^{-1}k - k$, $\xi\left(g_{s_i}^{(a)}, g_{s_j}^{(a)}\right) = 0$ for $0 \le \phi(g_{s_i}^{(a)}g_{s_j}^{(a)}) < 2\pi$ and $\xi\left(g_{s_i}^{(a)}, g_{s_j}^{(a)}\right) = 1$ for $2\pi \le \phi(g_{s_i}^{(a)}g_{s_j}^{(a)}) < 4\pi$. Therefore, it is only necessary to find out all projective irreps of $\tilde{G}_{SS}^k$. For simplicity, we use $g_i^{(a)}$ to represent $\left\{g_{s_i}^{(a)}\middle\|R_i\middle|\tau_i^{(a)}\right\}$ hereafter.

Next, CSCO method is employed to obtain projective irreps by decomposing the regular projective reps [67, 68]. In quantum mechanics, a set of commuting operators $(J^2, J_z)$ form the CSCO of the Hilbert space of angular momentum. The corresponding quantum numbers, $j$ and $m_j$, are sufficient to diagonalize the Hamiltonian and label all the resulting eigenstates. Similarly, the basic idea of CSCO approach used here is to decompose the projective reps of $\tilde{G}_{SS}^k$ into blocks to distinguish all the irreps of $\tilde{G}_{SS}^k$. This is done by constructing a series of class operators analogous to the set of $(J^2, J_z)$, which commute with each other and commute with every group element. The specific procedure is described in Appendix E.

For any antiunitary little group $G_{SS}^k$, one can first decompose it with respect to its maximal unitary subgroups $L_{SS}^k$: $G_{SS}^k = L_{SS}^k \cup TAL_{SS}^k$, where $TA$ is the antiunitary coset representative element. The projective irreps $M_k^l\left(g_i^{(a)}\right)$ and the corresponding irreps $d_k^l\left(g_i^{(a)}\right)$or $g_i^{(a)} \in \tilde{L}_{SS}^k = L_{SS}^k/\mathbb{T}$ can certainly be treated by using the CSCO method. Assuming that the basis set of $d_k^l\left(g_i^{(a)}\right)$ is $|\psi\rangle = |\psi_1, \psi_2, \dots, \psi_{n_l}\rangle$, then for the coset the basis set can be adopted as $|\phi\rangle = |\phi_1, \phi_2, \dots, \phi_{n_l}\rangle = TA|\psi_1, \psi_2, \dots, \psi_{n_l}\rangle$. Consequently, the co-representations (co-reps) for the full basis set $|\psi, \phi\rangle$ are:

$$D_k^l(g_i^{(a)}) = \begin{bmatrix} d_k^l\left(g_i^{(a)}\right) & 0 \\ 0 & d_k^l\left(A^{-1}g_i^{(a)}A\right)^* \end{bmatrix},$$

$$D_k^l(TAg_i^{(a)}) = \begin{bmatrix} 0 & d_k^l\left(TAg_i^{(a)}TA\right) \\ d_k^l\left(g_i^{(a)}\right)^* & 0 \end{bmatrix}. \quad (7)$$

Then we can prove the following modified Dimmock and Wheeler's character sum rule (see Appendix E), which helps to identify whether the co-reps are irreducible or not:

$$\sum_{g_i^{(a)} \in TA\tilde{L}_{SS}^k} \chi^l((g_i^{(a)})^2) = \begin{cases} +|\tilde{L}_{SS}^k| & (a) \\ -|\tilde{L}_{SS}^k| & (b) \\ 0 & (\mathrm{c}) \end{cases}. \quad (8)$$

For case ($a$), the co-rep matrices $D_k^l$ in Eq. (7) are reducible and have the same dimension as the irrep $d_k^l$. For cases ($b$) and ($c$), the dimension of $D_k^l$ is doubled compared with that of $d_k^l$. Following the approach presented here, we can obtain the projective co-irreps of any arbitrary $k$-points for all finite SSGs.

The abovementioned procedure of obtaining the co-irreps is exemplified by two realistic materials, i.e., $CeAuAl_3$ and $CoNb_3S_6$, as shown in Appendix G and Appendix H. The corresponding electronic band structures and additional exotic properties of these two materials are discussed in Sec. IV.

## C. Representation of collinear SSGs

The CSCO method cannot be applied to collinear SSGs because the $SO(2)$ spin rotation symmetry renders the SSG a continuously infinite group. Interestingly, the incorporation of spin-only group $G_{SO}^l = Z_2^K \ltimes SO(2)$ ensures that the co-irreps of collinear SSGs can be obtained by considering the single-valued co-irreps of 230 type-II MSGs plus the extra degeneracies in spin space originated from some crucial symmetries beyond MSGs. We next discuss collinear SSGs in two categories, i.e., those describing ferromagnets and antiferromagnets.

For SSGs describing collinear ferromagnets and ferrimagnets, it is straightforward that $G_0 = L_0$, $G^S = 1$, and $G_{SO}^l = Z_2^K \ltimes SO(2)$. The corresponding SSG can be written as

$$G_{SS} = \{E||L_0\} \times Z_2^K \ltimes SO(2). \quad (9)$$

While $Z_2^K \ltimes SO(2)$ itself cannot contribute extra degeneracy in spin space (see Appendix F), the conjugate symmetry operator $K$ can combine two conjugated single-valued irreps in real space. Therefore, the co-irreps of this type of SSGs (230 in total) are the same as spinless grey SGs ($L_0 \times Z_2^K$), or single-valued co-irreps of type-II MSGs.

For SSGs describing collinear antiferromagnets, $L_0$ is the sublattice group, which is an index-2 normal subgroup of $G_0$. In addition, we have $G^S = -1$ and a real-space operation $A$ connecting the two sublattices with opposite magnetic moments. The corresponding SSG can be expressed as

$$G_{SS} = (\{E\|L_0\} \cup \{T\|AL_0\}) \times Z_2^K \ltimes SO(2). \tag{10}$$

The critical symmetries beyond MSGs include spin $SO(2)$, $\{T\|A\}$ and $\{U_{\boldsymbol{n}}(\pi)\|A\}$. Among these, the $SO(2)$ group provides conjugated one-dimensional (1D) irreps $\Gamma_{+1/2}^{\mathrm{S}}$ and $\Gamma_{-1/2}^{\mathrm{S}}$ in spin space, leading to the following degeneracy doubling mechanisms: if the little group of $k$ has $\{U_{\boldsymbol{n}}(\pi)\|A\}$ or $\{T\|A\}$ symmetry, the combination of $SO(2)$ and $\{U_{\boldsymbol{n}}(\pi)\|A\}$ or $\{T\|A\}$ pairs $\Gamma_{+1/2}^{\mathrm{S}}$ and $\Gamma_{-1/2}^{\mathrm{S}}$ into a two-dimensional (2D) irrep or co-irrep, respectively (see Appendix F).

Overall, the general procedure to obtain co-irreps of collinear SSGs can be summarized as the co-irreps of the single-valued co-irreps of the type-II MSG $L_0 \times Z_2^K$ (real space) plus additional double degeneracy caused by $\{U_{\boldsymbol{n}}(\pi)\|A\}$ or $\{T\|A\}$ symmetry (spin space). As an example, we perform calculations of the co-irreps for a collinear antiferromagnet $RuO_2$, as shown in Appendix F. The corresponding electronic band structures and the extra degeneracies in SSG are discussed in Sec. IV.

## IV. Realistic materials

While SOC ultimately exists in realistic magnetic systems, under the circumstances when SOC effect is weak or when considering the SOC-induced effects (e.g., orbital polarization) [69], it is useful to analyze the wavefunction properties of a SOC-free Hamiltonian [19]. SSGs provide a comprehensive symmetry description of such Hamiltonians. Given a realistic material with specific atomic positions and local moments, a practical need is to identify its SSG with all possible symmetry operations.

In the following, we first introduce our online program for identifying the SSG symmetries of a magnetic crystal [59]. We then perform density functional theory (DFT) calculations on several material candidates to exemplify their SSG symmetries, the electronic band degeneracies, spin textures, geometric Hall effects, and the distinctions from MSG. Importantly, we show that the spin-space part and real-space part of a SSG directly lead to the features of spin polarization and geometric Hall effect, respectively. Our representative examples include the prototypical altermagnet $RuO_2$ (t-type SSG), spiral antiferromagnet $CeAuAl_3$ (k-type SSG) with coplanar configuration, and noncoplanar antiferromagnet $CoNb_3S_6$ (g-type SSG).

**A. SSG identification for a given magnetic structure**

The aim of the online program is to identify the SSG and all the symmetry operations herein of a magnetic material, given the lattice parameters, the atomic positions, and the local moments in a magnetic cell. The required input is a .mcif or .cif file (with magnetic moments). The outputs of our program are the details of the SSG, including $L_0$, $G_0$, $i_t$, $i_k$, $G^s$, type of magnetic moments, the international notation, standard magnetic cell, and all the symmetry operations written based on the lattice vectors of the standard magnetic cell. The specific procedure for identifying the SSG is as follows (see Fig. 4):

*Step 1*: Obtain the SG $G_R$ of the input structure (without considering the magnetic moments) using the module SPGLIB [70], which is an open-source Python package for searching crystal symmetries.

*Step 2*: Determine the type of magnetic moments, i.e., noncoplanar, coplanar, or collinear, by calculating the rank of all the vectors of magnetic moments.

*Step 3*: Extract all the magnetic moments and obtain the allowed PG $G_S$ for spin space using PYMATGEN [71], which is an open-source Python library for materials analysis. In order to construct the $G_{NS}$, the spin-only part in $G_S$ should be excluded. For collinear magnetic configurations, set $G_S$ to be $\{1, \bar{1}\}$ ; for coplanar magnetic configurations, add a constant small canting to each moment to exclude the mirror operation ($Z_2^K$) in $G_S$.

*Step 4*: Traverse all the symmetry operations of the direct product group $G_R \times G_S$ and apply them to the magnetic structure. The operations that keep the magnetic structure invariant form the nontrivial SSG $G_{NS}$ of the given material.

*Step 5*: Within the symmetry operations of $G_{NS}$, all elements of real-space operations form $G_0$ by dropping the operations in spin space; all elements of $G_0$ that map the identity operation in spin space form the sublattice group $L_0$; all elements of spin-space operations form PG $G^s$; $i_k$, $i_t$ and the International notation of $G_{NS}$ can also be obtained straightforwardly.

*Step 6*: Add spin-only group to $G_{NS}$ to obtain $G_{SS}$. For collinear configurations, we output the symmetry operations in $G_{NS}$ because the spin-only group is continuously infinite; for coplanar configurations, we output the symmetry operations in $G_{SS} = G_{NS} \times Z_2^K$.

We provide in Appendix I an example of .mcif file ($RuO_2$) and the resultant SSG identification output to illustrate the functionality of our program. Furthermore, we have also identified the SSGs for all the ~1970 experimentally reported magnetic structures provided in MAGNDATA database on the Bilbao Crystallographic Server. The results are also listed on the website of our online program.

## B. Extra band degeneracies in altermagnet $RuO_2$

$RuO_2$ is an altermagnet with a spin-polarized Fermi surface and thus holds great potential to realize various spintronic effects, including spin-polarized current, giant magnetoresistance, and spin-splitting torque [35-43]. It has a rutile structure with an out-of-plane collinear antiferromagnetic order, with Ru and O ions occupying 2a and 4f Wyckoff positions, respectively [Fig. 5(a)]. The resulting nonmagnetic SG and MSG are $P4_2/mnm$ (No. 136) and $P4_2'/mnm'$ (136.499), respectively. To elucidate its SSG, we first determine its sublattice SG $L_0$ by considering the subgroup of SG that preserves the moment of Ru, i.e., $L_0$ = *Cmmm* (No. 65). It is a *t*-type normal subgroup of $G_0$ = $P4_2/mnm$, with the subgroup indices $i_t = 2$, $i_k = 1$. The coset representative element, which is the symmetry connecting the sublattices with opposite magnetic moments, is a nonsymmorphic four-fold rotation $\{4_{001}^1|1/2\ 1/2\ 1/2\}$, and the coupled spin-space PG is $G^s = \bar{1} = \{E, T\}$. Consequently, the nontrivial SSG has the form $G_{NS} =$ $\{E||L_0\} \cup \{\bar{1}||4^1_{001}|1/2\ 1/2\ 1/2\}L_0$ , labeled as (65.136.1.1); the corresponding International notation is written as $P^{-1}4_2/^1m^{-1}n^1m$. The full SSG $G_{SS}$ taking into account the spin-only group, is $P^{-1}4_2/^1m^{-1}n^1m^{\infty m}1$, which is a continuously infinite group.

Fig. 5(c) shows the band structure of $RuO_2$ calculated by DFT (see Appendix J for DFT methods). It can be found that the bands along several high-symmetry lines, such as Γ-X-M, Γ-Z-R-A, have two-fold band degeneracies, while significant spin splittings occur along Γ-M and A-Z. The dimensions of the projective co-irreps for these wavevectors are shown in Fig. 5(d), demonstrating that the rep theory successfully reproduces the calculated band degeneracies. In comparison, we also plot in Fig. 5(e) the band structure with SOC, where the co-irreps of MSG dictate the band degeneracy [Fig. 5(f)]. It is shown that the high-symmetric line Z-R-A manifests band splitting with SOC, which is consistent with the MSG co-irreps. While in MSG, the little groups of *k* along Z-R-A (*mm*2 or *mmm*) only support 1D co-irreps, in SSG the little groups of these *k*-points have the operation $\{\bar{1}||m_{010}|1/2\ 1/2\ 1/2\}$, which can stick two conjugate 1D irreps of $SO(2)$ ($\Gamma^{S}_{+1/2}$ and $\Gamma^{S}_{-1/2}$) forming extra degeneracies. The details of the derivation of co-irreps in SSG and the comparison of co-irreps in SSG and MSG are provided in Appendix F. Similar extra degeneracies protected by SSG have also been discussed in $CoNb_3S_6$ with collinear order [49, 50], and $Mn_3Sn$ with coplanar magnetic order [19]. On the other hand, in SSG, the little group $^1m^mm^1m^{\infty}1$ at Σ and S points does not support 2D co-irreps, thus leading to spin splitting even without SOC.

### C. Spiral spin polarization in helimagnet $CeAuAl_3$

Spiral magnets, or helimagnets, present a type of magnetic ordering where the neighboring magnetic moments are arranged in a spiral pattern, with a characteristic turn angle between 0 and 180 degrees. Such magnetic orders generally originated from the competition between ferromagnetic and antiferromagnetic exchange interactions. Spiral magnets usually manifest combined spin rotation and fractional translation symmetry operations, which are not allowed within the MSG framework. However, for commensurate magnetic configurations, they serve as a nice platform to illustrate k-type and g-type SSGs. $CeAuAl_3$ has a tetragonal crystal structure with in-plane coplanar antiferromagnetic order [72], with magnetic Ce ion occupying 4a Wyckoff positions [Fig. 6(a)]. The spin of the Ce layer rotates $\pi/2$ when moving to its neighboring Ce layer, resulting in a spiral configuration with a four-fold cell expansion. While its nonmagnetic SG is *I*4*mm* (No. 107), the MSG is degraded to $P_c4_1$ (76.10), with only 8 symmetry operations left.

Within the regime of SSG, its sublattice group is $L_0$ = $P4mm$ (No. 99), a k-type normal subgroup of $G_0$ = *I*4*mm*, with the subgroup indices $i_t = 1$, $i_k = 4$. The coset representative element connecting the neighboring sublattices (Ce layers) is a fractional translation $\{1|1/2\ 1/2\ 1/4\}$. Therefore, there are two possibilities of $G^s$ being isomorphic to $G_0/L_0$ i.e., 4 and $\bar{4}$, resulting in two different nontrivial SSGs (99.107.4.1) and (99.107.4.2). $CeAuAl_3$ adopts the former one (99.107.4.1), which is constructed by mapping the element $4^1_{001}$ in spin space to $\{1|1/2\ 1/2\ 1/4\}$, i.e., $G_{NS} = L_0 \cup \{4^1_{001}||1|1/2\ 1/2\ 1/4\}L_0 \cup \{2_{001}||1|0\ 0\ 1/2\}L_0 \cup \{4^3_{001}||1|1/2\ 1/2\ 3/4\}L_0$. Specifically, it has a screw axis in spin space, i.e., $\{4^1_{001}||1|1/2\ 1/2\ 1/4\}$, served as the generator of the spin translation group $G^S_T$ isomorphic to PG 4. Thus, the corresponding International notation is written as $P^1 4^1 m^1 m^{4^1_{001}}(1/2\ 1/2\ 1/4)$ (Appendix C). The full SSG is written as $G_{SS} = G_{NS} \times Z_2^K = P^1 4^1 m^1 m^{4^1_{001}}(1/2\ 1/2\ 1/4)^m 1$, which has 64 symmetry operations. A similar procedure applies to (99.107.4.2) except that the coset representative of the nontrivial SSG is $\{\bar{4}^3_{001}||1|1/2\ 1/2\ 1/4\}$.

The DFT-calculated local moment for each Ce ion is 0.98 $\mu B$, which is close to the experimental value (1.05 $\mu B$ [72]). Fig. 6(b) exhibits the DFT-calculated band structure of $CeAuAl_3$. We find that all the bands along the X-M line are four-fold degenerate, and some of these bands split into two branches of doubly degenerate bands along the Γ-X line. For Γ and Z points, both Dirac and Weyl nodes exist. These node features are well explained by our CSCO projective rep theory, as comprehensively shown in Appendix G. As shown in Fig. 6(e), while X-M line only supports 4D co-irreps, Γ and Z points also allow 2D co-irreps. Note that while the full SSG is isomorphic to the type II MSG $I4mm1'$, the spin translation group generated by $\{4^1_{001}||1|1/2\ 1/2\ 1/4\}$ contributes to the factor system of projective reps, resulting in the emergence of 4D co-irreps. In comparison, the type II MSG $I4mm1'$ only supports 1D and 2D co-irreps. Importantly, while it is believed that the existence of "$U\tau$" symmetry (here is $\{2_{001}||1|0\ 0\ 1/2\}$) protects spin degeneracy throughout the Brillouin zone [32, 73], in noncollinear magnet $CeAuAl_3$ there is spin splitting along Γ-Z, indicating 1D co-irreps for the corresponding little groups with $\{U_{\boldsymbol{n}}(\pi)\|\tau\}$. Our results demonstrate that spin degeneracy enforced by $\{U_{\boldsymbol{n}}(\pi)\|\tau\}$ symmetry is indeed present in collinear antiferromagnets, as $\{U_{\boldsymbol{n}}(\pi)\|\tau\}$ sticks two conjugate 1D irreps of $SO(2)$.

More remarkably, such spiral magnets exhibit a new type of spiral spin polarization, for which the spin component aligns the spiral axis, rather than the direction of local moments. Fig. 6(d) shows the spin texture of $CeAuAl_3$ at the $k_z = \pi/2$ plane for the band marked in Fig. 6(b). While all the local moments are in-plane, the $S_x$ and $S_y$ components of the extended Bloch states are enforced to be zero, leaving significant and continuous $S_z$ distribution along the spiral direction ($k_z$). Such spiral spin polarization is in sharp contrast to the conventional Rashba and Dresselhaus spin polarization (along the $k$-dependent effective magnetic field) in nonmagnetic materials and Zeeman spin polarization (along the direction of local moments) in ferromagnetic materials. To explain this, we note that the spin little group at an arbitrary $k$-point is 4, indicating that the spin texture has only $S_z$ component and forms a four-fold symmetric pattern, consistent with our calculation shown in Fig. 6(d). Therefore, while such spin polarization can survive with moderate SOC, its physical mechanism is apparently beyond the scenario of MSG.

**D. Geometric Hall effect in nonplanar $CoNb_3S_6$**

$CoNb_3S_6$ has attracted great interest due to its surprisingly large anomalous Hall effect and controversial magnetic configurations [50, 74-78]. While its magnetic order was historically determined as collinear antiferromagnet by neutron diffraction measurements [79], a recent study reported an all-in–all-out noncoplanar antiferromagnetic order and accompanied topological Hall effect [78]. Here, we discuss the SSG symmetry of $CoNb_3S_6$ with the all-in–all-out antiferromagnetic order and conclude that such a noncoplanar order generates anomalous Hall effect even without the assistance of SOC. As shown in Fig. 7(a), $CoNb_3S_6$ has a hexagonal crystal structure (nonmagnetic SG is No. 182, $P6_322$) with magnetic Co ion occupying 2d Wyckoff positions. Interestingly, its all-in–all-out antiferromagnetic order forms a cubic structure in spin space, i.e., $G^s = \bar{4}3m$. However, the corresponding MSG is $P32'$ with only six symmetry elements because within the MSG framework, the spin rotation must compromise the spatial rotation, implying that the cubic nature of spin symmetry is lost.

With $G_0 = P6_322$, the sublattice group $L_0 = P2_1$ (No. 4) forms a g-type normal subgroup with the subgroup indices $i_t = 6$ and $i_k = 2 \times 2 \times 1 = 4$, the latter of which corresponds to two-fold cell expansions along the $x$ and $y$ directions, i.e., $\{1|1\ 0\ 0\}$ and $\{1|0\ 1\ 0\}$. In addition, the three-fold rotation $\{3_{001}^1|0\}$ and two-fold rotation $\{2_{110}|0\}$, also belong to $G_0/L_0$. Overall, there are two possibilities of 24-element $G^s$ being isomorphic to $G_0/L_0$ i.e., 432 and $\bar{4}3m$, with each corresponding to one inequivalent mapping. Therefore, there are two nontrivial SSGs in total, labeled by (4.182.4.1) and (4.182.4.2), respectively. In the case of $CoNb_3S_6$, real space $\{2_{100}|0\}$ operation couples spin space $m_{110}$ operation, rendering $G^s = \bar{4}3m$ and thus a nontrivial SSG (4.182.4.2). It can be generated using $\{3^2_{-11-1}||6^1_{001}|0\ 0\ 1/2\}$, $\{m_{110}||2_{100}|0\}$, $\{m_{011}||2_{210}|0\ 0\ 1/2\}$, and spin translation group $G_T^S$ is generated by $\{2_{001}||1|1\ 0\ 0\}$ and $\{2_{100}||1|0\ 1\ 0\}$. We note that the PG part of $G_T^S$ forms a $D_2$ group in spin space. Therefore, the corresponding international notation is written as $P^{3^2_{-11-1}}6_3{}^{m_{110}}2^{m_{011}}2|(2_{001}, 2_{100}, 1)$ (Appendix C). Since $G_{SO}$ is the identity group, the full SSG has 48 symmetry operations. Interestingly, such SSGs contain real space and spin space symmetries from totally incompatible crystal systems, a scenario that is impossible within the scope of MSG.

Fig. 7(b) shows the band structure of $CoNb_3S_6$ calculated using DFT. Notably, while $PT$ is absent, the bands exhibit double degeneracy throughout the Brillouin zone. Such a new form of spin degeneracy appears only in the framework of SSG, attributed to the double-valued 2D co-irreps of the little co-group $D_2$ in spin space (the elements in $G_T^S$ leave $k$ invariant) at any arbitrary $k$-points. Furthermore, the three perpendicular twofold spin rotation axes enforce any spin polarization components to be zero. Therefore, the spin degeneracy/spin splitting induced by magnetic order in noncollinear magnetic structures extends beyond the scope of MSGs, as only $T\tau$ symmetry in the spin translation group can be maintained when considering the MSGs. Consequently, we summarize the complete criteria of spin splitting for noncollinear antiferromagnetic orders, i.e., (i) the absence of $PT$, and (ii) the spin-space PG of the spin translation group does not contain $D_n$. This further highlights the importance of the enumeration and representation theory of SSGs.

The noncoplanar magnetic order could lead to responses without the assistance of SOC. For example, the anomalous Hall effect has long been attributed as a consequence of SOC under time-reversal breaking [80]. On the other hand, the exchange field induced by noncoplanar order could also induce anomalous Hall effect without SOC. Instead of the topological Hall effect, we use the terminology of geometric Hall effect to emphasize the sole origin of magnetic geometry. While the characteristic quantities of noncoplanar magnets, e.g., scalar spin chirality [81] or band-resolved Berry curvature [82], cannot fully describe the existence of the geometric Hall effect, we note that due to the absence of SOC, the geometric Hall conductivity tensor can be fully determined by SSG. For collinear and coplanar orders, the spin-only symmetry $TU_{\boldsymbol{n}}(\pi)$ behaves similarly to $T$, which enforces the geometric Hall conductivity $\sigma_{xy}^G$ to be zero when integrating the Berry curvature over the Brillouin zone. Therefore, the anomalous Hall effect in collinear or coplanar magnets (even for ferromagnets) must be a SOC effect. However, for noncoplanar $CoNb_3S_6$, the PG symmetries of real-space part of the SSG ($62'2'$) leave the $z$ component of the magnetization invariant. Thus, an orbital magnetization along the $z$ direction is allowed, leading to nonzero $\sigma_{xy}^G$. As shown in Fig. 7(d), our DFT calculation obtains nonzero geometric Hall conductivity throughout the energy window, consistent with our symmetry analysis. Remarkably, $\sigma_{xy}^G$ reaches 47 $\Omega^{-1}$cm$^{-1}$ at the Fermi level, which is comparable to the largest anomalous Hall conductivities in antiferromagnets with the assistance of SOC [83].

## V. Summary

To conclude, we present a systematic study of the enumeration and the representation theory of SSG applied to describe different magnetic configurations. By using a group extension approach, our enumeration constructs a collection of over 100000 nontrivial SSGs, named by a four-index nomenclature. The International notation system for SSGs is also developed. Furthermore, we derive the irreducible projective co-representations of the little groups of the wavevectors within the SSG framework, which is the foundation for understanding the symmetry-enforced degeneracies in band spectra. To facilitate the search for SSG symmetries, we develop an online program (https://findspingroup.com/) that can identify the SSG symmetries of any given magnetic crystal. The SSG identification for all ~1970 experimentally reported magnetic structures provided in the MAGNDATA database is also provided. We then show representative material examples, including altermagnet $RuO_2$, helimagnets $CeAuAl_3$, and noncoplanar antiferromagnet $CoNb_3S_6$ to illustrate their SSG symmetries and the emergent properties beyond the MSG framework. Our work further develops the group theory in describing materials with magnetic order, thereby unlocking possibilities for future exploration into exotic phenomena within magnetic materials.

*Note added.* During the finalization of our work, we became aware of related studies on the classification of spin space groups from Chen Fang's and Zhi-Da Song's groups [84, 85]. In addition, we also noticed recent studies on symmetry analysis and search of spin space groups [86, 87], as well as the representations of spin point groups [88].

## Acknowledgements

We thank Chen Fang and Zhi-Da Song for helpful discussions. This work was supported by National Key R&D Program of China under Grant No. 2020YFA0308900, the National Natural Science Foundation of China under Grant No. 12274194, Guangdong Provincial Key Laboratory for Computational Science and Material Design under Grant No. 2019B030301001, Shenzhen Science and Technology Program under Grant No. RCJC20221008092722009, the Science, Technology and Innovation Commission of Shenzhen Municipality (No. ZDSYS20190902092905285), Special Funds for the Cultivation of Guangdong College Students' Scientific and Technological Innovation (No. pdjh2024c10202) and Center for Computational Science and Engineering of Southern University of Science and Technology.

**Table I:** Nontrivial SSGs for different crystal systems and different magnetic configurations. "Collinear only" indicates nontrivial SSGs that only support collinear magnetic orders, i.e., $G^S = 1$ and $\bar{1}$. Similarly, "Noncoplanar only" requires that $G^S$ is a polyhedral PG ($T$, $T_d$, $T_h$, $O$, and $O_h$). "Coplanar" contains nontrivial SSGs that support coplanar magnetic orders.

| Crystal system | Collinear only | Coplanar | Noncoplanar only | Total |
|---|---|---|---|---|
| Triclinic (2) | 5 | 55 | 0 | 60 |
| Monoclinic (13) | 78 | 3540 | 0 | 3618 |
| Orthorhombic (59) | 503 | 53734 | 0 | 54237 |
| Tetragonal (68) | 502 | 31185 | 0 | 31687 |
| Trigonal (25) | 83 | 2331 | 62 | 2476 |
| Hexagonal (27) | 137 | 7149 | 111 | 7397 |
| Cubic (36) | 113 | 840 | 184 | 1137 |
| Total (230) | 1421 | 98834 | 357 | 100612 |

**Table II.** The PG $G^S$ in spin space for all the SSGs with the cell expansion limited to $i_k \leq 8$. H-M symbol stands for Hermann-Mauguin symbol.

| H-M symbol | Point group |
|---|---|
| n | 1, 2, 3, 4, 5, 6, 7, 8, 9, 10, 12, 14, 15, 16, 18, 20, 21, 24, 28, 30, 42 |
| nm | 3m, 5m, 7m, 9m, (15)m, (21)m |
| nmm | mm2, 4mm, 6mm, 8mm, (10)mm, (12)mm, (14)mm, (16)mm, (18)mm, (20)mm, (24)mm, (28)mm, (30)mm, (42)mm |
| -n | -1, m, -3, -4, -5, -6, -7, -8, -9, -10, -12, -14. -15, -16, -18, -20. -21, -24, -28, -30, -42, |
| n/m | 2/m, 4/m, 6/m, 8/m, (10)/m, (12)/m, (14)/m, (16)/m |
| n2 | 32, 52, 72, 92, (15)2, (21)2 |
| n22 | 222, 422, 622, 822, (10)22, (12)22, (14)22, (16)22, (18)22, (20)22, (24)22, (28)22, (30)22, (42)22 |
| -nm | -3m, -5m, -7m, -9m, (-15)m, (-21)m, |
| -n2m | -42m, -62m, -82m, (-10)2m, (-12)2m, (-14)2m, (-16)2m, (-18)2m, (-20)2m, (-24)2m, (-28)2m, (-30)2m, (-42)2m, |
| n/mmm | mmm, 4/mmm, 6/mmm, 8/mmm, (10)/mmm, (12)/mmm, (14)/mmm, (16)/mmm, |
| cubic | 23, 432, m-3, -43m, m-3m |

**Table III.** Symmetry operations and general positions of SSG (13.54.1.1) $P^{-1}c^{-1}c^{1}a$. In this table, a, b, and c represent the coordinates of a general position using the lattice basis of the SSG; x, y, and z denote the corresponding components of the magnetic moment using Cartesian coordinate.

| Operation | Coordinates | Operation | Coordinates |
|---|---|---|---|
| $\{1\|\|1\|0\}$ | a, b, c,<br>x, y, z | $\{-1\|\|2_{100}\|1/2\ 0\ 1/2\}$ | a+1/2, -b, -c+1/2,<br>-x, -y, -z |
| $\{1\|\|-1\|0\}$ | -a, -b, -c,<br>x, y, z | $\{-1\|\|m_{100}\|1/2\ 0\ 1/2\}$ | -a+1/2, b, c+1/2,<br>-x, -y, -z |
| $\{1\|\|2_{001}\|1/2\ 0\ 0\}$ | -a+1/2, -b, c,<br>x, y, z | $\{-1\|\|2_{010}\|0\ 0\ 1/2\}$ | -a, b, -c+1/2,<br>-x, -y, -z |
| $\{1\|\|m_{001}\|1/2\ 0\ 0\}$ | a+1/2, b, -c,<br>x, y, z | $\{-1\|\|m_{010}\|0\ 0\ 1/2\}$ | a, -b, c+1/2, -x,<br>-y, -z |

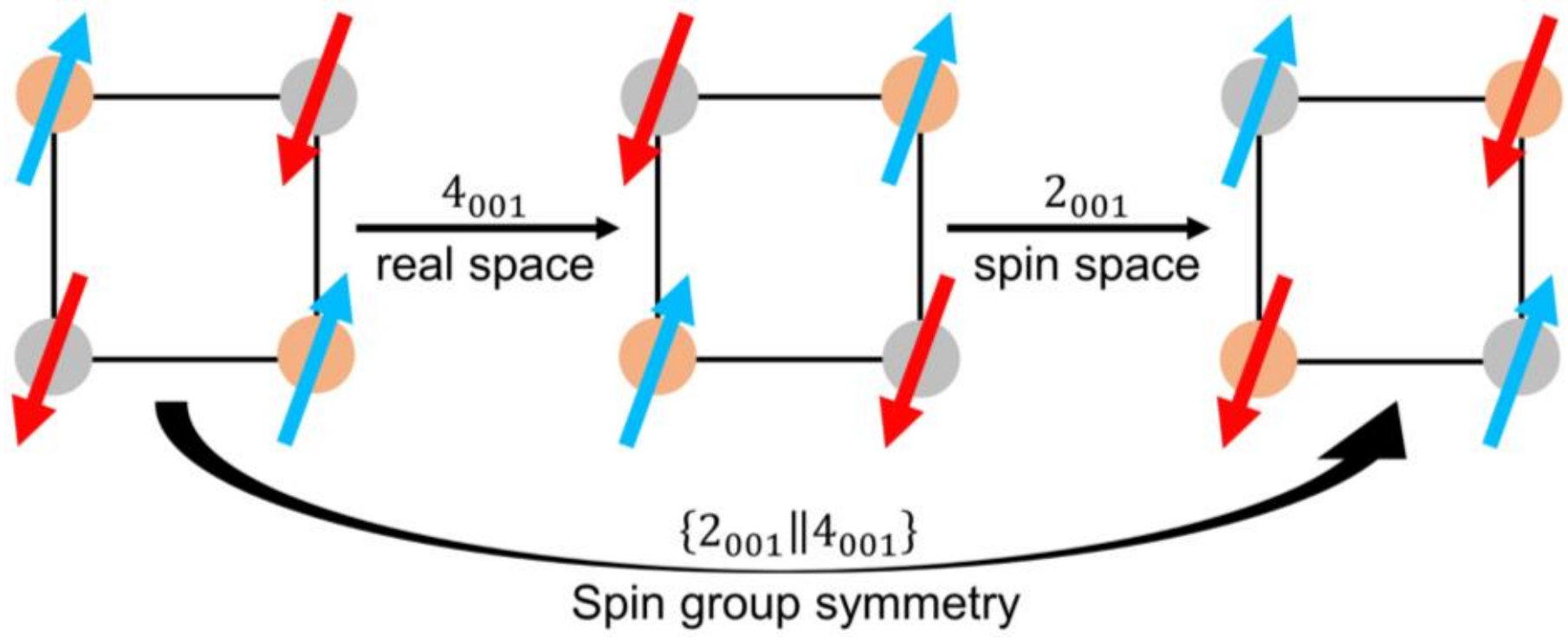

**Fig. 1:** Schematic plot of a spin group symmetry of an antiferromagnetic structure. It takes a four-fold rotation ($4_{001}$) in real space followed by a two-fold rotation ($2_{001}$) in spin space, constituting a spin group symmetry $\{2_{001}||4_{001}\}$. Such a symmetry operation contains separated lattice and spin rotations and is thus beyond the framework of magnetic groups.

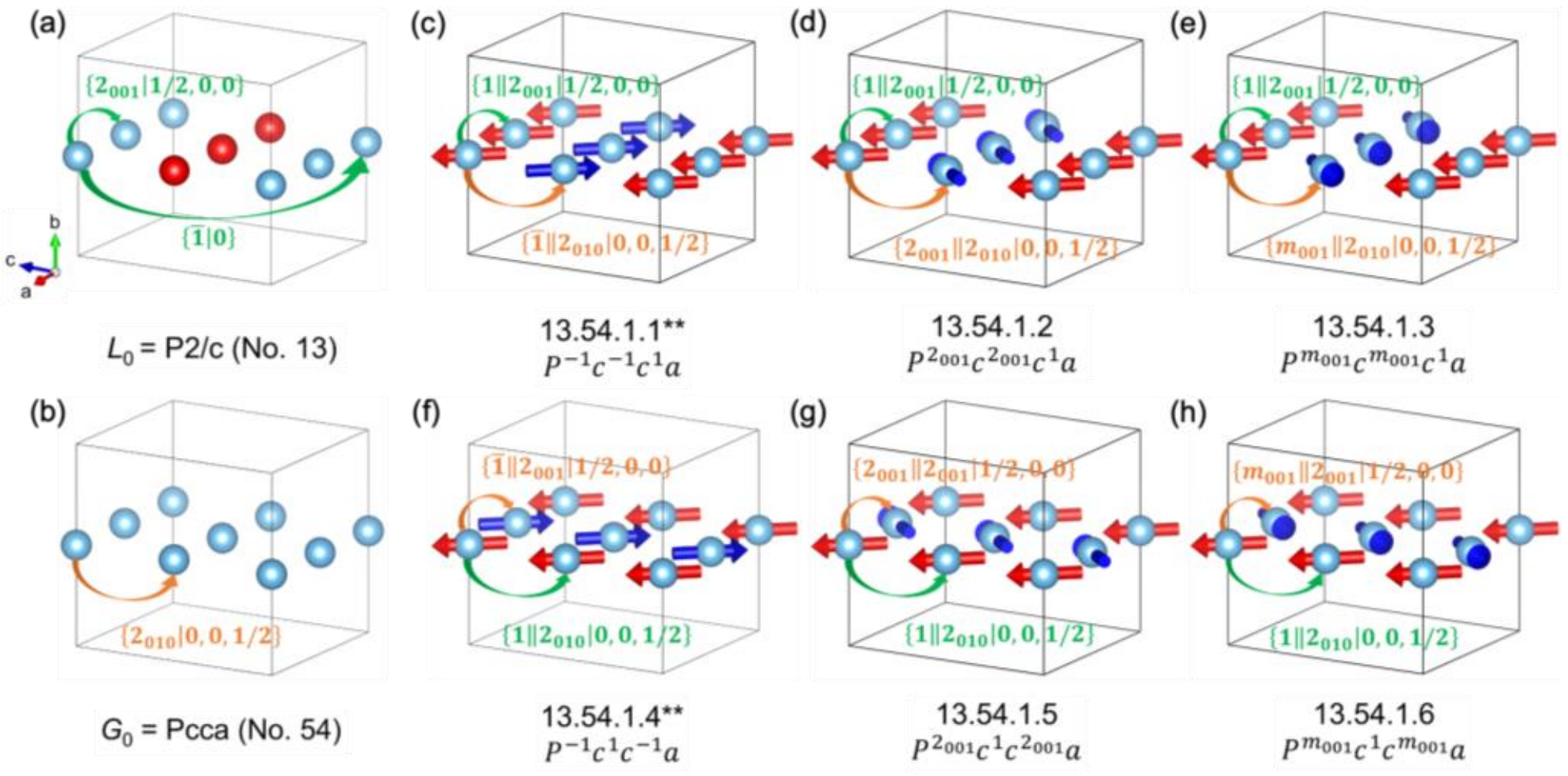

**Fig. 2:** Nontrivial SSGs generated from a $t$-type $(L_0, G_0)$ subgroup pair and the corresponding magnetic configurations. In this case, the coordinate system $xyz$ in spin space is coincide with the coordinate system $abc$ in real space. (a) Structure with sublattice group $L_0 = P2/c$ (No. 13). The group generators are indicated by green arrows. The balls with different colors denote different sublattices. (b) Structure with nonmagnetic group $G_0 = Pcca$ (No. 54). The group generators that connect different sublattices are indicated by orange arrows. (c-h) Magnetic configurations and the corresponding International notations for different nontrivial SSGs. Green and orange arrows indicate the generators connecting the same and the different sublattices, respectively. Double star denotes SSGs for collinear magnetic configurations. Note that when considering spin-only group, the SSGs generated by $G^s = 2$ and $m$ (e.g., panels (d) and (e); panels (g) and (h)) are equivalent.

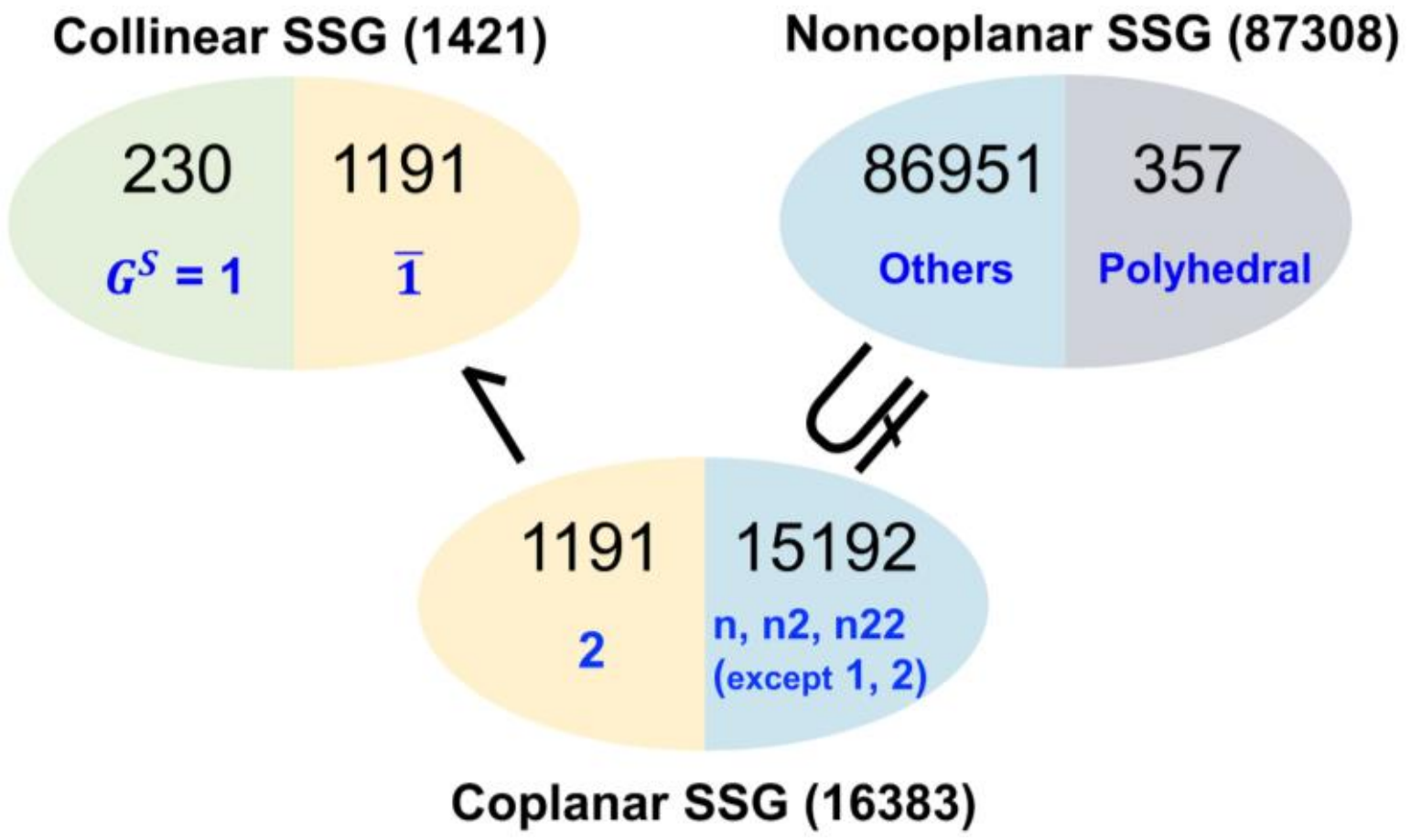

**Fig. 3:** Summary of inequivalent SSGs for collinear, coplanar, and noncoplanar magnetic configurations when considering their spin-only groups. The blue fonts denote the corresponding $G^s$. "Polyhedral" indicates polyhedral PGs, i.e., $T$, $T_d$, $T_h$, $O$, and $O_h$. "Others" indicates other $G^s$ except 1, $\bar{1}$, 2, $m$, 2/$m$, and polyhedral PGs. "⇀" means that when considering collinear spin-only group $G_{SO}^{l}$, $G^s = \bar{1}$ and 2 yield equivalent collinear SSGs. "⊊" means that when considering coplanar spin-only group $G_{SO}^{p}$, the 86951 nontrivial SSGs (supporting both coplanar and noncoplanar magnetic configurations) is reduced to 15192 inequivalent coplanar SSGs.

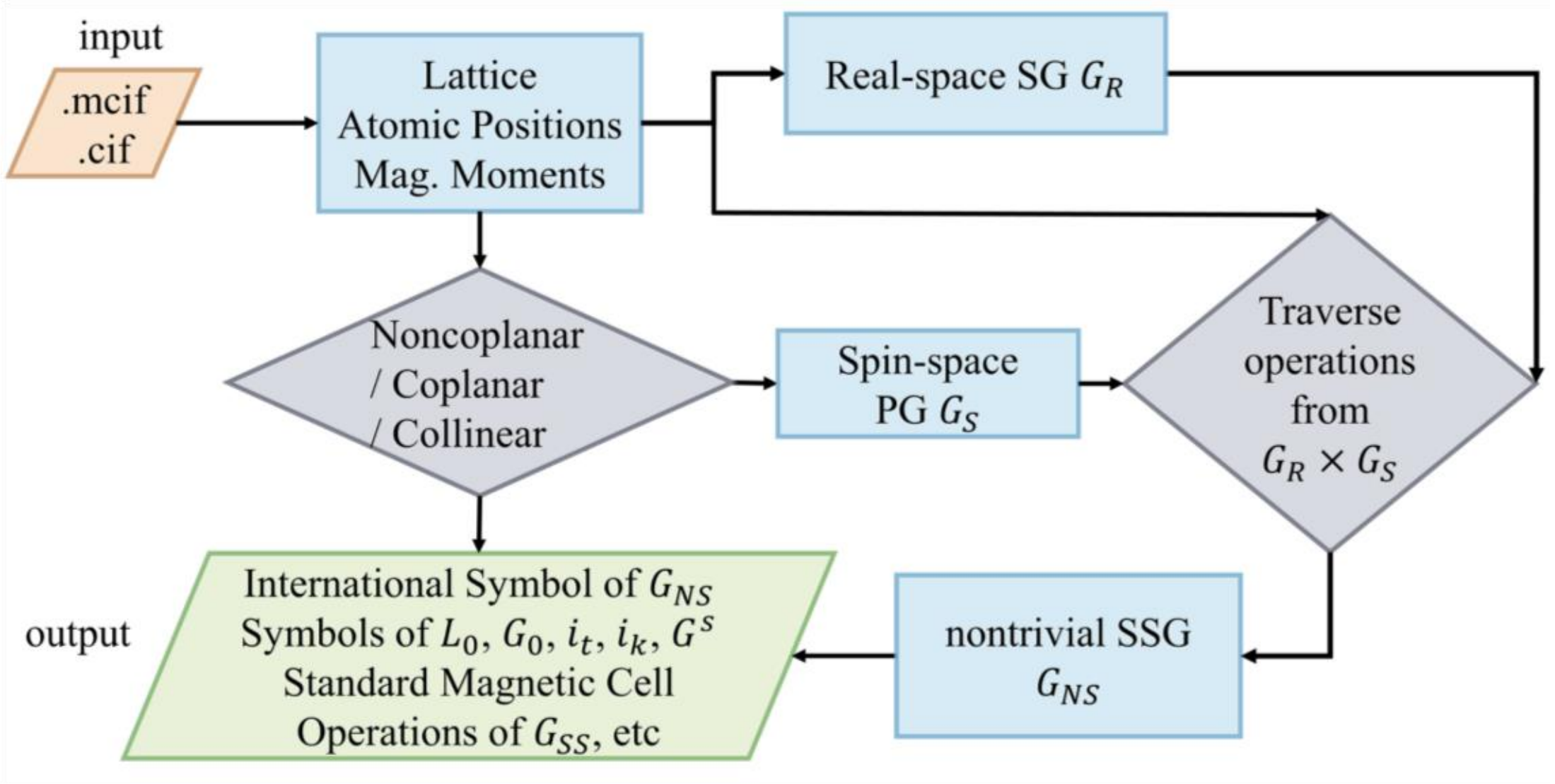

**Fig. 4:** Procedure of the identification of SSG from a given magnetic structure.

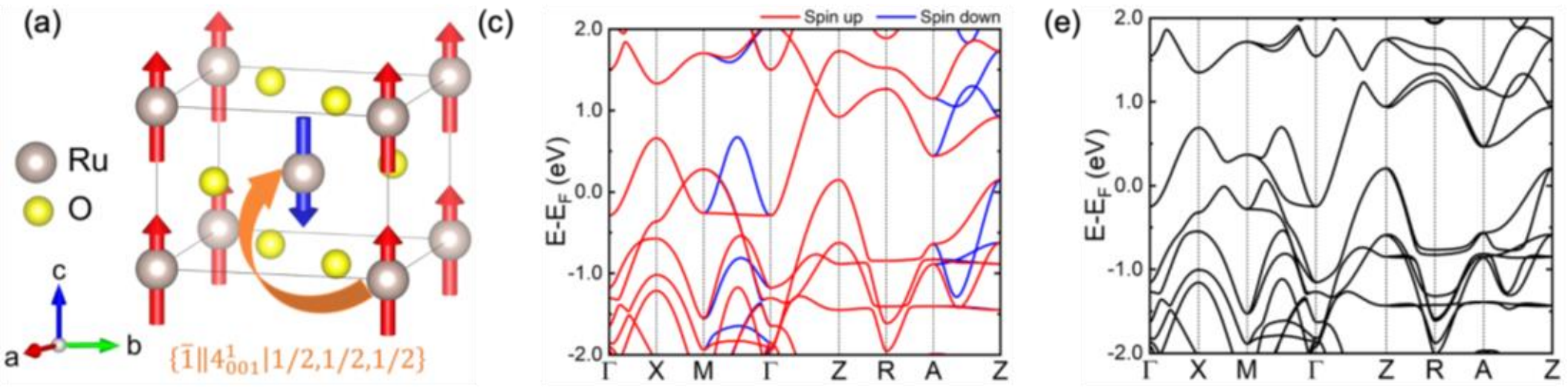

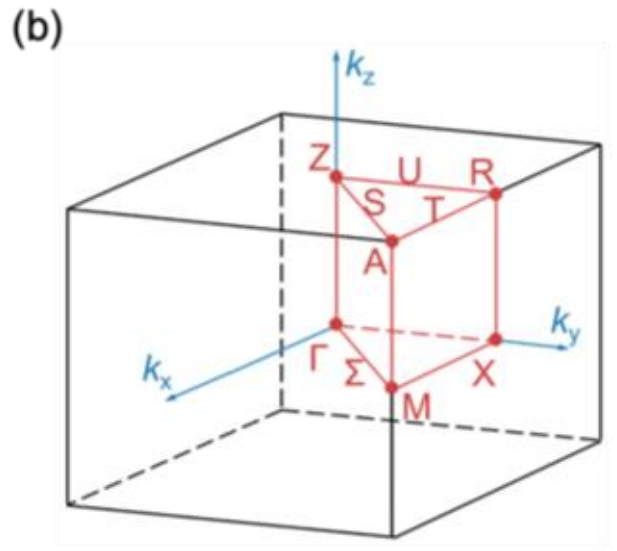

(d)

| k point | Little group | Co-Irreps |
|---|---|---|
| Γ (0, 0, 0) | $^{\bar{1}}4/^{1}m^{\bar{1}}m^{1}m^{\infty m}1$ | $\Gamma^{+,-}_{1,2,3,4}(2)$ |
| M(1/2, 1/2, 0) | $^{\bar{1}}4/^{1}m^{\bar{1}}m^{1}m^{\infty m}1$ | $M^{+,-}_{1,2,3,4}(2)$ |
| Σ (u, u, 0) | $^{1}m^{m}m^{1}m^{\infty}1$ | $\Sigma_{1,2,3,4}(1)$ |
| Z (0, 0, 1/2) | $^{\bar{1}}4/^{1}m^{\bar{1}}m^{1}m^{\infty m}1$ | $Z^{+,-}_{1,2,3,4}(2)$ |
| U (0, v, 1/2) | $^{2}m^{\bar{1}}m^{1}m^{\infty}1$ | $U_{1,2}(2)$ |
| R (0, 1/2, 1/2) | $^{\bar{1}}m^{\bar{1}}m^{1}m^{\infty m}1$ | $R^{+,-}_{1,2}(2)$ |
| T (u, 1/2, 1/2) | $^{2}m^{\bar{1}}m^{1}m^{\infty}1$ | $T_{1,2}(2)$ |
| A (1/2, 1/2, 1/2) | $^{\bar{1}}4/^{1}m^{\bar{1}}m^{1}m^{\infty m}1$ | $A^{+,-}_{1,2,3,4}(2)$ |
| S (u, u, 1/2) | $^{1}m^{m}m^{1}m^{\infty}1$ | $S_{1,2,3,4}(1)$ |

(f)

| k point | Little group | Co-Irreps |
|---|---|---|
| Γ (0, 0, 0) | $4'/mm'm$ | $\Gamma_{5,6}(2)$ |
| M(1/2, 1/2, 0) | $4'/mm'm$ | $M_{5,6}(2)$ |
| Σ (u, u, 0) | $m'm2'$ | $\Sigma_{3,4}(1)$ |
| Z (0, 0, 1/2) | $4'/mm'm$ | $Z_{3,4}(2)$ |
| U (0, v, 1/2) | $mm2$ | $U_{2,3,4,5}(1)$ |
| R (0, 1/2, 1/2) | $mmm$ | $R_{2\sim 9}(1)$ |
| T (u, 1/2, 1/2) | $mm2$ | $T_{2,3,4,5}(1)$ |
| A (1/2, 1/2, 1/2) | $4'/mm'm$ | $A_{3,4}(2)$ |
| S (u, u, 1/2) | $m'm2'$ | $S_{3,4}(1)$ |

**Fig. 5:** Spin space group of altermagnet $RuO_2$. (a) Crystal structure. The symmetry connecting different sublattices is indicated. The coordinate system *xyz* in spin space is coincide with the coordinate system *abc* in real space. (b) Brillouin zone in momentum space. (c) SOC-free band structure with the projection of the spin components. (d) Little groups and the corresponding projective co-irreps for different *k*-points within the regime of SSG. Brown fonts denote the *k*-points manifesting spin splitting; Green fonts denote the *k*-points manifesting extra degeneracies compared with the band structure with SOC. (e) Same as (c) but with SOC. (f) Same as (d) but within the regime of MSG.

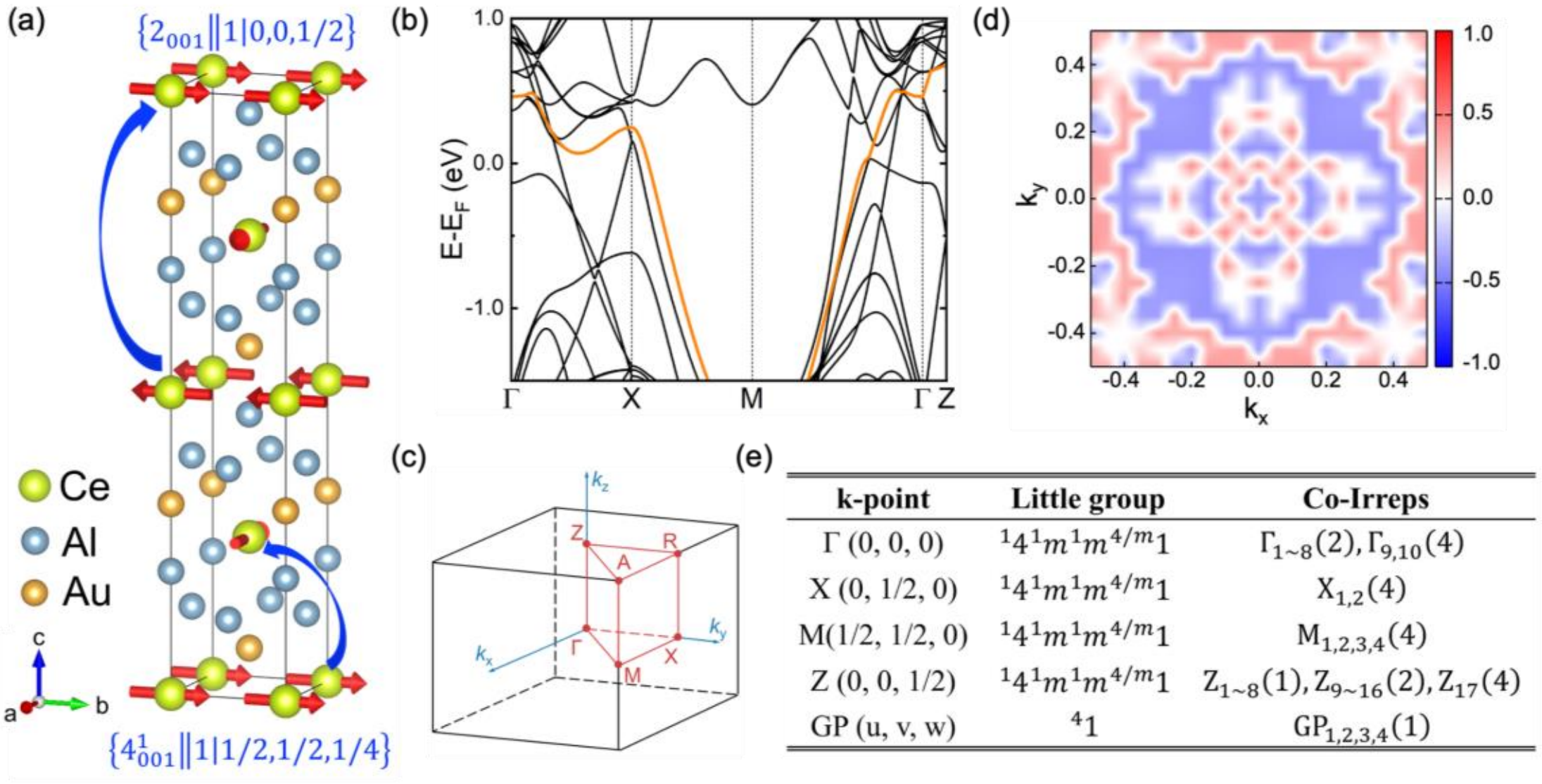

| k-point | Little group | Co-Irreps |
|---|---|---|
| Γ (0, 0, 0) | ${}^14{}^1m{}^1m{}^{4/m}1$ | $\Gamma_{1\sim8}(2), \Gamma_{9,10}(4)$ |
| X (0, 1/2, 0) | ${}^14{}^1m{}^1m{}^{4/m}1$ | $X_{1,2}(4)$ |
| M(1/2, 1/2, 0) | ${}^14{}^1m{}^1m{}^{4/m}1$ | $M_{1,2,3,4}(4)$ |
| Z (0, 0, 1/2) | ${}^14{}^1m{}^1m{}^{4/m}1$ | $Z_{1\sim8}(1), Z_{9\sim16}(2), Z_{17}(4)$ |
| GP (u, v, w) | ${}^41$ | $GP_{1,2,3,4}(1)$ |

**Fig. 6:** Spin space group of spiral antiferromagnet $CeAuAl_3$. (a) Crystal structure. The symmetry connecting different sublattices are indicated. The coordinate system *xyz* in spin space is coincide with the coordinate system *abc* in real space. (b) SOC-free electronic band structure. (c) Brillouin zone in momentum space. (d) Spin polarization $S_z$ at the $k_z = \pi/2$ plane for the band marked in panel (b). (e) Little groups and the corresponding projective co-irreps for different *k*-points.

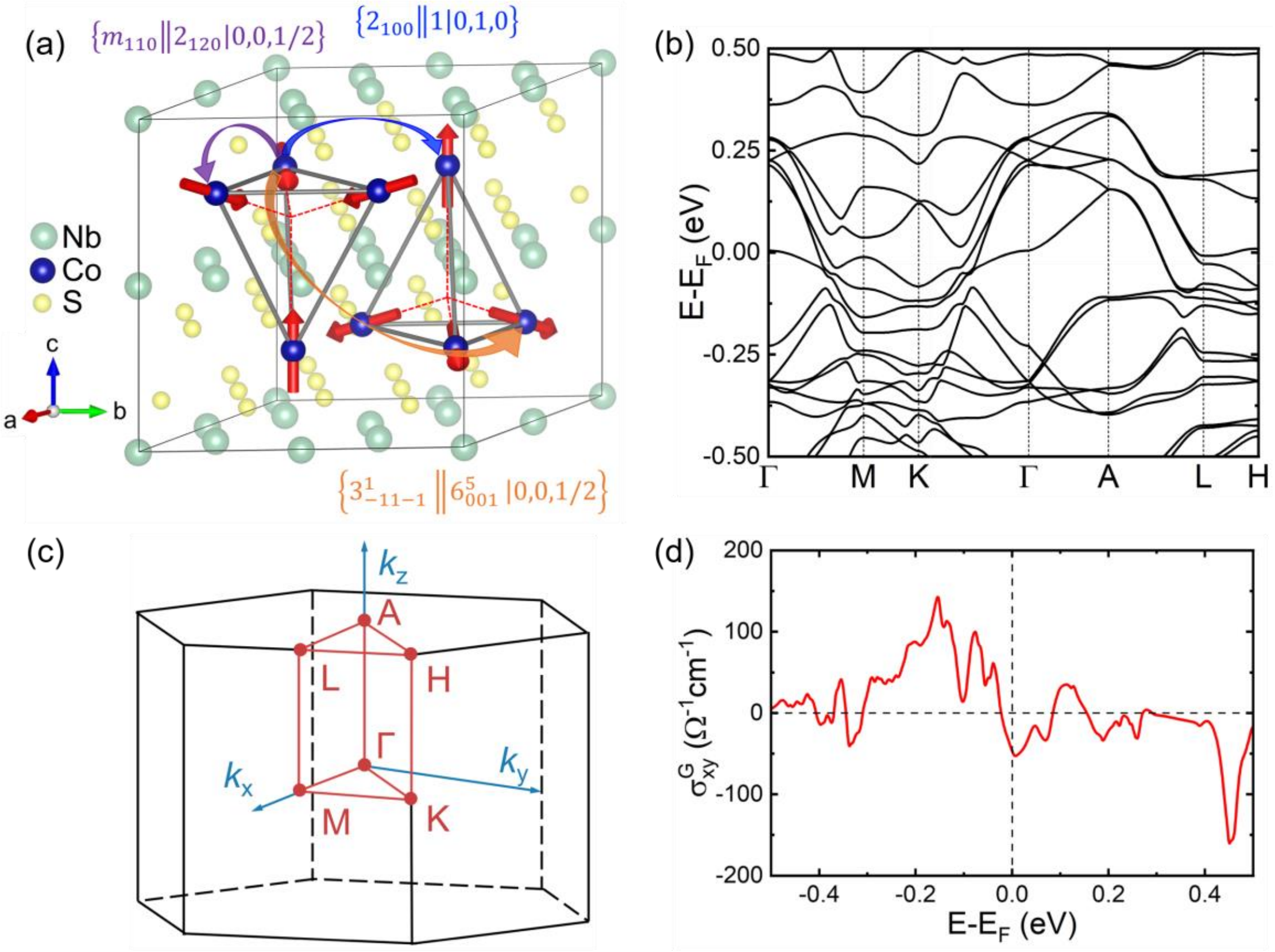

**Fig. 7:** Spin space group of noncoplanar antiferromagnet $CoNb_3S_6$. (a) Crystal structure. The symmetry connecting different sublattices are indicated. The cubic structure in spin space is illustrated by the two tetrahedrons. The relationship between the *xyz* coordinate system in spin space and the *abc* coordinate system in real space can be expressed as follows: $a = -x - y$, $b = y + z$, and $c = -x + y - z$. (b) Electronic band structure. (c) Brillouin zone in momentum space. (d) Geometric Hall conductivity as a function of energy. SOC is excluded in the calculations.

# Appendix

In Appendix, we begin in Sec. A by defining the SSGs, establishing connections with the more familiar nonmagnetic SGs and MSGs. In Sec. B, we review the International notation and Seitz notation for SGs and MSGs, and extend these notations to describe SSGs. Next, in Sec. C, we introduce the nomenclature for *t*-type, *k*-type and *g*-type SSGs and show how to obtain all group elements using International notation of SSGs. These nonmenclatures are also integrated in our online platform findspingroup.com. In Sec. D, the site symmetry groups and Wyckoff positions are introduced into SSGs. The construction of a magnetic structure using site symmetry groups are also explained. In Sec. E, we provide some details regarding the rep theory of SSG, including the projective reps for SSGs, the decomposition of regular projective reps using CSCO method and the modified Dimmock and Wheeler's character sum rule. In Sec. F, we derive the band reps of the little groups of *k* for collinear SSGs, exemplified by $RuO_2$. In Sec. G and H, we show the band reps of SSGs and the comparison of those between SSGs and MSGs for coplanar $CeAuAl_3$ and noncoplanar $CoNb_3S_6$, respectively. In Sec. I, we provide an example ($RuO_2$) of our online program for SSG identification. Lastly, in Sec. J, we provide the computational methods for the first-principles calculations.

# Nomenclature

| Abbreviation | Stands for |
|---|---|
| PG | point group |
| SG | space group |
| MPG | magnetic point group |
| MSG | magnetic space group |
| SPG | spin point group |
| SSG | spin space group |
| SOC | spin-orbit coupling |
| rep | representation |
| irrep | irreducible representation |
| co-rep | co-representation |
| co-irrep | irreducible co-representation |
| HM | Hermann-Mauguin Notation, International Notation |
| CSCO | complete set of commuting operators |
| DFT | density functional theory |
| $G_0$ | SG, supergroup of $L_0$ |
| $L_0$ | normal subgroup of $G_0$ |
| $g_s$ | operations in spin space |
| $g_l$ | operations in real space |
| $R$ | PG part of $g_l$ |
| $\tau$ | fractional translation part of $g_l$ |
| $\mathbb{T}$ | translation group |
| $P$ | space inversion |
| $T$ | time reversal |
| $U_\alpha(\theta)$ | spin rotation of $\theta$-degree along the $\alpha$ axis |
| $C_\beta(\theta)$ | spatial rotation of $\theta$-degree along the $\beta$ axis |
| $T(G)$ | translation subgroup of SG $G$ |
| $P(G)$ | quotient group of $G$ with respect to the $T(G)$ |
| $t$ | integral translation operation in translation group $\mathbb{T}$ or $T(G)$ |
| $\|G\|$ | order of group $G$ |
| $i_t$ | PG-index of $\|P(G_0)\|/\|P(L_0)\|$ |
| $i_k$ | supercell index of $\|T(G_0)\|/\|T(L_0)\|$ |
| $G^s$ | mapped PG in spin space |
| $G_{SG}$ | SG |
| $G_{MS}$ | MSG |
| $G_{SS}$ | SSG |
| $G_{SO}$ | spin-only group |
| $G_{NS}$ | nontrivial SSG |
| $L$ | maximal unitary subgroup of $G$ |
| $G^k$ | little group of $G$ at $k$ point |
| $\tilde{G}^k$ | the quotient group of $G^k$ with respect to translation group $\mathbb{T}$ |
| $\breve{G}^k$ | little co-group of $G$ at $k$ point |
| $\breve{G}^{k,*}$ | central extension group of the $\breve{G}^k$ |
| $Z_g$ | cyclic group of integers |
| $\overline{\tilde{G}^k}$ | intrinsic group of $\tilde{G}^k$ |
| $M_k$ | projective irrep of the little group $G^k$ |
| $d_k$ | irrep of $\tilde{G}^k$ |
| $D_k$ | rep matrix of co-rep |
| $G_R$ | SG of input structure without magnetic configurations |
| $G_S$ | the quotient group of the PG of magnetic moment vectors with respect to the spin-only group |
| $\left\{g_{s_i}^{(a)} \middle\| R_i \middle\| \tau_i^{(a)}\right\}$ | SSG symmtry with $i$ labels PG part $R_i$ in real space and $(a)$ labels the corresponding translations and spin operations |
| $\phi\left(g_{s_i}^{(a)}\right)$ | rotation angle of $g_{s_i}^{(a)}$ |
| $N_G(L)$ | normalizer of $L$ with respect to $G$ |
| $N_{A^+}(L)$ | chirality-preserving affine normalizer of $L$ |
| $G_T^S$ | spin translation group |
| $G^{\boldsymbol{q}}$ | site-symmetry group at $\boldsymbol{q}$ point |
| $M_T$ | transformation matrix |

# Table of contents

## A. SGs, MSGs, SSGs and their representations

### 1. Types of symmetry operations

We first discuss SGs, MSGs and SSGs in terms of their symmetry operations, as summarized in Table A1. SGs describe the geometry of a nonmagnetic material, or the Hamiltonian that does not contain the spin operator (nonmagnetic materials without SOC). Therefore, their symmetry operations include spatial symmorphic symmetry (pure spatial PG symmetry) $\{C_n(\theta), IC_n(\theta)|0\}$ and spatial non-symmorphic symmetry $\{C_n(\theta), IC_n(\theta)|\tau\}$, as shown in the fist two rows of Table A1.

On the other hand, while MSGs and SSGs contains all the operations listed in Table A1, the specific manifestations of each row for the two types of groups are quite different. If no additional symmetry operations is introduced compared to ordinary SGs, the MSGs are often regarded as type I MSG (colorless group). If a MSG contains pure spin PG symmetry $\{g_s||E|0\}$ where no spatial rotations or fractional translations $\tau$ are involved, $g_s$ can only be time-reversal $T$, corresponding to a type-II MSG (grey group). If a MSG contains general symmetry $\{g_s||C_n(\theta), IC_n(\theta)|\tau\}$, $g_s$ can only be proper or improper spin rotations $U_n(\theta)$ or $TU_n(\theta)$ compatible with the spatial rotation $C_n(\theta)$, corresponding to a type-III MSG (colored point group). If a MSG contains spin non-symmorphic symmetry $\{g_s||E|\tau\}$ where a spin operation is combined with a fractional translation $\tau$, again, $g_s$ can only be time-reversal $T$, corresponding to a type-IV MSG (colored lattice). In comparison, SSGs allow separated spatial and spin rotations, leading to much more symmetry operations that are not allowed by MSGs. Specifically, pure spin PG symmetries in SSGs form the spin-only group $G_{SO}$, including $G_{SO}^p = \{E, TU_n(\pi)\} = Z_2^K$ for coplanar configurations and $G_{SO}^l = Z_2^K \ltimes SO(2)$ for collinear configurations. In addition, spin non-symmorphic symmetry widely exists in spiral magnetic configurations with a *k*-type SSG. For example, the *k*-type SSG with $i_k = 4$ has spin non-symmorphic symmetry $\{C_{4z}||E|\tau_{1/4}\}$.

For general symmorphic symmetry and general non-symmorphic symmetry, MSG requires that $g_s$ can only be proper or improper spin rotations $U_n(\theta)$ or $TU_n(\theta)$ compatible with the spatial rotation $C_n(\theta)$. In comparison, within the framework of SSG, $g_s$ can be $U_m(\phi)$ or $TU_m(\phi)$ with the rotational angle and axis not necessarily compatible with those of the spatial rotation.

**Table A1:** Six types of operations in SGs, MSGs and SSGs. For pure spin PG symmetry and spin non-symmorphic symmetry in MSGs, $g_s$ can only be time-reversal $T$; for general symmorphic symmetry and general non-symmorphic symmetry in MSGs, $g_s$ can only be proper or improper spin rotations $U_n(\theta)$ or $TU_n(\theta)$ compatible with the spatial rotation $C_n(\theta)$.

| | Notation | Spin PG | Spatial PG | fractional translation |
|---|---|---|---|---|
| Spatial symmorphic symmetry | $\{C_n(\theta), IC_n(\theta)\|0\}$ | | √ | |
| Spatial non-symmorphic symmetry | $\{C_n(\theta), IC_n(\theta)\|\tau\}$ | | √ | √ |
| Pure spin symmetry | $\{g_s\|\|E\|0\}$ | √ | | |
| Spin non-symmorphic symmetry | $\{g_s\|\|E\|\tau\}$ | √ | | √ |
| General symmorphic symmetry | $\{g_s\|\|C_n(\theta), IC_n(\theta)\|0\}$ | √ | √ | |
| General non-symmorphic symmetry | $\{g_s\|\|C_n(\theta), IC_n(\theta)\|\tau\}$ | √ | √ | √ |

## 2. Construction of SSGs by finding inequivalent coset decompositions and inequivalent mappings

This part is devoted to explaining the mathematical procedure of finding inequivalent coset decompositions and inequivalent mappings between the quotient group $G_0/L_0$ and the spin-space PG $G^s$, which is an essential step to eliminate equivalent SSGs in the enumeration process. Before getting into the details of the mathematical procedure, we first introduce some necessary concepts.

*Chirality-preserving operations* are symmetry operations that preserve the chirality of the object. Specifically, in three-dimensional (3D) Euclidean space $\mathbb{R}^3$, the linear transformations form 3D orthogonal group $O(3) = SO(3) \cup PSO(3)$, where $P$ represents space inversion. Elements of $SO(3)$ are regarded as chirality-preserving operations because their transformation matrices have a determinant of 1. Therefore, the chirality-preserving mappings $\mathbb{R}^3$ includes all translations, proper rotations, and screw rotations, yet excluding reflections, inversions, rotoinversions, and glide reflections.

*Affine transformation*, or affine mapping, is a linear mapping method that perserves points, straight lines, and planes. After an affine transformation, although the Euclidean lengths and angles may change, sets of parallel lines remain parallel.

*Normalizer.* For a group $G$ and its subgroup $L$, the normalizer of $L$ with respect to $G$ denotes the set of elements $g \in G$ that leave the subgroup of $L$ unchanged by conjugation:

$$N_G(L) = \{g | g^{-1}Lg = L,\ g \in G\}. \tag{A1}$$

Let us review establishment of SG briefly. In the enumeration of SG, The group $G_0$ and $G_0'$ belong to the same affine space-group type if an affine transformation $a \in T(\mathbb{R}^3) \rtimes GL(3, R)$ exists, for which

$$G_0' = a^{-1}G_0a \tag{A2}$$

holds. Here $T(\mathbb{R}^3)$ contains all translations in $\mathbb{R}^3$ and $GL(3, R)$ denotes the 3D linear group whose entries of matrices are real numbers. However, there are only 219 affine space-group types in total as the rotation part of an affine transformation $a$ may have a negative determinant. In other words, it means that $a$ does not preserve the chirality of $G_0$. Only when the affine transformation $a$ is required to be chirality-preserving during the mapping process, we can distinguish eleven pairs of enantiomorphic SGs (left-handed versus right-handed structures), as shown in Table A2, and 230 = 219 + 11 crystallographic SG types are obtained [64].

To construct a SSG, the first step is to decompose a SG $G_0$ into its subgroups $L_0$.

$$G_0 = L_0 \cup g_1L_0 \cup \ldots \cup g_{n-1}L_0. \tag{A3}$$

For two pairs $(G_0, L_0)$ and $(G_0', L_0')$, it is obvious if $G_0$ and $G_0'$ belong to different SGs, as well as $L_0$ and $L_0'$, the coset decompositions and thus the resulting SSGs are different. On the other hand, even if both of $(G_0, L_0)$ and $(G_0', L_0')$ cannot be distinguished just by their SGs, the coset decompositions of the two pairs may still be different. To provide a more detailed explanation, one can firstly transform $L_0$ and $L_0'$ to the default setting of SGs so that they are identical ($L_0 = L_0'$). Then a chirality-preserving affine mapping $a$ in $\mathbb{R}^3$ is constructed to make the coset representatives of $G_0 = L_0 \cup g_1L_0 \cup \ldots \cup g_{n-1}L_0$ and $G_0' = L_0 \cup g_1'L_0 \cup \ldots \cup g_{n-1}'L_0$ become identical.

Overall, the chirality-preserving affine normalizer $N_{A^+}(L_0)$ was employed to find all inequivalent coset decompositions [64]. All affine mappings in $\mathbb{R}^3$ that map $L_0$ onto itself by conjugation and preserve its chirality form the chirality-preserving affine

normalizer of $L_0$:

$$N_{A^+}(L_0) = \{a | a^{-1} L_0 a = L_0,\ a \in T(\mathbb{R}^3) \rtimes SL(3,\mathbb{Z})\} \quad \text{(A4)}$$

Here, $SL(3,\mathbb{Z})$ denotes the 3D linear group whose entries of matrices are integers and determinant is of 1. Thus, two pairs $(G_0, L_0)$ and $(G_0', L_0)$ are equivalent if their supergroups are conjugated under $a \in N_{A^+}(L_0)$:

$$a^{-1} G_0' a = G_0,\ \exists a \in N_{A^+}(L_0) \quad \text{(A5)}$$

The conclusion can be directly extended when considering the inclusion of spin part, i.e., nontrivial SSGs. Two nontrivial SSGs $G_{NS}'$ and $G_{NS}$ with equivalent $(G_0, L_0)$ pairs are equivalent if they can be mapped onto each other by:

$$a^{-1} G_{NS}' a = G_{NS},\ \exists a \in \left(N_{A^+}(L_0) \cap N_{A^+}(G_0)\right) \times N_{O(3)}(G^s) \quad \text{(A6)}$$

where $N_{A^+}(G_0)$ is the chirality-preserving affine normalizer of $G_0$ and $N_{O(3)}(G^s)$ is the normalizer of $G^s$ in orthogonal group $O(3)$.

The abovementioned mathematical procedure of identifying all inequivalent coset decompositions and inequivalent mappings for constructing SSGs using Eq. (A4) to (A6) is performed by Magma, which is a large, well-supported software package designed for computations in abstract mathematical objects. It basically provides nearly all the significant algorithms for finite groups and finitely presented infinite groups.

**Table A2:** Eleven pairs of enantiomorphic SGs

| | |
|---|---|
| $P4_1$ (No. 76) | $P4_3$ (No. 78) |
| $P4_122$ (No. 91) | $P4_322$ (No. 95) |
| $P4_12_12$ ($No.$ 92) | $P4_32_12$ ($No.$ 96) |
| $P3_1$ ($No.$ 144) | $P3_2$ ($No.$ 145) |
| $P3_112$ ($No.$ 151) | $P3_212$ ($No.$ 153) |
| $P3_121$ ($No.$ 152) | $P3_221$ ($No.$ 154) |
| $P6_1$ ($No.$ 169) | $P6_5$ ($No.$ 170) |
| $P6_2$ ($No.$ 171) | $P6_4$ ($No.$ 172) |
| $P6_122$ ($No.$ 178) | $P6_522$ ($No.$ 179) |
| $P6_222$ ($No.180$) | $P6_422$ ($No.181$) |
| $P4_132$ ($No.213$) | $P4_332$ ($No.212$) |

## 3. Little group and little co-group

The translation generators of a SG $G$ determine its Bravais lattice and also its Brillouin zone (BZ). Points in the BZ are denoted by $k$ in momentum space. The action of $g = \{R|\tau\} \in G$ on $k$ is $gk = Rk$, where translation $\tau$ leaves $k$ invariant. All elements in a SG $G$ that keeps a reciprocal vector $k$ invariant form a group named the *little group* of $k$. Since we focus on the properties of Bloch states under band theory, where Bloch states are labeled by reciprocal vectors $k$, the little group of $k$ is of great importance. The PG part of a little group is called a *little co-group*. Such concept is introduced because the theory of projective rep maps the derivation of irreps of a little group to the derivation of projective irreps of the corresponding little co-group, in which lattice translation is neglected and the fractional translation is included in factor systems (the details will be discussed later).

## 4. Representation, irreducible representation and regular representation

A *representation* of group $G$ is formally defined as a homomorphism from $G$ to the group form by general linear transformations on a vector space $V$, i.e., the general linear group $GL(V)$. The matrix forms of those linear transformations are called *representation matrices*. In quantum physics, group rep is considered with a certain group of quantum states as a basis. Under this situation, a group rep can be simply understood as the matrix form of all symmetry operators. This method allows us to describe the symmetry of quantum systems with easily manipulated linear algebra. Generally, by basis transformation, a vector space can be separated into subspaces such that all vectors of one subspace are only transformed into each other. This allows us to block diagonalize the full rep matrices of all group elements simultaneously and view the full rep as the direct sum of reps on every subspace. Those reps which have rep matrices that cannot be block diagonalized by any basis transformation are called *irreducible representations*. In quantum physics, irreps describe the degeneracy of quantum states and encode basic symmetry information of quantum states, e.g., parity, angular momentum numbers, etc.

A *regular representation* is defined as the rep by using the group space as the rep space, where the group elements function as both the operators and as the bases for the space. It is a particularly interesting set of matrices that forms a reducible rep of the group $G$, which can be block-diagonalized to give all irreps of the group. The matrices of the regular rep are constructed with the following relation:

$$D_{jk}^{\mathrm{reg}}(R_i) = \begin{cases} 1 & if\ R_i R_k = R_j \\ 0 & otherwise \end{cases}. \tag{A7}$$

The multiplicity $a_j$ of each irrep in the regular rep can be calculated to be equal to its dimension $l_j$:

$$a_j = \frac{1}{g}\sum_{R\in G} \chi_j(R)^* \chi^{reg}(R) = \chi_j(E)^* = l_j. \tag{A8}$$

Namely,

$$X^{-1}D^{reg}(R)X = \bigoplus_j l_j D_j(R), \qquad \chi(R) = \sum_j l_j \chi_j^{reg}(R), \tag{A9}$$

where $\chi_j(R)$, $D_j(R)$ represent the character and the rep matrix for the irrep $j$ of the group $G$, respectively, while $\chi^{reg}(R)$, $D^{reg}(R)$ represent the character and the rep matrix for the reducible rep of the regular rep, repsectively. $E$, $g$, $X$ stand for the identity, the order of the group $G$ and the similarity transformation, respectively.

In summary, the regular rep contains each irrep a number of times equal to the dimensionality of the rep, i.e., the *n*-dimensional irreps appears *n* times in regular reps.

## 5. Projective representation in ordinary SGs

In SGs, the fractional translations usually bring complexities to the reps, leading to extra degeneracies, etc. The reps of any little groups of $k$ can be obtained by the projective reps of the corresponding little co-group. To elaborate on this, we first consider a little group $G^k$ with the coset form with respect to translation group $\mathbb{T}$ to be:

$$G^k = \{R_1|\tau_1\}\mathbb{T} \cup \{R_2|\tau_2\}\mathbb{T} \cup \ldots \cup \{R_n|\tau_n\}\mathbb{T}. \tag{A10}$$

The rep of $\mathbb{T}$ with a basis of Bloch functions at momentum $k$ is already known with the form $exp(-ik\cdot\tau)$. Thus, to get a full list of irreps of $G^k$, all we need to do is to derive the irreps of the coset representatives $\tilde{G}^k = G^k/\mathbb{T} = (\{R_1|\tau_1\}, \{R_2|\tau_2\}, \ldots, \{R_n|\tau_n\})$. One can naively think the rep of the coset representatives as the rep of the corresponding little co-group $\breve{G}^k = (R_1, R_2, \ldots, R_n)$. However, they are not always the same because of the possible nonsymmorphic translation $\tau_i$, i.e., if we assume $R_iR_j = R_k$, the product of $\{R_i|\tau_i\}$ and $\{R_j|\tau_j\}$ may not be $\{R_k|t_k\}$, but follow the below relation:

$$\{R_i|\tau_i\}\{R_j|\tau_j\} = \{R_iR_j|\tau_i + R_i\tau_j\} = \{E|\tau_{ij}\}\{R_k|\tau_k\}, \quad \text{(A11)}$$

where $\tau_{ij} = \tau_i + R_i\tau_j - \tau_k$. Therefore, the nonsymmorphic translation $\tau_i$ and $\tau_j$ could result in a lattice translation $\{E|\tau_{ij}\}$ beyond the $\{R_k|\tau_k\}$. Thus, the product relation of a rep $d_k$ at $k$ point should obey the following:

$$d_k(\{R_i|\tau_i\})d_k(\{R_j|\tau_j\}) = d_k(\{E|\tau_{ij}\})d_k(\{R_k|\tau_k\}) = exp(-ik \cdot \tau_{ij})d_k(\{R_k|\tau_k\}), \quad \text{(A12)}$$

which exhibit an additional phase factor $exp(-ik \cdot \tau_{ij})$.

Strictly speaking, the group of rep matrices could be much larger than the group of coset representatives $\tilde{G}^k$, or the little co-group $\breve{G}^k$. These rep matrices should include all reps of $\{R_iR_j|\tau_i + R_i\tau_j\}$ in addition to those of $\{R_i|\tau_i\}$. In other words, the group of rep matrices could cover little co-group $\breve{G}^k$ multiple times because one $R_k$ could combine with different translations. To simplify the problem, we can use the theory of projective rep to "mod" the influence of nonsymmorphic translations and get all irreps of $G^k$ by deriving the corresponding irreducible projective reps.

*Projective representation* is formally defined as follows: For a group $H$ consisting of elements $h_i$, where $i$ ranges from 1 to $|H|$, a non-singular matrix function $\Delta$ on group $H$ is a *projective representation* of $H$ if it satisfies the following rule:

For each group product $h_ih_j = h_k$, there exists a scalar function $\mu(h_i, h_j)$ on the group elements $h_i, h_j$, obeying that

$$\Delta(h_i)\Delta(h_j) = \mu(h_i, h_j)\Delta(h_k). \quad \text{(A13)}$$

The function $\mu(h_i, h_j)$ is called a *factor system* for the projective rep $\Delta$.

In the problem of multiple reps of $\{R_iR_j|\tau_i + R_i\tau_j\}$ discussed above, we can see that the product of rep $d_k$ does not follow the rule of Eq. (A13) because $\tau_{ij} = \tau_i + R_i\tau_j - \tau_k$ is dependent on $\tau_k$. To solve this, we define the projective rep corresponding to $d_k$ to be $M_k$:

$$d_k(\{R|\tau\}) = exp(-ik \cdot \tau)\, M_k(\{R|\tau\}). \quad \text{(A14)}$$

Under this definition, one can show that

$$M_k(\{R_i|\tau_i\})M_k(\{R_j|\tau_j\}) = exp\,(-iK_i \cdot \tau_j)\, M_k(\{R_k|\tau_k\})\,, \quad \text{(A15)}$$

where $K_i = R_i^{-1}k - k$. The set of $M_k$ together with a factor system $\mu(R_i, R_j) = exp(-iK_i \cdot \tau_j)$ form a projective irrep of the little group $G^k$. The set $M_k$ forms an irrep of the isomorphic PG, where the factor system $\mu(R_i, R_j) = exp(-iK_i \cdot \tau_j)$ is completely dependent on $\{R_i|\tau_i\}$ and $\{R_j|\tau_j\}$. Consequently, to get all irreps $d_k$ of $\tilde{G}^k$, one needs to derive all the corresponding projective irreps $M_k$ of $\breve{G}^k$. In practice, $M_k$ is obtained by central extension method, which is introduced in the following.

## 6. Central extension

We now encounter two possibilities regarding the multiplier factor system. One case is all the factor system $\mu(R_i, R_j) = 1$, and hence the irreps of little group $G^k$ and the little co-group $\breve{G}^k$ coincide. If one of following three conditions is reached, the factor system maps into unity: (1) $G$ is a symmorphic SG; (2) $G^k$ is a symmorphic group; (3) $k$ lies inside the first Brillouin zone.

Another case is that the factor system $\mu(R_i, R_j)$ can take multiple values, when $G^k$ is a non-symmorphic group and $k$ lies on the Brillouin zone boundary. In such situation, a typical treatment, named *central extension*, to obtain the projective reps is to find the reps of a new abstract group $\breve{G}^{k,*} = \breve{G}^k \times Z_g$, where $Z_g$ denotes the cyclic group of integers ( $0,1,\dots,g-1$ ). Here $g$ comes from the factor system $\mu(R_i, R_j) = exp(2\pi i a(R_i, R_j)/g)$ with the condition $0 \leq a(R_i, R_j) \leq g-1$. The multiplication table of the operations in $\breve{G}_k^*$ is determined by the following relationship:

$$(R_i, \alpha)(R_j, \beta) = (R_i R_j, a(R_i, R_j) + \alpha + \beta), \tag{A16}$$

where $(R_i, \alpha)$ are group elements of the central extension of little cogroup $\breve{G}^{k,*}$ and $\alpha \in \{0,1,\dots,g-1\}$.

Following this approach, $\breve{G}^{k,*}$ can be identified with one isomorphic abstract group and the irreps can be determined. Of all irreps, we only focus on the right irreps with $M_k(E, \alpha) = exp(2\pi i\alpha/g)\mathbb{I}$, where $\mathbb{I}$ is the identity matrix. Following above, we can construct the character table of the projective irreps of $G^k$. More details and specific examples can be found in Ref. [2].

## 7. Co-representation

In magnetic materials, group description usually involves antiunitary operations that include time-reversal operator *T*. These operations contain a complex conjugation operation that transforms any complex number into its complex conjugate. A famous example of band degeneracy induced by antiunitary groups is the so-called Kramers degeneracy. However, a comprehensive understanding of the transformation of quantum states under antiunitary groups requires more than just knowledge of Kramers degeneracy. To achive this, we must first know the unitary rep matrices under antiunitary groups. Owing to the properties of antiunitary operations, the matrix representatatives $D$ do not obey the ordinary multiplication relations associated with unitary groups, but rather satisfy the following equations:

$$D(u_1)D(u_2) = D(u_1u_2), D(u)D(a) = D(ua), \tag{A17}$$

$$D(a)D^*(u) = D(au), D(a_1)D^*(a_2) = D(a_1a_2), \tag{A18}$$

where $u, u_1, u_2$ represent any unitary operation, while $a, a_1, a_2$ represent any antiunitary operation of the same magnetic group, $D^*$ is the complex conjugate rep of $D$. This kind of rep is named as "*co-representation*" by Wigner. The theory of *co-rep* is useful for analyzing extra degeneracies caused by antiunitary operations and the corresponding degenerate states.

## 8. Dimmock and Wheeler's character sum rule

Following the above, the co-irreps of MSGs are constructed. For any little group $G_{MS}^k$ of a MSG $G_{MS}$: (1) if $G_{MS}^k$ contains only unitary group elements, then the irreps of $G_{MS}^k$ are given as those in SGs; (2) if $G_{MS}^k$ contains antiunitary group elements, then $G_{MS}^k$ can be written as $G_{MS}^k = L_{MS}^k \cup TAL_{MS}^k$, where $L_{MS}^k$ is a unitary subgroup of index 2 in $G_{MS}^k$, $TA \notin L_{MS}^k$ is an antiunitary group element of $G_{MS}^k$. Now we define the quotient group $\tilde{G}_{MS}^k = G_{MS}^k/\mathbb{T} = \tilde{L}_{MS}^k \cup TA\tilde{L}_{MS}^k$. Since the irreps of $\tilde{L}_{MS}^k$ are already given in SGs, we only need to classify the induced co-irreps from the irreps of the maximal unitary subgroup $\tilde{L}_{MS}^k$ using the Dimmock and Wheeler's sum rule as summarized in three cases:

(a) the co-rep $D^{(i)}$ of $\tilde{G}_{MS}^k$ correspond to a single rep $d^{(i)}$ of $\tilde{L}_{MS}^k$ and has the same dimension, in this case, no extra degeneracy is introduced.
(b) the co-rep $D^{(i)}$ of $\tilde{G}_{MS}^k$ correspond to a single rep $d^{(i)}$ of $\tilde{L}_{MS}^k$ but with the doubled dimension, in this case, the degeneracy is doubled.

(c) the co-rep $D^{(i)}$ of $\tilde{G}_{MS}^{k}$ correspond to two inequivalent reps of $\tilde{L}_{MS}^{k}$, but these two can be stick together under the antiunitary operators.

We summarize the Dimmock and Wheeler's sum rule as

$$\sum_{g\in TA\tilde{L}_{MS}^{k}} \chi(g^2) = \begin{cases} +\left|\tilde{L}_{MS}^{k}\right| & (a) \\ -\left|\tilde{L}_{MS}^{k}\right| & (b) \\ 0 & (c) \end{cases}, \tag{A19}$$

where $\chi$ is the character of $d$.

## B. International notation and Seitz symbol

In International notation, SGs are designated by a symbol $Bg_1g_2g_3$ that combines the SG symmetry with a uppercase letter describing the centering of the Bravais lattice. Specifically, $B$ is denoted by P for primitive lattice, A, B, C for base-centered lattice, I for body-centered lattice, F for face-centered lattice and R for rhombohedral lattice. The subsequent three symbols $g_1g_2g_3$ represent the representative symmetry operations when projected along one of the high symmetry directions of the crystal. The high symmetry directions of different crystal systems are shown as follows:

### 1. Viewing direction of SGs

**Table B1:** The viewing direction of SGs in International notations.

| order | Triclinic | Monoclinic | Orthorhombic | Tetragonal | Trigonal | Hexagonal | Cubic |
|---|---|---|---|---|---|---|---|
| 1 | | [010] | [100] | [001] | [001] | [001] | [100] |
| 2 | | | [010] | [100] | [100] | [100] | [111] |
| 3 | | | [001] | [110] | [210] | [210] | [110] |

For example, the four letters of $P3_212$ (153) represents a primitive lattice, consisting of a 3-fold rotation along the [001] direction with a 2/3$c$ translation, identity along the [100] direction and a 2-fold rotation along the [210] direction.

### 2. International symbols of SGs

Here we list International symbols and their corresponding symmetry operations used for SGs, such as rotation, mirror, rotation-inversion in the PG, as well as glide, screw in the SG.

**Table B2:** Symbols of SG using International notation, including rotation axes, mirror planes, screw axes and glide planes.

| Name | Notation | Remark |
|---|---|---|
| identity | 1 | |
| inversion | -1 | |
| rotation | n | n-fold rotation |
| rotation-inversion | -n | n-fold rotation followed by an inversion |
| mirror | m | reflection in a plane |
| mirror plane ⊥ to n-fold axes | n/m | |
| screw | $n_i$ | n-fold rotation followed by *i*/n fraction translation (*i* = 1, 2, .. n-1) |
| glide mirror (combination of a mirror and a fraction translation) | | |
| axial glide | a | tranlation of a/2 |
| | b | tranlation of b/2 |
| | c | tranlation of c/2 |
| diagnonal glide | n | tranlation of (a+b)/2, (b+c)/2, (a+c)/2 |
| diamond glide | d | tranlation of (a+b)/4, (b+c)/4, (a+c)/4 |
| two ⊥ glides* | e | |

* SGs with two perpendicular glides are attributed to the centering type and the glide with perpendicular direction. For instance, the SG Aem2 (39) has a mirror parallel to *a* with a translation of b/2 and a centering fraction translation of (b+c)/2, resulting in the mirror parallel to *a* with a translation of c/2. As a consequence, Aem2 can be referred as Abm2 and Acm2 simultaneously. Therefore, the symbol "e" is employed for such planes. Similar situations can be found in Aea2 (41), Cmce (64), Cmme (67) and Ccce (68).

### 3. Seitz symbols in real space

International notations only provide description of representative group elements for denoting SGs. In order to denote all group elements, Seitz symbols are introduced. We generalized the Seitz symbol $\{R|\tau\}$ in SGs to $\{g_s||R|\tau\}$ in SSGs, where $g_s$, $R$, and $\tau$ represent spin rotation, space rotation and translation, respectively. However, due to the complexity of translational supercell, the PG in spin space will not always have the same crystal systems as the space part. For example, the space part of 4.182.4.2 ($CoNb_3S_6$) belongs to the hexagonal system ($P6_322$), while the spin part belongs to the cubic system (-43m). Therefore, we define the space part of SSG symmetry in the lattice basis, i.e., *a*, *b*, *c*, while the spin part of SSG symmetry is defined in Cartesian coordinate system, i.e., *x*, *y*, *z*. The nomenclature for the symbol *R* in real space is shown below:

The symbol *R* signifies the operations of identity and space inversion with 1 and -1, respectively. The symbol *m* represents mirror operations, while the symbols *n* (*n* = 2, 3, 4, 6) are utilized for *n*-fold rotations, and -*n* (n = 3, 4, 6) denote rotation-inversion operations. For rotations and rotation-inversions of order higher than 2, a superscript *i* is used to indicate the number of applications of rotations and the subscript *hkl* is used to denote the characteristic direction of the operation. For example, the $4^3_{001}$ represents an anticlockwise rotation of 270-degree along the [001] crystallographic direction.

The following two tables list all Seitz symbols for the SG operations in the seven crystal systems. Here $\vec{a}$, $\vec{b}$ and $\vec{c}$ denote the lattice basis vectors, while a, b and c represent the coordination using the lattice basis *a*, *b*, *c*.

**Table B3:** Seitz symbols for PG part of space-group symmetry operations of the cubic, tetragonal, orthorhombic, monoclinic and triclinic crystal systems.

| No. | Seitz symbol | Axes | Coordination |
|---|---|---|---|
| 1 | 1 | | a,b,c |
| 2 | $2_{001}$ | $\vec{c}$ | -a,-b,c |
| 3 | $2_{010}$ | $\vec{b}$ | -a,b,-c |
| 4 | $2_{100}$ | $\vec{a}$ | a,-b,-c |
| 5 | $3^1_{111}$ | $\vec{a}+\vec{b}+\vec{c}$ | c,a,b |
| 6 | $3^1_{-11-1}$ | $-\vec{a}+\vec{b}-\vec{c}$ | c,-a,-b |
| 7 | $3^1_{1-1-1}$ | $\vec{a}-\vec{b}-\vec{c}$ | -c,-a,b |
| 8 | $3^1_{-1-11}$ | $-\vec{a}-\vec{b}+\vec{c}$ | -c,a,-b |
| 9 | $3^2_{111}$ | $\vec{a}+\vec{b}+\vec{c}$ | b,c,a |
| 10 | $3^2_{1-1-1}$ | $\vec{a}-\vec{b}-\vec{c}$ | -b,c,-a |
| 11 | $3^2_{-1-11}$ | $-\vec{a}-\vec{b}+\vec{c}$ | b,-c,-a |
| 12 | $3^2_{-11-1}$ | $-\vec{a}+\vec{b}-\vec{c}$ | -b,-c,a |
| 13 | $2_{110}$ | $\vec{a}+\vec{b}$ | b,a,-c |
| 14 | $2_{1-10}$ | $\vec{a}-\vec{b}$ | -b,-a,-c |
| 15 | $4^3_{001}$ | $\vec{c}$ | b,-a,c |
| 16 | $4^1_{001}$ | $\vec{c}$ | -b,a,c |
| 17 | $4^3_{100}$ | $\vec{a}$ | a,c,-b |
| 18 | $2_{011}$ | $\vec{b}+\vec{c}$ | -a,c,b |
| 19 | $2_{01-1}$ | $\vec{b}-\vec{c}$ | -a,-c,-b |
| 20 | $4^1_{100}$ | $\vec{a}$ | a,-c,b |
| 21 | $4^1_{010}$ | $\vec{b}$ | c,b,-a |
| 22 | $2_{101}$ | $\vec{a}+\vec{c}$ | c,-b,a |
| 23 | $4^3_{010}$ | $\vec{b}$ | -c,b,a |
| 24 | $2_{-101}$ | $\vec{a}-\vec{c}$ | -c,-b,-a |
| 25 | -1 | | -a,-b,-c |
| 26 | $m_{001}$ | $\vec{c}$ | a,b,-c |
| 27 | $m_{010}$ | $\vec{b}$ | a,-b,c |
| 28 | $m_{100}$ | $\vec{a}$ | -a,b,c |
| 29 | $-3^1_{111}$ | $\vec{a}+\vec{b}+\vec{c}$ | -c,-a,-b |
| 30 | $-3^1_{-11-1}$ | $-\vec{a}+\vec{b}-\vec{c}$ | -c,a,b |
| 31 | $-3^1_{1-1-1}$ | $\vec{a}-\vec{b}-\vec{c}$ | c,a,-b |
| 32 | $-3^1_{-1-11}$ | $-\vec{a}-\vec{b}+\vec{c}$ | c,-a,b |
| 33 | $-3^2_{111}$ | $\vec{a}+\vec{b}+\vec{c}$ | -b,-c,-a |
| 34 | $-3^2_{1-1-1}$ | $\vec{a}-\vec{b}-\vec{c}$ | b,-c,a |
| 35 | $-3^2_{-1-11}$ | $-\vec{a}-\vec{b}+\vec{c}$ | -b,c,a |
| 36 | $-3^2_{-11-1}$ | $-\vec{a}+\vec{b}-\vec{c}$ | b,c,-a |
| 37 | $m_{110}$ | $\vec{a}+\vec{b}$ | -b,-a,c |
| 38 | $m_{1-10}$ | $\vec{a}-\vec{b}$ | b,a,c |
| 39 | $-4^3_{001}$ | $\vec{c}$ | -b,a,-c |
| 40 | $-4^1_{001}$ | $\vec{c}$ | b,-a,-c |
| 41 | $-4^3_{100}$ | $\vec{a}$ | -a,-c,b |
| 42 | $m_{011}$ | $\vec{b}+\vec{c}$ | a,-c,-b |
| 43 | $m_{01-1}$ | $\vec{b}-\vec{c}$ | a,c,b |
| 44 | $-4^{1}_{100}$ | $\vec{a}$ | -a,c,-b |
| 45 | $-4^{1}_{010}$ | $\vec{b}$ | -c,-b,a |
| 46 | $m_{101}$ | $\vec{a}+\vec{c}$ | -c,b,-a |
| 47 | $-4^{3}_{010}$ | $\vec{b}$ | c,-b,-a |
| 48 | $m_{-101}$ | $\vec{a}-\vec{c}$ | c,b,a |

**Table B4:** Seitz symbols for PG part of space-group symmetry operations of the hexagonal and trigonal crystal systems.

| No. | Seitz symbol | Axes | Coordination |
|---|---|---|---|
| 1 | 1 | | a,b,c |
| 2 | $3^{1}_{001}$ | $\vec{c}$ | -b,a - b,c |
| 3 | $3^{2}_{001}$ | $\vec{c}$ | -a + b,-a,c |
| 4 | $2_{001}$ | $\vec{c}$ | -a,-b,c |
| 5 | $6^{5}_{001}$ | $\vec{c}$ | b,-a + b,c |
| 6 | $6^{1}_{001}$ | $\vec{c}$ | a - b,a,c |
| 7 | $2_{110}$ | $\vec{a}+\vec{b}$ | b,a,-c |
| 8 | $2_{100}$ | $\vec{a}$ | a - b,-b,-c |
| 9 | $2_{010}$ | $\vec{b}$ | -a,-a + b,-c |
| 10 | $2_{1-10}$ | $\vec{a}-\vec{b}$ | -b,-a,-c |
| 11 | $2_{120}$ | $\vec{a}$+2$\vec{b}$ | -a + b,b,-c |
| 12 | $2_{210}$ | 2$\vec{a}$+$\vec{b}$ | a,a - b,-c |
| 13 | -1 | | -a,-b,-c |
| 14 | $-3^{1}_{001}$ | $\vec{c}$ | b,-a + b,-c |
| 15 | $-3^{2}_{001}$ | $\vec{c}$ | a - b,a,-c |
| 16 | $m_{001}$ | $\vec{c}$ | a,b,-c |
| 17 | $-6^{5}_{001}$ | $\vec{c}$ | -b,a - b,-c |
| 18 | $-6^{1}_{001}$ | $\vec{c}$ | -a + b,-a,-c |
| 19 | $m_{110}$ | $\vec{a}+\vec{b}$ | -b,-a,c |
| 20 | $m_{100}$ | $\vec{a}$ | -a + b,b,c |
| 21 | $m_{010}$ | $\vec{b}$ | a,a - b,c |
| 22 | $m_{1-10}$ | $\vec{a}-\vec{b}$ | b,a,c |
| 23 | $m_{120}$ | $\vec{a}$+2$\vec{b}$ | a - b,-b,c |
| 24 | $m_{210}$ | 2$\vec{a}$+$\vec{b}$ | -a,-a + b,c |

## 4. Seitz symbols in spin space

We extend the nomenclature for the symbol $g_s$ in spin space:

The symbol $g_s$ is 1 and -1 for identity and time reversal, respectively. The symbols *n* (*n* = 2, 3, 4, 5, 6, 7, 8 …) are used for *n*-fold spin rotations. The improper spin rotation represents the time reversal followed by proper spin rotations, including *m* for the combination of the time reversal and 2-fold rotation and -*n* for the combination of the time reversal and *n*-fold rotation. Similarly, a superscript *i* is used to indicate the number of applications of rotations and the subscript *hkl* denote the characteristic direction of the operation. For example, $5^2_{001}$ stands for 144-degree rotation along the [001] direction. We include three axial directions [100], [010], [001], six face-diagonal directions [110], [1-10], [101], [10-1], [011], [01-1] and four body-diagonal direction [111], [-11-1], [1-1-1], [-1-11]. In addition, the subscript $\theta \in (0, \pi)$ is employed to label the specific direction in the *x*-*y* plane. For example, the $2_\theta$ label the in-plane 2-fold rotation around the in-plane $\theta$-direction axis with respect to *x* direction (shown in Fig. B1).

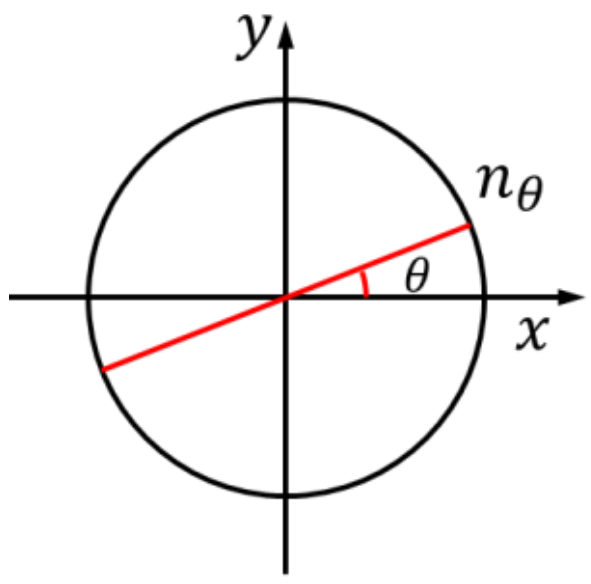

**Fig. B1:** Schematic of in-plane two-fold rotation $2_\theta$ or mirror $m_\theta$ in spin space.

## 5. Representative symmetries for SGs

Now we present the complete list of SG $Bg_1g_2g_3$ with International notation and their corresponding Seitz symbols, which are used to label SSGs according to their crystal systems.

First we list the detailed centering-type fractional translation $b_1, b_2, b_3$ for SG in various Bravais lattices, i.e., the corresponding symmetry operations of $B$.

**Table B5:** The centering-type fractional translation $b_1, b_2, b_3$ for SG in different Bravais lattices, which is used as default sequence in ***g***-type SSG.

| Bravais lattice | $b_1$ | $b_2$ | $b_3$ |
|---|---|---|---|
| P | | | |
| F | $\{1\|1/2\ 1/2\ 0\}$ | $\{1\|1/2\ 0\ 1/2\}$ | $\{1\|0\ 1/2\ 1/2\}$ |
| I | $\{1\|1/2\ 1/2\ 1/2\}$ | | |
| A | $\{1\|0\ 1/2\ 1/2\}$ | | |
| C | $\{1\|1/2\ 1/2\ 0\}$ | | |
| R | $\{1\|2/3\ 1/3\ 1/3\}$ | $\{1\|1/3\ 2/3\ 2/3\}$ | |

Secondly, we show the following three representative symbols, $g_1g_2g_3$, within 230 SGs $Bg_1g_2g_3$, categorized into triclinic, monoclinic, orthorhombic, tetragonal, trigonal, hexagonal and cubic crystal systems.

**Table B6:** Table of symbol-group element correspondence in SGs of triclinic crystal systems, with no viewing direction is used.

| No. | HM | $g_1$ | $g_2$ | $g_3$ |
|---|---|---|---|---|
| 1 | P1 | $\{1\|0\}$ | | |
| 2 | P-1 | $\{-1\|0\}$ | | |

**Table B7:** Table of symbol-group element correspondence in SGs of monoclinic crystal systems, with viewing direction of [010] in International notation.

| No. | HM | $g_1$ | $g_2$ | $g_3$ |
|---|---|---|---|---|
| 3 | P2 | $\{2_{010}\|0\}$ | | |
| 4 | P2₁ | $\{2_{010}\|0\ 1/2\ 0\}$ | | |
| 5 | C2 | $\{2_{010}\|0\}$ | | |
| 6 | Pm | $\{m_{010}\|0\}$ | | |
| 7 | Pc | $\{m_{010}\|0\ 0\ 1/2\}$ | | |
| 8 | Cm | $\{m_{010}\|0\}$ | | |
| 9 | Cc | $\{m_{010}\|0\ 0\ 1/2\}$ | | |
| 10 | P2/m | $\{2_{010}\|0\}$ | | |
| | | $\{m_{010}\|0\}$ | | |
| 11 | $P2_1/m$ | $\{2_{010}\|0\ 1/2\ 0\}$ | | |
| | | $\{m_{010}\|0\ 1/2\ 0\}$ | | |
| 12 | C2/m | $\{2_{010}\|0\}$ | | |
| | | $\{m_{010}\|0\}$ | | |
| 13 | P2/c | $\{2_{010}\|0\ 0\ 1/2\}$ | | |
| | | $\{m_{010}\|0\ 0\ 1/2\}$ | | |
| 14 | $P2_1/c$ | $\{2_{010}\|0\ 1/2\ 1/2\}$ | | |
| | | $\{m_{010}\|0\ 1/2\ 1/2\}$ | | |
| 15 | C2/c | $\{2_{010}\|0\ 0\ 1/2\}$ | | |
| | | $\{m_{010}\|0\ 0\ 1/2\}$ | | |

**Table B8:** Table of symbol-group element correspondence in SGs of orthorhombic crystal systems, with viewing direction of [100], [010] and [001] in International notation.

| No. | HM | $g_1$ | $g_2$ | $g_3$ |
|---|---|---|---|---|
| 16 | P222 | $\{2_{100}\|0\}$ | $\{2_{010}\|0\}$ | $\{2_{001}\|0\}$ |
| 17 | $P222_1$ | $\{2_{100}\|0\}$ | $\{2_{010}\|0\ 0\ 1/2\}$ | $\{2_{001}\|0\ 0\ 1/2\}$ |
| 18 | $P2_12_12$ | $\{2_{100}\|1/2\ 1/2\ 0\}$ | $\{2_{010}\|1/2\ 1/2\ 0\}$ | $\{2_{001}\|0\}$ |
| 19 | $P2_12_12_1$ | $\{2_{100}\|1/2\ 1/2\ 0\}$ | $\{2_{010}\|0\ 1/2\ 1/2\}$ | $\{2_{001}\|1/2\ 0\ 1/2\}$ |
| 20 | $C222_1$ | $\{2_{100}\|0\}$ | $\{2_{010}\|0\ 0\ 1/2\}$ | $\{2_{001}\|0\ 0\ 1/2\}$ |
| 21 | C222 | $\{2_{100}\|0\}$ | $\{2_{010}\|0\}$ | $\{2_{001}\|0\}$ |
| 22 | F222 | $\{2_{100}\|0\}$ | $\{2_{010}\|0\}$ | $\{2_{001}\|0\}$ |
| 23 | I222 | $\{2_{100}\|0\}$ | $\{2_{010}\|0\}$ | $\{2_{001}\|0\}$ |
| 24 | $I2_12_12_1$ | $\{2_{100}\|1/2\ 1/2\ 0\}$ | $\{2_{010}\|0\ 1/2\ 1/2\}$ | $\{2_{001}\|1/2\ 0\ 1/2\}$ |
| 25 | Pmm2 | $\{m_{100}\|0\}$ | $\{m_{010}\|0\}$ | $\{2_{001}\|0\}$ |
| 26 | $Pmc2_1$ | $\{m_{100}\|0\}$ | $\{m_{010}\|0\ 0\ 1/2\}$ | $\{2_{001}\|0\ 0\ 1/2\}$ |
| 27 | Pcc2 | $\{m_{100}\|0\ 0\ 1/2\}$ | $\{m_{010}\|0\ 0\ 1/2\}$ | $\{2_{001}\|0\}$ |
| 28 | Pma2 | $\{m_{100}\|1/2\ 0\ 0\}$ | $\{m_{010}\|1/2\ 0\ 0\}$ | $\{2_{001}\|0\}$ |
| 29 | $Pca2_1$ | $\{m_{100}\|1/2\ 0\ 1/2\}$ | $\{m_{010}\|1/2\ 0\ 0\}$ | $\{2_{001}\|0\ 0\ 1/2\}$ |
| 30 | Pnc2 | $\{m_{100}\|0\ 1/2\ 1/2\}$ | $\{m_{010}\|0\ 1/2\ 1/2\}$ | $\{2_{001}\|0\}$ |
| 31 | $Pmn2_1$ | $\{m_{100}\|0\}$ | $\{m_{010}\|1/2\ 0\ 1/2\}$ | $\{2_{001}\|1/2\ 0\ 1/2\}$ |
| 32 | Pba2 | $\{m_{100}\|1/2\ 1/2\ 0\}$ | $\{m_{010}\|1/2\ 1/2\ 0\}$ | $\{2_{001}\|0\}$ |
| 33 | $Pna2_1$ | $\{m_{100}\|1/2\ 1/2\ 1/2\}$ | $\{m_{010}\|1/2\ 1/2\ 0\}$ | $\{2_{001}\|0\ 0\ 1/2\}$ |
| 34 | Pnn2 | $\{m_{100}\|1/2\ 1/2\ 1/2\}$ | $\{m_{010}\|1/2\ 1/2\ 1/2\}$ | $\{2_{001}\|0\}$ |
| 35 | Cmm2 | $\{m_{100}\|0\}$ | $\{m_{010}\|0\}$ | $\{2_{001}\|0\}$ |
| 36 | $Cmc2_1$ | $\{m_{100}\|0\}$ | $\{m_{010}\|0\ 0\ 1/2\}$ | $\{2_{001}\|0\ 0\ 1/2\}$ |
| 37 | Ccc2 | $\{m_{100}\|0\ 0\ 1/2\}$ | $\{m_{010}\|0\ 0\ 1/2\}$ | $\{2_{001}\|0\}$ |
| 38 | Amm2 | $\{m_{100}\|0\}$ | $\{m_{010}\|0\}$ | $\{2_{001}\|0\}$ |
| 39 | Aem2 | $\{m_{100}\|0\ 1/2\ 0\}$ | $\{m_{010}\|0\ 1/2\ 0\}$ | $\{2_{001}\|0\}$ |
| 40 | Ama2 | $\{m_{100}\|1/2\ 0\ 0\}$ | $\{m_{010}\|1/2\ 0\ 0\}$ | $\{2_{001}\|0\}$ |
| 41 | Aea2 | $\{m_{100}\|1/2\ 1/2\ 0\}$ | $\{m_{010}\|1/2\ 1/2\ 0\}$ | $\{2_{001}\|0\}$ |
| 42 | Fmm2 | $\{m_{100}\|0\}$ | $\{m_{010}\|0\}$ | $\{2_{001}\|0\}$ |
| 43 | Fdd2 | $\{m_{100}\|1/4\ 1/4\ 1/4\}$ | $\{m_{010}\|1/4\ 1/4\ 1/4\}$ | $\{2_{001}\|0\}$ |
| 44 | Imm2 | $\{m_{100}\|0\}$ | $\{m_{010}\|0\}$ | $\{2_{001}\|0\}$ |
| 45 | Iba2 | $\{m_{100}\|1/2\ 1/2\ 0\}$ | $\{m_{010}\|1/2\ 1/2\ 0\}$ | $\{2_{001}\|0\}$ |
| 46 | Ima2 | $\{m_{100}\|1/2\ 0\ 0\}$ | $\{m_{010}\|1/2\ 0\ 0\}$ | $\{2_{001}\|0\}$ |
| 47 | Pmmm | $\{m_{100}\|0\}$ | $\{m_{010}\|0\}$ | $\{m_{001}\|0\}$ |
| 48 | Pnnn | $\{m_{100}\|0\ 1/2\ 1/2\}$ | $\{m_{010}\|1/2\ 0\ 1/2\}$ | $\{m_{001}\|1/2\ 1/2\ 0\}$ |
| 49 | Pccm | $\{m_{100}\|0\ 0\ 1/2\}$ | $\{m_{010}\|0\ 0\ 1/2\}$ | $\{m_{001}\|0\}$ |
| 50 | Pban | $\{m_{100}\|0\ 1/2\ 0\}$ | $\{m_{010}\|1/2\ 0\ 0\}$ | $\{m_{001}\|1/2\ 1/2\ 0\}$ |
| 51 | Pmma | $\{m_{100}\|1/2\ 0\ 0\}$ | $\{m_{010}\|0\}$ | $\{m_{001}\|1/2\ 0\ 0\}$ |
| 52 | Pnna | $\{m_{100}\|0\ 1/2\ 1/2\}$ | $\{m_{010}\|1/2\ 1/2\ 1/2\}$ | $\{m_{001}\|1/2\ 0\ 0\}$ |
| 53 | Pmna | $\{m_{100}\|0\}$ | $\{m_{010}\|1/2\ 0\ 1/2\}$ | $\{m_{001}\|1/2\ 0\ 1/2\}$ |
| 54 | Pcca | $\{m_{100}\|1/2\ 0\ 1/2\}$ | $\{m_{010}\|0\ 0\ 1/2\}$ | $\{m_{001}\|1/2\ 0\ 0\}$ |
| 55 | Pbam | $\{m_{100}\|1/2\ 1/2\ 0\}$ | $\{m_{010}\|1/2\ 1/2\ 0\}$ | $\{m_{001}\|0\}$ |
| 56 | Pccn | $\{m_{100}\|1/2\ 0\ 1/2\}$ | $\{m_{010}\|0\ 1/2\ 1/2\}$ | $\{m_{001}\|1/2\ 1/2\ 0\}$ |
| 57 | Pbcm | $\{m_{100}\|0\ 1/2\ 0\}$ | $\{m_{010}\|0\ 1/2\ 1/2\}$ | $\{m_{001}\|0\ 0\ 1/2\}$ |
| 58 | Pnnm | $\{m_{100}\|1/2\ 1/2\ 1/2\}$ | $\{m_{010}\|1/2\ 1/2\ 1/2\}$ | $\{m_{001}\|0\}$ |
| 59 | Pmmn | $\{m_{100}\|1/2\ 0\ 0\}$ | $\{m_{010}\|0\ 1/2\ 0\}$ | $\{m_{001}\|1/2\ 1/2\ 0\}$ |
| 60 | Pbcn | $\{m_{100}\|1/2\ 1/2\ 0\}$ | $\{m_{010}\|0\ 0\ 1/2\}$ | $\{m_{001}\|1/2\ 1/2\ 1/2\}$ |
| 61 | Pbca | $\{m_{100}\|1/2\ 1/2\ 0\}$ | $\{m_{010}\|0\ 1/2\ 1/2\}$ | $\{m_{001}\|1/2\ 0\ 1/2\}$ |
| 62 | Pnma | $\{m_{100}\|1/2\ 1/2\ 1/2\}$ | $\{m_{010}\|0\ 1/2\ 0\}$ | $\{m_{001}\|1/2\ 0\ 1/2\}$ |
| 63 | Cmcm | $\{m_{100}\|0\}$ | $\{m_{010}\|0\ 0\ 1/2\}$ | $\{m_{001}\|0\ 0\ 1/2\}$ |
| 64 | Cmce | $\{m_{100}\|0\}$ | $\{m_{010}\|0\ 1/2\ 1/2\}$ | $\{m_{001}\|0\ 1/2\ 1/2\}$ |
| 65 | Cmmm | $\{m_{100}\|0\}$ | $\{m_{010}\|0\}$ | $\{m_{001}\|0\}$ |
| 66 | Cccm | $\{m_{100}\|0\ 0\ 1/2\}$ | $\{m_{010}\|0\ 0\ 1/2\}$ | $\{m_{001}\|0\}$ |
| 67 | Cmme | $\{m_{100}\|0\}$ | $\{m_{010}\|0\ 1/2\ 0\}$ | $\{m_{001}\|0\ 1/2\ 0\}$ |
| 68 | Ccce | $\{m_{100}\|1/2\ 0\ 1/2\}$ | $\{m_{010}\|0\ 0\ 1/2\}$ | $\{m_{001}\|1/2\ 0\ 0\}$ |
| 69 | Fmmm | $\{m_{100}\|0\}$ | $\{m_{010}\|0\}$ | $\{m_{001}\|0\}$ |
| 70 | Fddd | $\{m_{100}\|0\ 1/4\ 1/4\}$ | $\{m_{010}\|1/4\ 0\ 1/4\}$ | $\{m_{001}\|1/4\ 1/4\ 0\}$ |
| 71 | Immm | $\{m_{100}\|0\}$ | $\{m_{010}\|0\}$ | $\{m_{001}\|0\}$ |
| 72 | Ibam | $\{m_{100}\|1/2\ 1/2\ 0\}$ | $\{m_{010}\|1/2\ 1/2\ 0\}$ | $\{m_{001}\|0\}$ |
| 73 | Ibca | $\{m_{100}\|1/2\ 1/2\ 0\}$ | $\{m_{010}\|0\ 1/2\ 1/2\}$ | $\{m_{001}\|1/2\ 0\ 1/2\}$ |
| 74 | Imma | $\{m_{100}\|0\}$ | $\{m_{010}\|0\ 1/2\ 0\}$ | $\{m_{001}\|0\ 1/2\ 0\}$ |

**Table B9:** Table of symbol-group element correspondence in SGs of tetragonal crystal systems, with viewing direction of [001], [100] and [110] in International notation.

| No. | HM | $g_1$ | $g_2$ | $g_3$ |
|---|---|---|---|---|
| 75 | P4 | $\{4^1_{001}\|0\}$ | | |
| 76 | $P4_1$ | $\{4^1_{001}\|0\ 0\ 1/4\}$ | | |
| 77 | $P4_2$ | $\{4^1_{001}\|0\ 0\ 1/2\}$ | | |
| 78 | $P4_3$ | $\{4^1_{001}\|0\ 0\ 3/4\}$ | | |
| 79 | I4 | $\{4^1_{001}\|0\}$ | | |
| 80 | $I4_1$ | $\{4^1_{001}\|0\ 1/2\ 1/4\}$ | | |
| 81 | P-4 | $\{-4^1_{001}\|0\}$ | | |
| 82 | I-4 | $\{-4^1_{001}\|0\}$ | | |
| 83 | P4/m | $\{4_{001}^{1}\|0\}$ | | |
| | | $\{m_{001}\|0\}$ | | |
| 84 | $P4_2/m$ | $\{4_{001}^{1}\|0\ 0\ 1/2\}$ | | |
| | | $\{m_{001}\|0\}$ | | |
| 85 | P4/n | $\{4_{001}^{1}\|1/2\ 0\ 0\}$ | | |
| | | $\{m_{001}\|1/2\ 1/2\ 0\}$ | | |
| 86 | $P4_2/n$ | $\{4_{001}^{1}\|0\ 1/2\ 1/2\}$ | | |
| | | $\{m_{001}\|1/2\ 1/2\ 0\}$ | | |
| 87 | I4/m | $\{4_{001}^{1}\|0\}$ | | |
| | | $\{m_{001}\|0\}$ | | |
| 88 | $I4_1/a$ | $\{4_{001}^{1}\|3/4\ 1/4\ 1/4\}$ | | |
| | | $\{m_{001}\|1/2\ 0\ 1/2\}$ | | |
| 89 | P422 | $\{4_{001}^{1}\|0\}$ | $\{2_{100}\|0\}$ | $\{2_{110}\|0\}$ |
| 90 | $P42_12$ | $\{4_{001}^{1}\|1/2\ 1/2\ 0\}$ | $\{2_{100}\|1/2\ 1/2\ 0\}$ | $\{2_{110}\|0\}$ |
| 91 | $P4_122$ | $\{4_{001}^{1}\|0\ 0\ 1/4\}$ | $\{2_{100}\|0\ 0\ 1/2\}$ | $\{2_{110}\|0\ 0\ 3/4\}$ |
| 92 | $P4_12_12$ | $\{4_{001}^{1}\|1/2\ 1/2\ 1/4\}$ | $\{2_{100}\|1/2\ 1/2\ 3/4\}$ | $\{2_{110}\|0\}$ |
| 93 | $P4_222$ | $\{4_{001}^{1}\|0\ 0\ 1/2\}$ | $\{2_{100}\|0\}$ | $\{2_{110}\|0\ 0\ 1/2\}$ |
| 94 | $P4_22_12$ | $\{4_{001}^{1}\|1/2\ 1/2\ 1/2\}$ | $\{2_{100}\|1/2\ 1/2\ 1/2\}$ | $\{2_{110}\|0\}$ |
| 95 | $P4_322$ | $\{4_{001}^{1}\|0\ 0\ 3/4\}$ | $\{2_{100}\|0\ 0\ 1/2\}$ | $\{2_{110}\|0\ 0\ 1/4\}$ |
| 96 | $P4_32_12$ | $\{4_{001}^{1}\|1/2\ 1/2\ 3/4\}$ | $\{2_{100}\|1/2\ 1/2\ 1/4\}$ | $\{2_{110}\|0\}$ |
| 97 | I422 | $\{4_{001}^{1}\|0\}$ | $\{2_{100}\|0\}$ | $\{2_{110}\|0\}$ |
| 98 | $I4_122$ | $\{4_{001}^{1}\|0\ 1/2\ 1/4\}$ | $\{2_{100}\|0\ 1/2\ 1/4\}$ | $\{2_{110}\|1/2\ 1/2\ 1/2\}$ |
| 99 | P4mm | $\{4_{001}^{1}\|0\}$ | $\{m_{100}\|0\}$ | $\{m_{110}\|0\}$ |
| 100 | P4bm | $\{4_{001}^{1}\|0\}$ | $\{m_{100}\|1/2\ 1/2\ 0\}$ | $\{m_{110}\|1/2\ 1/2\ 0\}$ |
| 101 | $P4_2cm$ | $\{4_{001}^{1}\|0\ 0\ 1/2\}$ | $\{m_{100}\|0\ 0\ 1/2\}$ | $\{m_{110}\|0\}$ |
| 102 | $P4_2nm$ | $\{4_{001}^{1}\|1/2\ 1/2\ 1/2\}$ | $\{m_{100}\|1/2\ 1/2\ 1/2\}$ | $\{m_{110}\|0\}$ |
| 103 | P4cc | $\{4_{001}^{1}\|0\}$ | $\{m_{100}\|0\ 0\ 1/2\}$ | $\{m_{110}\|0\ 0\ 1/2\}$ |
| 104 | P4nc | $\{4_{001}^{1}\|0\}$ | $\{m_{100}\|1/2\ 1/2\ 1/2\}$ | $\{m_{110}\|1/2\ 1/2\ 1/2\}$ |
| 105 | $P4_2mc$ | $\{4_{001}^{1}\|0\ 0\ 1/2\}$ | $\{m_{100}\|0\}$ | $\{m_{110}\|0\ 0\ 1/2\}$ |
| 106 | $P4_2bc$ | $\{4_{001}^{1}\|0\ 0\ 1/2\}$ | $\{m_{100}\|1/2\ 1/2\ 0\}$ | $\{m_{110}\|1/2\ 1/2\ 1/2\}$ |
| 107 | I4mm | $\{4_{001}^{1}\|0\}$ | $\{m_{100}\|0\}$ | $\{m_{110}\|0\}$ |
| 108 | I4cm | $\{4_{001}^{1}\|0\}$ | $\{m_{100}\|0\ 0\ 1/2\}$ | $\{m_{110}\|0\ 0\ 1/2\}$ |
| 109 | $I4_1md$ | $\{4_{001}^{1}\|0\ 1/2\ 1/4\}$ | $\{m_{100}\|1/2\ 1/2\ 1/2\}$ | $\{m_{110}\|0\ 1/2\ 1/4\}$ |
| 110 | $I4_1cd$ | $\{4_{001}^{1}\|0\ 1/2\ 1/4\}$ | $\{m_{100}\|1/2\ 1/2\ 0\}$ | $\{m_{110}\|0\ 1/2\ 3/4\}$ |
| 111 | P-42m | $\{-4_{001}^{1}\|0\}$ | $\{2_{100}\|0\}$ | $\{m_{110}\|0\}$ |
| 112 | P-42c | $\{-4_{001}^{1}\|0\}$ | $\{2_{100}\|0\ 0\ 1/2\}$ | $\{m_{110}\|0\ 0\ 1/2\}$ |
| 113 | $P\text{-}42_1m$ | $\{-4_{001}^{1}\|0\}$ | $\{2_{100}\|1/2\ 1/2\ 0\}$ | $\{m_{110}\|1/2\ 1/2\ 0\}$ |
| 114 | $P\text{-}42_1c$ | $\{-4_{001}^{1}\|0\}$ | $\{2_{100}\|1/2\ 1/2\ 1/2\}$ | $\{m_{110}\|1/2\ 1/2\ 1/2\}$ |
| 115 | P-4m2 | $\{-4_{001}^{1}\|0\}$ | $\{m_{100}\|0\}$ | $\{2_{110}\|0\}$ |
| 116 | P-4c2 | $\{-4_{001}^{1}\|0\}$ | $\{m_{100}\|0\ 0\ 1/2\}$ | $\{2_{110}\|0\ 0\ 1/2\}$ |
| 117 | P-4b2 | $\{-4_{001}^{1}\|0\}$ | $\{m_{100}\|1/2\ 1/2\ 0\}$ | $\{2_{110}\|1/2\ 1/2\ 0\}$ |
| 118 | P-4n2 | $\{-4_{001}^{1}\|0\}$ | $\{m_{100}\|1/2\ 1/2\ 1/2\}$ | $\{2_{110}\|1/2\ 1/2\ 1/2\}$ |
| 119 | I-4m2 | $\{-4_{001}^{1}\|0\}$ | $\{m_{100}\|0\}$ | $\{2_{110}\|0\}$ |
| 120 | I-4c2 | $\{-4_{001}^{1}\|0\}$ | $\{m_{100}\|0\ 0\ 1/2\}$ | $\{2_{110}\|0\ 0\ 1/2\}$ |
| 121 | I-42m | $\{-4_{001}^{1}\|0\}$ | $\{2_{100}\|0\}$ | $\{m_{110}\|0\}$ |
| 122 | I-42d | $\{-4_{001}^{1}\|0\}$ | $\{2_{100}\|1/2\ 0\ 3/4\}$ | $\{m_{110}\|1/2\ 0\ 3/4\}$ |
| 123 | P4/mmm | $\{4_{001}^{1}\|0\}$<br>$\{m_{001}\|0\}$ | $\{m_{100}\|0\}$ | $\{m_{110}\|0\}$ |
| 124 | P4/mcc | $\{4_{001}^{1}\|0\}$<br>$\{m_{001}\|0\}$ | $\{m_{100}\|0\ 0\ 1/2\}$ | $\{m_{110}\|0\ 0\ 1/2\}$ |
| 125 | P4/nbm | $\{4_{001}^{1}\|1/2\ 0\ 0\}$<br>$\{m_{001}\|1/2\ 1/2\ 0\}$ | $\{m_{100}\|0\ 1/2\ 0\}$ | $\{m_{110}\|0\}$ |
| 126 | P4/nnc | $\{4_{001}^{1}\|1/2\ 0\ 0\}$<br>$\{m_{001}\|1/2\ 1/2\ 0\}$ | $\{m_{100}\|0\ 1/2\ 1/2\}$ | $\{m_{110}\|0\ 0\ 1/2\}$ |
| 127 | P4/mbm | $\{4_{001}^{1}\|0\}$<br>$\{m_{001}\|0\}$ | $\{m_{100}\|1/2\ 1/2\ 0\}$ | $\{m_{110}\|1/2\ 1/2\ 0\}$ |
| 128 | P4/mnc | $\{4_{001}^{1}\|0\}$<br>$\{m_{001}\|0\}$ | $\{m_{100}\|1/2\ 1/2\ 1/2\}$ | $\{m_{110}\|1/2\ 1/2\ 1/2\}$ |
| 129 | P4/nmm | $\{4_{001}^{1}\|1/2\ 0\ 0\}$<br>$\{m_{001}\|1/2\ 1/2\ 0\}$ | $\{m_{100}\|1/2\ 0\ 0\}$ | $\{m_{110}\|1/2\ 1/2\ 0\}$ |
| 130 | P4/ncc | $\{4_{001}^{1}\|1/2\ 0\ 0\}$<br>$\{m_{001}\|1/2\ 1/2\ 0\}$ | $\{m_{100}\|1/2\ 0\ 1/2\}$ | $\{m_{110}\|1/2\ 1/2\ 1/2\}$ |
| 131 | $P4_2/mmc$ | $\{4_{001}^{1}\|0\ 0\ 1/2\}$ | $\{m_{100}\|0\}$ | $\{m_{110}\|0\ 0\ 1/2\}$ |
| | | $\{m_{001}\|0\}$ | | |
| 132 | $P4_2/mcm$ | $\{4^{1}_{001}\|0\ 0\ 1/2\}$<br>$\{m_{001}\|0\}$ | $\{m_{100}\|0\ 0\ 1/2\}$ | $\{m_{110}\|0\}$ |
| 133 | $P4_2/nbc$ | $\{4^{1}_{001}\|1/2\ 0\ 1/2\}$<br>$\{m_{001}\|1/2\ 1/2\ 0\}$ | $\{m_{100}\|0\ 1/2\ 0\}$ | $\{m_{110}\|0\ 0\ 1/2\}$ |
| 134 | $P4_2/nnm$ | $\{4^{1}_{001}\|1/2\ 0\ 1/2\}$<br>$\{m_{001}\|1/2\ 1/2\ 0\}$ | $\{m_{100}\|0\ 1/2\ 1/2\}$ | $\{m_{110}\|0\}$ |
| 135 | $P4_2/mbc$ | $\{4^{1}_{001}\|0\ 0\ 1/2\}$<br>$\{m_{001}\|0\}$ | $\{m_{100}\|1/2\ 1/2\ 0\}$ | $\{m_{110}\|1/2\ 1/2\ 1/2\}$ |
| 136 | $P4_2/mnm$ | $\{4^{1}_{001}\|1/2\ 1/2\ 1/2\}$<br>$\{m_{001}\|0\}$ | $\{m_{100}\|1/2\ 1/2\ 1/2\}$ | $\{m_{110}\|0\}$ |
| 137 | $P4_2/nmc$ | $\{4^{1}_{001}\|1/2\ 0\ 1/2\}$<br>$\{m_{001}\|1/2\ 1/2\ 0\}$ | $\{m_{100}\|1/2\ 0\ 0\}$ | $\{m_{110}\|1/2\ 1/2\ 1/2\}$ |
| 138 | $P4_2/ncm$ | $\{4^{1}_{001}\|1/2\ 0\ 1/2\}$<br>$\{m_{001}\|1/2\ 1/2\ 0\}$ | $\{m_{100}\|1/2\ 0\ 1/2\}$ | $\{m_{110}\|1/2\ 1/2\ 0\}$ |
| 139 | $I4/mmm$ | $\{4^{1}_{001}\|0\}$<br>$\{m_{001}\|0\}$ | $\{m_{100}\|0\}$ | $\{m_{110}\|0\}$ |
| 140 | $I4/mcm$ | $\{4^{1}_{001}\|0\}$<br>$\{m_{001}\|0\}$ | $\{m_{100}\|0\ 0\ 1/2\}$ | $\{m_{110}\|0\ 0\ 1/2\}$ |
| 141 | $I4_1/amd$ | $\{4^{1}_{001}\|1/4\ 3/4\ 1/4\}$<br>$\{m_{001}\|1/2\ 0\ 1/2\}$ | $\{m_{100}\|0\}$ | $\{m_{110}\|3/4\ 1/4\ 3/4\}$ |
| 142 | $I4_1/acd$ | $\{4^{1}_{001}\|1/4\ 3/4\ 1/4\}$<br>$\{m_{001}\|1/2\ 0\ 1/2\}$ | $\{m_{100}\|0\ 0\ 1/2\}$ | $\{m_{110}\|3/4\ 1/4\ 1/4\}$ |

**Table B10:** Table of symbol-group element correspondence in SGs of trigonal crystal systems, with viewing direction of [001], [100] and [210] in International notation.

| No. | HM | $g_1$ | $g_2$ | $g_3$ |
|---|---|---|---|---|
| 143 | $P3$ | $\{3^{1}_{001}\|0\}$ | | |
| 144 | $P3_1$ | $\{3^{1}_{001}\|0\ 0\ 1/3\}$ | | |
| 145 | $P3_2$ | $\{3^{1}_{001}\|0\ 0\ 2/3\}$ | | |
| 146 | $R3$ | $\{3^{1}_{001}\|0\}$ | | |
| 147 | P-3 | $\{-3^{1}_{001}\|0\}$ | | |
| 148 | R-3 | $\{-3^{1}_{001}\|0\}$ | | |
| 149 | P312 | $\{3^{1}_{001}\|0\}$ | $\{1\|0\}$ | $\{2_{210}\|0\}$ |
| 150 | P321 | $\{3^{1}_{001}\|0\}$ | $\{2_{100}\|0\}$ | $\{1\|0\}$ |
| 151 | P$3_1$12 | $\{3^{1}_{001}\|0\ 0\ 1/3\}$ | $\{1\|0\}$ | $\{2_{210}\|0\}$ |
| 152 | P$3_1$21 | $\{3^{1}_{001}\|0\ 0\ 1/3\}$ | $\{2_{100}\|0\ 0\ 2/3\}$ | $\{1\|0\}$ |
| 153 | P$3_2$12 | $\{3^{1}_{001}\|0\ 0\ 2/3\}$ | $\{1\|0\}$ | $\{2_{210}\|0\}$ |
| 154 | P$3_2$21 | $\{3^{1}_{001}\|0\ 0\ 2/3\}$ | $\{2_{100}\|0\ 0\ 1/3\}$ | $\{1\|0\}$ |
| 155 | R32 | $\{3^{1}_{001}\|0\}$ | $\{2_{100}\|0\}$ | |
| 156 | P3m1 | $\{3^{1}_{001}\|0\}$ | $\{m_{100}\|0\}$ | $\{1\|0\}$ |
| 157 | P31m | $\{3^{1}_{001}\|0\}$ | $\{1\|0\}$ | $\{m_{210}\|0\}$ |
| 158 | P3c1 | $\{3^{1}_{001}\|0\}$ | $\{m_{100}\|0\ 0\ 1/2\}$ | $\{1\|0\}$ |
| 159 | P31c | $\{3^{1}_{001}\|0\}$ | $\{1\|0\}$ | $\{m_{210}\|0\ 0\ 1/2\}$ |
| 160 | R3m | $\{3^{1}_{001}\|0\}$ | $\{m_{100}\|0\}$ | |
| 161 | R3c | $\{3^{1}_{001}\|0\}$ | $\{m_{100}\|0\ 0\ 1/2\}$ | |
| 162 | P-31m | $\{-3^{1}_{001}\|0\}$ | $\{1\|0\}$ | $\{m_{210}\|0\}$ |
| 163 | P-31c | $\{-3^{1}_{001}\|0\}$ | $\{1\|0\}$ | $\{m_{210}\|0\ 0\ 1/2\}$ |
| 164 | P-3m1 | $\{-3^{1}_{001}\|0\}$ | $\{m_{100}\|0\}$ | $\{1\|0\}$ |
| 165 | P-3c1 | $\{-3^{1}_{001}\|0\}$ | $\{m_{100}\|0\ 0\ 1/2\}$ | $\{1\|0\}$ |
| 166 | R-3m | $\{-3^{1}_{001}\|0\}$ | $\{m_{100}\|0\}$ | |
| 167 | R-3c | $\{-3^{1}_{001}\|0\}$ | $\{m_{100}\|0\ 0\ 1/2\}$ | |

**Table B11:** Table of symbol-group element correspondence in SGs of hexagonal crystal systems, with viewing direction of [001], [100] and [210] in International notation.

| No. | HM | $g_1$ | $g_2$ | $g_3$ |
|---|---|---|---|---|
| 168 | P6 | $\{6^{1}_{001}\|0\}$ | | |
| 169 | P$6_1$ | $\{6^{1}_{001}\|0\ 0\ 1/6\}$ | | |
| 170 | P$6_5$ | $\{6^{1}_{001}\|0\ 0\ 5/6\}$ | | |
| 171 | P$6_2$ | $\{6^{1}_{001}\|0\ 0\ 1/3\}$ | | |
| 172 | P$6_4$ | $\{6^{1}_{001}\|0\ 0\ 2/3\}$ | | |
| 173 | P$6_3$ | $\{6^{1}_{001}\|0\ 0\ 1/2\}$ | | |
| 174 | P-6 | $\{-6^1_{001}\|0\}$ | | |
| 175 | P6/m | $\{6^1_{001}\|0\}$ | $\{m_{001}\|0\}$ | |
| 176 | P6<sub></sub>3/m | $\{6^1_{001}\|0\ 0\ 1/2\}$ | $\{m_{001}\|0\ 0\ 1/2\}$ | |
| 177 | P622 | $\{6^1_{001}\|0\}$ | $\{2_{100}\|0\}$ | $\{2_{210}\|0\}$ |
| 178 | $P6_122$ | $\{6^1_{001}\|0\ 0\ 1/6\}$ | $\{2_{100}\|0\}$ | $\{2_{210}\|0\ 0\ 1/6\}$ |
| 179 | $P6_522$ | $\{6^1_{001}\|0\ 0\ 5/6\}$ | $\{2_{100}\|0\}$ | $\{2_{210}\|0\ 0\ 5/6\}$ |
| 180 | $P6_222$ | $\{6^1_{001}\|0\ 0\ 1/3\}$ | $\{2_{100}\|0\}$ | $\{2_{210}\|0\ 0\ 1/3\}$ |
| 181 | $P6_422$ | $\{6^1_{001}\|0\ 0\ 2/3\}$ | $\{2_{100}\|0\}$ | $\{2_{210}\|0\ 0\ 2/3\}$ |
| 182 | $P6_322$ | $\{6^1_{001}\|0\ 0\ 1/2\}$ | $\{2_{100}\|0\}$ | $\{2_{210}\|0\ 0\ 1/2\}$ |
| 183 | P6mm | $\{6^1_{001}\|0\}$ | $\{m_{100}\|0\}$ | $\{m_{210}\|0\}$ |
| 184 | P6cc | $\{6^1_{001}\|0\}$ | $\{m_{100}\|0\ 0\ 1/2\}$ | $\{m_{210}\|0\ 0\ 1/2\}$ |
| 185 | $P6_3cm$ | $\{6^1_{001}\|0\ 0\ 1/2\}$ | $\{m_{100}\|0\ 0\ 1/2\}$ | $\{m_{210}\|0\}$ |
| 186 | $P6_3mc$ | $\{6^1_{001}\|0\ 0\ 1/2\}$ | $\{m_{100}\|0\}$ | $\{m_{210}\|0\ 0\ 1/2\}$ |
| 187 | P-6m2 | $\{-6^1_{001}\|0\}$ | $\{m_{100}\|0\}$ | $\{2_{210}\|0\}$ |
| 188 | P-6c2 | $\{-6^1_{001}\|0\ 0\ 1/2\}$ | $\{m_{100}\|0\ 0\ 1/2\}$ | $\{2_{210}\|0\}$ |
| 189 | P-62m | $\{-6^1_{001}\|0\}$ | $\{2_{100}\|0\}$ | $\{m_{210}\|0\}$ |
| 190 | P-62c | $\{-6^1_{001}\|0\ 0\ 1/2\}$ | $\{2_{100}\|0\}$ | $\{m_{210}\|0\ 0\ 1/2\}$ |
| 191 | P6/mmm | $\{6^1_{001}\|0\}$<br>$\{m_{001}\|0\}$ | $\{m_{100}\|0\}$ | $\{m_{210}\|0\}$ |
| 192 | P6/mcc | $\{6^1_{001}\|0\}$<br>$\{m_{001}\|0\}$ | $\{m_{100}\|0\ 0\ 1/2\}$ | $\{m_{210}\|0\ 0\ 1/2\}$ |
| 193 | $P6_3/mcm$ | $\{6^1_{001}\|0\ 0\ 1/2\}$<br>$\{m_{001}\|0\ 0\ 1/2\}$ | $\{m_{100}\|0\ 0\ 1/2\}$ | $\{m_{210}\|0\}$ |
| 194 | $P6_3/mmc$ | $\{6^1_{001}\|0\ 0\ 1/2\}$<br>{m_001\|0 0 1/2} | $\{m_{100}\|0\}$ | $\{m_{210}\|0\ 0\ 1/2\}$ |

**Table B12:** Table of symbol-group element correspondence in SGs of hexagonal crystal systems, with viewing direction of [100], [111] and [110] in International notation.

| No. | HM | $g_1$ | $g_2$ | $g_3$ |
|---|---|---|---|---|
| 195 | P23 | $\{2_{100}\|0\}$ | $\{3^1_{111}\|0\}$ | |
| 196 | F23 | $\{2_{100}\|0\}$ | $\{3^1_{111}\|0\}$ | |
| 197 | I23 | $\{2_{100}\|0\}$ | $\{3^1_{111}\|0\}$ | |
| 198 | P213 | $\{2_{100}\|1/2\ 1/2\ 0\}$ | $\{3^1_{111}\|0\}$ | |
| 199 | I213 | $\{2_{100}\|1/2\ 1/2\ 0\}$ | $\{3^1_{111}\|0\}$ | |
| 200 | Pm-3 | $\{m_{100}\|0\}$ | $\{-3^1_{111}\|0\}$ | |
| 201 | Pn-3 | $\{m_{100}\|0\ 1/2\ 1/2\}$ | $\{-3^1_{111}\|0\}$ | |
| 202 | Fm-3 | $\{m_{100}\|0\}$ | $\{-3^1_{111}\|0\}$ | |
| 203 | Fd-3 | $\{m_{100}\|0\ 1/4\ 1/4\}$ | $\{-3^1_{111}\|0\}$ | |
| 204 | Im-3 | $\{m_{100}\|0\}$ | $\{-3^1_{111}\|0\}$ | |
| 205 | Pa-3 | $\{m_{100}\|1/2\ 1/2\ 0\}$ | $\{-3^1_{111}\|0\}$ | |
| 206 | Ia-3 | $\{m_{100}\|1/2\ 1/2\ 0\}$ | $\{-3^1_{111}\|0\}$ | |
| 207 | P432 | $\{4^1_{100}\|0\}$ | $\{3^1_{111}\|0\}$ | $\{2_{110}\|0\}$ |
| 208 | P4232 | $\{4^1_{100}\|1/2\ 1/2\ 1/2\}$ | $\{3^1_{111}\|0\}$ | $\{2_{110}\|1/2\ 1/2\ 1/2\}$ |
| 209 | F432 | $\{4^1_{100}\|0\}$ | $\{3^1_{111}\|0\}$ | $\{2_{110}\|0\}$ |
| 210 | F4132 | $\{4^1_{100}\|1/4\ 3/4\ 3/4\}$ | $\{3^1_{111}\|0\}$ | $\{2_{110}\|3/4\ 1/4\ 3/4\}$ |
| 211 | I432 | $\{4^1_{100}\|0\}$ | $\{3^1_{111}\|0\}$ | $\{2_{110}\|0\}$ |
| 212 | P4332 | $\{4^1_{100}\|3/4\ 3/4\ 1/4\}$ | $\{3^1_{111}\|0\}$ | $\{2_{110}\|1/4\ 3/4\ 3/4\}$ |
| 213 | P4132 | $\{4^1_{100}\|1/4\ 1/4\ 3/4\}$ | $\{3^1_{111}\|0\}$ | $\{2_{110}\|3/4\ 1/4\ 1/4\}$ |
| 214 | I4132 | $\{4^1_{100}\|1/4\ 1/4\ 3/4\}$ | $\{3^1_{111}\|0\}$ | $\{2_{110}\|3/4\ 1/4\ 1/4\}$ |
| 215 | P-43m | $\{-4^1_{100}\|0\}$ | $\{3^1_{111}\|0\}$ | $\{m_{110}\|0\}$ |
| 216 | F-43m | $\{-4^1_{100}\|0\}$ | $\{3^1_{111}\|0\}$ | $\{m_{110}\|0\}$ |
| 217 | I-43m | $\{-4^1_{100}\|0\}$ | $\{3^1_{111}\|0\}$ | $\{m_{110}\|0\}$ |
| 218 | P-43n | $\{-4^1_{100}\|1/2\ 1/2\ 1/2\}$ | $\{3^1_{111}\|0\}$ | $\{m_{110}\|1/2\ 1/2\ 1/2\}$ |
| 219 | F-43c | $\{-4^1_{100}\|1/2\ 1/2\ 1/2\}$ | $\{3^1_{111}\|0\}$ | $\{m_{110}\|1/2\ 1/2\ 1/2\}$ |
| 220 | I-43d | $\{-4^1_{100}\|3/4\ 3/4\ 1/4\}$ | $\{3^1_{111}\|0\}$ | $\{m_{110}\|1/4\ 3/4\ 3/4\}$ |
| 221 | Pm-3m | $\{m_{100}\|0\}$ | $\{-3^1_{111}\|0\}$ | $\{m_{110}\|0\}$ |
| 222 | Pn-3n | $\{m_{100}\|0\ 1/2\ 1/2\}$ | $\{-3^1_{111}\|0\}$ | $\{m_{110}\|0\ 0\ 1/2\}$ |
| 223 | Pm-3n | $\{m_{100}\|0\}$ | $\{-3^1_{111}\|0\}$ | $\{m_{110}\|1/2\ 1/2\ 1/2\}$ |
| 224 | Pn-3m | $\{m_{100}\|0\ 1/2\ 1/2\}$ | $\{-3^1_{111}\|0\}$ | $\{m_{110}\|1/2\ 1/2\ 0\}$ |
| 225 | Fm-3m | $\{m_{100}\|0\}$ | $\{-3^1_{111}\|0\}$ | $\{m_{110}\|0\}$ |
| 226 | Fm-3c | $\{m_{100}\|0\}$ | $\{-3^1_{111}\|0\}$ | $\{m_{110}\|1/2\ 1/2\ 1/2\}$ |
| 227 | Fd-3m | $\{m_{100}\|1/2\ 1/4\ 3/4\}$ | $\{-3^1_{111}\|0\}$ | $\{m_{110}\|1/4\ 3/4\ 1/2\}$ |
| 228 | Fd-3c | $\{m_{100}\|1/2\ 3/4\ 1/4\}$ | $\{-3^1_{111}\|0\}$ | $\{m_{110}\|1/4\ 3/4\ 0\}$ |
| 229 | Im-3m | $\{m_{100}\|0\}$ | $\{-3^1_{111}\|0\}$ | $\{m_{110}\|0\}$ |
| 230 | Ia-3d | $\{m_{100}\|1/2\ 1/2\ 0\}$ | $\{-3^{1}_{111}\|0\}$ | $\{m_{110}\|1/4\ 3/4\ 3/4\}$ |

## C. The nomenclature for t-type, k-type and g-type SSGs

Below we give the guidance for constructing nontrivial SSG $G_{NS}$ using the nomenclature of *t*-type, *k*-type and *g*-type SSG.

### 1. t-type SSG

*t*-type SSG $B^{g_{s_1}}g_1{}^{g_{s_2}}g_2{}^{g_{s_3}}g_3$ can be formed from $\{g_{s_1}\|g_1\}$, $\{g_{s_2}\|g_2\}$, $\{g_{s_3}\|g_3\}$ .

### I. 65.136.1.1

Following Table B9, $P^{-1}4_2/^{1}m^{-1}n^{1}m$ (65.136.1.1) can be generated using $\{-1\|4^1_{001}|1/2\ 1/2\ 1/2\}$, $\{1\|m_{001}|0\}$, $\{-1\|m_{100}|1/2\ 1/2\ 1/2\}$ and $\{1\|m_{110}|0\}$. All elements are listed below. Here a, b, c and x, y, z coordination correspond to space coordination and vector of magnetic moment, respectively, $\tau$ = (1/2 1/2 1/2).

**Table C1:** General positions of $P^{-1}4_2/^{1}m^{-1}n^{1}m$ (65.136.1.1).

| Seitz | Coordination | Seitz | Coordination |
|---|---|---|---|
| $\{1\|1|0\}$ | a, b, c,<br>x, y, z | $\{-1\|4^1_{001}|\tau\}$ | -b+1/2, a+1/2, c+1/2,<br>-x, -y, -z |
| $\{1\|2_{001}|0\}$ | -a, -b, c,<br>x, y, z | $\{-1\|4^3_{001}|\tau\}$ | b+1/2, -a+1/2, c+1/2,<br>-x, -y, -z |
| $\{1\|2_{110}|0\}$ | b, a, -c,<br>x, y, z | $\{-1\|2_{100}|\tau\}$ | a+1/2, -b+1/2, -c+1/2,<br>-x, -y, -z |
| $\{1\|2_{1-10}|0\}$ | -b, -a, -c,<br>x, y, z | $\{-1\|2_{010}|\tau\}$ | -a+1/2, b+1/2, -c+1/2,<br>-x, -y, -z |
| $\{1\|-1|0\}$ | -a, -b, -c,<br>x, y, z | $\{-1\|-4^1_{001}|\tau\}$ | b+1/2, -a+1/2, -c+1/2,<br>-x, -y, -z |
| $\{1\|m_{001}|0\}$ | a, b, -c,<br>x, y, z | $\{-1\|-4^3_{001}|\tau\}$ | -b+1/2, a+1/2, -c+1/2,<br>-x, -y, -z |
| $\{1\|m_{110}|0\}$ | -b, -a, c,<br>x, y, z | $\{-1\|m_{100}|\tau\}$ | -a+1/2, b+1/2, c+1/2,<br>-x, -y, -z |
| $\{1\|m_{1-10}|0\}$ | b, a, c, | $\{-1\|m_{010}|\tau\}$ | a+1/2, -b+1/2, c+1/2, |

| | x, y, z | | -x, -y, -z |
|---|---|---|---|

As shown in Table C1, we can conclude that SSG with $G^s = -1$ only supports collinear magnetic structures, supposing that no inequivalent magnetic ion in the magnetic cell. Take the collinear antiferromagnet $RuO_2$ as an example, whose nontrivial SSG is $P^{-1}4_2/^{1}m^{-1}n^{1}m$ (65.136.1.1). The spin only part is $G_{SO}^{l} = Z_K^2 \ltimes SO(2)$, which is named as ${}^{\infty m}1$ in International notation. As a result, for collinear antiferromagnet $RuO_2$, its complete SSG is $P^{-1}4_2/^{1}m^{-1}n^{1}m^{\infty m}1$.

## II. 47.221.1.2

Now we turn to a more complex one, $P^{1}m^{3_{001}^{1}}\text{-}3^{m_{010}}m$ (47.221.1.2). $\{m_{100}|0\}$, $\{-3_{111}^{1}|0\}$, $\{m_{110}|0\}$ are obtained following Table B12. Now we perform the coset decomposition using $G_0 = Pm-3m$ and $L_0 = Pmmm$.

$$G_0 = \bigcup_{\alpha} g_\alpha L_0 \tag{C1}$$

$$L_0 = \{\{1|0\},\{2_{100}|0\},\{2_{010}|0\},\{2_{001}|0\},\{-1|0\},\{m_{100}|0\},\{m_{010}|0\},\{m_{001}|0\}\} \tag{C2}$$

$$G_0/L_0 = \{g_\alpha\} = \{\{1|0\},\{3_{111}^{1}|0\},\{3_{111}^{2}|0\},\{2_{110}|0\},\{2_{101}|0\},\{2_{011}|0\}\} \tag{C3}$$

According to representative symmetry operations of 47.221.1.2, $\{1||m_{100}|0\}$, $\{3_{001}^{1}||-3_{111}^{1}|0\}$, $\{m_{010}||m_{110}|0\}$, we can obtain the mapping between $G^s = 3m$ and $G_0/L_0$:

$$\{g_S||g_\alpha\} = \left\{\begin{matrix} \{1||1|0\},\{3_{001}^{1}||3_{111}^{1}|0\},\{3_{001}^{2}||3_{111}^{2}|0\}, \\ \{m_{010}||2_{110}|0\},\{m_{\pi/6}||2_{101}|0\},\{m_{5\pi/6}||2_{011}|0\} \end{matrix}\right\}. \tag{C4}$$

Then we make the direct product of $\{E||L_0\}$ and $\{g_S||g_\alpha\}$, all the general positions are then obtained and listed in Table C2.

**Table C2:** General positions of $P^{1}m^{3_{001}^{1}}\text{-}3^{m_{010}}m$ (47.221.1.2).

| Seitz | Coordination |
|---|---|
| $\{1\|\|1\|0\}$ | a, b, c, x, y, z |
| $\{1\|\|2_{100}\|0\}$ | a, -b, -c, x, y, z |
| $\{1\|\|2_{010}\|0\}$ | -a, b, -c, x, y, z |
| $\{1\|\|2_{001}\|0\}$ | -a, -b, c, x, y, z |
| $\{1\|\|-1\|0\}$ | -a, -b, -c, x, y, z |
| $\{1\|\|m_{100}\|0\}$ | -a, b, c, x, y, z |
| $\{1\|\|m_{010}\|0\}$ | a, -b, -c, x, y, z |
| $\{1\|\|m_{001}\|0\}$ | a, b, -c, x, y, z |
| $\{3_{001}^{1}\|\|3_{111}^{1}\|0\}$ | c, a, b, -(x+√3y)/2, (√3x-y)/2, z |
| $\{3_{001}^{1}\|\|3_{1-11}^{1}\|0\}$ | c, -a, -b, -(x+√3y)/2, (√3x-y)/2, z |
| $\{3_{001}^{1}\|\|3_{1-1-1}^{1}\|0\}$ | -c, -a, b, -(x+√3y)/2, (√3x-y)/2, z |
| $\{3_{001}^{1}\|\|3_{11-1}^{1}\|0\}$ | -c, a, -b, -(x+√3y)/2, (√3x-y)/2, z |
| $\{3_{001}^{1}\|\|-3_{111}^{1}\|0\}$ | -c, -a, -b, -(x+√3y)/2, (√3x-y)/2, z |
| $\{3_{001}^{1}\|\|-3_{1-11}^{1}\|0\}$ | -c, a, b, -(x+√3y)/2, (√3x-y)/2, z |
| $\{3_{001}^{1}\|\|-3_{1-1-1}^{1}\|0\}$ | c, a, -b, -(x+√3y)/2, (√3x-y)/2, z |
| $\{3_{001}^{1}\|\|-3_{11-1}^{1}\|0\}$ | c, -a, b, -(x+√3y)/2, (√3x-y)/2, z |
| $\{3_{001}^{2}\|\|3_{111}^{2}\|0\}$ | b, c, a, (-x+√3y)/2, -(√3x+y)/2, z |
| $\{3_{001}^{2}\|\|3_{1-11}^{2}\|0\}$ | -b, -c, a, (-x+√3y)/2, -(√3x+y)/2, z |
| $\{3_{001}^{2}\|\|3_{1-1-1}^{2}\|0\}$ | -b, c, -a, (-x+√3y)/2, -(√3x+y)/2, z |
| $\{3_{001}^{2}\|\|3_{11-1}^{2}\|0\}$ | b, -c, -a, (-x+√3y)/2, -(√3x+y)/2, z |
| $\{3_{001}^{2}\|\|-3_{111}^{2}\|0\}$ | -b, -c, -a, (-x+√3y)/2, -(√3x+y)/2, z |
| $\{3_{001}^{2}\|\|-3_{1-11}^{2}\|0\}$ | b, c, -a, (-x+√3y)/2, -(√3x+y)/2, z |
| $\{3_{001}^{2}\|\|-3_{1-1-1}^{2}\|0\}$ | b, -c, a, (-x+√3y)/2, -(√3x+y)/2, z |
| $\{3_{001}^{2}\|\|-3_{11-1}^{2}\|0\}$ | -b, c, a, (-x+√3y)/2, -(√3x+y)/2, z |
| $\{m_{010}\|\|2_{110}\|0\}$ | b, a, -c, x, -y, z |
| $\{m_{010}\|\|2_{1-10}\|0\}$ | -b, -a, -c, x, -y, z |
| $\{m_{010}\|\|4_{001}^{1}\|0\}$ | -b, a, c, x, -y, z |
| $\{m_{010}\|\|4_{001}^{3}\|0\}$ | b, -a, c, x, -y, z |
| $\{m_{010}\|\|m_{110}\|0\}$ | -b, -a, c, x, -y, z |
| $\{m_{010}\|\|m_{1-10}\|0\}$ | b, a, c, x, -y, z |
| $\{m_{010}\|\|-4_{001}^{1}\|0\}$ | b, -a, -c, x, -y, z |
| $\{m_{010}\|\|-4_{001}^{3}\|0\}$ | -b, a, -c, x, -y, z |
| $\{m_{\pi/6}\|2_{011}\|0\}$ | -a, c, b, -(x+√3y)/2, (y-√3x)/2, z |
| $\{m_{\pi/6}\|2_{01-1}\|0\}$ | -a, -c, -b, -(x+√3y)/2, (y-√3x)/2, z |
| $\{m_{\pi/6}\|4^1_{100}\|0\}$ | a, -c, b, -(x+√3y)/2, (y-√3x)/2, z |
| $\{m_{\pi/6}\|4^3_{100}\|0\}$ | a, c, -b, -(x+√3y)/2, (y-√3x)/2, z |
| $\{m_{\pi/6}\|m_{011}\|0\}$ | a, -c, -b, -(x+√3y)/2, (y-√3x)/2, z |
| $\{m_{\pi/6}\|m_{01-1}\|0\}$ | a, c, b, -(x+√3y)/2, (y-√3x)/2, z |
| $\{m_{\pi/6}\|-4^1_{100}\|0\}$ | -a, c, -b, -(x+√3y)/2, (y-√3x)/2, z |
| $\{m_{\pi/6}\|-4^3_{100}\|0\}$ | -a, -c, b, -(x+√3y)/2, (y-√3x)/2, z |
| $\{m_{5\pi/6}\|2_{101}\|0\}$ | c, -b, a, (-x+√3y)/2, (√3x+y)/2, z |
| $\{m_{5\pi/6}\|2_{10-1}\|0\}$ | -c, -b, -a, (-x+√3y)/2, (√3x+y)/2, z |
| $\{m_{5\pi/6}\|4^1_{010}\|0\}$ | c, b, -a, (-x+√3y)/2, (√3x+y)/2, z |
| $\{m_{5\pi/6}\|4^3_{010}\|0\}$ | -c, b, a, (-x+√3y)/2, (√3x+y)/2, z |
| $\{m_{5\pi/6}\|m_{101}\|0\}$ | -c, b, -a, (-x+√3y)/2, (√3x+y)/2, z |
| $\{m_{5\pi/6}\|m_{10-1}\|0\}$ | c, b, a, (-x+√3y)/2, (√3x+y)/2, z |
| $\{m_{5\pi/6}\|-4^1_{010}\|0\}$ | -c, -b, a, (-x+√3y)/2, (√3x+y)/2, z |
| $\{m_{5\pi/6}\|-4^3_{010}\|0\}$ | c, -b, -a, (-x+√3y)/2, (√3x+y)/2, z |

**III.13.54.1.1~13.54.1.6**

All general positions of 13.54.1.1 to 13.54.1.6 are also listed below in Table C3 to C8. The International notations for 13.54.1.1 to 13.54.1.6 are $P^{-1}c^{-1}c^{1}a$, $P^{2_{001}}c^{2_{001}}c^{1}a$, $P^{m_{001}}c^{m_{001}}c^{1}a$, $P^{-1}c^{1}c^{-1}a$, $P^{2_{001}}c^{1}c^{2_{001}}a$ and $P^{m_{001}}c^{1}c^{m_{001}}a$, respectively.

**Table C3:** General positions of $P^{-1}c^{-1}c^{1}a$ (13.54.1.1).

| Seitz | Coordination | Seitz | Coordination |
|---|---|---|---|
| $\{1\|1\|0\}$ | a, b, c,<br>x, y, z | $\{-1\|2_{100}\|1/2\ 0\ 1/2\}$ | a+1/2, -b, -c+1/2,<br>-x, -y, -z |
| $\{1\|-1\|0\}$ | -a, -b, -c,<br>x, y, z | $\{-1\|m_{100}\|1/2\ 0\ 1/2\}$ | -a+1/2, b, c+1/2,<br>-x, -y, -z |
| $\{1\|\|2_{001}\|1/2\ 0\ 0\}$ | -a+1/2, -b, c,<br>x, y, z | $\{-1\|\|2_{010}\|0\ 0\ 1/2\}$ | -a, b, -c+1/2,<br>-x, -y, -z |
| $\{1\|\|m_{001}\|1/2\ 0\ 0\}$ | a+1/2, b, -c,<br>x, y, z | $\{-1\|\|m_{010}\|0\ 0\ 1/2\}$ | a, -b, c+1/2, -x,<br>-y, -z |

**Table C4:** General positions of $P^{2_{001}}c^{2_{001}}c^{1}a$ (13.54.1.2).

| Seitz | Coordination | Seitz | Coordination |
|---|---|---|---|
| $\{1\|\|1\|0\}$ | a, b, c,<br>x, y, z | $\{2_{001}\|\|2_{100}\|1/2\ 0\ 1/2\}$ | a+1/2, -b, -c+1/2,<br>-x, -y, z |
| $\{1\|\|-1\|0\}$ | -a, -b, -c,<br>x, y, z | $\{2_{001}\|\|m_{100}\|1/2\ 0\ 1/2\}$ | -a+1/2, b, c+1/2,<br>-x, -y, z |
| $\{1\|\|2_{001}\|1/2\ 0\ 0\}$ | -a+1/2, -b, c,<br>x, y, z | $\{2_{001}\|\|2_{010}\|0\ 0\ 1/2\}$ | -a, b, -c+1/2,<br>-x, -y, z |
| $\{1\|\|m_{001}\|1/2\ 0\ 0\}$ | a+1/2, b, -c,<br>x, y, z | $\{2_{001}\|\|m_{010}\|0\ 0\ 1/2\}$ | a, -b, c+1/2,<br>-x, -y, z |

**Table C5:** General positions of $P^{m_{001}}c^{m_{001}}c^{1}a$ (13.54.1.3).

| Seitz | Coordination | Seitz | Coordination |
|---|---|---|---|
| $\{1\|\|1\|0\}$ | a, b, c,<br>x, y, z | $\{m_{001}\|\|2_{100}\|1/2\ 0\ 1/2\}$ | a+1/2, -b, -c+1/2,<br>x, y, -z |
| $\{1\|\|-1\|0\}$ | -a, -b, -c,<br>x, y, z | $\{m_{001}\|\|m_{100}\|1/2\ 0\ 1/2\}$ | -a+1/2, b, c+1/2,<br>x, y, -z |
| $\{1\|\|2_{001}\|1/2\ 0\ 0\}$ | -a+1/2, -b, c,<br>x, y, z | $\{m_{001}\|\|2_{010}\|0\ 0\ 1/2\}$ | -a, b, -c+1/2,<br>x, y, -z |
| $\{1\|\|m_{001}\|1/2\ 0\ 0\}$ | a+1/2, b, -c,<br>x, y, z | $\{m_{001}\|\|m_{010}\|0\ 0\ 1/2\}$ | a, -b, c+1/2,<br>x, y, -z |

**Table C6:** General positions of $P^{-1}c^{1}c^{-1}a$ (13.54.1.4).

| Seitz | Coordination | Seitz | Coordination |
|---|---|---|---|
| $\{1\|\|1\|0\}$ | a, b, c,<br>x, y, z | $\{-1\|\|2_{100}\|1/2\ 0\ 1/2\}$ | a+1/2, -b, -c+1/2,<br>-x, -y, -z |
| $\{1\|\|-1\|0\}$ | -a, -b, -c, | $\{-1\|\|m_{100}\|1/2\ 0\ 1/2\}$ | -a+1/2, b, c+1/2, |
| | x, y, z | | -x, -y, -z |
| $\{1\|\|2_{010}\|0\ 0\ 1/2\}$ | -a, b, -c+1/2,<br>x, y, z | $\{-1\|\|2_{001}\|1/2\ 0\ 0\}$ | -a+1/2, -b, c,<br>-x, -y, -z |
| $\{1\|\|m_{010}\|0\ 0\ 1/2\}$ | a, -b, c+1/2,<br>x, y, z | $\{-1\|\|m_{001}\|1/2\ 0\ 0\}$ | a+1/2, b, -c,<br>-x, -y, -z |

**Table C7:** General positions of $P^{2_{001}}c^{1}c^{2_{001}}a$ (13.54.1.5).

| Seitz | Coordination | Seitz | Coordination |
|---|---|---|---|
| $\{1\|\|1\|0\}$ | a, b, c,<br>x, y, z | $\{2_{001}\|\|2_{100}\|1/2\ 0\ 1/2\}$ | a+1/2, -b, -c+1/2,<br>-x, -y, z |
| $\{1\|\|-1\|0\}$ | -a, -b, -c,<br>x, y, z | $\{2_{001}\|\|m_{100}\|1/2\ 0\ 1/2\}$ | -a+1/2, b, c+1/2,<br>-x, -y, z |
| $\{1\|\|2_{010}\|0\ 0\ 1/2\}$ | -a, b, -c+1/2,<br>x, y, z | $\{2_{001}\|\|2_{001}\|1/2\ 0\ 0\}$ | -a+1/2, -b, c,<br>-x, -y, z |
| $\{1\|\|m_{010}\|0\ 0\ 1/2\}$ | a, -b, c+1/2,<br>x, y, z | $\{2_{001}\|\|m_{001}\|1/2\ 0\ 0\}$ | a+1/2, b, -c,<br>-x, -y, z |

**Table C8:** General positions of $P^{m_{001}}c^{1}c^{m_{001}}a$ (13.54.1.6).

| Seitz | Coordination | Seitz | Coordination |
|---|---|---|---|
| $\{1\|\|1\|0\}$ | a, b, c,<br>x, y, z | $\{m_{001}\|\|2_{100}\|1/2\ 0\ 1/2\}$ | a+1/2, -b, -c+1/2,<br>x, y, -z |
| $\{1\|\|-1\|0\}$ | -a, -b, -c,<br>x, y, z | $\{m_{001}\|\|m_{100}\|1/2\ 0\ 1/2\}$ | -a+1/2, b, c+1/2,<br>x, y, -z |
| $\{1\|\|2_{010}\|0\ 0\ 1/2\}$ | -a, b, -c+1/2,<br>x, y, z | $\{m_{001}\|\|2_{001}\|1/2\ 0\ 0\}$ | -a+1/2, -b, c,<br>x, y, -z |
| $\{1\|\|m_{010}\|0\ 0\ 1/2\}$ | a, -b, c+1/2,<br>x, y, z | $\{m_{001}\|\|m_{001}\|1/2\ 0\ 0\}$ | a+1/2, b, -c,<br>x, y, -z |

## 2. k-type SSG

In the BNS setting, MSGs are constructed from space group $L_0$. This can be achieved by adding a "primed sublattice" generated by an operation that combines time-inversion with a fractional translation to describe the “reversal” of spin between different nonmagnetic unit cells. The type-IV group is constructed by $M_{IV} = L_0 + T\tau L_0$, where $\tau$ is the mentioned translation. The $k$-type SSG can be seen as the multi-color externsion of MSG. In this case, due to the decoupling of spin space and lattice space, a point group in spin space is used to describe the propagation vector in the enlarged magnetic unit cell. As a result, we use the nomenclature of $B^1g_1{}^1g_2{}^1g_3{}^{g_{s_1}}\tau_1{}^{g_{s_2}}\tau_2{}^{g_{s_3}}\tau_3$ for $k$-type SSG under $L_0$ basis. In this vein, the $k$-type SSG is directly constructed from the direct product of sublattice SG $\{E||Bg_1g_2g_3\}$ and additional spin translation group $G_T^S$, where $G_T^S$ can be generated by $\{g_{s_1}||\tau_1\}$, $\{g_{s_2}||\tau_2\}$ and $\{g_{s_3}||\tau_3\}$. Note that the integer translation of $L_0$ is always $t_a = (100), t_b = (010), t_c = (001)$.

## I. 99.107.4.1

$P^14^1m^1m^{4^1_{001}}(1/2\ 1/2\ 1/4)$ (99.107.4.1) can be constructed using $\{E||P4mm\}$ and $G_T^S = \{\{1||1|0\}, \{4^1_{001}||1|1/2\ 1/2\ 1/4\}, \{2_{001}||1|0\ 0\ 1/2\}, \{4^3_{001}||1|1/2\ 1/2\ 3/4\}\}$.

**Table C9:** General positions of $P^14^1m^1m^{4^1_{001}}(1/2\ 1/2\ 1/4)$ (99.107.4.1).

| Seitz | Coordination |
|---|---|
| $\{1\|\|1\|0\}$ | a, b, c, x, y, z |
| $\{1\|\|2_{001}\|0\}$ | -a, -b, c, x, y, z |
| $\{1\|\|4^1_{001}\|0\}$ | -b, a, c, x, y, z |
| $\{1\|\|4^3_{001}\|0\}$ | b, -a, c, x, y, z |
| $\{1\|\|m_{010}\|0\}$ | a, -b, c, x, y, z |
| $\{1\|\|m_{100}\|0\}$ | -a, b, c, x, y, z |
| $\{1\|\|m_{110}\|0\}$ | -b, -a, c, x, y, z |
| $\{1\|\|m_{1-10}\|0\}$ | b, a, c, x, y, z |
| $\{4^1_{001}\|\|1\|1/2\ 1/2\ 1/4\}$ | a+1/2, b+1/2, c+1/4, -y, x, z |
| $\{4^1_{001}\|\|2_{001}\|1/2\ 1/2\ 1/4\}$ | -a+1/2, -b+1/2, c+1/4, -y, x, z |
| $\{4^1_{001}\|\|4^1_{001}\|1/2\ 1/2\ 1/4\}$ | -b+1/2, a+1/2, c+1/4, -y, x, z |
| $\{4_{001}^{1}\|\|4_{001}^{3}\|1/2\ 1/2\ 1/4\}$ | b+1/2, -a+1/2, c+1/4, -y, x, z |
| $\{4_{001}^{1}\|\|m_{010}\|1/2\ 1/2\ 1/4\}$ | a+1/2, -b+1/2, c+1/2, -y, x, z |
| $\{4_{001}^{1}\|\|m_{100}\|1/2\ 1/2\ 1/4\}$ | -a+1/2, b+1/2, c+1/4, -y, x, z |
| $\{4_{001}^{1}\|\|m_{110}\|1/2\ 1/2\ 1/4\}$ | -b+1/2, -a+1/2, c+1/4, -y, x, z |
| $\{4_{001}^{1}\|\|m_{1-10}\|1/2\ 1/2\ 1/4\}$ | b+1/2, a+1/2, c+1/4, -y, x, z |
| $\{2_{001}\|\|1\|0\ 0\ 1/2\}$ | a, b, c+1/2, -x, -y, z |
| $\{2_{001}\|\|2_{001}\|0\ 0\ 1/2\}$ | -a, -b, c+1/2, -x, -y, z |
| $\{2_{001}\|\|4_{001}^{1}\|0\ 0\ 1/2\}$ | -b, a, c+1/2, -x, -y, z |
| $\{2_{001}\|\|4_{001}^{3}\|0\ 0\ 1/2\}$ | b, -a, c+1/2, -x, -y, z |
| $\{2_{001}\|\|m_{010}\|0\ 0\ 1/2\}$ | a, -b, c+1/2, -x, -y, z |
| $\{2_{001}\|\|m_{100}\|0\ 0\ 1/2\}$ | -a, b, c+1/2, -x, -y, z |
| $\{2_{001}\|\|m_{110}\|0\ 0\ 1/2\}$ | -b, -a, c+1/2, -x, -y, z |
| $\{2_{001}\|\|m_{1-10}\|0\ 0\ 1/2\}$ | b, a, c+1/2, -x, -y, z |
| $\{4_{001}^{3}\|\|1\|1/2\ 1/2\ 1/4\}$ | a+1/2, b+1/2, c+3/4, y, -x, z |
| $\{4_{001}^{3}\|\|2_{001}\|1/2\ 1/2\ 1/4\}$ | -a+1/2, -b+1/2, c+3/4, y, -x, z |
| $\{4_{001}^{3}\|\|4_{001}^{1}\|1/2\ 1/2\ 1/4\}$ | -b+1/2, a+1/2, c+3/4, y, -x, z |
| $\{4_{001}^{3}\|\|4_{001}^{3}\|1/2\ 1/2\ 1/4\}$ | b+1/2, -a+1/2, c+3/4, y, -x, z |
| $\{4_{001}^{3}\|\|m_{010}\|1/2\ 1/2\ 1/4\}$ | a+1/2, -b+1/2, c+3/4, y, -x, z |
| $\{4_{001}^{3}\|\|m_{100}\|1/2\ 1/2\ 1/4\}$ | -a+1/2, b+1/2, c+3/4, y, -x, z |
| $\{4_{001}^{3}\|\|m_{110}\|1/2\ 1/2\ 1/4\}$ | -b+1/2, -a+1/2, c+3/4, y, -x, z |
| $\{4_{001}^{3}\|\|m_{1-10}\|1/2\ 1/2\ 1/4\}$ | b+1/2, a+1/2, c+3/4, y, -x, z |

Similarly, $P^{1}4^{1}m^{1}m^{-4_{001}^{3}}(1/2\ 1/2\ 1/4)$(99.107.4.2) can be constructed by $G_T^S = \{\{1\|1\|0\},\ \{-4_{001}^{3}\|1\|1/2\ 1/2\ 1/4\}, \{2_{001}\|1\|0\ 0\ 1/2\}, \{-4_{001}^{1}\|1\|1/2\ 1/2\ 3/4\}\}$ and $\{E\|P4mm\}$.

## II. 174.174.3.1

Now we show why we use the $L_0$ basis, rather than the $G_0$ basis for *k*-type SSG. The $L_0$ basis depict the magnetic primitive cell, while the $G_0$ basis sometimes give a larger primitive cell. Now we take $P^{1}{-6}^{3_{001}^{1}}(2/3\ 1/3\ 0)$ (174.174.3.1) as an example.

Under $L_0$ basis, the sublattice SG is $L_0 = P-6$, the spin translation group is $G_T^S = \{\{1||1|0\}, \{3_{001}^1||1|2/3\ 1/3\ 0\}, \{3_{001}^2||1|1/3\ 2/3\ 0\}\}$. All general positions are listed in Table C10.

**Table C10:** General positions of $P^1{-}6^{3_{001}^1}(2/3\ 1/3\ 0)$ (174.174.3.1).

| Seitz | Coordination |
|---|---|
| $\{1\|\|1\|0\}$ | a, b, c, x, y, z |
| $\{1\|\|3_{001}^1\|0\}$ | -b, a-b, c, x, y, z |
| $\{1\|\|3_{001}^2\|0\}$ | -a+b, -a, c, x, y, z |
| $\{1\|\|m_{001}\|0\}$ | a, b, -c, x, y, z |
| $\{1\|\|-6_{001}^5\|0\}$ | -b, a-b, -c, x, y, z |
| $\{1\|\|-6_{001}^1\|0\}$ | -a+b, -a, -c, x, y, z |
| $\{3_{001}^1\|\|1\|2/3\ 1/3\ 0\}$ | a+2/3, b+1/3, c, -(x+√3y)/2, (√3x-y)/2, z |
| $\{3_{001}^1\|\|3_{001}^1\|2/3\ 1/3\ 0\}$ | -b+2/3, a-b+1/3, c, -(x+√3y)/2, (√3x-y)/2, z |
| $\{3_{001}^1\|\|3_{001}^2\|2/3\ 1/3\ 0\}$ | -a+b+2/3, -a+1/3, c, -(x+√3y)/2, (√3x-y)/2, z |
| $\{3_{001}^1\|\|m_{001}\|2/3\ 1/3\ 0\}$ | a+2/3, b+1/3, -c, -(x+√3y)/2, (√3x-y)/2, z |
| $\{3_{001}^1\|\|-6_{001}^5\|2/3\ 1/3\ 0\}$ | -b+2/3, a-b+1/3, -c, -(x+√3y)/2, (√3x-y)/2, z |
| $\{3_{001}^1\|\|-6_{001}^1\|2/3\ 1/3\ 0\}$ | -a+b+2/3, -a+1/3, -c, -(x+√3y)/2, (√3x-y)/2, z |
| $\{3_{001}^2\|\|1\|1/3\ 2/3\ 0\}$ | a+1/3, b+2/3, c, (-x+√3y)/2, -(√3x+y)/2, z |
| $\{3_{001}^2\|\|3_{001}^1\|1/3\ 2/3\ 0\}$ | -b+1/3, a-b+2/3, c, (-x+√3y)/2, -(√3x+y)/2, z |
| $\{3_{001}^2\|\|3_{001}^2\|1/3\ 2/3\ 0\}$ | -a+b+1/3, -a+2/3, c, (-x+√3y)/2, -(√3x+y)/2, z |
| $\{3_{001}^2\|\|m_{001}\|1/3\ 2/3\ 0\}$ | a+1/3, b+2/3, -c, (-x+√3y)/2, -(√3x+y)/2, z |
| $\{3_{001}^2\|\|-6_{001}^5\|1/3\ 2/3\ 0\}$ | -b+1/3, a-b+2/3, -c, (-x+√3y)/2, -(√3x+y)/2, z |
| $\{3_{001}^2\|\|-6_{001}^1\|1/3\ 2/3\ 0\}$ | -a+b+1/3, -a+2/3, -c, (-x+√3y)/2, -(√3x+y)/2, z |

Using the transformation matrix $M_T = \begin{pmatrix} 1 & 1 & 0 & 0 \\ -1 & 2 & 0 & 0 \\ 0 & 0 & 1 & 0 \end{pmatrix}$, which transform from $L_0$ basis to $G_0$ basis. Then three spin translation operation is transformed into $\{1||1|0\}$, $\{3_{001}^1||1|1\ 0\ 0\}$, $\{3_{001}^2||1|1\ 1\ 0\}$, and the full $G_T^S$ is

$$G_T^S = \left\{\begin{matrix} \{1\|1|0\}, \{1\|1|1\ 2\ 0\}, \{1\|1|2\ 1\ 0\}, \\ \{3_{001}^1\|1|1\ 0\ 0\}, \{3_{001}^2\|1|1\ 1\ 0\}, \{3_{001}^1\|1|0\ 1\ 0\}, \\ \{3_{001}^2\|1|2\ 0\ 0\}, \{3_{001}^2\|1|0\ 2\ 0\}, \{3_{001}^1\|1|2\ 2\ 0\} \end{matrix}\right\}. \quad \text{(C5)}$$

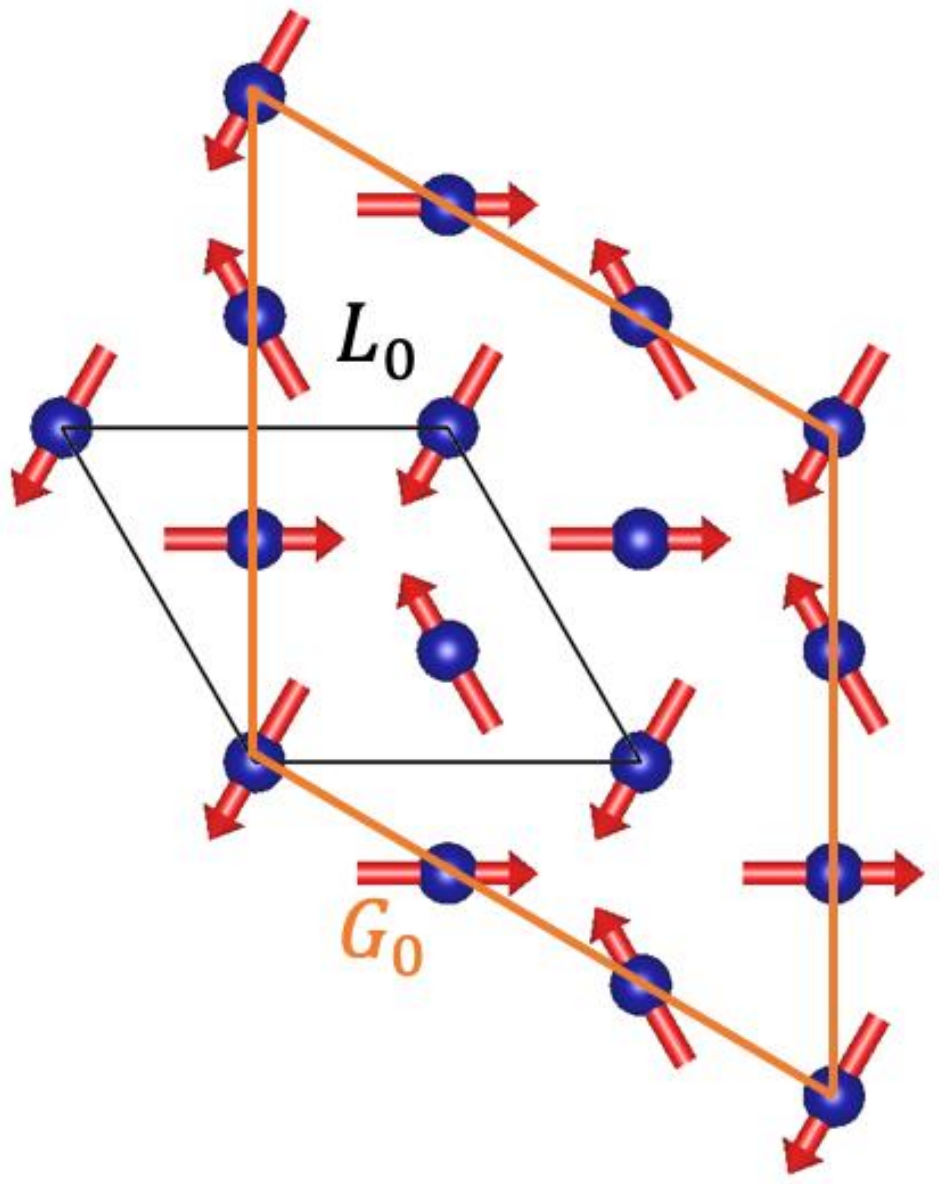

**Fig. C1:** Magnetic primitive cell of $P^1{-}6^{3_{001}^1}(2/3\ 1/3\ 0)$ (174.174.3.1) under $L_0$ basis (black line) and under $G_0$ basis (orange line).

It is clear that the magnetic cell under $G_0$ basis is three times larger than $L_0$ basis as shown in Fig. C1. In other words, the magnetic cell under $L_0$ basis is also the minimum periodic unit cell, while the magnetic cell under $G_0$ basis is sometimes larger than the previous, depending on the transform matrix.

### 3. g-type SSG

For $g$-type SSG with $B^{g_{s_1}}g_1{}^{g_{s_2}}g_2{}^{g_{s_3}}g_3|(g_{s_4}, g_{s_5}, g_{s_6}; g_{s_7}, g_{s_8}, g_{s_9})$ in $G_0$ basis, $g_{s_4}, g_{s_5}, g_{s_6}$ are combined with three integer translation $t_a, t_b, t_c$, while $g_{s_7}, g_{s_8}, g_{s_9}$ are combined with centering-type fractional translation $b_1, b_2, b_3$. The spin translation group can be constructed from $\{g_{s_4}\|t_a\}$, $\{g_{s_5}\|t_b\}$, $\{g_{s_6}\|t_c\}$, $\{g_{s_7}\|b_1\}$, $\{g_{s_8}\|b_2\}$ and $\{g_{s_7}\|b_3\}$. The sequence of $b_1, b_2, b_3$ for different Bravais lattices is given in Table B5.

When $B = P$ for primitive lattice, $g_{s_7}, g_{s_8}, g_{s_9}$ are omitted. Below we show several examples of $g$-type SSG.

## I. 4.182.4.2

The first is $P^{3^2_{-11-1}}6_3{}^{m_{110}}2^{m_{011}}2|(2_{001}, 2_{100}, 1)$ (4.182.4.2) of $CoNb_3S_6$. It can be generated using $\{3^2_{-11-1}||6^1_{001}|0\ 0\ 1/2\}$, $\{m_{110}||2_{100}|0\}$, $\{m_{011}||2_{210}|0\ 0\ 1/2\}$ and the spin translation group $G_T^S = \{\{1||1|0\}, \{2_{001}||1|1\ 0\ 0\}, \{2_{100}||1|0\ 1\ 0\}, \{2_{010}||1|1\ 1\ 0\}\}$.

**Table C11:** General positions of $P^{3^2_{-11-1}}6_3{}^{m_{110}}2^{m_{011}}2|(2_{001}, 2_{100}, 1)$ (4.182.4.2).

| Seitz | Coordination |
|---|---|
| $\{1\|\|1\|0\}$ | a, b, c, x, y, z |
| $\{3^1_{-11-1}\|\|3^1_{001}\|0\}$ | -b, a-b, c, z, -x, -y |
| $\{3^2_{-11-1}\|\|3^2_{001}\|0\}$ | -a+b, -a, c, -y, -z, x |
| $\{1\|\|2_{001}\|0\ 0\ 1/2\}$ | -a, -b, c+1/2, x, y, z |
| $\{3^1_{-11-1}\|\|6^5_{001}\|0\ 0\ 1/2\}$ | b, -a+b, c+1/2, z, -x, -y |
| $\{3^2_{-11-1}\|\|6^1_{001}\|0\ 0\ 1/2\}$ | a-b, a, c+1/2, -y, -z, x |
| $\{m_{-101}\|\|2_{110}\|0\}$ | b, a, -c, z, y, x |
| $\{m_{110}\|\|2_{100}\|0\}$ | a-b, -b, -c, -y, -x, z |
| $\{m_{011}\|\|2_{010}\|0\}$ | -a, -a+b, -c, x, -z, -y |
| $\{m_{-101}\|\|2_{1-10}\|0\ 0\ 1/2\}$ | -b, -a, -c+1/2, z, y, x |
| $\{m_{110}\|\|2_{120}\|0\ 0\ 1/2\}$ | -a+b, b, -c+1/2, -y, -x, z |
| $\{m_{011}\|\|2_{210}\|0\ 0\ 1/2\}$ | a, a-b, -c+1/2, x, -z, -y |
| $\{2_{001}\|\|1\|1\ 0\ 0\}$ | 1+a, b, c, -x, -y, z |
| $\{3^1_{-1-11}\|\|3^1_{001}\|1\ 0\ 0\}$ | 1-b, a-b, c, -z, x, -y |
| $\{3^2_{111}\|\|3^2_{001}\|1\ 0\ 0\}$ | 1-a+b, -a, c, y, z, x |
| $\{2_{001}\|\|2_{001}\|1\ 0\ 1/2\}$ | 1-a, -b, c+1/2, -x, -y, z |
| $\{3^1_{-1-11}\|\|6^5_{001}\|1\ 0\ 1/2\}$ | 1+b, -a+b, c+1/2, -z, x, -y |
| $\{3^2_{111}\|\|6^1_{001}\|1\ 0\ 1/2\}$ | 1+a-b, a, c+1/2, y, z, x |
| $\{-4^1_{010}\|\|2_{110}\|1\ 0\ 0\}$ | 1+b, a, -c, -z, -y, x |
| $\{m_{1-10}\|\|2_{100}\|1\ 0\ 0\}$ | 1+a-b, -b, -c, y, x, z |
| $\{-4^1_{100}\|\|2_{010}\|1\ 0\ 0\}$ | 1-a, -a+b, -c, -x, z, -y |
| $\{-4^1_{010}\|\|2_{1-10}\|1\ 0\ 1/2\}$ | 1-b, -a, -c+1/2, -z, -y, x |
| $\{m_{1-10}\|\|2_{120}\|1\ 0\ 1/2\}$ | 1-a+b, b, -c+1/2, y, x, z |
| $\{-4^1_{100}\|\|2_{210}\|1\ 0\ 1/2\}$ | 1+a, a-b, -c+1/2, -x, z, -y |
| $\{2_{100}\|\|1\|0\ 1\ 0\}$ | a, 1+b, c, x, -y, -z |
| $\{3^1_{111}\|\|3^1_{001}\|0\ 1\ 0\}$ | -b, 1+a-b, c, z, x, y |
| $\{3^2_{1-1-1}\|\|3^2_{001}\|0\ 1\ 0\}$ | -a+b, 1-a, c, -y, z, -x |
| $\{2_{100}\|\|2_{001}\|0\ 1\ 1/2\}$ | -a, 1-b, c+1/2, x, -y, -z |
| $\{3^1_{111}\|\|6^5_{001}\|0\ 1\ 1/2\}$ | b, 1-a+b, c+1/2, z, x, y |
| $\{3^2_{1-1-1}\|\|6^1_{001}\|0\ 1\ 1/2\}$ | a-b, 1+a, c+1/2, -y, z, -x |
| $\{-4^3_{010}\|\|2_{110}\|0\ 1\ 0\}$ | b, 1+a, -c, z, -y, -x |
| $\{-4^3_{001}\|\|2_{100}\|0\ 1\ 0\}$ | a-b, 1-b, -c, -y, x, -z |
| $\{m_{01-1}\|\|2_{010}\|0\ 1\ 0\}$ | -a, 1-a+b, -c, x, z, y |
| $\{-4^3_{010}\|\|2_{1-10}\|0\ 1\ 1/2\}$ | -b, 1-a, -c+1/2, z, -y, -x |
| $\{-4^3_{001}\|\|2_{120}\|0\ 1\ 1/2\}$ | -a+b, 1+b, -c+1/2, -y, x, -z |
| $\{m_{01-1}\|\|2_{210}\|0\ 1\ 1/2\}$ | a, 1+a-b, -c+1/2, x, z, y |
| $\{2_{010}\|\|1\|1\ 1\ 0\}$ | a, b, c, -x, y, -z |
| $\{3^1_{1-1-1}\|\|3^1_{001}\|1\ 1\ 0\}$ | -b, a-b, c, -z, -x, y |
| $\{3^2_{-1-11}\|\|3^2_{001}\|1\ 1\ 0\}$ | -a+b, -a, c, y, -z, -x |
| $\{2_{010}\|\|2_{001}\|1\ 1\ 1/2\}$ | -a, -b, c+1/2, -x, y, -z |
| $\{3^1_{1-1-1}\|\|6^5_{001}\|1\ 1\ 1/2\}$ | b, -a+b, c+1/2, -z, -x, y |
| $\{3^2_{-1-11}\|\|6^1_{001}\|1\ 1\ 1/2\}$ | a-b, a, c+1/2, y, -z, -x |
| $\{m_{101}\|\|2_{110}\|1\ 1\ 0\}$ | b, a, -c, -z, y, -x |
| $\{-4^1_{001}\|\|2_{100}\|1\ 1\ 0\}$ | a-b, -b, -c, y, -x, -z |
| $\{-4^3_{100}\|\|2_{010}\|1\ 1\ 0\}$ | -a, -a+b, -c, -x, -z, y |
| $\{m_{101}\|\|2_{1-10}\|1\ 1\ 1/2\}$ | -b, -a, -c+1/2, -z, y, -x |
| $\{-4^1_{001}\|\|2_{120}\|1\ 1\ 1/2\}$ | -a+b, b, -c+1/2, y, -x, -z |
| $\{-4^3_{100}\|\|2_{210}\|1\ 1\ 1/2\}$ | a, a-b, -c+1/2, -x, -z, y |

## II. 3.22.8.1

In $G_0$ basis, the Bravais-centering fractional translation $b_1 b_2 b_3$ can also combine with nontrivial spin operation. Below we show how to construct the full SSG from the nonmenclature of $F^{m_{010}}2^{m_{010}}2^{1}2|(1,1,2_{001}; m_{001}, 4^{1}_{001}, -4^{3}_{001})$ (3.22.8.1).

The spin translation group $G_T^S$ for 3.22.8.1 can be constructed from $\{2_{001}||1|0\ 0\ 1\}$, $\{m_{001}||1|1/2\ 1/2\ 0\}$, $\{4^{1}_{001}||1|1/2\ 0\ 1/2\}$, $\{-4^{3}_{001}||1|0\ 1/2\ 1/2\}$, and the full $G_T^S$ is

$$G_T^S = \left\{\begin{array}{c} \{1||1|0\}, \{-1||1|1/2\ 1/2\ 1\}, \\ \{-4^{3}_{001}||1|0\ 1/2\ 1/2\}, \{-4^{1}_{001}||1|0\ 1/2\ 3/2\}, \\ \{2_{001}||1|0\ 0\ 1\}, \{m_{001}||1|1/2\ 1/2\ 0\}, \\ \{4^{1}_{001}||1|1/2\ 0\ 1/2\}, \{4^{3}_{001}||1|1/2\ 0\ 3/2\} \end{array}\right\}. \tag{C6}$$

The integral translations along three basis vector of $G_0$ for 3.22.8.1 are $\{1||1|1\ 0\ 0\}$, $\{1||1|0\ 1\ 0\}$ and $\{1||1|0\ 0\ 2\}$, respectively. Then we add $\{m_{010}||2_{100}|0\}$, $\{m_{010}||2_{010}|0\}$, $\{1||2_{001}|0\}$, and all general positions of SSG is shown below:

**Table C12:** General positions of $F^{m_{010}}2^{m_{010}}2^{1}2|(1,1,2_{001}; m_{001}, 4^{1}_{001}, -4^{3}_{001})$ (3.22.8.1).

| Seitz | Coordination |
|---|---|
| $\{1\|\|1\|0\}$ | a, b, c, x, y, z |
| $\{2_{001}\|\|1\|0\ 0\ 1\}$ | a, b, c+1, -x, -y, z |
| $\{4^{1}_{001}\|\|1\|1/2\ 0\ 1/2\}$ | a+1/2, b, c+1/2, -y, x, z |
| $\{4^{3}_{001}\|\|1\|1/2\ 0\ 3/2\}$ | a+1/2, b, c+3/2, y, -x, z |
| $\{-1\|\|1\|1/2\ 1/2\ 1\}$ | a+1/2, b+1/2, c+1, -x, -y, -z |
| $\{m_{001}\|\|1\|1/2\ 1/2\ 0\}$ | a+1/2, b+1/2, c, x, y, -z |
| $\{-4^{1}_{001}\|\|1\|0\ 1/2\ 3/2\}$ | a, b+1/2, c+3/2, y, -x, -z |
| $\{-4^{3}_{001}\|\|1\|0\ 1/2\ 1/2\}$ | a, b+1/2, c+1/2, -y, x, -z |
| $\{1\|\|2_{001}\|0\}$ | -a, -b, c, x, y, z |
| $\{2_{001}\|\|2_{001}\|0\ 0\ 1\}$ | -a, -b, c+1, -x, -y, z |
| $\{4^{1}_{001}\|\|2_{001}\|1/2\ 0\ 1/2\}$ | -a+1/2, -b, c+1/2, -y, x, z |
| $\{4^{3}_{001}\|\|2_{001}\|1/2\ 0\ 3/2\}$ | -a+1/2, -b, c+3/2, y, -x, z |
| $\{-1\|\|2_{001}\|1/2\ 1/2\ 1\}$ | -a+1/2, -b+1/2, c+1, -x, -y, -z |
| $\{m_{001}\|\|2_{001}\|1/2\ 1/2\ 0\}$ | -a+1/2, -b+1/2, c, x, y, -z |
| $\{-4^{1}_{001}\|\|2_{001}\|0\ 1/2\ 3/2\}$ | -a, -b+1/2, c+3/2, y, -x, -z |
| $\{-4^3_{001}\|\|2_{001}\|0\ 1/2\ 1/2\}$ | -a, -b+1/2, c+1/2, -y, x, -z |
| $\{m_{010}\|\|2_{010}\|0\}$ | -a, b, -c, x, -y, z |
| $\{m_{100}\|\|2_{010}\|0\ 0\ 1\}$ | -a, b, -c+1, -x, y, z |
| $\{m_{110}\|\|2_{010}\|1/2\ 0\ 1/2\}$ | -a+1/2, b, -c+1/2, -y, -x, z |
| $\{m_{-110}\|\|2_{010}\|1/2\ 0\ 3/2\}$ | -a+1/2, b, -c+3/2, y, x, z |
| $\{2_{010}\|\|2_{010}\|1/2\ 1/2\ 1\}$ | -a+1/2, b+1/2, -c+1, -x, y, -z |
| $\{2_{100}\|\|2_{010}\|1/2\ 1/2\ 0\}$ | -a+1/2, b+1/2, -c, x, -y, -z |
| $\{2_{110}\|\|2_{010}\|0\ 1/2\ 3/2\}$ | -a, b+1/2, -c+3/2, y, x, -z |
| $\{2_{-110}\|\|2_{010}\|0\ 1/2\ 1/2\}$ | -a, b+1/2, -c+1/2, -y, -x, -z |
| $\{m_{010}\|\|2_{100}\|0\}$ | a, -b, -c, x, -y, z |
| $\{m_{100}\|\|2_{100}\|0\ 0\ 1\}$ | a, -b, -c+1, -x, y, z |
| $\{m_{110}\|\|2_{100}\|1/2\ 0\ 1/2\}$ | a+1/2, -b, -c+1/2, -y, -x, z |
| $\{m_{-110}\|\|2_{100}\|1/2\ 0\ 3/2\}$ | a+1/2, -b, -c+3/2, y, x, z |
| $\{2_{010}\|\|2_{100}\|1/2\ 1/2\ 1\}$ | a+1/2, -b+1/2, -c+1, -x, y, -z |
| $\{2_{100}\|\|2_{100}\|1/2\ 1/2\ 0\}$ | a+1/2, -b+1/2, -c, x, -y, -z |
| $\{2_{110}\|\|2_{100}\|0\ 1/2\ 3/2\}$ | a, -b+1/2, -c+3/2, y, x, -z |
| $\{2_{-110}\|\|2_{100}\|0\ 1/2\ 1/2\}$ | a, -b+1/2, -c+1/2, -y, -x, -z |

## D. Introduction to site-symmetry groups and Wyckoff positions of SSGs

This section aims to define the site-symmetry group $G^{\boldsymbol{q}}$ at a point **q** in a crystal that is invariant under a SSG $G$, which comprises spin-space symmetry operations $g_i$:

$$g_i = \{g_{s_i} \| R_i | \tau_i\}, \tag{D1}$$

where $g_{s_i}$ represents a spin-space unitary/antiunitary symmetry operation encompassing the identity, a rotation, time reversal, or time reversal followed by rotation; $R_i$ denotes a real-space symmetry operation that is either the identity, inversion, a rotation or a rotation-inversion; and $\tau_i$ is a real-space pure translation.

Given a point **q** in a crystal, the action of $g_i$ on **q** can be expressed as:

$$g_i \boldsymbol{q} = R_i \boldsymbol{q} + \tau_i, \tag{D2}$$

where only $R_i$ and $\tau_i$ act on **q** due to the invariance of spatial coordinates under both time revesal and spin-space rotation. Those operations $g_i^q$ that leave the point invariant form the site-symmetry group $G^{\boldsymbol{q}}$, i.e., $g_i^q \boldsymbol{q} = \boldsymbol{q}$. Additionally, the elements $g_\alpha \in G$ but $g_\alpha \notin G^{\boldsymbol{q}}$ in magnetic primitive cell form the cosets $G/G^{\boldsymbol{q}}$, which is the set of points $\{\boldsymbol{q}_\alpha = g_\alpha \boldsymbol{q} | g_\alpha \notin G^{\boldsymbol{q}}\}$, where $\alpha = 1,2,\dots,n$. These points $\boldsymbol{q}_\alpha$ is also called as Wyckoff positions, and *n* represents the multiplicity. Therefore, the site-symmetry group $G^{\boldsymbol{q}}$ is a finite subgroup of the SSG $G$, i.e., $G^{\boldsymbol{q}} \subset G$. It is important to note that $G^{\boldsymbol{q}}$ does not contain any element in the form of $\{g_{s_i} \| 1 | \tau_i\}$ or $\{1\|1|\tau_i\}$, where 1 is the identity and $\tau_i$ is a translation.

In the following discussion, we explore the relationship between nontrivial site-SSG and nontrivial SPG. For *t*-type SSG, $G^{\boldsymbol{q}}$ is found to be isomorphic to one of the 598 nontivial SPGs detailed in Litvin's work [18]. Specially, for *k*-type and *g*-type SSGs, the spin translation group $G_T^S = \left\{\{1\|1|0\}, \{g_{s_1}\|1|\tau_1\}, \dots, \{g_{s_{n-1}}\|1|\tau_{n-1}\}\right\}$ is unable to stablize any point in position space. Thus, $G^{\boldsymbol{q}}$ can not contain any operation in the form of $\{g_{s_i} \| 1 | \tau_i\}$. In summary, for *t*-type, *k*-type and *g*-type SSG, $G^{\boldsymbol{q}}$ always exhibits isomorphism to one of the 598 nontivial SPGs. In contrast, the spin translation group $G_T^S$ always present in the coset $G/G^{\boldsymbol{q}}$, which tells the difference of volumn between magnetic unit cell and nonmangnetic unit cell.

Next we show how to obtain Wyckoff positions in *t*-type, *k*-type and *g*-type SSG.

## 1. t-type SSG

As shown in the main text, the nontrivial SSG $B^{g_{s_1}}g_1{}^{g_{s_2}}g_2{}^{g_{s_3}}g_3$ can be obtained using the PG $G^s = \{1, g_{s_1}, g_{s_2} \dots g_{s_{n-1}}\}$ in spin space and the quotient group $G_0/L_0$ using Eq. (3). Since the site $\boldsymbol{q}$ is invariant under any spin-space operation $g_{s_i}$, and the magnetic cell of *t*-type SSG is the same with nonmagnetic cell. Then Eq. (3) can be viewed as

$$G_0 = L_0 \cup \{E||g_1\}L_0 \cup \{E||g_2\}L_0 \cup \dots \cup \{E||g_{\mathrm{n}-1}\}L_0, \quad \text{(D3)}$$

when act on the site $\boldsymbol{q}$. In other word, the *t*-type SSG shares the same Wyckoff-position multiplicities and coordinates with SG $G_0$. Besides, the site-symmetry group $G^{\boldsymbol{q}}$ is also isomorphic to the site-symmetry group $G_0^{\boldsymbol{q}}$ of SG $G_0$.

## I. 13.54.1.1~13.54.1.6

Here we use 13.54.1.1~13.54.1.6 as an example, telling about the construction of Wyckoff positions, site-symmetry groups and construct realistic magnetic structures using these SSGs.

As stated before, the *t*-type SSG shares the same Wyckoff-position multiplicities and coordinates with SG $G_0$. Below is these information about $G_0 = Pcca$ (No. 54).

**Table D1:** Wyckoff positions for SG $G_0 = Pcca$ (No. 54) from WYCKPOS at Bilbao Crystallography Server [89].

| WP | Site-SG | Coordinates |
|---|---|---|
| 4a | -1 | (0, 0, 0), (1/2, 0, 0), (0, 0, 1/2), (1/2, 0, 1/2) |
| 4b | -1 | (0, 1/2, 0), (1/2, 1/2, 0), (0, 1/2, 1/2), (1/2, 1/2, 1/2) |
| 4c | 2 | (0, b, 1/4), (1/2, -b, 1/4), (0, -b, 3/4), (1/2, b, 3/4) |
| 4d | 2 | (1/4, 0, c), (3/4, 0, -c+1/2), (3/4, 0, -c), (1/4, 0, c+1/2) |
| 4e | 2 | (1/4, 1/2, c), (3/4, 1/2, -c+1/2), (3/4, 1/2, -c), (1/4, 1/2, c+1/2) |
| 8f | 1 | (a, b, c), (-a+1/2, -b, c), (-a, b, -c+1/2), (a+1/2, -b, -c+1/2),<br>(-a, -b, -c), (a+1/2, b, -c), (a, -b, c+1/2), (-a+1/2, b, c+1/2) |

For the *t*-type SSG, only the site-symmetry group and the magnetic moment on each coordinate should be added for constructing Wyckoff-positions of SSG. Here we choose some representative sites to show the construction of site-SSG of $P^{-1}c^{-1}c^{1}a$ (13.54.1.1).

**Table D2:** Representative sites for all Wyckoff positions for $G_0 = Pcca$ (No. 54).

| WP | Representative | Site-symmetry group |
|---|---|---|
| 4a | (0, 0, 0) | $\{1\|0\},\{-1\|0\}$ |
| 4b | (0, 1/2, 0) | $\{1\|0\},\{-1\|0\ 1\ 0\}$ |
| 4c | (0, b, 1/4) | $\{1\|0\},\{2_{010}\|0\ 0\ 1/2\}$ |
| 4d | (1/4, 0, c) | $\{1\|0\},\{2_{001}\|1/2\ 0\ 0\}$ |
| 4e | (1/4, 1/2, c) | $\{1\|0\},\{2_{001}\|1/2\ 1\ 0\}$ |
| 8f | (a, b, c) | $\{1\|0\}$ |

Here we show all symmetry operations of 13.54.1.1

$$P^{-1}c^{-1}c^{1}a = \left\{ \begin{array}{c} \{1\|1|0\},\{1\|-1|0\}, \\ \{1\|2_{001}|1/2\ 0\ 0\},\{1\|m_{001}|1/2\ 0\ 0\} \\ \{-1\|2_{100}|1/2\ 0\ 1/2\},\{-1\|m_{100}|1/2\ 0\ 1/2\} \\ \{-1\|2_{010}|0\ 0\ 1/2\},\{-1\|m_{010}|0\ 0\ 1/2\} \end{array} \right\}, \tag{D4}$$

Following above, we map the spin operation with space part to obtain Wyckoff positions of $P^{-1}c^{-1}c^{1}a$ (13.54.1.1), and we calculate the constrain on magnetic moments.

**Table D3:** Representative sites for all Wyckoff positions for $P^{-1}c^{-1}c^{1}a$ (13.54.1.1).

| 13.54.1.1<br>$P^{-1}c^{-1}c^{1}a$ | Representative | Site-symmetry group | notation | magnetic moments |
|---|---|---|---|---|
| 4a | (0, 0, 0) | $\{1\|1\|0\},\{1\|-1\|0\}$ | ${}^{1}-1$ | (x, y, z) |
| 4b | (0, 1/2, 0) | $\{1\|\|1\|0\},\{1\|\|-1\|0\ 1\ 0\}$ | $^{1}-1$ | (x, y, z) |
| 4c | (0, b, 1/4) | $\{1\|\|1\|0\},\{-1\|\|2_{010}\|0\ 0\ 1/2\}$ | $^{-1}2$ | (0, 0, 0) |
| 4d | (1/4, 0, c) | $\{1\|\|1\|0\},\{1\|\|2_{001}\|1/2\ 0\ 0\}$ | $^{1}2$ | (x, y, z) |
| 4e | (1/4, 1/2, c) | $\{1\|\|1\|0\},\{1\|\|2_{001}\|1/2\ 1\ 0\}$ | $^{1}2$ | (x, y, z) |
| 8f | (a, b, c) | $\{1\|\|1\|0\}$ | $^{1}1$ | (x, y, z) |

Following these procedure, the Wyckoff positions of $P^{-1}c^{-1}c^{1}a$ (13.54.1.1) are constructed and shown in Table D4. Similarly, Wyckoff positions of 13.54.1.2~13.54.1.6 are also constructed and shown in Table D5 to D9.

**Table D4:** Wyckoff positions for $P^{-1}c^{-1}c^{1}a$ (13.54.1.1).

| WP | Site-SSG | Coordinates |
|---|---|---|
| 4a | $^{1}-1$ | (0, 0, 0 \| x, y, z), (1/2, 0, 0 \| x, y, z),<br>(0, 0, 1/2 \| -x, -y, -z), (1/2, 0, 1/2 \| -x, -y, -z) |
| 4b | $^{1}-1$ | (0, 1/2, 0 \| x, y, z), (1/2, 1/2, 0 \| x, y, z),<br>(0, 1/2, 1/2 \| -x, -y, -z), (1/2, 1/2, 1/2 \| -x, -y, -z) |
| 4c | $^{-1}2$ | (0, b, 1/4 \| 0, 0, 0), (1/2, -b, 1/4 \| 0, 0, 0),<br>(0, -b, 3/4 \| 0, 0, 0), (1/2, b, 3/4 \| 0, 0, 0) |
| 4d | $^{1}2$ | (1/4, 0, c \| x, y, z), (3/4, 0, -c+1/2 \| -x, -y, -z),<br>(3/4, 0, -c \| x, y, z), (1/4, 0, c+1/2 \| -x, -y, -z) |
| 4e | $^{1}2$ | (1/4, 1/2, c \| x, y, z), (3/4, 1/2, -c+1/2 \| -x, -y, -z),<br>(3/4, 1/2, -c \| x, y, z), (1/4, 1/2, c+1/2 \| -x, -y, -z) |
| 8f | $^{1}1$ | (a, b, c \| x, y, z), (-a+1/2, -b, c \| x, y, z),<br>(-a, b, -c+1/2 \| -x, -y, -z), (a+1/2, -b, -c+1/2 \| -x, -y, -z),<br>(-a, -b, -c \| x, y, z), (a+1/2, b, -c \| x, y, z),<br>(a, -b, c+1/2 \| -x, -y, -z), (-a+1/2, b, c+1/2 \| -x, -y, -z) |

**Table D5:** Wyckoff positions for $P^{2_{001}}c^{2_{001}}c^{1}a$ (13.54.1.2).

| WP | Site-SSG | Coordinates |
|---|---|---|
| 4a | $^{1}-1$ | (0, 0, 0 \| x, y, z), (1/2, 0, 0 \| x, y, z),<br>(0, 0, 1/2 \| -x, -y, z), (1/2, 0, 1/2 \| -x, -y, z) |
| 4b | $^{1}-1$ | (0, 1/2, 0 \| x, y, z), (1/2, 1/2, 0 \| x, y, z), |
| | | (0, 1/2, 1/2 \| -x, -y, z), (1/2, 1/2, 1/2 \| -x, -y, z) |
| 4c | ${}^{2_{001}}2$ | (0, b, 1/4 \| 0, 0, z), (1/2, -b, 1/4 \| 0, 0, z),<br>(0, -b, 3/4 \| 0, 0, z), (1/2, b, 3/4 \| 0, 0, z) |
| 4d | ${}^{1}2$ | (1/4, 0, c \| x, y, z), (3/4, 0, -c+1/2 \| -x, -y, z),<br>(3/4, 0, -c \| x, y, z), (1/4, 0, c+1/2 \| -x, -y, z) |
| 4e | ${}^{1}2$ | (1/4, 1/2, c \| x, y, z), (3/4, 1/2, -c+1/2 \| -x, -y, z),<br>(3/4, 1/2, -c \| x, y, z), (1/4, 1/2, c+1/2 \| -x, -y, z) |
| 8f | ${}^{1}1$ | (a, b, c \| x, y, z), (-a+1/2, -b, c \| x, y, z),<br>(-a, b, -c+1/2 \| -x, -y, z), (a+1/2, -b, -c+1/2 \| -x, -y, z),<br>(-a, -b, -c \| x, y, z), (a+1/2, b, -c \| x, y, z),<br>(a, -b, c+1/2 \| -x, -y, z), (-a+1/2, b, c+1/2 \| -x, -y, z) |

**Table D6:** Wyckoff positions for $P^{m_{001}}c^{m_{001}}c^{1}a$ (13.54.1.3).

| WP | Site-SSG | Coordinates |
|---|---|---|
| 4a | ${}^{1}-1$ | (0, 0, 0 \| x, y, z), (1/2, 0, 0 \| x, y, z),<br>(0, 0, 1/2 \| x, y, -z), (1/2, 0, 1/2 \| x, y, -z) |
| 4b | ${}^{1}-1$ | (0, 1/2, 0 \| x, y, z), (1/2, 1/2, 0 \| x, y, z),<br>(0, 1/2, 1/2 \| x, y, -z), (1/2, 1/2, 1/2 \| x, y, -z) |
| 4c | ${}^{m_{001}}2$ | (0, b, 1/4 \| x, y, 0), (1/2, -b, 1/4 \| x, y, 0),<br>(0, -b, 3/4 \| x, y, 0), (1/2, b, 3/4 \| x, y, 0) |
| 4d | ${}^{1}2$ | (1/4, 0, c \| x, y, z), (3/4, 0, -c+1/2 \| x, y, -z),<br>(3/4, 0, -c \| x, y, z), (1/4, 0, c+1/2 \| x, y, -z) |
| 4e | ${}^{1}2$ | (1/4, 1/2, c \| x, y, z), (3/4, 1/2, -c+1/2 \| x, y, -z),<br>(3/4, 1/2, -c \| x, y, z), (1/4, 1/2, c+1/2 \| x, y, -z) |
| 8f | ${}^{1}1$ | (a, b, c \| x, y, z), (-a+1/2, -b, c \| x, y, z),<br>(-a, b, -c+1/2 \| x, y, -z), (a+1/2, -b, -c+1/2 \| x, y, -z),<br>(-a, -b, -c \| x, y, z), (a+1/2, b, -c \| x, y, z),<br>(a, -b, c+1/2 \| x, y, -z), (-a+1/2, b, c+1/2 \| x, y, -z) |

**Table D7:** Wyckoff positions for $P^{-1}c^{1}c^{-1}a$ (13.54.1.4).

| WP | Site-SSG | Coordinates |
|---|---|---|
| 4a | ${}^{1}-1$ | (0, 0, 0 \| x, y, z), (1/2, 0, 0 \| -x, -y, -z), |
| | | (0, 0, 1/2 \| x, y, z), (1/2, 0, 1/2 \| -x, -y, -z) |
| 4b | ${}^{1}{-1}$ | (0, 1/2, 0 \| x, y, z), (1/2, 1/2, 0 \| -x, -y, -z),<br>(0, 1/2, 1/2 \| x, y, z), (1/2, 1/2, 1/2 \| -x, -y, -z) |
| 4c | ${}^{1}2$ | (0, b, 1/4 \| x, y, z), (1/2, -b, 1/4 \| -x, -y, -z),<br>(0, -b, 3/4 \| x, y, z), (1/2, b, 3/4 \| -x, -y, -z) |
| 4d | ${}^{-1}2$ | (1/4, 0, c \| 0, 0, 0), (3/4, 0, -c+1/2 \| 0, 0, 0),<br>(3/4, 0, -c \| 0, 0, 0), (1/4, 0, c+1/2 \| 0, 0, 0) |
| 4e | ${}^{-1}2$ | (1/4, 1/2, c \| 0, 0, 0), (3/4, 1/2, -c+1/2 \| 0, 0, 0),<br>(3/4, 1/2, -c \| 0, 0, 0), (1/4, 1/2, c+1/2 \| 0, 0, 0) |
| 8f | ${}^{1}1$ | (a, b, c \| x, y, z), (-a+1/2, -b, c \| -x, -y, -z),<br>(-a, b, -c+1/2 \| x, y, z), (a+1/2, -b, -c+1/2 \| -x, -y, -z),<br>(-a, -b, -c \| x, y, z), (a+1/2, b, -c \| -x, -y, -z),<br>(a, -b, c+1/2 \| x, y, z), (-a+1/2, b, c+1/2 \| -x, -y, -z) |

**Table D8:** Wyckoff positions for $P^{2_{001}}c^{1}c^{2_{001}}a$ (13.54.1.5).

| WP | Site-SSG | Coordinates |
|---|---|---|
| 4a | ${}^{1}{-1}$ | (0, 0, 0 \| x, y, z), (1/2, 0, 0 \| -x, -y, z),<br>(0, 0, 1/2 \| x, y, z), (1/2, 0, 1/2 \| -x, -y, z) |
| 4b | ${}^{1}{-1}$ | (0, 1/2, 0 \| x, y, z), (1/2, 1/2, 0 \| -x, -y, z),<br>(0, 1/2, 1/2 \| x, y, z), (1/2, 1/2, 1/2 \| -x, -y, z) |
| 4c | ${}^{1}2$ | (0, b, 1/4 \| x, y, z), (1/2, -b, 1/4 \| -x, -y, z),<br>(0, -b, 3/4 \| x, y, z), (1/2, b, 3/4 \| -x, -y, z) |
| 4d | ${}^{2_{001}}2$ | (1/4, 0, c \| 0, 0, z), (3/4, 0, -c+1/2 \| 0, 0, z),<br>(3/4, 0, -c \| 0, 0, z), (1/4, 0, c+1/2 \| 0, 0, z) |
| 4e | ${}^{2_{001}}2$ | (1/4, 1/2, c \| 0, 0, z), (3/4, 1/2, -c+1/2 \| 0, 0, z),<br>(3/4, 1/2, -c \| 0, 0, z), (1/4, 1/2, c+1/2 \| 0, 0, z) |
| 8f | ${}^{1}1$ | (a, b, c \| x, y, z), (-a+1/2, -b, c \| -x, -y, z),<br>(-a, b, -c+1/2 \| x, y, z), (a+1/2, -b, -c+1/2 \| -x, -y, z),<br>(-a, -b, -c \| x, y, z), (a+1/2, b, -c \| -x, -y, z),<br>(a, -b, c+1/2 \| x, y, z), (-a+1/2, b, c+1/2 \| -x, -y, z) |

**Table D9:** Wyckoff positions for $P^{m_{001}}c^{1}c^{m_{001}}a$ (13.54.1.6).

| WP | Site-SSG | Coordinates |
|---|---|---|
| 4a | ${}^{1}{-1}$ | (0, 0, 0 \| x, y, z), (1/2, 0, 0 \| x, y, -z),<br>(0, 0, 1/2 \| x, y, z), (1/2, 0, 1/2 \| x, y, -z) |
| 4b | ${}^{1}{-1}$ | (0, 1/2, 0 \| x, y, z), (1/2, 1/2, 0 \| x, y, -z),<br>(0, 1/2, 1/2 \| x, y, z), (1/2, 1/2, 1/2 \| x, y, -z) |
| 4c | ${}^{1}2$ | (0, b, 1/4 \| x, y, z), (1/2, -b, 1/4 \| x, y, -z),<br>(0, -b, 3/4 \| x, y, z), (1/2, b, 3/4 \| x, y, -z) |
| 4d | ${}^{m_{001}}2$ | (1/4, 0, c \| x, y, 0), (3/4, 0, -c+1/2 \| x, y, 0),<br>(3/4, 0, -c \| x, y, 0), (1/4, 0, c+1/2 \| x, y, 0) |
| 4e | ${}^{m_{001}}2$ | (1/4, 1/2, c \| x, y, 0), (3/4, 1/2, -c+1/2 \| x, y, 0),<br>(3/4, 1/2, -c \| x, y, 0), (1/4, 1/2, c+1/2 \| x, y, 0) |
| 8f | ${}^{1}1$ | (a, b, c \| x, y, z), (-a+1/2, -b, c \| x, y, -z),<br>(-a, b, -c+1/2 \| x, y, z), (a+1/2, -b, -c+1/2 \| x, y, -z),<br>(-a, -b, -c \| x, y, z), (a+1/2, b, -c \| x, y, -z),<br>(a, -b, c+1/2 \| x, y, z), (-a+1/2, b, c+1/2 \| x, y, -z) |

## 2. k-type SSG

Unlike *t*-type SSGs, the Wyckoff positions in *k*-type SSGs have more complicated dependencies on the Wyckoff positions in the SG $G_0$, which is similar to the Type-IV MSGs in the BNS setting. This complication arise from the spin translation operations $\{g_{s_i}||1|\tau_i\}$ in *k*-type SSG $B^1g_1{}^1g_2{}^1g_3{}^{g_{s_1}}\tau_1{}^{g_{s_2}}\tau_2{}^{g_{s_3}}\tau_3$, which enlarges the magnetic unit cell of a crystal compared to the nonmagnetic unit cell of SG $G_0$. To construct the Wyckoff positions of *k*-type SSGs, the Wyckoff positions of SG $G_0$, the transformation matrix $M_{\mathrm{T}}$ between $G_0$ and $L_0$, and the spin translation group $G_T^S = \{\{1||1|0\}, \{g_{s_1}||1|\tau_1\}, \dots, \{g_{s_{n-1}}||1|\tau_{n-1}\}\}$ are employed. The multiplicity of each Wyckoff position in k-type SSG will be an integer multiple of the index $i_k$ when compared to the nonmagnetic primitive cell of SG $G_0$. Here the transformation matrix $M_{\mathrm{T}}$ is introduced in order to suit the $L_0$ basis of *k*-type SSGs.

## I. 99.107.4.1

First, we list the Wyckoff positions of SG $G_0 = I4$mm (No. 107) in Table D10.

**Table D10:** Wyckoff positions for SG $G_0 = I4$mm (No. 107) from WYCKPOS at Bilbao Crystallography Server [89].

| WP | Site-SG | Coordinates |
|---|---|---|
| | | (0, 0, 0) + (1/2, 1/2, 1/2) |
| 2a | 4mm | (0, 0, c) |
| 4b | 2mm | (0, 1/2, c), (1/2, 0, c) |
| 8c | m | (a, a, c), (-a, -a, c), (-a, a, c), (a, -a, c) |
| 8d | m | (a, 0, c), (-a, 0, c), (0, a, c), (0, -a, c) |
| 16e | 1 | (a, b, c), (-a, -b, c), (-b, a, c), (b, -a, c)<br>(a, -b, c), (-a, b, c), (-b, -a, c), (b, a, c) |

Next, Wyckoff positions and site-SSGs for $P^{1}4^{1}m^{1}m^{4_{001}^{1}}(1/2\ 1/2\ 1/4)$ (99.107.4.1) are constructed using the transformation matrix

$$M_{\mathrm{T}} = \begin{pmatrix} 1 & 0 & 0 & 0 \\ 0 & 1 & 0 & 0 \\ 0 & 0 & 2 & 0 \end{pmatrix}, \qquad \text{(D5)}$$

which tells the volumn ratio between $G_0$ and $L_0$, $\mathrm{V}(G_0){:}\,\mathrm{V}(L_0) = 1{:}\,2$ from $\det(M_{\mathrm{T}}) = 2$. Therefore, the multiplicities of Wyckoff-positions is doubled in comparison with $G_0 = I4$mm (No. 107) , and the Wyckoff positions for $P^{1}4^{1}m^{1}m^{4_{001}^{1}}(1/2\ 1/2\ 1/4)$ (99.107.4.1) are shown in Table D11.

**Table D11:** Wyckoff positions for $P^{1}4^{1}m^{1}m^{4_{001}^{1}}(1/2\ 1/2\ 1/4)$ (99.107.4.1).

| WP | Site-SSG | Coordinates |
|---|---|---|
| | | **(0, 0, 0 \| x, y, z) + (1/2, 1/2, 1/4 \| -y, x, z)<br>+ (0, 0, 1/2 \| -x, -y, z) + (1/2, 1/2, 3/4 \| y, -x, z)** |
| 4a | $^{1}4^{1}m^{1}m$ | (0, 0, c) |
| 8b | $^{1}2^{1}m^{1}m$ | (0, 1/2, c), (1/2, 0, c) |
| 16c | $^{1}m$ | (a, a, c), (-a, -a, c), (-a, a, c), (a, -a, c) |
| 16d | ${}^{1}m$ | (a, 0, c), (-a, 0, c), (0, a, c), (0, -a, c) |
| 32e | ${}^{1}1$ | (a, b, c), (-a, -b, c), (-b, a, c), (b, -a, c)<br>(a, -b, c), (-a, b, c), (-b, -a, c), (b, a, c) |

## 3. g-type SSG

Similar to the *k*-type SSGs, the Wyckoff positions in *g*-type SSGs have more complicated dependencies on the Wyckoff positions in the SG $G_0$, which is similar to the Type-IV MSGs in the OG setting. Since now we must use the $G_0$ basis to maintain the most spatial rotation part. This complication arise from the spin translation operations $\{g_{s_i}||1|\tau_i\}$ in *g*-type SSG $B^{g_{s_1}}g_1{}^{g_{s_2}}g_2{}^{g_{s_3}}g_3|(g_{s_4}, g_{s_5}, g_{s_6}; g_{s_7}, g_{s_8}, g_{s_9})$, which enlarges the magnetic unit cell of a crystal compared to the nonmagnetic unit cell of SG $G_0$. To construct the Wyckoff positions of *k*-type SSGs, the Wyckoff positions of SG $G_0$ and the spin translation group $G_T^S = \left\{\{1||1|0\},\ \{g_{s_1}||1|\tau_1\}, \dots, \{g_{s_{n-1}}||1|\tau_{n-1}\}\right\}$ are employed. Here the transformation matrix $M_{\mathrm{T}}$ is no more necessary since the *g*-type SSGs are under $G_0$ basis.

## I. 4.182.4.2

Wyckoff positions of the *g*-type SSG $P^{3^2_{-11-1}}6_3{}^{m_{110}}2^{m_{011}}2|(2_{001}, 2_{100}, 1)$ (4.182.4.2) are constructed from Wyckoff positions of SG $G_0 = P6_322$ (No. 182) shown in Table D12.

**Table D12:** Wyckoff positions for SG $G_0 = P6_322$ (No. 182) from WYCKPOS at Bilbao Crystallography Server [89].

| WP | Site-SG | Coordinates |
|---|---|---|
| 2a | 321 | (0, 0, 0), (0, 0, 1/2) |
| 2b | 312 | (0, 0, 1/4), (0, 0, 3/4) |
| 2c | 312 | (1/3, 2/3, 1/4), (2/3, 1/3, 3/4) |
| 2d | 312 | (1/3, 2/3, 3/4), (2/3, 1/3, 1/4) |
| 4e | 3 | (0, 0, c), (0, 0, c+1/2), (0, 0, -c), (0, 0, -c+1/2) |
| 4f | 3 | (1/3, 2/3, c), (2/3, 1/3, c+1/2), (2/3, 1/3, -c), (1/3, 2/3, -c+1/2) |
| 6g | 2 | (a, 0, 0), (0, a, 0), (-a, -a, 0),<br>(-a, 0, 1/2), (0, -a, 1/2), (-a, -a, 1/2) |
| 6h | 2 | (a, 2a, 1/4), (-2a, -a, 1/4), (a, -a, 1/4),<br>(-a, -2a, 3/4), (2a, a, 3/4), (-a, a, 3/4) |
| 12i | 1 | (a, b, c), (-b, a-b, c), (-a+b, -a, c), (-a, -b, c+1/2),<br>(b, -a+b, c+1/2), (a-b, a, c+1/2), (b, a, -c), (a-b, -b, -c),<br>(-a, -a+b, -c), (-b, -a, -c+1/2), (-a+b, b, -c+1/2), (a, a-b, -c+1/2) |

For *g*-type SSGs, it is necessary to establish constraints on the magnetic moment and the spatial coordinate of SG $G_0$ using site-SSG. Additionally, the multiplicities arising from the volume difference between the magnetic unit cell and the nonmagnetic cell must be taken into account when constructing the Wyckoff-positions of SSG. Here the volumn of magnetic cell is four times larger than the nonmagnetic cell, resulting in the muliplicities being four times larger than those of the nonmangnetic SG 182. Here we choose some representative sites to show the construction of site-SSG of the *g*-type SSG $P^{3^2_{-11-1}}6_3{}^{m_{110}}2^{m_{011}}2|(2_{001}, 2_{100}, 1)$ (4.182.4.2).

**Table D13:** Representative sites for all Wyckoff positions for SSG 4.182.4.2.

| SSG | Representative | Generators of Site-symmetry group |
|---|---|---|
| 8a | (0, 0, 0) | $\{3^1_{-11-1} \| 3^1_{001} \| 0\}, \{m_{110} \| 2_{100} \| 0\}$ |
| 8b | (0, 0, 1/4) | $\{3^1_{-11-1} \| 3^1_{001} \| 0\}, \{m_{011} \| 2_{210} \| 0\ 0\ 1/2\}$ |
| 8c | (1/3, 2/3, 1/4) | $\{3^1_{-11-1} \| 3^1_{001} \| 1\ 1\ 0\}, \{m_{011} \| 2_{210} \| 0\ 1\ 1/2\}$ |
| 8d | (1/3, 2/3, 3/4) | $\{3^1_{-11-1} \| 3^1_{001} \| 1\ 1\ 0\}, \{m_{011} \| 2_{210} \| 0\ 1\ 3/2\}$ |
| 16e | (0, 0, c) | $\{3^1_{-11-1} \| 3^1_{001} \| 0\}$ |
| 16f | (1/3, 2/3, c) | $\{3^1_{-11-1} \| 3^1_{001} \| 1\ 1\ 0\}$ |
| 24g | (a, 0, 0) | $\{m_{110} \| 2_{100} \| 0\}$ |
| 24h | (a, 2a, 1/4) | $\{m_{110} \| 2_{120} \| 0\ 0\ 1/2\}$ |
| 48i | (a, b, c) | $\{1 \| 0\}$ |

**Table D14:** Representative sites, site-SSGs and the corresponding magnetic moments for all Wyckoff positions for SSG 4.182.4.2.

| SSG | Representative | Site-SSG | magnetic moment |
|---|---|---|---|
| 8a | (0, 0, 0) | ${}^{3^{1}_{-11-1}}3\,{}^{m_{110}}2\,{}^{1}1$ | (x, -x, x) |
| 8b | (0, 0, 1/4) | ${}^{3^{1}_{-11-1}}3\,{}^{1}1\,{}^{m_{011}}2$ | (x, -x, x) |
| 8c | (1/3, 2/3, 1/4) | ${}^{3^{1}_{-11-1}}3\,{}^{1}1\,{}^{m_{011}}2$ | (x, -x, x) |
| 8d | (1/3, 2/3, 3/4) | ${}^{3^{1}_{-11-1}}3\,{}^{1}1\,{}^{m_{011}}2$ | (x, -x, x) |
| 16e | (0, 0, c) | ${}^{3^{1}_{-11-1}}3$ | (x, -x, x) |
| 16f | (1/3, 2/3, c) | ${}^{3^{1}_{-11-1}}3$ | (x, -x, x) |
| 24g | (a, 0, 0) | ${}^{m_{110}}2$ | (x, -x, z) |
| 24h | (a, 2a, 1/4) | ${}^{m_{110}}2$ | (x, -x, z) |
| 48i | (a, b, c) | ${}^{1}1$ | (x, y, z) |

Using the elements of site-SSG. the Wyckoff Positions are determined as follows:

**Table D15:** Wyckoff positions for $P^{3^{2}_{-11-1}}6_3{}^{m_{110}}2{}^{m_{011}}2|(2_{001}, 2_{100}, 1)$ (4.182.4.2).

| WP | Site-SSG | Coordinates |
|---|---|---|
| | | (0, 0, 0 \| x, y, z) + (1, 0, 0 \| -x, -y, z)<br>+ (0, 1, 0 \| x, -y, -z) + (1, 1, 0 \| -x, y, -z) |
| 8a | ${}^{3^{1}_{-11-1}}3\,{}^{m_{110}}2\,{}^{1}1$ | (0, 0, 0 \| x, -x, x), (0, 0, 1/2 \| x, -x, x) |
| 8b | ${}^{3^{1}_{-11-1}}3\,{}^{1}1\,{}^{m_{011}}2$ | (0, 0, 1/4 \| x, -x, x), (0, 0, 3/4 \| x, -x, x) |
| 8c | ${}^{3^{1}_{-11-1}}3\,{}^{1}1\,{}^{m_{011}}2$ | (1/3, 2/3, 1/4 \| x, -x, x), (5/3, 4/3, 3/4 \| x, -x, x) |
| 8d | ${}^{3^{1}_{-11-1}}3\,{}^{1}1\,{}^{m_{011}}2$ | (1/3, 2/3, 3/4 \| x, -x, x), (5/3, 4/3, 1/4 \| x, -x, x) |
| 16e | ${}^{3^{1}_{-11-1}}3$ | (0, 0, c \| x, -x, x), (0, 0, c+1/2 \| x, -x, x),<br>(0, 0, -c \| x, -x, x), (0, 0, -c+1/2 \| x, -x, x) |
| 16f | ${}^{3^{1}_{-11-1}}3$ | (1/3, 2/3, c \| x, -x, x), (5/3, 4/3, c+1/2 \| x, -x, x),<br>(2/3, 1/3, -c \| x, -x, x), (4/3, 5/3, -c+1/2 \| x, -x, x) |
| 24g | ${}^{m_{110}}2$ | (a, 0, 0 \| x, -x, z), (0, a, 0 \| z, -x, x),<br>(-a, -a, 0 \| x, -z, x), (-a, 0, 1/2 \| x, -x, z),<br>(0, -a, 1/2 \| z, -x, x), (a, a, 1/2 \| x, -z, x) |
| 24h | $m_{110}2$ | (a, 2a, 1/4 \| x, -x, z), (-2a, -a, 1/4 \| z, -x, x), (a, -a, 1/4 \| x, -z, x), (-a, -2a, 3/4 \| x, -x, z), (2a, a, 3/4 \| z, -x, x), (-a, a, 3/4 \| x, -z, x) |
| 48i | $^{1}1$ | (a, b, c \| x, y, z), (-b, a-b, c \| z, -x, -y), (-a+b, -a, c \| -y, -z, x), (-a, -b, c+1/2 \| x, y, z), (b, -a+b, c+1/2 \| z, -x, -y), (a-b, a, c+1/2 \| -y, -z, x), (b, a, -c \| z, y, x), (a-b, -b, -c \| -y, -x, z), (-a, -a+b, -c \| x, -z, -y), (-b, -a, -c+1/2 \| z, y, x), (-a+b, b, -c+1/2 \| -y, -x, z), (a, a-b, -c+1/2 \| x, -z, -y) |

It is important to note that the translation period is now 2a, 2b, c for three crystallographic axes *a*, *b*, *c*, respectively. For example, for Wyckoff positions 8c of SSG 4.182.4.2, the eight coordinates are constrained as (1/3, 2/3, 1/4 | x, -x, x), (5/3, 4/3, 3/4 | x, -x, x), (4/3, 2/3, 1/4 | -x, x, x), (2/3, 4/3, 3/4 | -x, x, x), (1/3, 5/3, 1/4 | x, x, -x), (5/3, 1/3, 3/4 | x, x, -x), (4/3, 5/3, 1/4 | -x, -x, -x) and (2/3, 1/3, 3/4 | -x, -x, -x), respectively. The magnetic structure can be constructed using parameters below:

G = 182, *a* = *b* = 5.749 Å, *c* = 11.886 Å, α = β = 90°, γ = 120°, Co (2d) (1/3, 2/3, 3/4), Nb (2a) (0, 0, 0), Nb (4f) (1/3, 2/3, 0.5056) and S (12i) (0.3322, 0.3313, 0.1306). Then we perform the supercell using 2×2×1 and add magnetic moments: Co (8a) (1/3, 2/3, 3/4 | x, -x, x) with x = -1 (shown in Figs. D1 (a)~(c)).

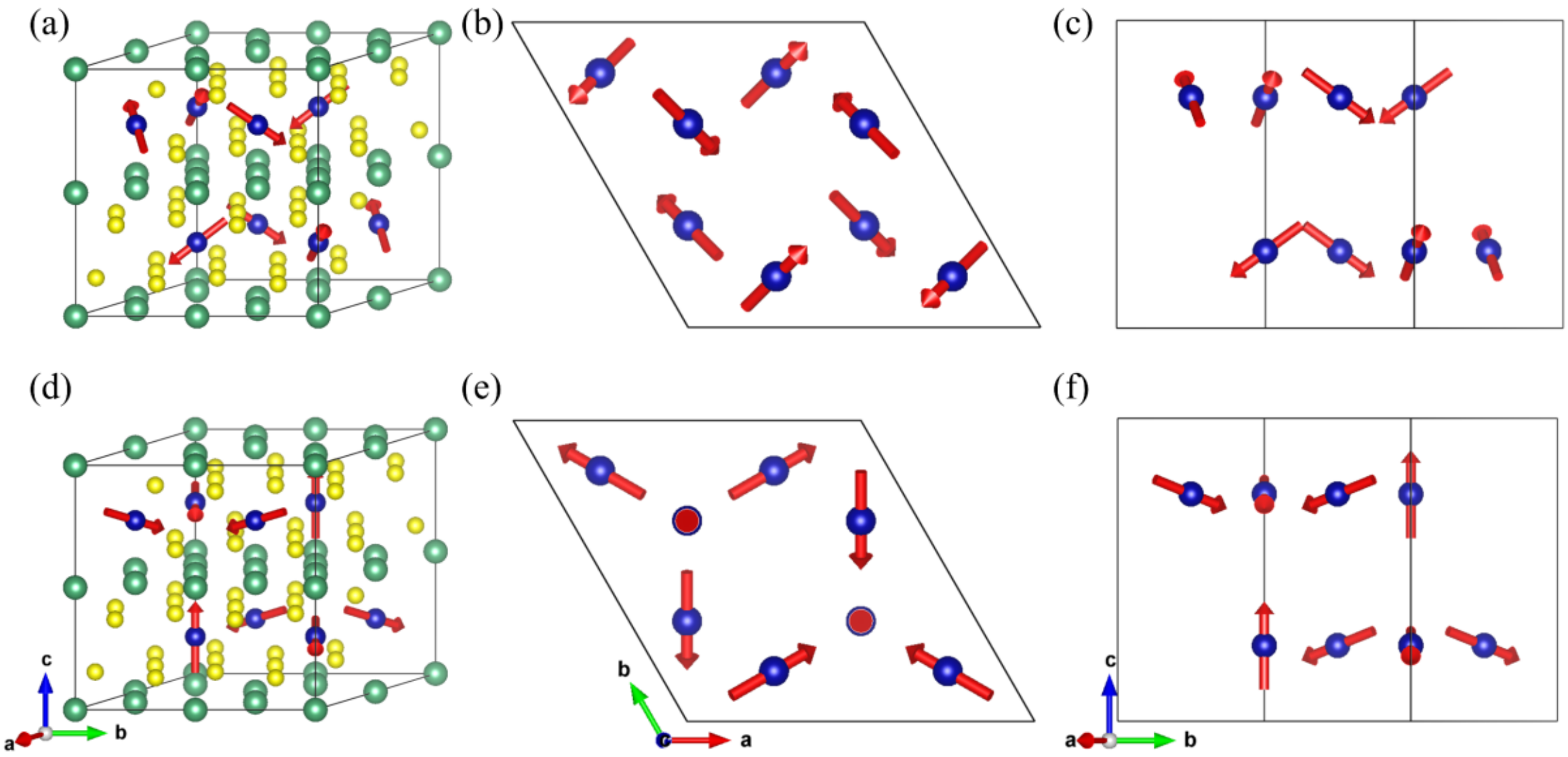

**Fig. D1:** (a)~(c) Magnetic structures constructed by $P^{3^{2}_{-11-1}}6_3{}^{m_{110}}2^{m_{011}}2|(2_{001},2_{100},1)$ using Table D15. (d)~(e) The real magnetic structure of triple-Q $CoNb_3S_6$. For clearity, we move the magnetic cell with $\tau_a = 1\ in\ G_0\ or\ 1/2\ in\ L_0$.

After transform [-11-1] to [001] direction in spin space using transformation matrix *M*, we can obtain Figs. D1 (d)~(f). The transformation matrix $M_T$ is

$$M_T = \begin{pmatrix} -\frac{1}{\sqrt{2}} & -\frac{1}{\sqrt{2}} & 0 \\ -\frac{1}{\sqrt{6}} & \frac{1}{\sqrt{6}} & \frac{\sqrt{6}}{3} \\ -\frac{1}{\sqrt{3}} & \frac{1}{\sqrt{3}} & -\frac{1}{\sqrt{3}} \end{pmatrix}. \tag{D6}$$

After transformation, $P^{3^{2}_{-11-1}}6_3{}^{m_{110}}2^{m_{011}}2|(2_{001},2_{100},1)$ (4.182.4.2) can be written as $P^{3^{2}_{001}}6_3{}^{m_{100}}2^{m_{010}}2|(2_{24-3},2_{423},1)$, which is equivalent with the magnetic structure of $Co_3NbS_6$ as shown in Figs. D1 (d)~(f).

From above, we can derive magnetic structures based on the Wyckoff positions and site symmetry groups of SSGs. It is important to emphasize that once detailed magnetic moments are imposed, the magnetic structure can exhibit certain global symmetry, which is called spin only group in the main text. For collinear, coplanar, non-coplanar magnetic structure, the spin only groups $G_{SO}$ are $Z_2^K \ltimes SO(2)$, $Z_2^K$ and $E$, respectively, where $Z_2^K = \{E, TU_n(\pi)\}$, $SO(2) = U_z(\phi)$, $\phi \in [0,2\pi)$. Here $TU_n(\pi)$ represents time reversal followed by a two-fold spin rotation along any axis perpendicular to the spin axis *z*, $U_z(\phi)$ denotes an any-fold spin rotation along the spin axis *z* and $E$ is the identity. These elements leave the spatial coordinates invariant.

## E. Representation theory of SSG

### 1. Projective representation for spin rotation

In SSG, the projective rep is similar to that of MSG, where spin rotation is included. Based on SG, in SSG we introduce spin operation here with the full group of spin operations being $SO(3) \times Z_2^T$. Since we focus on spin-1/2 systems, the full spin rotation group $SO(3)$ should be represented by $SU(2)$ double-valued rep, which is also a projective rep. Specifically, for any two elements $g_{s_i}$ and $g_{s_j}$ in spin rotation group $SO(3)$, the multiplication of these two elements could have rotation angle $\phi(g_{s_i} g_{s_j})$ between 0 and $2\pi$ or between $2\pi$ and $4\pi$. We use ${}^{d}1$ to represent a rotation of $2\pi$. Then we have:

$$g_{s_i} g_{s_j} = \begin{cases} g_{s_l} & 0 \le \phi\left(g_{s_i} g_{s_j}\right) < 2\pi \\ {}^{d}1 g_{s_l} & 2\pi \le \phi\left(g_{s_i} g_{s_j}\right) < 4\pi \end{cases}, \tag{E1}$$

where $g_{s_l} \in SO(3)$.

Then, we define $\rho(g_{s_i})$ to be the $SU(2)$ rep of $SO(3)$ with rep matrices belong to $SU(2)$ group. We define the rotation angle and rotation axis of $\rho(g_{s_i})$ in $SU(2)$ group to be the same as those of $g_{s_i}$ in $SO(3)$ group. Accordingly, we will have $\rho({}^{d}1) = -1$.

Therefore, the product of any two rep matrices follows the below relation:

$$\rho(g_{s_i})\rho\left(g_{s_j}\right) = \begin{cases} \rho(g_{s_k}) & 0 \le \phi\left(g_{s_i} g_{s_j}\right) < 2\pi \\ -\rho(g_{s_k}) & 2\pi \le \phi\left(g_{s_i} g_{s_j}\right) < 4\pi \end{cases}, \tag{E2}$$

Since the the coefficients of the above product results completely dependent on $g_{s_i}$ and $g_{s_j}$. This $SU(2)$ rep is naturally a projective rep of $SO(3)$ with factor system $\pm 1$.

Then, for a spin-1/2 rep $d_k(\{g_s \| R|\tau\})$ of little group $G_k$ of a SSG, the corresponding projective rep $M_k(\{g_s \| R|\tau\})$ has the same definition as that of SG:

$$d_k^l(\{g_s \| R|\tau\}) = exp(-ik \cdot \tau)\, M_k^l(\{g_s \| R|\tau\}). \tag{E3}$$

The product of two projective rep matrices follows the below relation:

$$M_k^l(\{g_{s_i} \| R_i|\tau_i\}) M_k^l\left(\left\{g_{s_j} \middle\| R_j \middle| \tau_j\right\}\right) = (-1)^{\xi\left(g_{s_i}, g_{s_j}\right)} exp\left(-iK_i \cdot \tau_j\right) M_k^l(\{g_{s_l} \| R_l|\tau_l\}). \tag{E4}$$

Here $\{g_{s_l} \| R_l|\tau_l\} = \left\{g_{s_i} g_{s_j} \middle\| R_i R_j \middle| \tau_i + R_i \tau_j \ mod\ T(L_0)\right\}$ , $K_i = R_i^{-1} k - k$ , $\xi\left(g_{s_i}, g_{s_j}\right) = 0$ for $0 \le \phi(g_{s_i} g_{s_j}) < 2\pi$ and $\xi\left(g_{s_i}, g_{s_j}\right) = 1$ for $2\pi \le \phi(g_{s_i} g_{s_j}) <$

$4\pi$.

## 2. Decomposition of regular projective representations using CSCO method

In quantum mechanics, the commuting operators $(J^2, J_z)$ form the CSCO of the Hilbert space of angular momentum, effectively characterizing the systems in terms of these observables. Note that $J^2$ is the invariant quantity (Casimir element) of $SO(3)$ and $J_z$ is that of its subgroup $SO(2)$. Similarly, the projective irreps in a certain unitary group $G$ can be identified through CSCO. The core principle of the CSCO method lies in the fact that every projective irrep, reduced from the regular projective reps, is labeled by eigenvalues of the CSCO. Consequently, all projective irreps can be extracted from the left regular projective reps.

The overall strategy to decompose the regular projective reps can be summarized as follows, with detailed explanations of some terminologies provided later

1) Construct the left regular projective reps for a given factor system;

2) Construct the CSCO-I formed by the class operators of $G$ to label the different rep spaces;

3) Construct the CSCO-II formed by the class operators of the canonical subgroup chain of the group *G* to distinguish each of the irreducible bases as the eigenvalues of CSCO-I are often degenerate (as shown in Fig. E1(b)).

4) Since the *n*-dimensional irreps appears *n* times in the regular reps. The next step is to construct the CSCO-III formed by the class operators of the canonical subgroup chain of the intrinsic group $\bar{G}$ to distinguish the projective irreps that occur more than once (as shown in Fig. E1(c));

5) Change the class and repeat above procedure until the projective irreps of all classes are obtained.

It is important to note that the character of a certain unitary group elements is invariant under unitary transformation and the character if each group element in a class is the same. Therefore, the CSCO method is applied to obtain the projective irreps of $\tilde{L}_{SS}^{k}$, which is the maximal unitary subgroup of the little group $\tilde{G}_{SS}^{k}$.

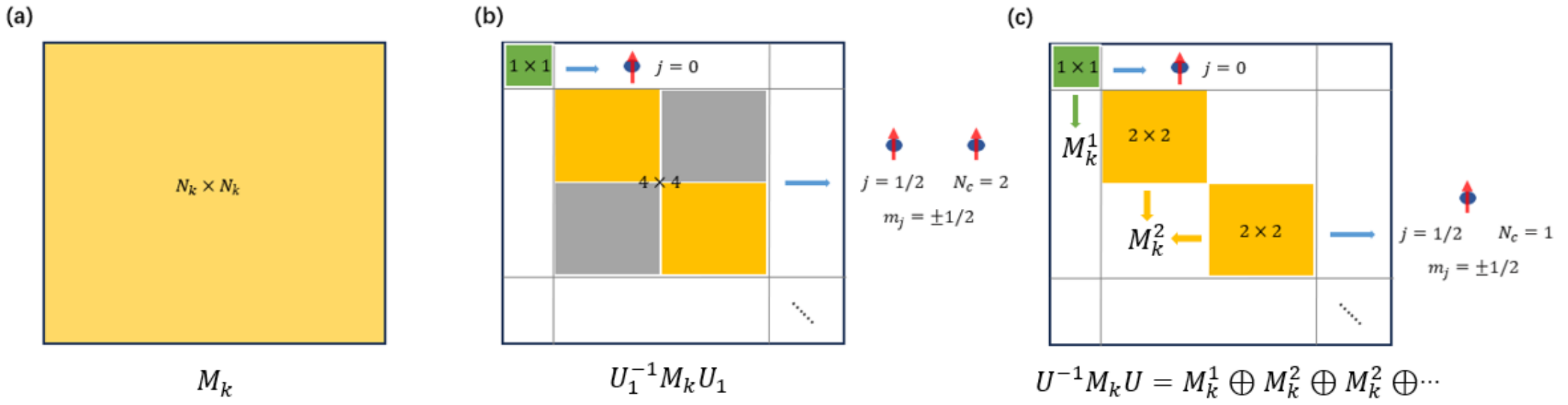

**Fig. E1:** The overall procedure of reducing left regular projective reps $M_k(g_i^{(a)})$. (a) the left regular projective reps are $N_k \times N_k$ rep matrices. (b) the transformation matrices $U_1$ formed by common eigenvectors of $(C, C(s))$ can partly diagonalize $M_k(g_i^{(a)})$ but fail to distinguish the same irreps. (c) the transformation matrices $U$ formed by common eigenvectors of $(C, C(s), \bar{C}(s))$ can be used to completely lift the remaining degeneracy and find all inequivalent irreps.

Several essential concepts of the projective reps are introduced in the following:

1) *Left regular projective representation*: In mathematics, the regular rep refers to a particular way of representing a group $G$ as a set of linear transformations on a vector space V, whose bases are nothing but the elements of $G$. In addition, the linear transformation for each group element that acts on each basis of V is defined as the group operation. The reason we use regular projective rep to obtain the irreps of a unitary SSG is because a regular rep is a reducible rep that contains all the possible irreps. The left regular projective reps for a given factor system can be constructed using the group space as the rep space, where the group elements $g_i^{(a)} \in \tilde{L}_{SS}^k$ are also bases $|g_i^{(a)}\rangle$. A group element $g_i^{(a)}$ acts on $|g_j^{(a)}\rangle$:

$$g_i^{(a)}|g_j^{(a)}\rangle = (-1)^{\xi\left(\phi\left(g_{s_l}^{(a)}\right)\right)} exp\left(-iK_i \cdot \tau_j^{(a)}\right)|g_l^{(a)}\rangle, \quad \text{(E5)}$$

the matrix element

$$M_k\left(g_i^{(a)}\right)_{g,g_j^{(a)}} = \left\langle g\middle|g_i^{(a)}\middle|g_j^{(a)}\right\rangle = (-1)^{\xi\left(\phi\left(g_{s_l}^{(a)}\right)\right)} \exp\left(-iK_i \cdot \tau_j^{(a)}\right)\delta_{g,g_i^{(a)}g_j^{(a)}}. \quad \text{(E6)}$$

where $g_i$ have the form of $\left\{g_{s_i}^{(a)} \middle\| R_i \middle| \tau_i^{(a)}\right\}$, $K_i = R_i^{-1}k - k$ and the superscript $(a)$ labels different translations and spin operations accompanied with one $R_i$ .

$(-1)^{\xi(\phi(g_{s_l}^{(a)}))}$ equals 1 or -1 when $0 \le \phi(g_{s_l}^{(a)}) < 2\pi$ or $2\pi \le \phi(g_{s_l}^{(a)}) < 4\pi$, respectively.

2) *Class operator*: For a regular projetive rep of a certain finite group $\tilde{L}_{SS}^{k}$, the class operator is defined as:

$$C_i^{(a)} = \sum_{g_j^{(a)} \in \tilde{L}_{SS}^{k}} M_k\left(g_j^{(a)}\right)^{-1} M_k\left(g_i^{(a)}\right) M_k\left(g_j^{(a)}\right), \tag{E7}$$

where the class operator $C_i^{(a)}$ commutes with all the regular projective rep $M_k(g_j^{(a)})$.

3) Canonical subgroup chain: The canonical subgroup chain has a close relationship with the restricted reps. In the mathematical sense, an irrep $d_k^l$ of the group $\tilde{L}_k$ is certainly a rep of $\tilde{L}_k$'s subgroup $\tilde{L}_{k1}$, denoted by $d_k^l \downarrow \tilde{L}_{k1}$. ($\tilde{L}_k$ is the maximal unitary subgroup of $\tilde{G}_k$) In general, the restricted rep $d_k^l \downarrow \tilde{L}_{k1}$ is a reducible rep of $\tilde{L}_{k1}$, which can be reduced to a direct sum of the irreps of $\tilde{L}_{k1}$,

$$d_k^l \downarrow \tilde{L}_{k1} = \bigoplus_{v} a_v^l d_{k1}^v, \tag{E8}$$

where $a_v^l$ is the multiplicity that irrep $d_{k1}^v$ of $\tilde{L}_{k1}$ occurs in the restricted rep $d_k^l \downarrow \tilde{L}_{k1}$. When $a_v^l \le 1$ for all $l$ and $v$, $\tilde{L}_{k1}$ is said to be a canonical subgroup of $\tilde{L}_k$. A group chain $\tilde{L}_k > \tilde{L}_{k1} > \tilde{L}_{k2} > \cdots > \tilde{L}_{kn}$ is a canonical subgroup chain if $\tilde{L}_{k(i+1)}$ is a canonical subgroup of $\tilde{L}_{ki}$ for $i = 0, 1, \ldots, n-1$ and $\tilde{L}_{kn}$ is an Abelian group.

Although we do not know the irreps $d_k^l$ of the group $\tilde{L}_k$ in advance, we can still find the canonical subgroup by using certain empirical strategies in practical terms. Taking $\tilde{L}_k > \tilde{L}_{k1}$ as an example, $\tilde{L}_{k1}$ contains all the translations and pure spin symmetries of $\tilde{L}_k$ while spatial PG part $P(\tilde{L}_{k1})$ is one of the maximal subgroup of spatial PG part $P(\tilde{L}_k)$. In this way, $\tilde{L}_{k1}$ usually serves as a canonical subgroup of $\tilde{L}_k$ and the canonical subgroup chain could be constructed.

4) *Intrinsic group:* The intrinsic group $\bar{G}$ of a group $G$ is the group of elements with multiplication on basis defined as right multiplication. Specifically, for each element $R$ of a group $G$, we can define a corresponding operator $\bar{R}$ in the group rep space $L_g$ through the right-muliplication rule with:

$$\bar{R}S = SR \text{ for all } S \in L_g, \tag{E9}$$

The group formed by the collection of operators $\bar{R}$ is called the *intrinsic group* of $\bar{G}$.

5) *Intrinsic regular projective reps*: each group element $g_i^{(a)} \in \tilde{L}_{SS}^k$ corresponds to an element $\bar{g}_i^{(a)}$ in the intrinsic group $\overline{\tilde{L}_{SS}^k}$, which obey right multiplication rule with $\bar{g}_i^{(a)} g_j^{(a)} = g_j^{(a)} g_i^{(a)}$. Distinct from the typical right regular reps, $\overline{\tilde{L}_{SS}^k}$ is anti-isomorphic to $\tilde{L}_{SS}^k$. Thus, one can define the intrinsic regular projective reps:

$$\bar{g}_i^{(a)} \left| g_j^{(a)} \right\rangle = (-1)^{\xi(\phi(g_{s_m}^{(a)}))} exp(-iK_j \cdot \tau_i^{(a)}) \left| g_m^{(a)} \right\rangle, \tag{E10}$$

in which $g_m^{(a)} = \{g_{s_m}^{(a)} || R_m | \tau_m^{(a)}\} = \{g_{s_j}^{(a)} g_{s_i}^{(a)} || R_j R_i | (\tau_j^{(a)} + R_j \tau_i^{(a)})\}$. Note that regular projective reps of $\tilde{G}_{SS}^k$ commutes with its intrinsic regular projective reps.

6) CSCO-I: the linear combination of class operators $C(\tilde{L}_{SS}^k) = \sum_i \lambda_i C_i^{(a)}$ form CSCO-I of the left regular projective rep, where $\lambda_i$ is any constant.

7) CSCO-II: For the canonical subgroup chain of $\tilde{L}_{SS}^k$, $\tilde{L}_{SS}^k(s) = \tilde{L}_{SS1}^k > \tilde{L}_{SS2}^k > \cdots$, repeat the procedure in CSCO-I:

$$C_{i1}^{(a)} = \sum_{g_j^{(a)} \in \tilde{L}_{SS1}^k} M_k\left(g_{j1}^{(a)}\right)^{-1} M_k\left(g_{i1}^{(a)}\right) M_k\left(g_{j1}^{(a)}\right), \tag{E11}$$

where the linear combination of class operators $C(\tilde{L}_{SS1}^k) = \sum_i \lambda_i C_{i1}^{(a)}$ form CSCO-Is for the subgroup chain. The collection of the corresponding class operators $C(s) = (C(\tilde{L}_{SS1}^k), C(\tilde{L}_{SS2}^k), \dots)$ commute with the $C$, forming the operator set $(C, C(s))$, also known as CSCO-II.

8) CSCO-III: For the canonical subgroup chain of $\overline{\tilde{L}_{SS}^k}$, $\overline{\tilde{L}_{SS}^k}(s) = \overline{\tilde{L}_{SS1}^k} > \overline{\tilde{L}_{SS2}^k} > \cdots$, repeat the procedure in CSCO-I:

$$\bar{C}_{i1}^{(a)} = \sum_{g_j^{(a)} \in \overline{\tilde{L}_{SS1}^k}} M_k\left(\bar{g}_{j1}^{(a)}\right)^{-1} M_k\left(\bar{g}_{i1}^{(a)}\right) M_k\left(\bar{g}_{j1}^{(a)}\right), \tag{E12}$$

where the linear combination of class operators $C(\overline{\tilde{L}_{SS1}^k}) = \sum_i \lambda_i \bar{C}_{i1}^{(a)}$ form CSCO-Is for the subgroup chain of the intrinsic group $\overline{\tilde{L}_{SS}^k}$. The collection of the corresponding class operators $\bar{C}(s)=(\bar{C}(\tilde{L}_{SS1}^{k}),\bar{C}(\tilde{L}_{SS2}^{k}),\ldots)$ commute with the $C$ and $C(s)$. This complete set of class operators $(C,C(s),\bar{C}(s))$ is named as CSCO-III.

Now we show the decomposition of the regular projective reps of the little group $\tilde{G}_{SS}^{k}$ using the CSCO method.

1. The left regular projective reps $M_k(g_i^{(a)})$ are block-diagonalized to obtain all the projective irreps, where the projective reps $M_k(g_i^{(a)})$ are $N_k\times N_k$ reducible matrices (shown in Fig. E1 (a)).

Starting from the unitary $\tilde{L}_{SS}^{k}$, in which the character $\chi_i^l$ of a projective irrep is a function of class operators $C_i^{(a)}$ as defined in Eq. (E7). The number of independent classes $N_c$ equals to the number of projective irreps; and a $n_l$-dimensional projective irrep will appear $n_l$ times in $M_k(g_i^{(a)})$. By constructing the linear combination of class operators $C=\sum_i k_i C_i^{(a)}$ ($k_i$ are arbitrary constant numbers), we can decompose $M_k(g_i^{(a)})$ into $N_c$ blocks. The number of eigenvalues of $C$ is equal to the number $N_c$ of inequivalent irreps. Analogous to quantum mechanics, we obtain the principal quantum number $|j\rangle$.

2. Generally, these principle quantum numbers are degenerate and cannot distinguish each of the irreducible bases. Therefore, the magnetic quantum number is introduced from the canonical subgroup chain of $\tilde{L}_{SS}^{k}$, $\tilde{L}_{SS}^{k}(s)=\tilde{L}_{SS1}^{k}>\tilde{L}_{SS2}^{k}>\cdots$ [68, 90]. The collection of the corresponding class operators $C(s)=(C(\tilde{L}_{SS1}^{k}),C(\tilde{L}_{SS2}^{k}),\ldots)$ commute with the $C$, forming a CSCO-II set of $(C,C(s))$ together with $C$ and provide the required magnetic quantum numbers $m_j$ (shown in Fig. E1(b)). Such a set helps us to distinguish the rows of a $n_l$-dimensional block of $M_k(g_i^{(a)})$. The CSCO-II can distinguish all the bases only if each irrep is one-dimensional in the reduced regular rep.

3. However, in a left regular projective rep, an $n_l$-dimensional projective irrep occurs $n_l$ times (See Appendix A4). Therefore, we need another canonical subgroup chain of the intrinsic group $\overline{\tilde{L}_{SS}^{k}}$, $\overline{\tilde{L}_{SS}^{k}}(s)=\overline{\tilde{L}_{SS1}^{k}}>\overline{\tilde{L}_{SS2}^{k}}>\cdots$ and the set of corresponding class operators, $\bar{C}(s)=(\bar{C}(\tilde{L}_{SS1}^{k}),\bar{C}(\tilde{L}_{SS2}^{k}),\ldots)$ to lift the remaining degeneracy. Finally, we have a CSCO set of $(C,C(s),\bar{C}(s))$ and a corresponding collection of eigenvalues, say $|j,m_j,\bar{m}_j\rangle$, of group $\tilde{L}_{SS}^{k}$. By rearranging their eigenvectors, we can construct a transformation matrix to block diagonalize $M_k(g_i^{(a)})$ and get all projective irreps (shown in Fig. E1(c)).

### 3. Modified Dimmock and Wheeler's character sum rule for SSG

For SSGs, the anitiunitary little group can also be decomposed into its maximal unitary subgroups $G_{SS}^k = L_{SS}^k \cup TAL_{SS}^k$, where $L_{SS}^k$ is a unitary subgroup of index 2 in $G_{SS}^k$. For calculating the co-irreps, we define the quotient group $\tilde{G}_{SS}^k = \tilde{L}_{SS}^k \cup TA\tilde{L}_{SS}^k$. Now we prove the modified Dimmock and Wheeler's character sum rule to account for the co-irreps in SSGs.

The basis set of irrep $d_k^l(g_i^{(a)})$ of the maximal unitary group $\tilde{L}_{SS}^k$ is defined as $|\psi\rangle = |\psi_1, \psi_2, \ldots, \psi_{n_l}\rangle$. When the anti-unitary group element is taken into consideration, the basis set can be denoted as $|\phi\rangle = |\phi_1, \phi_2, \ldots, \phi_{n_l}\rangle = TA|\psi_1, \psi_2, \ldots, \psi_{n_l}\rangle$. Consequently, the co-reps for the full basis set $|\psi, \phi\rangle$ are:

$$D_k^l\left(g_i^{(a)}\right) = \begin{bmatrix} d_k^l\left(g_i^{(a)}\right) & 0 \\ 0 & d_k^l\left(A^{-1}g_i^{(a)}A\right)^* \end{bmatrix}, \tag{E13}$$

$$D_k^l\left(TAg_i^{(a)}\right) = \begin{bmatrix} 0 & d_k^l\left(TAg_i^{(a)}TA\right) \\ d_k^l\left(g_i^{(a)}\right)^* & 0 \end{bmatrix}. \tag{E14}$$

Two co-reps $D_k^{l_2}$ and $D_k^l$ are unitarily equivalent if there exists a unitary matrix $U$ such that:

$$D_k^{l_2}\left(g_i^{(a)}\right) = U^{-1}D_k^l\left(g_i^{(a)}\right)U \text{ for all } g_i^{(a)} \in \tilde{L}_{SS}^k, \tag{E15}$$

$$D_k^{l_2}\left(B_i^{(a)}\right) = U^{-1}D_k^l\left(B_i^{(a)}\right)U^* \text{ for all } B_i^{(a)} \in TA\tilde{L}_{SS}^k, \tag{E16}$$

For the modified Dimmock and Wheeler's character sum rule, which helps us calculate the co-irreps $D_k^l$ of $\tilde{G}_{SS}^k$ from the irreps of $\tilde{L}_{SS}^k$, or the Eq. (8) in our main text,

$$\sum_{g_i^{(a)} \in \tilde{L}_{SS}^k} \chi\left(\left(TAg_i^{(a)}\right)^2\right) = \begin{cases} +\left|\tilde{L}_{SS}^k\right| & (a) \\ -\left|\tilde{L}_{SS}^k\right| & (b) \\ 0 & \text{(c)} \end{cases}, \tag{E17}$$

where $\chi$ is the character of $d_k^l$.

For the case (a), the co-rep matrices $D_k^l$ have identical dimensions with the irrep $d_k^l$:

$$D_k^l\left(g_i^{(a)}\right) = d_k^l\left(g_i^{(a)}\right), \tag{E18}$$

$$D_k^l\left(TAg_i^{(a)}\right) = \pm d_k^l\left(Ag_i^{(a)}A^{-1}\right)C, \tag{E19}$$

where $C$ is an unitary matrix with $CC^* = +d_k^l((TA)^2)$. That means the irrep $d_k^l(g_i^{(a)})$ is equivalent to $d_k^l(A^{-1}g_i^{(a)}A)^*$ and they are associated by the matrix transformation:

$$d_k^l\left(g_i^{(a)}\right) = Cd_k^l\left(A^{-1}g_i^{(a)}A\right)^* C^{-1}. \tag{E20}$$

For the case (b), the dimension of $D_k^l$ is doubled compared with $d_k^l$:

$$D_k^l\left(g_i^{(a)}\right) = \begin{bmatrix} d_k^l\left(g_i^{(a)}\right) & 0 \\ 0 & d_k^l\left(g_i^{(a)}\right) \end{bmatrix}, \tag{E21}$$

$$D_k^l\left(TAg_i^{(a)}\right) = \begin{bmatrix} 0 & -d_k^l\left(TAg_i^{(a)}(TA)^{-1}\right)C \\ d_k^l\left(TAg_i^{(a)}(TA)^{-1}\right)C & 0 \end{bmatrix}, \tag{E22}$$

where $C$ is an unitary matrix with $CC^* = +d_k^l((TA)^2)$. $d_k^l(g_i^{(a)})$ and $d_k^l(A^{-1}g_i^{(a)}A)^*$ still satisfy the Eq. (E20).

For the case (c), the dimension of $D_k^l$ is also doubled with respect to $d_k^l$:

$$D_k^l\left(g_i^{(a)}\right) = \begin{bmatrix} d_k^l\left(g_i^{(a)}\right) & 0 \\ 0 & d_k^l\left(A^{-1}g_i^{(a)}A\right)^* \end{bmatrix}, \tag{E23}$$

$$D_k^l\left(TAg_i^{(a)}\right) = \begin{bmatrix} 0 & d_k^l\left(TAg_i^{(a)}TA\right) \\ d_k^l\left(g_i^{(a)}\right)^* & 0 \end{bmatrix}. \tag{E24}$$

However, $d_k^l(g_i^{(a)})$ and $d_k^l(A^{-1}g_i^{(a)}A)^*$ are inequivalent irreps of $\tilde{L}_{SS}^k$.

The anti-unitary operation $A_T$ can be expressed in terms of the time reversal operator $T$ and its unitary part $A$:

$$A_T = TA. \tag{E25}$$

In this way:

$$\begin{aligned}
&\sum\nolimits_{B_i^{(a)}\in A_T\tilde{L}_{SS}^k}\chi\left((B_i^{(a)})^2\right)\\
&=\sum\nolimits_{B_i^{(a)}\in A_T\tilde{L}_{SS}^k}\sum\nolimits_i d_k^l\left((B_i^{(a)})^2\right)_{ii}\\
&=\sum\nolimits_i\sum\nolimits_{g_i^{(a)}\in\tilde{L}_{SS}^k}\left[d_k^l({A_T}^2)d_k^l\left({A_T}^{-1}g_i^{(a)}A_T\right)d_k^l\left(g_i^{(a)}\right)\right]_{ii}\\
&=\sum\nolimits_{s,t}\sum\nolimits_i\sum\nolimits_{g_i^{(a)}\in\tilde{L}_{SS}^k}d_k^l({A_T}^2)_{is}d_k^l\left({A_T}^{-1}g_i^{(a)}A_T\right)_{st}d_k^l\left(g_i^{(a)}\right)_{ti}
\end{aligned}\tag{E26}$$

When $d_k^l(g_i^{(a)})$ and $d_k^l({A_T}^{-1}g_i^{(a)}A_T)^*$ are inequivalent

$$\sum\nolimits_{g_i^{(a)}\in\tilde{L}_{SS}^k}d_k^l\left({A_T}^{-1}g_i^{(a)}A_T\right)_{st}d_k^l\left(g_i^{(a)}\right)_{ti}=0.\tag{E27}$$

When $d_k^l(g_i^{(a)})$ and $d_k^l({A_T}^{-1}g_i^{(a)}A_T)^*$ are related by a unitary transformation matrix $C$ with $d_k^l\left(g_i^{(a)}\right)=Cd_k^l({A_T}^{-1}g_i^{(a)}A_T)^*C^{-1}$ and $CC^*=\pm d_k^l({A_T}^2)$:

$$\begin{aligned}
&\sum\nolimits_{B_i^{(a)}\in A_T\tilde{L}_{SS}^k}\chi\left(\left(B_i^{(a)}\right)^2\right)\\
&=\sum\nolimits_{s,t}\sum\nolimits_i d_k^l({A_T}^2)_{is}\sum\nolimits_{g_i^{(a)}\in\tilde{L}_{SS}^k}\left(C^{-1*}d_k^l\left(g_i^{(a)}\right)^*C^*\right)_{st}d_k^l\left(g_i^{(a)}\right)_{ti}\\
&=\sum\nolimits_{s,t}\sum\nolimits_i\sum\nolimits_{p,q}d_k^l({A_T}^2)_{is}\sum\nolimits_{g_i^{(a)}\in\tilde{L}_{SS}^k}{C_{sp}}^{-1*}d_k^l\left(g_i^{(a)}\right)_{pq}^*{C_{qt}}^*d_k^l\left(g_i^{(a)}\right)_{ti}\\
&=\sum\nolimits_{s,t}\sum\nolimits_i\sum\nolimits_{p,q}d_k^l({A_T}^2)_{is}{C_{sp}}^{-1*}{C_{qt}}^*\sum\nolimits_{g_i^{(a)}\in\tilde{L}_{SS}^k}d_k^l\left(g_i^{(a)}\right)_{pq}^*d_k^l\left(g_i^{(a)}\right)_{ti}
\end{aligned}\tag{E28}$$

Here we transform the irreps into projective ones:

$$\begin{aligned}
&\sum\nolimits_{g_i^{(a)}\in\tilde{L}_{SS}^k}d_k^l\left(g_i^{(a)}\right)_{pq}^*d_k^l\left(g_i^{(a)}\right)_{ti}\\
&\qquad=\sum\nolimits_{g_i^{(a)}\in\tilde{L}_{SS}^k}\exp\left(ik_1\cdot\left({\tau_i}_i^{(a)}\right)\right)M_k^l\left(g_i^{(a)}\right)_{pq}^*\exp\left(-ik_1\right.\\
&\qquad\qquad\left.\cdot\left({\tau_i}_i^{(a)}\right)\right)M_k^l\left(g_i^{(a)}\right)_{ti}\\
&=\sum\nolimits_{g_i^{(a)}\in\tilde{L}_{SS}^k}M_k^l\left(g_i^{(a)}\right)_{pq}^*M_k^l\left(g_i^{(a)}\right)_{ti}.
\end{aligned}\tag{E29}$$

Using the orthogonality relationships for unitary projective irreps (equation (3.7.17) in Ref. [2]):

$$\sum\nolimits_{g_i^{(a)}\in\tilde{L}_{SS}^k}M_k^l\left(g_i^{(a)}\right)_{pq}^*M_k^l\left(g_i^{(a)}\right)_{ti}=\frac{N_k}{n_l}\delta_{p,t}\delta_{q,i}.\tag{E30}$$

Where $n_l$ is the dimension of projective irrep $M_k^l$.we finally obtain:

$$\begin{aligned}\sum\nolimits_{B_i^{(a)}\in A_T\tilde{L}_{SS}^k}\chi\left((B_i^{(a)})^2\right) &= \sum\nolimits_{s,t}\sum\nolimits_{i}\sum\nolimits_{p,q} d_k^l({A_T}^2)_{is}{C_{sp}}^{-1*}{C_{qt}}^{*}\frac{N_k}{n_l}\delta_{p,t}\delta_{q,i} \\ &= \frac{N_k}{n_l}\sum\nolimits_{s,t}\sum\nolimits_{i} d_k^l({A_T}^2)_{is}{C_{st}}^{-1*}{C_{it}}^{*} \\ &= \frac{N_k}{n_l}\sum\nolimits_{s,i} d_k^l({A_T}^2)_{is}(C^*C)_{is} \\ &= \frac{N_k}{n_l}\sum\nolimits_{s,i} d_k^l({A_T}^2)_{is}(\pm d_k^l({A_T}^2)^*)_{is} \\ &= \frac{N_k}{n_l}\sum\nolimits_{s,i} d_k^l({A_T}^2)_{is}(\pm d_k^l({A_T}^{-2}))_{si} \\ &= \pm\frac{N_k}{n_l}\sum\nolimits_{i} d_k^l(E)_{ii} \\ &= \pm N_k\end{aligned} \tag{E31}$$

When $CC^* = +d_k^l({A_T}^2)$, $\sum_{B_i^{(a)}\in A_T\tilde{L}_{SS}^k}\chi\left((B_i^{(a)})^2\right) = +N_k$ corresponding to Eq. (E17)-(a).

When $CC^* = -d_k^l({A_T}^2)$, $\sum_{B_i^{(a)}\in A_T\tilde{L}_{SS}^k}\chi\left((B_i^{(a)})^2\right) = -N_k$ corresponding to Eq. (E17)-(b).

## F. Band representations in collinear SSGs

Now we turn to the derivation of co-irreps in collinear SSGs. Considering the spin-only group $G_{SO}^{l} = Z_2^K \ltimes SO(2) = {}^{\infty m}1$, the SSGs describing collinear ferromagnets and ferrimagnets with $G_0 = L_0, G^s = 1$ are expressed by

$$G_{SS} = \{E||L_0\} \times Z_2^K \ltimes SO(2), \tag{F1}$$

while for SSGs describing collinear antiferromagnets, $L_0$ is an index 2 normal subgroup of $G_0$ and $G^s = -1$, they can be written by

$$G_{SS} = (\{E||L_0\} \cup \{T||AL_0\}) \times Z_2^K \ltimes SO(2). \tag{F2}$$

Note that $G_{SS}$ actually is the quotient group with respect to the translation group $\mathbb{T}$.

### 1. Double-valued representation

Since we are focus on the SOC-free electronic band structure, then the spinor rep or double-valued rep are necessary for constructing the co-irreps of collinear SSGs. The spinor rep of rotations read

$$D\big(U_{\hat{\boldsymbol{n}}}(\phi)\big) = exp\left(-i\sigma \cdot \frac{\hat{\boldsymbol{n}}\phi}{2}\right)$$
$$= \begin{pmatrix} cos\left(\frac{\phi}{2}\right) - in_z sin\left(\frac{\phi}{2}\right) & (-in_x - n_y)sin\left(\frac{\phi}{2}\right) \\ (-in_x + n_y)sin\left(\frac{\phi}{2}\right) & cos\left(\frac{\phi}{2}\right) + in_z sin\left(\frac{\phi}{2}\right) \end{pmatrix}, \tag{F3}$$

where $\hat{\boldsymbol{n}} = \left(n_x, n_y, n_z\right)$ is the direction vector of the rotation axis $\hat{\boldsymbol{n}}$, $\phi$ is the rotation angle.

The character table of the double group $SO(2)$ are given in Table F1.

**Table F1:** Character table of the double group of $SO(2)$. ($2m \in \mathbb{Z}, 0 < \phi < 2\pi$)

| $SO(2)$ | $E$ | $U_z(\phi)$ |
|---|---|---|
| $\Gamma_0$ | 1 | 1 |
| $\Gamma_m$ | 1 | $e^{-im\phi}$ |

In collinear SSG, there are only nontrivial spin-space symmetries $T$, $U_{\boldsymbol{x}}(\pi)$, $U_z(\phi)$ and their combinations. (Here we define the spin principle axis and any axis perpendicular to the spin principle axis as the *z* and *x* direction, respectively). For electron systems with $m = \pm 1/2$,

$$D(U_z(\phi)) = \begin{pmatrix} e^{-\frac{i\phi}{2}} & 0 \\ 0 & e^{\frac{i\phi}{2}} \end{pmatrix}, \tag{F4}$$

$$D(T) = \begin{pmatrix} 0 & 1 \\ -1 & 0 \end{pmatrix} K, \tag{F5}$$

$$D(U_x(\pi)) = \begin{pmatrix} 0 & -i \\ -i & 0 \end{pmatrix}, \tag{F6}$$

where $K$ denotes complex conjugation.

In addition, we can find that

$$(U_x(\pi))^2 = T^2 = -E, \tag{F7}$$

$$(TU_x(\pi))^2 = E, \tag{F8}$$

$$(TU_x(\pi)U_z(\phi))^2 = E, \tag{F9}$$

$$(TU_z(\phi))^2 = U_z(2\phi), \tag{F10}$$

## 2. General rules for doubled degeneracy in SSGs

(1) The combination of $SO(2)$ with $U_x(\pi)A$ symmetry pairs two conjugated one-dimensional irreps $\Gamma^S_{+1/2}(1)$ and $\Gamma^S_{-1/2}(1)$ into a two-dimensional irrep $\Gamma^S_{\pm 1/2}(2)$ as shown in Table F2.

**Table F2:** Character table of the double group of ∞2. ($2m \in \mathbb{Z}, 0 < \phi < 2\pi$)

| ∞2 | $E$ | $U_z(\phi)$ | $U_x(\pi)$ |
|---|---|---|---|
| $\Gamma_{0g}$ | 1 | 1 | 1 |
| $\Gamma_{0u}$ | 1 | 1 | -1 |
| $\Gamma_m$ | 2 | $2cos(m\phi)$ | 0 |

(2) The combination of $SO(2)$ with $TA$ symmetry pairs two conjugated one-dimensional irreps $\Gamma^S_{+1/2}(1)$ and $\Gamma^S_{-1/2}(1)$ into a two-dimensional co-irrep $\Gamma^S_{\pm 1/2}(2)$ according to the Dimmock and Wheeler's character sum rule ($L = \infty$):

$$\sum_{\beta \in TAL} \chi(\beta^2) = \sum_{\phi} \chi\left((TAU_z(\phi))^2\right) = \sum_{\phi} \chi(U_z(2\phi)) = \sum_{\phi} \chi\begin{pmatrix} e^{-i\phi} & 0 \\ 0 & e^{i\phi} \end{pmatrix}$$

$$= \int_0^{2\pi} 2cos(2\phi) = 0 \cdots\ (c). \quad \text{(F11)}$$

(3) The combination of $SO(2)$ with $TU_x(\pi)$ symmetry cannot contribute to extra degeneracy in spin space according to the Dimmock and Wheeler's character sum rule ($L = \infty$):

$$\sum_{\beta \in TU_x(\pi)L} \chi(\beta^2) = \sum_{\phi} \chi\left(\left(TU_x(\pi)U_z(\phi)\right)^2\right)$$

$$= \sum_{\phi} \chi(E) = g \cdots\ (a). \quad \text{(F12)}$$

(4) The $TU_x(\pi)$ can combine two conjugated single-valued irreps in real space, similar to that in SGs.

(5) The concurrence of $SO(2)$ and $TU_x(\pi)$, $U_x(\pi)A$ (also $TA$) cannot contribute to quadruple degeneracy in spin space. The combination of $TU_x(\pi)$ and $\infty 2$ cannot pair the real two-dimensional irrep $\Gamma^S_{\pm 1/2}(2)$ of $\infty 2$, as calculated from the character sum rule ($L = \infty 2$):

$$\sum_{\beta \in TU_x(\pi)L} \chi(\beta^2) = \sum_{\phi} \left\{\chi\left(\left(TU_z(\phi)\right)^2\right) + \chi\left(\left(TU_x(\pi)U_z(\phi)\right)^2\right)\right\}$$

$$= 0 + g = g \qquad \cdots\ (a). \quad \text{(F13)}$$

**Table F3:** Some typical little co-group in collinear SSG.

| little cogroup | sum rule | type | Degeneracy in spin space |
|---|---|---|---|
| ${}^{\infty}1$ | | | |
| ${}^{2}1^{\infty}1$ | | | double |
| ${}^{-1}2^{\infty}1$ | $\sum_{\beta \in TAL} \chi(\beta^2) = 0$ | (c)* | double |
| ${}^{-1}{-1}^{\infty}1$ | $\sum_{\beta \in TAL} \chi(\beta^2) = 0$ | (c)* | double |
| ${}^{m}1^{\infty}1$ | $\sum_{\beta \in TAL} \chi(\beta^2) = g$ | (a) | no |
| ${}^{m}-1^{\infty}1$ | $\sum_{\beta \in TAL} \chi(\beta^2) = g$ | (a) | no |
| ${}^{-1}-1^{2}1^{\infty}1$ | $\sum_{\beta \in TAL} \chi(\beta^2) = g$ | (a) | double* |

* This double degeneracy in spin space arise from the unitary irrep of $\infty 2$, rather than the extra degeneracy when introducing the antiunitary element.

## 3. Little (co)groups of altermagnet $RuO_2$

This section presents the derivation of the co-irreps of little groups at high-symmetry points in the collinear SSG for the altermagnet $RuO_2$. First, the Brillouin zone and the little group for collinear antiferromagnet $RuO_2$ are given in Fig. F1 and Table F5, respectively.

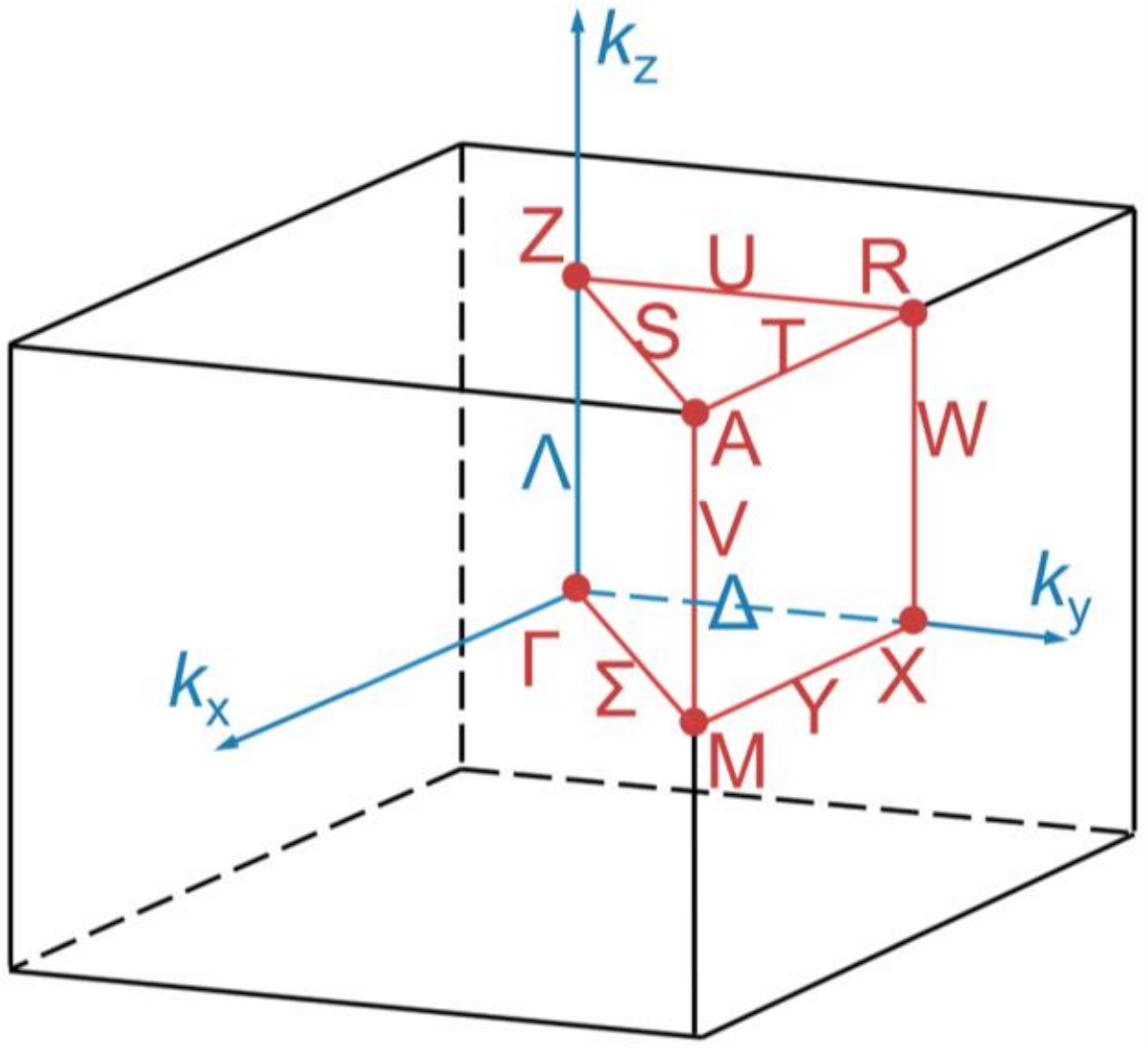

**Fig. F1:** Schematic of spin Brillouin zone for SSG $P^{-1}4_2/{}^{1}m^{-1}n^{1}m^{\infty m}1$, sharing the same magnetic Brillouin zone for MSG $P4_2'/mnm'$.

For collinear SSG describing antiferromagnets, they can be expressed by

$$G_{SS} = (\{E||L_0\} \cup \{T||AL_0\}) \times Z_2^K \ltimes SO(2)$$

$$= (\{E||L_0\} \cup \{T||AL_0\} \cup \{TU_n(\pi)||L_0\} \cup \{U_n(\pi)||AL_0\}) \ltimes SO(2). \quad \text{(F14)}$$

The nontrivial SSG of antiferromagnetic $RuO_2$ is $P^{-1}4_2/{}^{1}m^{-1}n^{1}m$ (65.136.1.1)

and $L_0 \cong C$mmm (65). Considering the spin-only group $G_{SO}^{l} = Z_2^K \ltimes SO(2) = {}^{\infty m}1$, it turns into $P^{-1}4_2/{}^{1}m^{-1}n^{1}m^{\infty m}1$. The complete list of 32 symmetry operations in $\{E||L_0\} \cup \{T||AL_0\} \cup \{TU_n(\pi)||L_0\} \cup \{U_n(\pi)||AL_0\}$ is provided in Table F1. Here $m_{\boldsymbol{n}}$, $2_{\boldsymbol{n}}$ denote $TU_{\boldsymbol{n}}(\pi)$ and $U_{\boldsymbol{n}}(\pi)$, respectively. In addition, $\tau = (1/2,1/2,1/2)$ denotes a body-centered half integer translation.

**Table F4:** Symmetry operations in $\{E||L_0\} \cup \{T||AL_0\} \cup \{TU_n(\pi)||L_0\} \cup \{U_n(\pi)||AL_0\}$.

| | Symmetry operations |
|---|---|
| $\{E\|\|L_0\}$ | $\{1\|\|1\|0\}$, $\{1\|\|2_{001}\|0\}$, $\{1\|\|2_{110}\|0\}$, $\{1\|\|2_{1-10}\|0\}$, $\{1\|\|-1\|0\}$, $\{1\|\|m_{001}\|0\}$, $\{1\|\|m_{110}\|0\}$, $\{1\|\|m_{1-10}\|0\}$ |
| $\{T\|\|AL_0\}$ | $\{-1\|\|4_{001}^1\|\tau\}$, $\{-1\|\|4_{001}^3\|\tau\}$, $\{-1\|\|2_{100}\|\tau\}$, $\{-1\|\|2_{010}\|\tau\}$, $\{-1\|\|-4_{001}^1\|\tau\}$, $\{-1\|\|-4_{001}^3\|\tau\}$, $\{-1\|\|m_{100}\|\tau\}$, $\{-1\|\|m_{010}\|\tau\}$ |
| $\{TU_n(\pi)\|\|L_0\}$ | $\{m_{\boldsymbol{n}}\|\|1\|0\}$, $\{m_{\boldsymbol{n}}\|\|2_{001}\|0\}$, $\{m_{\boldsymbol{n}}\|\|2_{110}\|0\}$, $\{m_{\boldsymbol{n}}\|\|2_{1-10}\|0\}$, $\{m_{\boldsymbol{n}}\|\|-1\|0\}$, $\{m_{\boldsymbol{n}}\|\|m_{001}\|0\}$, $\{m_{\boldsymbol{n}}\|\|m_{110}\|0\}$, $\{m_{\boldsymbol{n}}\|\|m_{1-10}\|0\}$ |
| $\{U_n(\pi)\|\|AL_0\}$ | $\{2_{\boldsymbol{n}}\|\|4_{001}^1\|\tau\}$, $\{2_{\boldsymbol{n}}\|\|4_{001}^3\|\tau\}$, $\{2_{\boldsymbol{n}}\|\|2_{100}\|\tau\}$, $\{2_{\boldsymbol{n}}\|\|2_{010}\|\tau\}$, $\{2_{\boldsymbol{n}}\|\|-4_{001}^1\|\tau\}$, $\{2_{\boldsymbol{n}}\|\|-4_{001}^3\|\tau\}$, $\{2_{\boldsymbol{n}}\|\|m_{100}\|\tau\}$, $\{2_{\boldsymbol{n}}\|\|m_{010}\|\tau\}$ |

The reciprocal SG of $P^{-1}4_2/{}^{1}m^{-1}n^{1}m^{\infty m}1$ is the same as $P4/mmm$ (123). The Brillouin zone of $P^{-1}4_2/{}^{1}m^{-1}n^{1}m^{\infty m}1$ is depicted in Fig. F1.

The little groups for $P^{-1}4_2/{}^{1}m^{-1}n^{1}m^{\infty m}1$ at every high symmetry points (HSPs) are shown in Table F5. Since ${}^{\infty}1$ is always invariant for any *k* points, we only list the elements in $\{E||L_0\} \cup \{T||AL_0\} \cup \{TU_n(\pi)||L_0\} \cup \{U_n(\pi)||AL_0\}$.

**Table F5:** Little group of $P^{-1}4_2/{}^{1}m^{-1}n^{1}m^{\infty m}1$ at each *k* point.

| HSP | Little group |
|---|---|
| Γ (0, 0, 0) | All 32 elements |
| X (0, 1/2, 0) | $\{1\|\|1\|0\}$, $\{1\|\|2_{001}\|0\}$, $\{1\|\|-1\|0\}$, $\{1\|\|m_{001}\|0\}$, $\{-1\|\|2_{100}\|\tau\}$, $\{-1\|\|2_{010}\|\tau\}$, $\{-1\|\|m_{100}\|\tau\}$, $\{-1\|\|m_{010}\|\tau\}$, $\{m_{\boldsymbol{n}}\|\|1\|0\}$, $\{m_{\boldsymbol{n}}\|\|2_{001}\|0\}$, $\{m_{\boldsymbol{n}}\|\|-1\|0\}$, $\{m_{\boldsymbol{n}}\|\|m_{001}\|0\}$, $\{2_{\boldsymbol{n}}\|\|2_{100}\|\tau\}$, $\{2_{\boldsymbol{n}}\|\|2_{010}\|\tau\}$, $\{2_{\boldsymbol{n}}\|\|m_{100}\|\tau\}$, $\{2_{\boldsymbol{n}}\|\|m_{010}\|\tau\}$ |
| M (1/2, 1/2, 0) | All 32 elements |
| Z (0, 0, 1/2) | All 32 elements |
| R (0, 1/2, 1/2) | $\{1\|\|1\|0\}, \{1\|\|2_{001}\|0\}, \{1\|\|-1\|0\}, \{1\|\|m_{001}\|0\}, \{-1\|\|2_{100}\|\tau\}, \{-1\|\|2_{010}\|\tau\}, \{-1\|\|m_{100}\|\tau\}, \{-1\|\|m_{010}\|\tau\}, \{m_{\boldsymbol{n}}\|\|1\|0\}, \{m_{\boldsymbol{n}}\|\|2_{001}\|0\}, \{m_{\boldsymbol{n}}\|\|-1\|0\}, \{m_{\boldsymbol{n}}\|\|m_{001}\|0\}, \{2_{\boldsymbol{n}}\|\|2_{100}\|\tau\}, \{2_{\boldsymbol{n}}\|\|2_{010}\|\tau\}, \{2_{\boldsymbol{n}}\|\|m_{100}\|\tau\}, \{2_{\boldsymbol{n}}\|\|m_{010}\|\tau\}$ |
| A (1/2, 1/2, 1/2) | All 32 elements |
| Δ (0, v, 0) | $\{1\|\|1\|0\}, \{1\|\|m_{001}\|0\}, \{-1\|\|2_{100}\|\tau\}, \{-1\|\|m_{010}\|\tau\}, \{m_{\boldsymbol{n}}\|\|2_{001}\|0\}, \{m_{\boldsymbol{n}}\|\|-1\|0\}, \{2_{\boldsymbol{n}}\|\|2_{010}\|\tau\}, \{2_{\boldsymbol{n}}\|\|m_{100}\|\tau\}$ |
| Y (u, 1/2, 0) | $\{1\|\|1\|0\}, \{1\|\|m_{001}\|0\}, \{-1\|\|2_{010}\|\tau\}, \{-1\|\|m_{100}\|\tau\}, \{m_{\boldsymbol{n}}\|\|2_{001}\|0\}, \{m_{\boldsymbol{n}}\|\|-1\|0\}, \{2_{\boldsymbol{n}}\|\|2_{100}\|\tau\}, \{2_{\boldsymbol{n}}\|\|m_{010}\|\tau\}$ |
| Σ (u, u, 0) | $\{1\|\|1\|0\}, \{1\|\|2_{110}\|0\}, \{1\|\|m_{001}\|0\}, \{1\|\|m_{1-10}\|0\}, \{m_{\boldsymbol{n}}\|\|2_{001}\|0\}, \{m_{\boldsymbol{n}}\|\|2_{1-10}\|0\}, \{m_{\boldsymbol{n}}\|\|-1\|0\}, \{m_{\boldsymbol{n}}\|\|m_{110}\|0\}$ |
| Λ (0, 0, w) | $\{1\|\|1\|0\}, \{1\|\|2_{001}\|0\}, \{1\|\|m_{110}\|0\}, \{1\|\|m_{1-10}\|0\}, \{-1\|\|2_{100}\|\tau\}, \{-1\|\|2_{010}\|\tau\}, \{-1\|\|-4^{1}_{001}\|\tau\}, \{-1\|\|-4^{3}_{001}\|\tau\}, \{m_{\boldsymbol{n}}\|\|2_{110}\|0\}, \{m_{\boldsymbol{n}}\|\|2_{1-10}\|0\}, \{m_{\boldsymbol{n}}\|\|-1\|0\}, \{m_{\boldsymbol{n}}\|\|m_{001}\|0\}, \{2_{\boldsymbol{n}}\|\|4^{1}_{001}\|\tau\}, \{2_{\boldsymbol{n}}\|\|4^{3}_{001}\|\tau\}, \{2_{\boldsymbol{n}}\|\|m_{100}\|\tau\}, \{2_{\boldsymbol{n}}\|\|m_{010}\|\tau\}$ |
| V (1/2, 1/2, w) | $\{1\|\|1\|0\}, \{1\|\|2_{001}\|0\}, \{1\|\|m_{110}\|0\}, \{1\|\|m_{1-10}\|0\}, \{-1\|\|2_{100}\|\tau\}, \{-1\|\|2_{010}\|\tau\}, \{-1\|\|-4^{1}_{001}\|\tau\}, \{-1\|\|-4^{3}_{001}\|\tau\}, \{m_{\boldsymbol{n}}\|\|2_{110}\|0\}, \{m_{\boldsymbol{n}}\|\|2_{1-10}\|0\}, \{m_{\boldsymbol{n}}\|\|-1\|0\}, \{m_{\boldsymbol{n}}\|\|m_{001}\|0\}, \{2_{\boldsymbol{n}}\|\|4^{1}_{001}\|\tau\}, \{2_{\boldsymbol{n}}\|\|4^{3}_{001}\|\tau\}, \{2_{\boldsymbol{n}}\|\|m_{100}\|\tau\}, \{2_{\boldsymbol{n}}\|\|m_{010}\|\tau\}$ |
| W (0, 1/2, w) | $\{1\|\|1\|0\}, \{1\|\|2_{001}\|0\}, \{-1\|\|2_{100}\|\tau\}, \{-1\|\|2_{010}\|\tau\}, \{m_{\boldsymbol{n}}\|\|-1\|0\}, \{m_{\boldsymbol{n}}\|\|m_{001}\|0\}, \{2_{\boldsymbol{n}}\|\|m_{100}\|\tau\}, \{2_{\boldsymbol{n}}\|\|m_{010}\|\tau\}$ |
| S (u, u, 1/2) | $\{1\|\|1\|0\}, \{1\|\|2_{110}\|0\}, \{1\|\|m_{001}\|0\}, \{1\|\|m_{1-10}\|0\}, \{m_{\boldsymbol{n}}\|\|2_{001}\|0\}, \{m_{\boldsymbol{n}}\|\|2_{1-10}\|0\}, \{m_{\boldsymbol{n}}\|\|-1\|0\}, \{m_{\boldsymbol{n}}\|\|m_{110}\|0\}$ |
| T (u, 1/2, 1/2) | $\{1\|\|1\|0\}, \{1\|\|m_{001}\|0\}, \{-1\|\|2_{010}\|\tau\}, \{-1\|\|m_{100}\|\tau\}, \{m_{\boldsymbol{n}}\|\|2_{001}\|0\}, \{m_{\boldsymbol{n}}\|\|-1\|0\}, \{2_{\boldsymbol{n}}\|\|2_{100}\|\tau\}, \{2_{\boldsymbol{n}}\|\|m_{010}\|\tau\}$ |
| U (0, v, 1/2) | $\{1\|\|1\|0\}, \{1\|\|m_{001}\|0\}, \{-1\|\|2_{100}\|\tau\}, \{-1\|\|m_{010}\|\tau\}, \{m_{\boldsymbol{n}}\|\|2_{001}\|0\}, \{m_{\boldsymbol{n}}\|\|-1\|0\}, \{2_{\boldsymbol{n}}\|\|2_{010}\|\tau\}, \{2_{\boldsymbol{n}}\|\|m_{100}\|\tau\}$ |
| B (0, v, w) | $\{1\|\|1\|0\}, \{-1\|\|2_{100}\|\tau\}, \{m_{\boldsymbol{n}}\|\|-1\|0\}, \{2_{\boldsymbol{n}}\|\|m_{010}\|\tau\}$ |
| C (u, u, w) | $\{1\|\|1\|0\}, \{1\|\|m_{1-10}\|0\}, \{m_{\boldsymbol{n}}\|\|2_{1-10}\|0\}, \{m_{\boldsymbol{n}}\|\|-1\|0\}$ |
| D (u, v, 0) | $\{1\|\|1\|0\}, \{1\|\|m_{001}\|0\}, \{m_{\boldsymbol{n}}\|\|2_{001}\|0\}, \{m_{\boldsymbol{n}}\|\|-1\|0\}$ |
| E (u, v, 1/2) | $\{1\|\|1\|0\}, \{1\|\|m_{001}\|0\}, \{m_{\boldsymbol{n}}\|\|2_{001}\|0\}, \{m_{\boldsymbol{n}}\|\|-1\|0\}$ |
| F (u, 1/2, w) | $\{1\|\|1\|0\}, \{-1\|\|2_{010}\|\tau\}, \{m_{\boldsymbol{n}}\|\| - 1\|0\}, \{2_{\boldsymbol{n}}\|\|m_{010}\|\tau\}$ |
| GP (u, v, w) | $\{1\|\|1\|0\}, \{m_{\boldsymbol{n}}\|\| - 1\|0\}$ |

## 4. Irreducible co-representations of $RuO_2$

(1) GP (u, v, w) with little group: $\tilde{G}_{SS}^{GP} = {}^{m}{-1}{}^{\infty}1$

The unitary subgroup is ${}^{\infty}1$, where it provides two conjugated one-dimensional irreps for electron. The introduction of $\{m_x\|| - 1|0\}$ follows the character sum rule: ($\boldsymbol{n} = x$, $\infty_z$ stands for $U_z(\phi)$)

$$\sum_{\beta \in TA\tilde{L}_{SS}^{GP}} \chi((\{m_x\infty_z\|| - 1|0\})^2) = g \cdots (a). \tag{F15}$$

No extra degeneracy in spin space is obtained when introducing the antiunitary part. The coirreps at GP are $GP_1^R(1)GP_{\pm 1/2}^S(1)$, which is abbreviated as $GP_1(1)$ for brief.

(2) B (0, v, w) with little group: $\tilde{G}_{SS}^{B} = {}^{-1}2/{}^{2}m{}^{\infty}1$

The combination of $\{2_{\boldsymbol{n}}\||m_{010}|\tau\}$ and ${}^{\infty}1$ pairs two conjugated one-dimensional irreps $B_{+1/2}^S(1)$ and $B_{-1/2}^S(1)$ into a two-dimensional irrep $B_{\pm 1/2}^S(2)$. Then perform the character sum rule:

$$\sum_{\beta \in TA\tilde{L}_{SS}^{B}} \chi(\beta^2) = \sum_{\infty_z} \{\chi((\{T\infty_z\||2_{100}|\tau\})^2) + \chi((\{m_x\infty_z\||2_{100}|\tau\})^2)\}$$

$$= 0 + g = g \qquad \cdots (a), \tag{F16}$$

no extra degeneracy in spin space is obtained when the antiunitary part is introduced. Then coirreps at B are $B_1(2)$.

(3) C (u, u, w) with little group: $\tilde{G}_{SS}^{C} = {}^{m}2/{}^{1}m{}^{\infty}1$

The character table of maximal unitary subgroup $\tilde{L}_{SS}^{C} = {}^{1}m{}^{\infty}1$ are listed in Table F6.

**Table F6:** Character table of ${}^{1}m{}^{\infty}1$.

| ${}^{1}m{}^{\infty}1$ | $\{1\|\|1\|0\}$ | $\{1\|\|m_{1-10}\|0\}$ | $\{\infty_z\|\|1\|0\}$ | $\{\infty_z\|\|m_{1-10}\|0\}$ |
|---|---|---|---|---|
| $C_1^R C_0^S$ | 1 | 1 | 1 | 1 |
| $C_2^R C_0^S$ | 1 | -1 | 1 | -1 |
| $C_1^R C_m^S$ | 1 | 1 | $e^{-im\phi}$ | $e^{-im\phi}$ |
| $C_2^R C_m^S$ | 1 | -1 | $e^{-im\phi}$ | $-e^{-im\phi}$ |

The irreps of maximal unitary subgroups describing electron with $m = \pm 1/2$ are $C_1(1)C_{+1/2}^S(1)$, $C_1(1)C_{-1/2}^S(1)$, $C_2(1)C_{+1/2}^S(1)$ and $C_2(1)C_{-1/2}^S(1)$.

When introducing antiunitary operation $\{m_x \| 2_{1-10}|0\}$, the character sum rule guarantee that no no extra degeneracy is obtained.

$$\sum_{\beta \in TA\tilde{L}_{SS}^C} \chi(\beta^2) = \chi((\{m_x \| 2_{1-10}|0\})^2) + \chi((\{m_x \| -1|0\})^2)$$
$$+\chi((\{m_x \infty_z \| 2_{1-10}|0\})^2) + \chi((\{m_x \infty_z \| -1|0\})^2) = g \cdots (a). \qquad \text{(F17)}$$

Then coirreps at C are $C_1(1)$ and $C_2(1)$, or abbreviated into $C_{1,2}(1)$.

(4) S (u, u, 1/2) with little group: $\tilde{G}_{SS}^S = {}^1m^m m^1 m^\infty 1$

The character table of the maximal unitary subgroup $\tilde{L}_{SS}^S = {}^1m^1 2^1 m^\infty 1$ can be obtained by Table F1 and F8.

**Table F8:** Character table of ${}^1m^1 2^1 m$.

| ${}^1m^1 2^1 m$ | $\{1\|1\|0\}$ | $\{1\|2_{110}\|0\}$ | $\{1\|m_{001}\|0\}$ | $\{1\|m_{1-10}\|0\}$, |
|---|---|---|---|---|
| $S_1$ | 1 | 1 | 1 | 1 |
| $S_2$ | 1 | 1 | -1 | -1 |
| $S_3$ | 1 | -1 | 1 | -1 |
| $S_4$ | 1 | -1 | -1 | 1 |

When introducing antiunitary operation $\{m_x \| m_{110}|0\}$, no extra degeneracy is obtained according to the sum rule:

$$\sum_{\beta \in TA\tilde{L}_{SS}^S} \chi(\beta^2) = \chi((\{m_x \| 2_{001}|0\})^2) + \chi((\{m_x \| 2_{1-10}|0\})^2) + \chi((\{m_x \| -1|0\})^2)$$
$$+\chi((\{m_x \| m_{110}|0\})^2) + \chi((\{m_x \infty_z \| 2_{001}|0\})^2) + \chi((\{m_x \infty_z \| 2_{1-10}|0\})^2)$$
$$+\chi((\{m_x \infty_z \| -1|0\})^2) + \chi((\{m_x \infty_z \| m_{110}|0\})^2) = g \cdots (a). \qquad \text{(F18)}$$

Then coirreps at S are $S_{1,2,3,4}(1)$.

(5) U (0, v, 1/2) with little group $\tilde{G}_{SS}^{U} = {}^{2}m^{-1}m^{1}m^{\infty}1$.

The little group $\tilde{G}_{SS}^{U}$ can be expressed as

$$\tilde{G}_{SS}^{U} = \tilde{L}_{SS}^{U} \cup A_1\tilde{L}_{SS}^{U}, \quad \text{(F19)}$$

$$\tilde{L}_{SS}^{U} = \left(\tilde{S}_{SS}^{U} \cup A_2\tilde{S}_{SS}^{U}\right) \ltimes SO(2), \quad \text{(F20)}$$

where $\tilde{L}_{SS}^{U}$ is the maximal unitary subgroup of $\tilde{G}_{SS}^{U}$, the little group of the sublattice group $\tilde{S}_{SS}^{U} = \{\{1\|1|0\}, \{1\|m_{001}|0\}\}$, $A_1 = \{-1\|2_{100}|\tau\}$, $A_2 = \{2_{\boldsymbol{n}}\|2_{010}|\tau\}$.

$\tilde{S}_{SS}^{U}$ provide $U_{1,2}^{R}(1)$ irreps in real space using the PG *m*; the combination of $A_1$ and $SO(2)$ provide the $U_{\pm1/2}^{S}(2)$ in spin space. The character table of $\tilde{L}_{SS}^{U}$ is list in Table F9. Since there are no conjugated irreps pair in real or spin space, the introduction of $A_2$ can not bring in new extra degeneracy. Therefore, the co-irreps at U point is $U_{1,2}(2)$.

**Table F9:** Character table of $\tilde{L}_{SS}^{U} = {}^{2}m^{2}2^{1}m^{\infty}1$.

| ${}^{2}m^{2}2^{1}m^{\infty}1$ | $\{1\|1\|0\}$ | $\{1\|m_{001}\|0\}$ | $\{U_z(\phi)\|1\|0\}$ | $\{U_z(\phi)\|m_{001}\|0\}$ | $\{U_x(\pi)\|1\|0\}$ | $\{U_x(\pi)\|m_{001}\|0\}$ |
|---|---|---|---|---|---|---|
| $U_1^R U_{0g}^S$ | 1 | 1 | 1 | 1 | 1 | 1 |
| $U_2^R U_{0g}^S$ | 1 | -1 | 1 | -1 | 1 | -1 |
| $U_1^R U_{0u}^S$ | 1 | 1 | 1 | 1 | -1 | -1 |
| $U_2^R U_{0u}^S$ | 1 | -1 | 1 | -1 | -1 | 1 |
| $U_1^R U_m^S$ | 2 | 2 | $2cos(m\phi)$ | $2cos(m\phi)$ | 0 | 0 |
| $U_2^R U_m^S$ | 2 | -2 | $2cos(m\phi)$ | $-2cos(m\phi)$ | 0 | 0 |

Following the same procedure, the complete coirreps of $P^{-1}4_2/{}^{1}m^{-1}n^{1}m^{\infty m}1$ at each *k* point are calculated and tabulated in Table F10.

**Table F10:** Little groups and co-irreps of $P^{-1}4_2/{}^{1}m^{-1}n^{1}m^{\infty m}1$ at each *k* point.

| HSP | Little group | Coirreps |
|---|---|---|
| Γ (0, 0, 0) | ${}^{-1}4/{}^{1}m^{-1}m^{1}m^{\infty m}1$ | $\Gamma_{1,2,3,4}^{+,-}(2)$ |
| X (0, 1/2, 0) | ${}^{-1}m^{-1}m^{1}m^{\infty m}1$ | $X_{1,2}^{+,-}(2)$ |
| M (1/2, 1/2, 0) | ${}^{-1}4/{}^{1}m^{-1}m^{1}m^{\infty m}1$ | $M_{1,2,3,4}^{+,-}(2)$ |
| Z (0, 0, 1/2) | ${}^{-1}4/{}^{1}m{}^{-1}m{}^{1}m{}^{\infty m}1$ | $Z^{+,-}_{1,2,3,4}(2)$ |
| R (0, 1/2, 1/2) | ${}^{-1}m{}^{-1}m{}^{1}m{}^{\infty m}1$ | $R^{+,-}_{1,2}(2)$ |
| A (1/2, 1/2, 1/2) | ${}^{-1}4/{}^{1}m{}^{-1}m{}^{1}m{}^{\infty m}1$ | $A^{+,-}_{1,2,3,4}(2)$ |
| Δ (0, v, 0) | ${}^{2}m{}^{-1}m{}^{1}m{}^{\infty}1$ | $\Delta_{1,2}(2)$ |
| Y (u, 1/2, 0) | ${}^{-1}m{}^{2}m{}^{1}m{}^{\infty}1$ | $Y_{1,2}(2)$ |
| Σ (u, u, 0) | ${}^{1}m{}^{m}m{}^{1}m{}^{\infty}1$ | $\Sigma_{1,2,3,4}(1)$ |
| Λ (0, 0, w) | ${}^{2}4/{}^{m}m{}^{2}m{}^{1}m{}^{\infty}1$ | $\Lambda_{1,2,3,4}(2)$ |
| V (1/2, 1/2, w) | ${}^{2}4/{}^{m}m{}^{2}m{}^{1}m{}^{\infty}1$ | $V_{1,2,3,4}(2)$ |
| W (0, 1/2, w) | ${}^{2}m{}^{2}m{}^{\mathrm{m}}m{}^{\infty}1$ | $W_{1,2}(2)$ |
| S (u, u, 1/2) | ${}^{1}m{}^{m}m{}^{1}m{}^{\infty}1$ | $S_{1,2,3,4}(1)$ |
| T (u, 1/2, 1/2) | ${}^{-1}m{}^{2}m{}^{1}m{}^{\infty}1$ | $T_{1,2}(2)$ |
| U (0, v, 1/2) | ${}^{2}m{}^{-1}m{}^{1}m{}^{\infty}1$ | $U_{1,2}(2)$ |
| B (0, v, w) | ${}^{-1}2/{}^{2}m{}^{\infty}1$ | $B_1(2)$ |
| C (u, u, w) | ${}^{m}2/{}^{1}m{}^{\infty}1$ | $C_{1,2}(1)$ |
| D (u, v, 0) | ${}^{m}2/{}^{1}m{}^{\infty}1$ | $D_{1,2}(1)$ |
| E (u, v, 1/2) | ${}^{m}2/{}^{1}m{}^{\infty}1$ | $E_{1,2}(1)$ |
| F (u, 1/2, w) | ${}^{-1}2/{}^{2}m{}^{\infty}1$ | $F_1(2)$ |
| GP (u, v, w) | ${}^{m}{-1}{}^{\infty}1$ | $GP_1(1)$ |

## 5. Comparison with the co-irreps of MSG

The single-valued and double-valued coirreps of MSG $P4_2'/mnm'$ (136.499) are also listed in Table F11 for comparison with the SOC-free band structure of $RuO_2$ in Fig. F2. In Table F11, the co-irreps that do not conform to the band degeneracy at the high-symmetry points/lines/planes in Fig. F2 are highlighted in red font.

**Table F11:** Comparison of coirreps of SSG $P^{-1}4_2/{}^{1}m{}^{-1}n{}^{1}m{}^{\infty m}1$ (65.136.1.1) and MSG $P4_2'/mnm'$ (136.499). The co-irreps of MSG are obtained from Bilbao, where the bar of double-valued co-irreps are left out. The highlighted co-irreps are those contracted with band structure of $RuO_2$ without SOC. The fifth column stands for the band degeneracy at different HSPs.

| HSP | SSG | MSG | | band |
|---|---|---|---|---|
| | | *single- valued* | *double-valued* | |
| Γ (0, 0, 0) | $\Gamma_{1,2,3,4}^{+,-}(2)$ | $\Gamma_{1,2}^{+,-}(1), \Gamma_3^{+,-}\Gamma_4^{+,-}(2)$ | $\Gamma_{5,6}(2)$ | 2 |
| X (0, 1/2, 0) | $X_{1,2}^{+,-}(2)$ | $X_{1,2}(2)$ | $X_{3,4}(2)$ | 2 |
| M (1/2, 1/2, 0) | $M_{1,2,3,4}^{+,-}(2)$ | $M_1^{+,-}M_2^{+,-}(2), M_{3,4}^{+,-}(1)$ | $M_{5,6}(2)$ | 2 |
| Z (0, 0, 1/2) | $Z_{1,2,3,4}^{+,-}(2)$ | $Z_{1,2}(2)$ | $Z_{3,4}(2)$ | 2 |
| R (0, 1/2, 1/2) | $R_{1,2}^{+,-}(2)$ | $R_{1,2}^{+,-}(2)$ | $R_{2\sim9}(1)$ | 2 |
| A (1/2, 1/2, 1/2) | $A_{1,2,3,4}^{+,-}(2)$ | $A_{1,2}(2)$ | $A_{3,4}(2)$ | 2 |
| Δ (0, v, 0) | $\Delta_{1,2}(2)$ | $\Delta_{1,2,3,4}(1)$ | $\Delta_5(2)$ | 2 |
| Y (u, 1/2, 0) | $Y_{1,2}(2)$ | $Y_{1,2,3,4}(1)$ | $Y_5(2)$ | 2 |
| Σ (u, u, 0) | $\Sigma_{1,2,3,4}(1)$ | $\Sigma_{1,2}(1)$ | $\Sigma_{3,4}(1)$ | 1 |
| Λ (0, 0, w) | $\Lambda_{1,2,3,4}(2)$ | $\Lambda_{1,2}(1), \Lambda_3\Lambda_4(2)$ | $\Lambda_5(2)$ | 2 |
| V (1/2, 1/2, w) | $V_{1,2,3,4}(2)$ | $V_1V_2(2), V_{3,4}(1)$ | $V_5(2)$ | 2 |
| W (0, 1/2, w) | $W_{1,2}(2)$ | $W_1(2)$ | $W_{2,3,4,5}(1)$ | 2 |
| S (u, u, 1/2) | $S_{1,2,3,4}(1)$ | $S_{1,2}(1)$ | $S_{3,4}(1)$ | 1 |
| T (u, 1/2, 1/2) | $T_{1,2}(2)$ | $T_1(2)$ | $T_{2,3,4,5}(1)$ | 2 |
| U (0, v, 1/2) | $U_{1,2}(2)$ | $U_1(2)$ | $U_{2,3,4,5}(1)$ | 2 |
| B (0, v, w) | $B_1(2)$ | $B_{1,2}(1)$ | $B_{3,4}(1)$ | 2 |
| C (u, u, w) | $C_{1,2}(1)$ | $C_1(1)$ | $C_2(1)$ | 1 |
| D (u, v, 0) | $D_{1,2}(1)$ | $D_{1,2}(1)$ | $D_{3,4}(1)$ | 1 |
| E (u, v, 1/2) | $E_{1,2}(1)$ | $E_{1,2}(1)$ | $E_{3,4}(1)$ | 1 |
| F (u, 1/2, w) | $F_1(2)$ | $F_{1,2}(1)$ | $F_{3,4}(1)$ | 2 |
| GP (u, v, w) | $GP_1(1)$ | $GP_1(1)$ | $GP_2(1)$ | 1 |

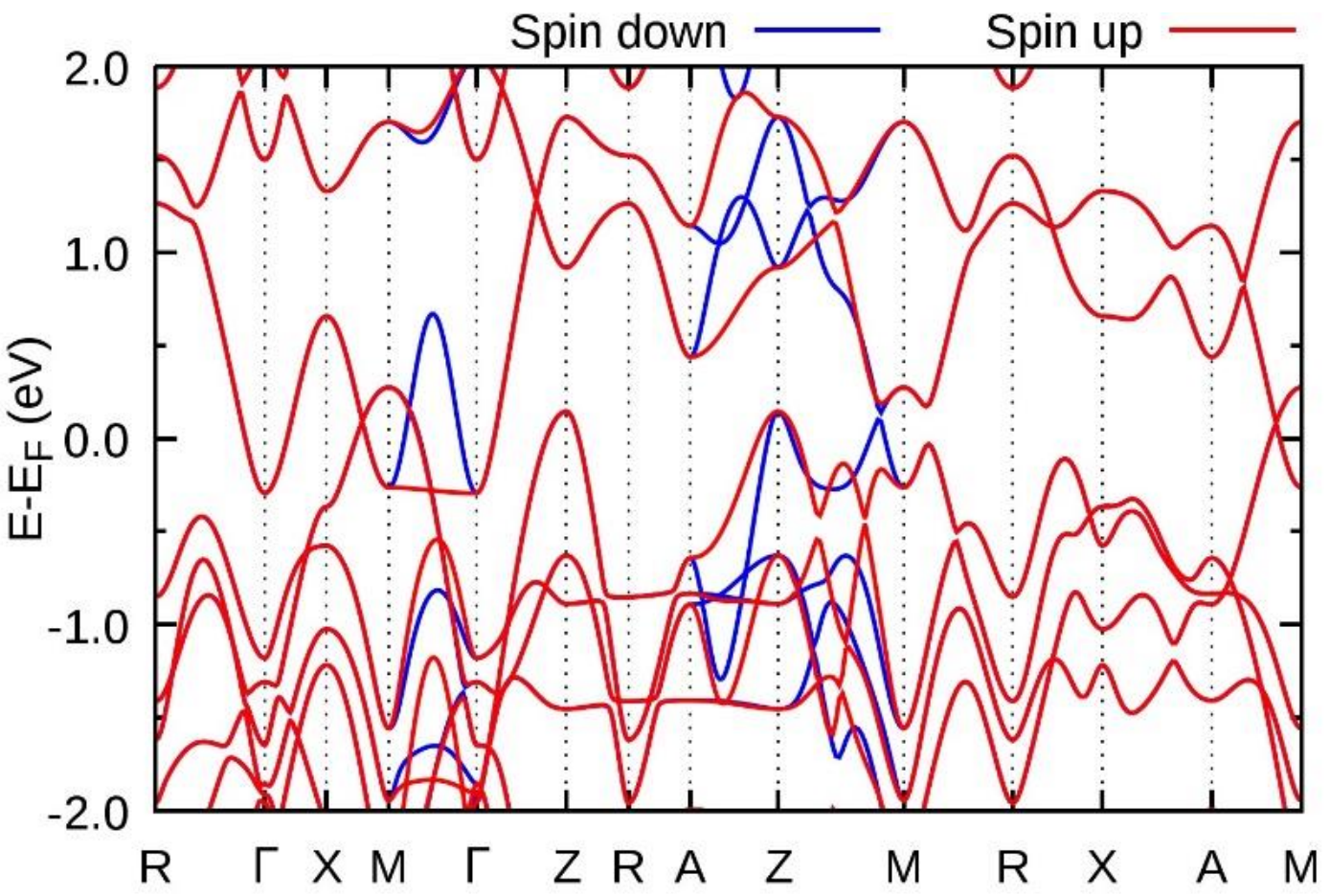

**Fig. F2:** SOC-free band structure of $RuO_2$ with projection of spin components following high-symmetry lines $R\overset{B}{-}\Gamma\overset{\Delta}{-}X\overset{Y}{-}M\overset{\Sigma}{-}\Gamma\overset{\Lambda}{-}Z\overset{U}{-}R\overset{T}{-}A\overset{S}{-}Z\overset{C}{-}M\overset{F}{-}R\overset{W}{-}X\overset{F}{-}A\overset{V}{-}M$ . Only $\Sigma$, $S$, $C$, $D$, $E$ have anti-ferromagnetic spin splitting (one-dimensional co-irreps), $\Sigma$ and $S$ line are at the $D$ and $E$ plane, respectively.

## G. Band representations in coplanar SSGs

In this section, we provide the derivation of the co-irrep matrices and character tables of the projective co-irreps of the little groups at high-symmetry points and lines in coplanar SSG. The general procedure, as exemplified by $CeAuAl_3$, is as follows: (1) identify the maximal unitary subgroup (MUSGs) $\tilde{L}_{SS}^{k}$ of the quotient group $\tilde{G}_{SS}^{k}$ of the little group $G_{SS}^{k}$; (2) construct the complete set of projective irreps of MUSGs by the CSCO approach introduced in Appendix E; (3) obtain co-irreps from those projective irreps based on the modified Dimmock and Wheeler's character sum rule, i.e., Eq. (E17).

The nontrivial SSG of $CeAuAl_3$ is $P^{1}4^{1}m^{1}m^{4_{001}^{1}}(1/2\ 1/2\ 1/4)$ (99.107.4.1). Considering the spin-only group $G_{SO}^{p} = Z_2^K = {}^{m}1$ for coplanar magnetic structure, it turns into $P^{1}4^{1}m^{1}m^{4_{001}^{1}}(1/2\ 1/2\ 1/4)^{m}1$, where ${}^{m}1 = T\{2_{001}||1|0\}$. The $Y(u,1/2,0)$ point of $P^{1}4^{1}m^{1}m^{4_{001}^{1}}(1/2\ 1/2\ 1/4)^{m}1$ is taken as an example to show the process of calculating projective irreps of coplanar SSG using the CSCO method.

First, the spin Brillouin zone and the magnetic Brillouin zone with all high-symmetric points/lines for coplanar antiferromagnet $CeAuAl_3$ are given in Figs. G1 and G2, respectively.

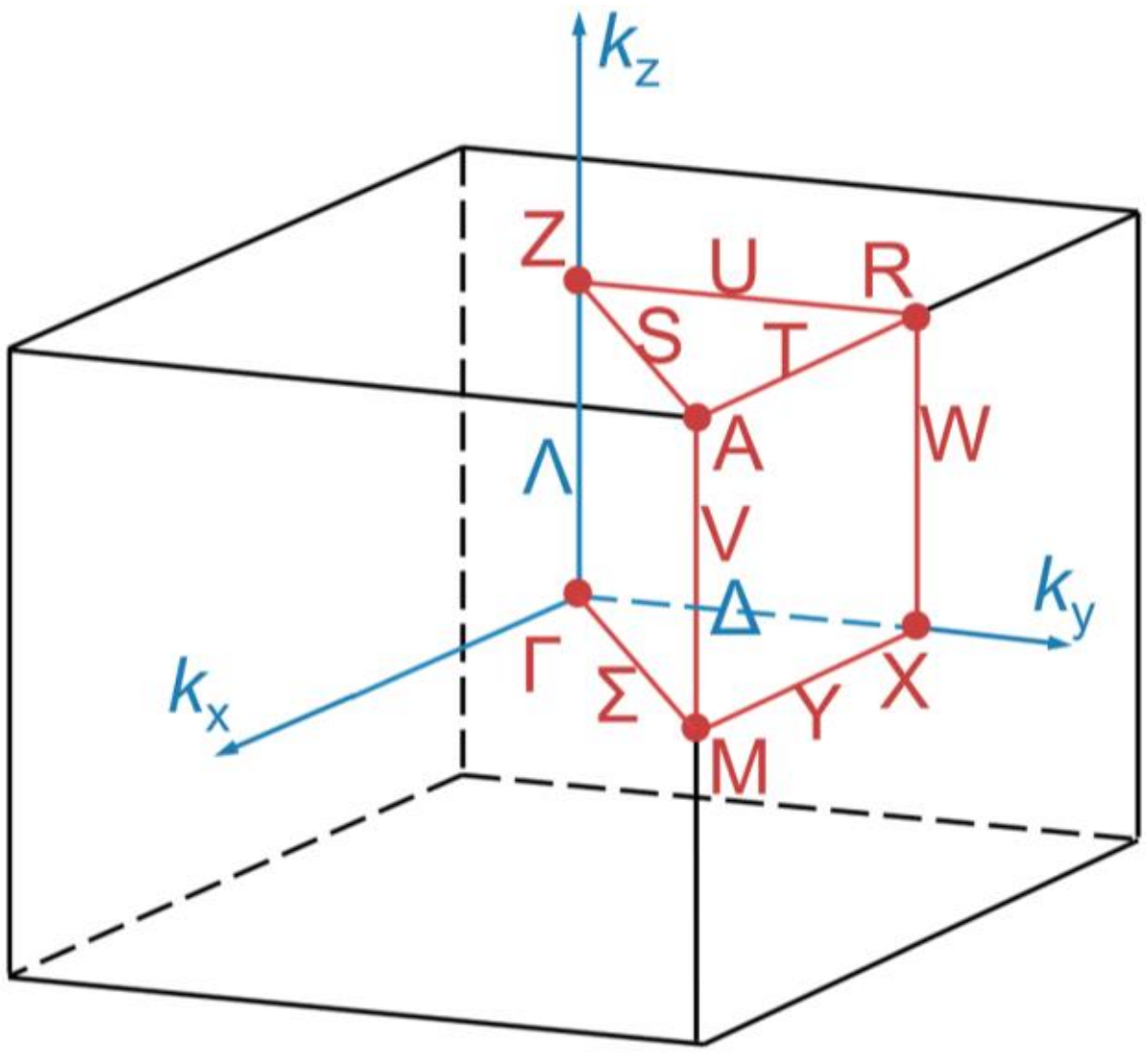

**Fig. G1:** Schematic of spin Brillouin zone for SSG $P^{1}4^{1}m^{1}m^{4_{001}^{1}}(1/2\ 1/2\ 1/4)^{m}1$.

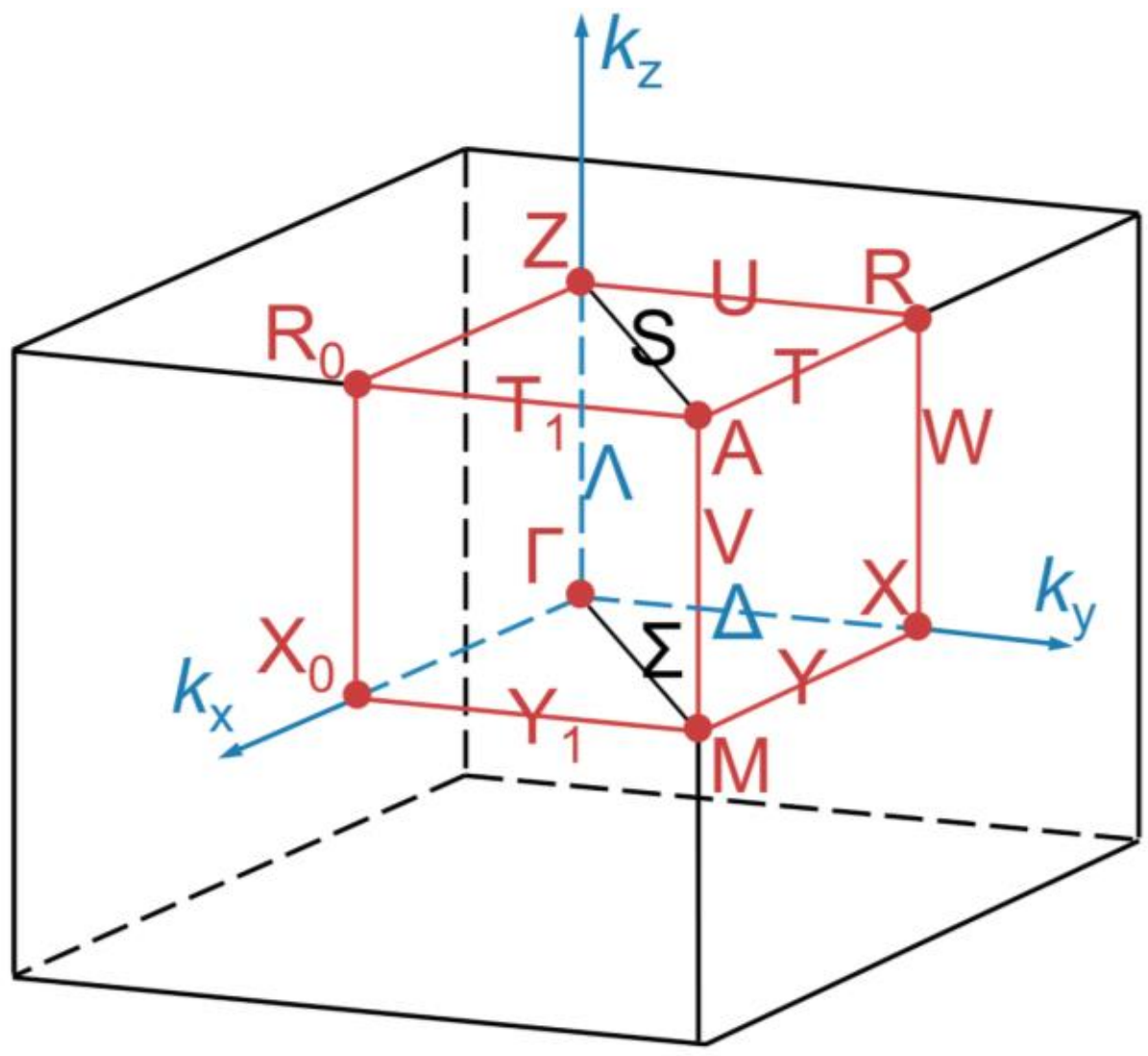

**Fig. G2:** Schematic of magnetic Brillouin zone for MSG $P_c4_1$.

## 1. Irreducible co-representations of CeAuAl3 at *Y*(u,1/2,0)

The little group $G_{SS}^{Y}$ of CeAuAl3 at $Y(u,$1/2,0) belongs to k-type SSGs. The quotient group $\tilde{G}_{SS}^{Y} = G_{SS}^{Y}/T(L_0)$, where $T(L_0)$ is the translational subgroup of $L_0 =$ P4mm (No. 99), contains eight unitary symmetry elements:

$$\begin{aligned} g_1^{(1)} &= \{1||1|0\}, & g_1^{(2)} &= \{4_{001}^1||1|1/2\ 1/2\ 1/4\}, \\ g_1^{(3)} &= \{2_{001}||1|0\ 0\ 1/2\}, & g_1^{(4)} &= \{4_{001}^3||1|1/2\ 1/2\ 3/4\}, \\ g_2^{(1)} &= \{1||m_{010}|0\}, & g_2^{(2)} &= \{4_{001}^1||m_{010}|1/2\ 1/2\ 1/4\}, \\ g_2^{(3)} &= \{2_{001}||m_{010}|0\ 0\ 1/2\}, & g_2^{(4)} &= \{4_{001}^3||m_{010}|1/2\ 1/2\ 3/4\} \end{aligned} \quad \text{(G1)}$$

and an anti-unitary generator:

$$g_3^{(1)} = TA = T\{2_{001}||m_{100}|0\}, \quad \text{(G2)}$$

Therefore, its MUSG $\tilde{L}_{SS}^{Y}$ is isomorphic to $4/m$. According to Eq. (E6), we next construct the left regular projective reps of $\tilde{L}_{SS}^{Y}$. The corresponding matrix element is written as:

$$M_k\left(g_i^{(a)}\right)_{g,g_j^{(a)}} = \left\langle g\left|g_i^{(a)}\right|g_j^{(a)}\right\rangle = (-1)^{\xi\left(\phi\left(g_{s_l}^{(a)}\right)\right)} exp\left(-iK_i\cdot\tau_j^{(a)}\right)\delta_{g,g_i^{(a)}g_j^{(a)}}. \quad \text{(G3)}$$

For instance,

$$g_1^{(2)}\left|g_1^{(4)}\right\rangle = \{(4_{001}^1)^4||1|1\ 1\ 1\} = (-1)^{\xi\left(\phi\left((4_{001}^1)^4\right)\right)} exp\left(-iK_1\cdot\tau_1^{(4)}\right)\left|g_1^{(1)}\right\rangle. \quad \text{(G4)}$$

where $\xi\left(\phi((4_{001}^1)^4)\right) = 1$ and $K_1 = 0$, resulting in $M_k(g_1^{(2)})_{1,4} = -1$.

For another example,

$$g_2^{(1)}\left|g_1^{(2)}\right\rangle = \{4_{001}^1||m_{010}|\ 1/2\ 1/2\ 1/4\} = (-1)^{\xi(\phi(4_{001}^1))}exp(-iK_2\cdot\tau_1^{(2)})\left|g_2^{(2)}\right\rangle \quad \text{(G5)}$$

where $\xi\left(\phi(4_{001}^1)\right) = 0$, $K_2 = 2\pi(0,-1,0)$ and $\tau_1^{(2)} = (1/2\ 1/2\ 1/4)$, resulting in $M_k(g_2^{(1)})_{6,2} = -1$.

Therefore, the left regular projective reps for generators of $\tilde{L}_{SS}^Y$ can be written as:

$$M_Y\left(g_1^{(2)}\right) = \begin{pmatrix} 0 & 0 & 0 & -1 & 0 & 0 & 0 & 0 \\ 1 & 0 & 0 & 0 & 0 & 0 & 0 & 0 \\ 0 & 1 & 0 & 0 & 0 & 0 & 0 & 0 \\ 0 & 0 & 1 & 0 & 0 & 0 & 0 & 0 \\ 0 & 0 & 0 & 0 & 0 & 0 & 0 & -1 \\ 0 & 0 & 0 & 0 & 1 & 0 & 0 & 0 \\ 0 & 0 & 0 & 0 & 0 & 1 & 0 & 0 \\ 0 & 0 & 0 & 0 & 0 & 0 & 1 & 0 \end{pmatrix}, \quad \text{(G6)}$$

$$M_Y\left(g_2^{(1)}\right) = \begin{pmatrix} 0 & 0 & 0 & 0 & 1 & 0 & 0 & 0 \\ 0 & 0 & 0 & 0 & 0 & -1 & 0 & 0 \\ 0 & 0 & 0 & 0 & 0 & 0 & 1 & 0 \\ 0 & 0 & 0 & 0 & 0 & 0 & 0 & -1 \\ 1 & 0 & 0 & 0 & 0 & 0 & 0 & 0 \\ 0 & -1 & 0 & 0 & 0 & 0 & 0 & 0 \\ 0 & 0 & 1 & 0 & 0 & 0 & 0 & 0 \\ 0 & 0 & 0 & -1 & 0 & 0 & 0 & 0 \end{pmatrix}. \quad \text{(G7)}$$

For each group element $g_i^{(a)} \in \tilde{L}_{SS}^Y$, there is a corresponding element $\bar{g}_i^{(a)}$ that obeys right multiplication rule $\bar{g}_i^{(a)}g_j^{(a)} = g_j^{(a)}g_i^{(a)}$. Remarkably, it has been proved that the collection of $\bar{g}_i^{(a)}$ forms a group named intrinsic group $\overline{\tilde{L}_{SS}^Y}$ [68]. Similarly, the matrix element is

$$M_k\left(\bar{g}_i^{(a)}\right)_{g,g_j^{(a)}} = \left\langle g\middle|\bar{g}_i^{(a)}\middle|g_j^{(a)}\right\rangle = (-1)^{\xi\left(\phi\left(g_{s_m}^{(a)}\right)\right)}exp\left(-iK_j\cdot\tau_i^{(a)}\right)\delta_{g,g_j^{(a)}g_i^{(a)}}, \quad \text{(G8)}$$

where $g_{s_m}^{(a)} = g_{s_j}^{(a)}g_{s_i}^{(a)}$, $K_j = R_j^{-1}k - k$.

For instance,

$$\bar{g}_1^{(2)}\left|g_1^{(4)}\right\rangle = \left|g_1^{(4)}g_1^{(2)}\right\rangle = \{(4_{001}^1)^4||1|1\ 1\ 1\}$$
$$= (-1)^{\xi\left(\phi\left((4_{001}^1)^4\right)\right)}exp\left(-iK_1\cdot\tau_1^{(2)}\right)\left|g_1^{(1)}\right\rangle, \quad \text{(G9)}$$

where $\xi\left(\phi((4_{001}^1)^4)\right) = 1$ and $K_1 = 0$, resulting in $M_k(\bar{g}_1^{(2)})_{1,4} = -1$.

For another example,

$$\bar{g}_2^{(1)}\left|g_1^{(2)}\right\rangle = \left|g_1^{(2)}g_2^{(1)}\right\rangle = \{4_{001}^1||m_{010}|\ 1/2\ 1/2\ 1/4\}$$

$$= (-1)^{\xi\left(\phi(4_{001}^1)\right)} exp\left(-iK_1 \cdot \tau_2^{(1)}\right)\left|g_2^{(2)}\right\rangle, \quad \text{(G10)}$$

where $\xi\left(\phi(4_{001}^1)\right) = 0$, $K_1 = 0$, resulting in $M_k(\bar{g}_2^{(1)})_{6,2} = 1$.

The matrix forms of generators of the intrinsic group $\overline{\tilde{L}_{SS}^Y}$ can be expressed by:

$$M_Y\left(\bar{g}_1^{(2)}\right) = \begin{pmatrix} 0 & 0 & 0 & -1 & 0 & 0 & 0 & 0 \\ 1 & 0 & 0 & 0 & 0 & 0 & 0 & 0 \\ 0 & 1 & 0 & 0 & 0 & 0 & 0 & 0 \\ 0 & 0 & 1 & 0 & 0 & 0 & 0 & 0 \\ 0 & 0 & 0 & 0 & 0 & 0 & 0 & 1 \\ 0 & 0 & 0 & 0 & -1 & 0 & 0 & 0 \\ 0 & 0 & 0 & 0 & 0 & -1 & 0 & 0 \\ 0 & 0 & 0 & 0 & 0 & 0 & -1 & 0 \end{pmatrix}, \quad \text{(G11)}$$

$$M_Y\left(\bar{g}_2^{(1)}\right) = \begin{pmatrix} 0 & 0 & 0 & 0 & 1 & 0 & 0 & 0 \\ 0 & 0 & 0 & 0 & 0 & 1 & 0 & 0 \\ 0 & 0 & 0 & 0 & 0 & 0 & 1 & 0 \\ 0 & 0 & 0 & 0 & 0 & 0 & 0 & 1 \\ 1 & 0 & 0 & 0 & 0 & 0 & 0 & 0 \\ 0 & 1 & 0 & 0 & 0 & 0 & 0 & 0 \\ 0 & 0 & 1 & 0 & 0 & 0 & 0 & 0 \\ 0 & 0 & 0 & 1 & 0 & 0 & 0 & 0 \end{pmatrix}, \quad \text{(G12)}$$

It should be noted that the dimension of the left regular projective rep $M_Y\left(g_i^{(a)}\right)$ is $N_k = \left|\tilde{L}_{SS}^Y\right| = 8$. Such $N_k \times N_k$ rep matrices are reducible, containing all projective irreps. In order to obtain all the projective irreps, we apply the method of CSCO to block diagonalize these matrices.

The class operators could be built up by

$$C_i^{(a)} = \sum_{g_j^{(a)} \in \tilde{L}_{SS}^Y} M_Y\left(g_j^{(a)}\right)^{-1} M_Y\left(g_i^{(a)}\right) M_Y\left(g_j^{(a)}\right), \quad \text{(G13)}$$

By constructing the linear combination of class operators $C = \sum_{i,a} k_i^{(a)} C_i^{(a)}$, where $k_i^{(a)}$ stand for the arbitrary real coefficient to avoid accidental elimination of the same class operators, the CSCO-I of $\tilde{L}_{SS}^Y$ can be obtained. In this case, it is convenient to set $k_i^{(a)} = 1$ for all $i = 1,2$ and $a = 1,2,3,4$. Therefore:

$$C = \sum_{\substack{i=1,2\\a=1,2,3,4}} C_i^{(a)} = \begin{pmatrix} 8 & 0 & -8 & 0 & 0 & 0 & 0 & 0 \\ 0 & 8 & 0 & -8 & 0 & 0 & 0 & 0 \\ 8 & 0 & 8 & 0 & 0 & 0 & 0 & 0 \\ 0 & 8 & 0 & 8 & 0 & 0 & 0 & 0 \\ 0 & 0 & 0 & 0 & 8 & 0 & -8 & 0 \\ 0 & 0 & 0 & 0 & 0 & 8 & 0 & -8 \\ 0 & 0 & 0 & 0 & 8 & 0 & 8 & 0 \\ 0 & 0 & 0 & 0 & 0 & 8 & 0 & 8 \end{pmatrix}, \tag{G14}$$

The number of different eigenvalues of $C$ equals the number of inequivalent irreps. The matrix $C$ in Eq. (G14) has two distinct eigenvalues, i.e., $8+8i$ and $8-8i$, each with a multiplicity of four. This implies that there are two inequivalent 2-dimensional irreps. The eigenvectors of $C$ could be used to partly diagonalize the left regular projective reps $M_Y\left(g_i^{(a)}\right)$. However, a $n_l$-dimensional projective irreps will appear $n_l$ times in $M_Y\left(g_i^{(a)}\right)$ and they are still degenerate at this stage.

To further decompose the left regular projective irreps, we find a subgroup of $\tilde{L}_{SS}^{Y}$, i.e., $\tilde{L}_{SS1}^{Y}$ with $g_1^{(1)}$, $g_1^{(2)}$, $g_1^{(3)}$ and $g_1^{(4)}$ forming a group ismorphic to PG 4. The corresponding class operator $C(s)$, which is also named CSCO-I of the subgroup chain $\tilde{L}_{SS1}^{Y} < \tilde{L}_{SS}^{Y}$, is formed by:

$$C(s) = \sum_{\substack{i=1,2\\a=1,2,3,4}} C_{i1}^{(a)} = \begin{pmatrix} 4 & -4 & -4 & -4 & 0 & 0 & 0 & 0 \\ 4 & 4 & -4 & -4 & 0 & 0 & 0 & 0 \\ 4 & 4 & 4 & -4 & 0 & 0 & 0 & 0 \\ 4 & 4 & 4 & 4 & 0 & 0 & 0 & 0 \\ 0 & 0 & 0 & 0 & 4 & -4 & -4 & -4 \\ 0 & 0 & 0 & 0 & 4 & 4 & -4 & -4 \\ 0 & 0 & 0 & 0 & 4 & 4 & 4 & -4 \\ 0 & 0 & 0 & 0 & 4 & 4 & 4 & 4 \end{pmatrix}, \tag{G15}$$

where the summation of $C_{i1}^{(a)} = \sum_{g_{j1}^{(a)} \in \tilde{L}_{SS1}^{Y}} M_Y(g_{j1}^{(a)})^{-1} M_Y(g_i^{(a)}) M_Y(g_{j1}^{(a)})$.

Furthermore, the corresponding intrinsic subgroup $\overline{\tilde{L}_{SS1}^{Y}} \cong C_4$ consists of $\bar{g}_1^{(1)}$, $\bar{g}_1^{(2)}$, $\bar{g}_1^{(3)}$ and $\bar{g}_1^{(4)}$. The CSCO-I $\bar{C}(s)$ of the intrinsic subgroup chain $\overline{\tilde{L}_{SS1}^{Y}} < \overline{\tilde{L}_{SS}^{Y}}$ can be written as:

$$\bar{C}(s)=\sum_{\substack{i=1,2\\a=1,2,3,4}}\bar{C}_{i1}^{(a)}=\begin{pmatrix}4&-4&-4&-4&0&0&0&0\\4&4&-4&-4&0&0&0&0\\4&4&4&-4&0&0&0&0\\4&4&4&4&0&0&0&0\\0&0&0&0&4&4&-4&4\\0&0&0&0&-4&4&4&-4\\0&0&0&0&4&-4&4&4\\0&0&0&0&-4&4&-4&4\end{pmatrix}, \tag{G16}$$

where $\bar{C}_{i1}^{(a)}=\sum_{\bar{g}_{j1}^{(a)}\in\overline{\tilde{L}_{SS1}^{Y}}}M_Y(\bar{g}_{j1}^{(a)})^{-1}M_Y(\bar{g}_i^{(a)})M_Y(\bar{g}_{j1}^{(a)})$.

We then use the complete set $(C,C(s),\bar{C}(s))$ to lift the remaining degeneracy of the left regular projective reps. By finding the common eigenvectors of $(C,C(s),\bar{C}(s))$, a unitary transformation matrix $U$ is obtained to block diagonalize the left regular projective reps of all group elements simultaneously:

$$U^{-1}M_Y\left(g_1^{(2)}\right)U=\begin{pmatrix}e^{i\frac{\pi}{4}}&0&0&0&0&0&0&0\\0&e^{-i\frac{3\pi}{4}}&0&0&0&0&0&0\\0&0&e^{-i\frac{\pi}{4}}&0&0&0&0&0\\0&0&0&e^{i\frac{3\pi}{4}}&0&0&0&0\\0&0&0&0&e^{-i\frac{3\pi}{4}}&0&0&0\\0&0&0&0&0&e^{i\frac{\pi}{4}}&0&0\\0&0&0&0&0&0&e^{i\frac{3\pi}{4}}&0\\0&0&0&0&0&0&0&e^{-i\frac{\pi}{4}}\end{pmatrix}, \tag{G17}$$

$$U^{-1}M_Y\left(g_2^{(1)}\right)U=\begin{pmatrix}0&-1&0&0&0&0&0&0\\-1&0&0&0&0&0&0&0\\0&0&0&-1&0&0&0&0\\0&0&-1&0&0&0&0&0\\0&0&0&0&0&e^{i\frac{\pi}{4}}&0&0\\0&0&0&0&e^{-i\frac{\pi}{4}}&0&0&0\\0&0&0&0&0&0&0&e^{-i\frac{\pi}{4}}\\0&0&0&0&0&0&e^{i\frac{\pi}{4}}&0\end{pmatrix}, \tag{G18}$$

The character table of projective irreps is shown in Table G1, where we can find that there are two different projective irreps $M_Y^1$:

$$M_Y^1\left(g_1^{(2)}\right)=\begin{pmatrix}e^{i\frac{3\pi}{4}}&0\\0&e^{-i\frac{\pi}{4}}\end{pmatrix},\quad M_Y^1\left(g_2^{(1)}\right)=\begin{pmatrix}0&e^{i\frac{3\pi}{4}}\\e^{-i\frac{3\pi}{4}}&0\end{pmatrix}, \tag{G19}$$

and $M_Y^2$:

$$M_Y^2\left(g_1^{(2)}\right)=\begin{pmatrix}e^{-i\frac{3\pi}{4}}&0\\0&e^{i\frac{\pi}{4}}\end{pmatrix},\quad M_Y^2\left(g_2^{(1)}\right)=\begin{pmatrix}0&e^{-i\frac{3\pi}{4}}\\e^{i\frac{3\pi}{4}}&0\end{pmatrix}. \tag{G20}$$

Finally, the irreps could be obtained by $d_Y^l\left(\left\{g_{s_i}^{(a)}\middle\|R_i\middle|\tau_i^{(a)}\right\}\right)=exp(-iY\cdot\tau_i^{(a)})M_Y^l\left(\left\{g_{s_i}^{(a)}\middle\|R_i\middle|\tau_i^{(a)}\right\}\right)$:

$$d_Y^1\left(g_1^{(2)}\right)=\begin{pmatrix} e^{i\left(\frac{\pi}{4}-u\pi\right)} & 0 \\ 0 & e^{-i\left(\frac{3\pi}{4}+u\pi\right)} \end{pmatrix},\quad d_Y^1\left(g_2^{(1)}\right)=\begin{pmatrix} 0 & e^{i\frac{3\pi}{4}} \\ e^{-i\frac{3\pi}{4}} & 0 \end{pmatrix}, \qquad \text{(G21)}$$

and:

$$d_Y^2\left(g_1^{(2)}\right)=\begin{pmatrix} e^{-i(\frac{5\pi}{4}+u\pi)} & 0 \\ 0 & e^{-i(\frac{\pi}{4}+u\pi)} \end{pmatrix},\quad d_Y^2\left(g_2^{(1)}\right)=\begin{pmatrix} 0 & e^{-i\frac{3\pi}{4}} \\ e^{i\frac{3\pi}{4}} & 0 \end{pmatrix}. \qquad \text{(G22)}$$

**Table G1:** Character table of projective irreps $M_Y^1$ and $M_Y^2$ of $\tilde{L}_{SS}^Y$.

| | $g_1^{(1)}$ | $g_1^{(2)}$ | $g_1^{(3)}$ | $g_1^{(4)}$ | $g_2^{(1)}$ | $g_2^{(2)}$ | $g_2^{(3)}$ | $g_2^{(4)}$ |
|---|---|---|---|---|---|---|---|---|
| $M_Y^1$ | 2 | 0 | $-2i$ | 0 | 0 | 0 | 0 | 0 |
| $M_Y^2$ | 2 | 0 | $2i$ | 0 | 0 | 0 | 0 | 0 |

We now introduce the anti-unitary generator $g_3^{(1)}=TA=T\{2_{001}||m_{100}|0\}$. Assuming that the basis set of $d_k^l(g_i^{(a)})$ is $|\psi\rangle=|\psi_1,\psi_2,\dots,\psi_{n_l}\rangle$, then for the coset the basis set can be adopted as $|\phi\rangle=|\phi_1,\phi_2,\dots,\phi_{n_l}\rangle=TA|\psi_1,\psi_2,\dots,\psi_{n_l}\rangle$ when anti-unitary elemented is considered. Consequently, the co-reps for the full basis set $|\psi,\phi\rangle$ are expressed by Eq. (E13) and (E14), respectively.

Then we can use the following modified Dimmock and Wheeler's character sum rule in Eq. (E17), which helps to identify whether the co-reps are irreducible or not.

By applying the character sum rule (note that $T^2=-1$):

$$\sum_{g_i^{(a)}\in\tilde{L}_{SS}^Y}\chi^1\left(\left(TAg_i^{(a)}\right)^2\right)=2+2i-2-2i+2-2i-2+2i=0\cdots(c), \qquad \text{(G23)}$$

$$\sum_{g_i^{(a)}\in\tilde{L}_{SS}^Y}\chi^2\left(\left(TAg_i^{(a)}\right)^2\right)=2-2i-2+2i+2+2i-2-2i=0\cdots(c). \qquad \text{(G24)}$$

Two-dimensional projective irreps are paired into four-dimensional co-irreps with the introduction of the anti-unitary elements, and the co-irreps are shown in Table G5.

## 2. Irreducible co-representations of CeAuAl3

Furthermore, co-irreps of the SSG $P^{1}4^{1}m^{1}m^{4^{1}_{001}}(1/2\ 1/2\ 1/4)^{m}1$ are tabulated in Tables G2 to G7. The compatibility relation for the momenta Δ(0,v,0), X(0,1/2,0) and Y(u,1/2,0) (along X-M) are shown in Table G8, which is consistent with the electronic band structure in Fig. 5(b) in the maintext.

**Table G2:** Co-irreps of SSG $P^{1}4^{1}m^{1}m^{4^{1}_{001}}(1/2\ 1/2\ 1/4)^{m}1$ at Γ(0,0,0). The number in the parentheses in the first column denotes the dimension of the co-irrep.

| | $\{4^{1}_{001}\|\|1\|1/2\ 1/2\ 1/4\}$ | $\{1\|\|m_{010}\|0\}$ | $\{1\|\|4^{1}_{001}\|0\}$ | $\{T2_{001}\|\|1\|0\}$ |
|---|---|---|---|---|
| $D_{\Gamma}^{1}(2)$ | $\begin{pmatrix} e^{i\frac{\pi}{4}} & 0 \\ 0 & e^{-i\frac{\pi}{4}} \end{pmatrix}$ | $\begin{pmatrix} 1 & 0 \\ 0 & 1 \end{pmatrix}$ | $\begin{pmatrix} 1 & 0 \\ 0 & 1 \end{pmatrix}$ | $\begin{pmatrix} 0 & 1 \\ 1 & 0 \end{pmatrix}$ |
| $D_{\Gamma}^{2}(2)$ | $\begin{pmatrix} e^{i\frac{3\pi}{4}} & 0 \\ 0 & e^{-i\frac{3\pi}{4}} \end{pmatrix}$ | $\begin{pmatrix} 1 & 0 \\ 0 & 1 \end{pmatrix}$ | $\begin{pmatrix} 1 & 0 \\ 0 & 1 \end{pmatrix}$ | $\begin{pmatrix} 0 & 1 \\ 1 & 0 \end{pmatrix}$ |
| $D_{\Gamma}^{3}(2)$ | $\begin{pmatrix} e^{-i\frac{\pi}{4}} & 0 \\ 0 & e^{i\frac{\pi}{4}} \end{pmatrix}$ | $\begin{pmatrix} -1 & 0 \\ 0 & -1 \end{pmatrix}$ | $\begin{pmatrix} 1 & 0 \\ 0 & 1 \end{pmatrix}$ | $\begin{pmatrix} 0 & 1 \\ 1 & 0 \end{pmatrix}$ |
| $D_{\Gamma}^{4}(2)$ | $\begin{pmatrix} e^{-i\frac{3\pi}{4}} & 0 \\ 0 & e^{i\frac{3\pi}{4}} \end{pmatrix}$ | $\begin{pmatrix} -1 & 0 \\ 0 & -1 \end{pmatrix}$ | $\begin{pmatrix} 1 & 0 \\ 0 & 1 \end{pmatrix}$ | $\begin{pmatrix} 0 & 1 \\ 1 & 0 \end{pmatrix}$ |
| $D_{\Gamma}^{5}(2)$ | $\begin{pmatrix} e^{-i\frac{\pi}{4}} & 0 \\ 0 & e^{i\frac{\pi}{4}} \end{pmatrix}$ | $\begin{pmatrix} 1 & 0 \\ 0 & 1 \end{pmatrix}$ | $\begin{pmatrix} -1 & 0 \\ 0 & -1 \end{pmatrix}$ | $\begin{pmatrix} 0 & 1 \\ 1 & 0 \end{pmatrix}$ |
| $D_{\Gamma}^{6}(2)$ | $\begin{pmatrix} e^{-i\frac{3\pi}{4}} & 0 \\ 0 & e^{i\frac{3\pi}{4}} \end{pmatrix}$ | $\begin{pmatrix} 1 & 0 \\ 0 & 1 \end{pmatrix}$ | $\begin{pmatrix} -1 & 0 \\ 0 & -1 \end{pmatrix}$ | $\begin{pmatrix} 0 & 1 \\ 1 & 0 \end{pmatrix}$ |
| $D_{\Gamma}^{7}(2)$ | $\begin{pmatrix} e^{i\frac{3\pi}{4}} & 0 \\ 0 & e^{-i\frac{3\pi}{4}} \end{pmatrix}$ | $\begin{pmatrix} -1 & 0 \\ 0 & -1 \end{pmatrix}$ | $\begin{pmatrix} -1 & 0 \\ 0 & -1 \end{pmatrix}$ | $\begin{pmatrix} 0 & 1 \\ 1 & 0 \end{pmatrix}$ |
| $D_{\Gamma}^{8}(2)$ | $\begin{pmatrix} e^{i\frac{\pi}{4}} & 0 \\ 0 & e^{-i\frac{\pi}{4}} \end{pmatrix}$ | $\begin{pmatrix} -1 & 0 \\ 0 & -1 \end{pmatrix}$ | $\begin{pmatrix} -1 & 0 \\ 0 & -1 \end{pmatrix}$ | $\begin{pmatrix} 0 & 1 \\ 1 & 0 \end{pmatrix}$ |
| $D_{\Gamma}^{9}(4)$ | $\begin{pmatrix} e^{-i\frac{\pi}{4}} & 0 & 0 & 0 \\ 0 & e^{-i\frac{\pi}{4}} & 0 & 0 \\ 0 & 0 & e^{i\frac{\pi}{4}} & 0 \\ 0 & 0 & 0 & e^{i\frac{\pi}{4}} \end{pmatrix}$ | $\begin{pmatrix} 0 & 1 & 0 & 0 \\ 1 & 0 & 0 & 0 \\ 0 & 0 & 0 & 1 \\ 0 & 0 & 1 & 0 \end{pmatrix}$ | $\begin{pmatrix} -i & 0 & 0 & 0 \\ 0 & i & 0 & 0 \\ 0 & 0 & i & 0 \\ 0 & 0 & 0 & -i \end{pmatrix}$ | $\begin{pmatrix} 0 & 0 & 1 & 0 \\ 0 & 0 & 0 & 1 \\ 1 & 0 & 0 & 0 \\ 0 & 1 & 0 & 0 \end{pmatrix}$ |
| $D_{\Gamma}^{10}(4)$ | $\begin{pmatrix} e^{-i\frac{3\pi}{4}} & 0 & 0 & 0 \\ 0 & e^{-i\frac{3\pi}{4}} & 0 & 0 \\ 0 & 0 & e^{i\frac{3\pi}{4}} & 0 \\ 0 & 0 & 0 & e^{i\frac{3\pi}{4}} \end{pmatrix}$ | $\begin{pmatrix} 0 & 1 & 0 & 0 \\ 1 & 0 & 0 & 0 \\ 0 & 0 & 0 & 1 \\ 0 & 0 & 1 & 0 \end{pmatrix}$ | $\begin{pmatrix} -i & 0 & 0 & 0 \\ 0 & i & 0 & 0 \\ 0 & 0 & i & 0 \\ 0 & 0 & 0 & -i \end{pmatrix}$ | $\begin{pmatrix} 0 & 0 & 1 & 0 \\ 0 & 0 & 0 & 1 \\ 1 & 0 & 0 & 0 \\ 0 & 1 & 0 & 0 \end{pmatrix}$ |

**Table G3:** Co-irrep of the SSG $P^{1}4^{1}m^{1}m^{4_{001}^{1}}(1/2\ 1/2\ 1/4)^{m}1$ at $\Delta(0,\mathrm{v},0)$. The number in the parentheses in the first column denotes the dimension of the co-irrep.

| | $\{4_{001}^{1}\|\|1\|1/2\ 1/2\ 1/4\}$ | $\{1\|\|m_{100}\|0\}$ | $\{T2_{001}\|\|m_{010}\|0\}$ |
|---|---|---|---|
| $D_{\Delta}^{1}(2)$ | $\begin{pmatrix} e^{-i(\frac{-\pi}{4}+v\pi)} & 0 \\ 0 & e^{-i(\frac{\pi}{4}+v\pi)} \end{pmatrix}$ | $\begin{pmatrix} 1 & 0 \\ 0 & 1 \end{pmatrix}$ | $\begin{pmatrix} 0 & 1 \\ 1 & 0 \end{pmatrix}$ |
| $D_{\Delta}^{2}(2)$ | $\begin{pmatrix} e^{-i(\frac{-3\pi}{4}+v\pi)} & 0 \\ 0 & e^{-i(\frac{3\pi}{4}+v\pi)} \end{pmatrix}$ | $\begin{pmatrix} 1 & 0 \\ 0 & 1 \end{pmatrix}$ | $\begin{pmatrix} 0 & 1 \\ 1 & 0 \end{pmatrix}$ |
| $D_{\Delta}^{3}(2)$ | $\begin{pmatrix} e^{-i(\frac{\pi}{4}+v\pi)} & 0 \\ 0 & e^{-i(\frac{-\pi}{4}+v\pi)} \end{pmatrix}$ | $\begin{pmatrix} -1 & 0 \\ 0 & -1 \end{pmatrix}$ | $\begin{pmatrix} 0 & 1 \\ 1 & 0 \end{pmatrix}$ |
| $D_{\Delta}^{4}(2)$ | $\begin{pmatrix} e^{-i(\frac{3\pi}{4}+v\pi)} & 0 \\ 0 & e^{-i(\frac{-3\pi}{4}+v\pi)} \end{pmatrix}$ | $\begin{pmatrix} -1 & 0 \\ 0 & -1 \end{pmatrix}$ | $\begin{pmatrix} 0 & 1 \\ 1 & 0 \end{pmatrix}$ |

**Table G4:** Co-irreps of the SSG $P^{1}4^{1}m^{1}m^{4_{001}^{1}}(1/2\ 1/2\ 1/4)^{m}1$ at X(0,1/2,0). The number in the parentheses in the first column denotes the dimension of the co-irrep.

| | $\{4_{001}^{1}\|\|1\|1/2\ 1/2\ 1/4\}$ | $\{1\|\|m_{010}\|0\}$ | $\{1\|\|2_{001}\|0\}$ | $\{T2_{001}\|\|1\|0\}$ |
|---|---|---|---|---|
| $D_{X}^{1}(4)$ | $\begin{pmatrix} 0 & i & 0 & 0 \\ -1 & 0 & 0 & 0 \\ 0 & 0 & 0 & -i \\ 0 & 0 & -1 & 0 \end{pmatrix}$ | $\begin{pmatrix} 1 & 0 & 0 & 0 \\ 0 & -1 & 0 & 0 \\ 0 & 0 & 1 & 0 \\ 0 & 0 & 0 & -1 \end{pmatrix}$ | $\begin{pmatrix} 1 & 0 & 0 & 0 \\ 0 & -1 & 0 & 0 \\ 0 & 0 & 1 & 0 \\ 0 & 0 & 0 & -1 \end{pmatrix}$ | $\begin{pmatrix} 0 & 0 & 1 & 0 \\ 0 & 0 & 0 & 1 \\ 1 & 0 & 0 & 0 \\ 0 & 1 & 0 & 0 \end{pmatrix}$ |
| $D_{X}^{2}(4)$ | $\begin{pmatrix} 0 & -i & 0 & 0 \\ -1 & 0 & 0 & 0 \\ 0 & 0 & 0 & i \\ 0 & 0 & -1 & 0 \end{pmatrix}$ | $\begin{pmatrix} 1 & 0 & 0 & 0 \\ 0 & -1 & 0 & 0 \\ 0 & 0 & 1 & 0 \\ 0 & 0 & 0 & -1 \end{pmatrix}$ | $\begin{pmatrix} -1 & 0 & 0 & 0 \\ 0 & 1 & 0 & 0 \\ 0 & 0 & -1 & 0 \\ 0 & 0 & 0 & 1 \end{pmatrix}$ | $\begin{pmatrix} 0 & 0 & 1 & 0 \\ 0 & 0 & 0 & 1 \\ 1 & 0 & 0 & 0 \\ 0 & 1 & 0 & 0 \end{pmatrix}$ |

**Table G5:** Co-irreps of the SSG $P^{1}4^{1}m^{1}m^{4_{001}^{1}}(1/2\ 1/2\ 1/4)^{m}1$ at Y(u,1/2,0). The number in the parentheses in the first column denotes the dimension of the co-irrep.

| | $\{4_{001}^{1}\|\|1\|1/2\ 1/2\ 1/4\}$ | $\{1\|\|m_{010}\|0\}$ | $\{T2_{001}\|\|m_{100}\|0\}$ |
|---|---|---|---|
| $D_{Y}^{1}(4)$ | $\begin{pmatrix} e^{i(\frac{\pi}{4}-u\pi)} & 0 & 0 & 0 \\ 0 & e^{-i(\frac{3\pi}{4}+u\pi)} & 0 & 0 \\ 0 & 0 & e^{-i(\frac{\pi}{4}+u\pi)} & 0 \\ 0 & 0 & 0 & e^{i(\frac{3\pi}{4}-u\pi)} \end{pmatrix}$ | $\begin{pmatrix} 0 & e^{i\frac{3\pi}{4}} & 0 & 0 \\ e^{-i\frac{3\pi}{4}} & 0 & 0 & 0 \\ 0 & 0 & 0 & e^{-i\frac{3\pi}{4}} \\ 0 & 0 & e^{i\frac{3\pi}{4}} & 0 \end{pmatrix}$ | $\begin{pmatrix} 0 & 0 & 1 & 0 \\ 0 & 0 & 0 & 1 \\ 1 & 0 & 0 & 0 \\ 0 & 1 & 0 & 0 \end{pmatrix}$ |

**Table G6:** Co-irreps of the SSG $P^{1}4^{1}m^{1}m^{4_{001}^{1}}(1/2\ 1/2\ 1/4)^{m}1$ at M(1/2,1/2,0). The number in the parentheses in the first column denotes the dimension of the co-irrep.

| | $\{4^1_{001}\|\|1\|1/2\ 1/2\ 1/4\}$ | $\{1\|\|m_{010}\|0\}$ | $\{1\|\|4^1_{001}\|0\}$ | $\{T2_{001}\|\|1\|0\}$ |
|---|---|---|---|---|
| $D^1_M(4)$ | $\begin{pmatrix}0&1&0&0\\i&0&0&0\\0&0&0&1\\0&0&-i&0\end{pmatrix}$ | $\begin{pmatrix}1&0&0&0\\0&-1&0&0\\0&0&1&0\\0&0&0&-1\end{pmatrix}$ | $\begin{pmatrix}1&0&0&0\\0&-1&0&0\\0&0&1&0\\0&0&0&-1\end{pmatrix}$ | $\begin{pmatrix}0&0&1&0\\0&0&0&1\\1&0&0&0\\0&1&0&0\end{pmatrix}$ |
| $D^2_M(4)$ | $\begin{pmatrix}0&i&0&0\\-1&0&0&0\\0&0&0&-i\\0&0&-1&0\end{pmatrix}$ | $\begin{pmatrix}1&0&0&0\\0&-1&0&0\\0&0&1&0\\0&0&0&-1\end{pmatrix}$ | $\begin{pmatrix}-1&0&0&0\\0&1&0&0\\0&0&-1&0\\0&0&0&1\end{pmatrix}$ | $\begin{pmatrix}0&0&1&0\\0&0&0&1\\1&0&0&0\\0&1&0&0\end{pmatrix}$ |
| $D^3_M(4)$ | $\begin{pmatrix}0&i&0&0\\-1&0&0&0\\0&0&0&-i\\0&0&-1&0\end{pmatrix}$ | $\begin{pmatrix}0&e^{-i\frac{3\pi}{4}}&0&0\\e^{i\frac{3\pi}{4}}&0&0&0\\0&0&0&e^{i\frac{3\pi}{4}}\\0&0&e^{-i\frac{3\pi}{4}}&0\end{pmatrix}$ | $\begin{pmatrix}-1&0&0&0\\0&1&0&0\\0&0&1&0\\0&0&0&-1\end{pmatrix}$ | $\begin{pmatrix}0&0&1&0\\0&0&0&1\\1&0&0&0\\0&1&0&0\end{pmatrix}$ |
| $D^4_M(4)$ | $\begin{pmatrix}0&e^{i\frac{\pi}{4}}&0&0\\e^{i\frac{\pi}{4}}&0&0&0\\0&0&0&e^{-i\frac{\pi}{4}}\\0&0&e^{-i\frac{\pi}{4}}&0\end{pmatrix}$ | $\begin{pmatrix}0&-i&0&0\\i&0&0&0\\0&0&0&i\\0&0&-i&0\end{pmatrix}$ | $\begin{pmatrix}-i&0&0&0\\0&i&0&0\\0&0&i&0\\0&0&0&-i\end{pmatrix}$ | $\begin{pmatrix}0&0&1&0\\0&0&0&1\\1&0&0&0\\0&1&0&0\end{pmatrix}$ |

**Table G7:** Co-irreps of the SSG $P^14^1m^1m^{4^1_{001}}(1/2\ 1/2\ 1/4)^m1$ at Z(0,0,1/2). The number in the parentheses in the first column denotes the dimension of the co-irrep.

| | $\{4^1_{001}\|\|1\|1/2\ 1/2\ 1/4\}$ | $\{1\|\|m_{010}\|0\}$ | $\{1\|\|4^1_{001}\|0\}$ | $\{T2_{001}\|\|1\|0\}$ |
|---|---|---|---|---|
| $D^1_Z(1)$ | 1 | 1 | 1 | 1 |
| $D^2_Z(1)$ | −1 | 1 | 1 | 1 |
| $D^3_Z(1)$ | 1 | −1 | 1 | 1 |
| $D^4_Z(1)$ | −1 | −1 | 1 | 1 |
| $D^5_Z(1)$ | 1 | 1 | −1 | 1 |
| $D^6_Z(1)$ | −1 | 1 | −1 | 1 |
| $D^7_Z(1)$ | −1 | −1 | −1 | 1 |
| $D^8_Z(1)$ | 1 | −1 | −1 | 1 |
| $D^9_Z(2)$ | $\begin{pmatrix}-i&0\\0&i\end{pmatrix}$ | $\begin{pmatrix}1&0\\0&1\end{pmatrix}$ | $\begin{pmatrix}1&0\\0&1\end{pmatrix}$ | $\begin{pmatrix}0&1\\1&0\end{pmatrix}$ |
| $D^{10}_Z(2)$ | $\begin{pmatrix}-i&0\\0&i\end{pmatrix}$ | $\begin{pmatrix}-1&0\\0&-1\end{pmatrix}$ | $\begin{pmatrix}1&0\\0&1\end{pmatrix}$ | $\begin{pmatrix}0&1\\1&0\end{pmatrix}$ |
| $D^{11}_Z(2)$ | $\begin{pmatrix}-i&0\\0&i\end{pmatrix}$ | $\begin{pmatrix}1&0\\0&1\end{pmatrix}$ | $\begin{pmatrix}-1&0\\0&-1\end{pmatrix}$ | $\begin{pmatrix}0&1\\1&0\end{pmatrix}$ |
| $D^{12}_Z(2)$ | $\begin{pmatrix}i&0\\0&-i\end{pmatrix}$ | $\begin{pmatrix}-1&0\\0&-1\end{pmatrix}$ | $\begin{pmatrix}-1&0\\0&-1\end{pmatrix}$ | $\begin{pmatrix}0&1\\1&0\end{pmatrix}$ |
| $D^{13}_Z(2)$ | $\begin{pmatrix}1&0\\0&1\end{pmatrix}$ | $\begin{pmatrix}0&1\\1&0\end{pmatrix}$ | $\begin{pmatrix}i&0\\0&-i\end{pmatrix}$ | $\begin{pmatrix}0&1\\1&0\end{pmatrix}$ |
| $D^{14}_Z(2)$ | $\begin{pmatrix}-1&0\\0&-1\end{pmatrix}$ | $\begin{pmatrix}0&1\\1&0\end{pmatrix}$ | $\begin{pmatrix}-i&0\\0&i\end{pmatrix}$ | $\begin{pmatrix}0&1\\1&0\end{pmatrix}$ |
| $D_Z^{15}(2)$ | $\begin{pmatrix}1 & 0\\0 & 1\end{pmatrix}$ | $\begin{pmatrix}0 & -1\\-1 & 0\end{pmatrix}$ | $\begin{pmatrix}-i & 0\\0 & i\end{pmatrix}$ | $\begin{pmatrix}0 & 1\\1 & 0\end{pmatrix}$ |
| $D_Z^{16}(2)$ | $\begin{pmatrix}-1 & 0\\0 & -1\end{pmatrix}$ | $\begin{pmatrix}0 & e^{i\frac{3\pi}{4}}\\e^{-i\frac{3\pi}{4}} & 0\end{pmatrix}$ | $\begin{pmatrix}i & 0\\0 & -i\end{pmatrix}$ | $\begin{pmatrix}0 & 1\\1 & 0\end{pmatrix}$ |
| $D_Z^{17}(4)$ | $\begin{pmatrix}-i & 0 & 0 & 0\\0 & -i & 0 & 0\\0 & 0 & i & 0\\0 & 0 & 0 & i\end{pmatrix}$ | $\begin{pmatrix}0 & 1 & 0 & 0\\1 & 0 & 0 & 0\\0 & 0 & 0 & 1\\0 & 0 & 1 & 0\end{pmatrix}$ | $\begin{pmatrix}-i & 0 & 0 & 0\\0 & i & 0 & 0\\0 & 0 & i & 0\\0 & 0 & 0 & -i\end{pmatrix}$ | $\begin{pmatrix}0 & 0 & 1 & 0\\0 & 0 & 0 & 1\\1 & 0 & 0 & 0\\0 & 1 & 0 & 0\end{pmatrix}$ |

**Table G8:** Compatibility relation of the SSG $P^{1}4^{1}m^{1}m^{4^{1}_{001}}(1/2\ 1/2\ 1/4)^{m}1$ between coirreps for k-vectors $\Delta$, X and Y. The little groups at different k-vectors are listed in the second row for the corresponding momenta.

| $\Delta(0,\mathrm{v},0)$ | X(0,1/2,0) | Y(u,1/2,0) |
|---|---|---|
| ${}^{1}m_{100}{}^{4/m}1$ | ${}^{1}4^{1}m^{1}m^{4/m}1$ | ${}^{1}m_{010}{}^{4/m}1$ |
| $D_\Delta^1(2)\oplus D_\Delta^2(2)$ | $D_X^1(4)$ | $D_Y^1(4)$ |
| $D_\Delta^3(2)\oplus D_\Delta^4(2)$ | $D_X^2(4)$ | $D_Y^1(4)$ |

## 3. General applicability of CSCO method in CeAuAl3's MSG

The reps of MSGs can be naturally incorporated into our theory framework. Compared with SSG, the symmetry of the system is reduced to the MSG $P_C4_1$ (No. 76.10) after the introduction of SOC. Although the co-rep theories of MSGs have been firmly established on textbooks, we provide the deduction of the co-irrep matrices of CeAuAl3' s magnetic little group at high-symmetry point $A(1/2,1/2,1/2)$ through the CSCO method to demonstrate its general applicability.

The quotient group $\tilde{G}_{MS}^{A}=G_{MS}^{A}/T(L_0)\cong 4/m$ of CeAuAl3's magnetic little group contains four unitary symmetry elements:

$$\begin{aligned} g_1^{(1)} &= \{1||1|0\}, & g_2^{(1)} &= \{4_{001}^1||4_{001}^1|0\ 0\ 1/4\}, \\ g_3^{(1)} &= \{2_{001}||2_{001}|0\ 0\ 1/2\}, & g_4^{(1)} &= \{4_{001}^3||4_{001}^3|0\ 0\ 3/4\}, \end{aligned} \tag{G25}$$

and an anti-unitary generator:

$$g_5^{(1)} = TA = T\{1||1|0\ 0\ 1/2\}. \tag{G26}$$

We firstly consider only the MUSG $\tilde{L}_{MS}^{A} \cong 4$ of $\tilde{G}_{MS}^{A}$. Then according to the matrix element of the left regular projective reps:

$$M_k\left(g_i^{(a)}\right)_{g,g_j^{(a)}} = \left\langle g\middle|g_i^{(a)}\middle|g_j^{(a)}\right\rangle = (-1)^{\xi\left(\phi\left(g_{s_l}^{(a)}\right)\right)} exp\left(-iK_i \cdot \tau_j^{(a)}\right)\delta_{g,g_i^{(a)}g_j^{(a)}}, \quad \text{(G27)}$$

the rep matrices of the generators of MUSG $\tilde{L}_{MS}^{A}$ can be expressed as follows:

$$M_A\left(g_2^{(1)}\right) = \begin{pmatrix} 0 & 0 & 0 & -1 \\ 1 & 0 & 0 & 0 \\ 0 & 1 & 0 & 0 \\ 0 & 0 & 1 & 0 \end{pmatrix}, \quad \text{(G28)}$$

The CSCO-I $C$ of $\tilde{L}_{MS}^{A}$, for $k_i^{(a)} = 1$ for all $i = 1,2,3,4$ and $a = 1$ are constructed using the linear combination of class operators $C = \sum_{i,a} k_i^{(a)} C_i^{(a)}$:

$$C = \sum_{\substack{i=1,2,3,4 \\ a=1}} C_i^{(a)} = \begin{pmatrix} 4 & -4 & -4 & -4 \\ 4 & 4 & -4 & -4 \\ 4 & 4 & 4 & -4 \\ 4 & 4 & 4 & 4 \end{pmatrix}, \quad \text{(G29)}$$

where $C_i^{(a)} = \sum_{g_j^{(a)} \in \tilde{L}_{MS}^{A}} M_A(g_j^{(a)})^{-1} M_A(g_i^{(a)}) M_A(g_j^{(a)})$ are class operators and $k_i^{(1)}$ are arbitrary real coefficients to avoid accidental elimination of the same class operator.

It is noteworthy that in this particular example, we observe the existence of four distinct eigenvalues of $C$, namely $4 + 9.6569i$, $4 - 9.6569i$, $4 + 1.6569i$, and $4 - 1.6569i$, which implies that the CSCO-1 $C$ can sufficiently serve as the CSCO of the MUSG $\tilde{L}_{MS}^{A}$ . In addition, $\tilde{L}_{MS}^{A}$ possesses four inequivalent one-dimensional irreps (shown in Table G9). The irreps $d_A^l$ can be obtained by $d_A^l(\{g_{s_i}^{(a)}||R_i|\tau_i^{(a)}\}) = exp(-iA \cdot \tau_i^{(a)}) M_A^l(\{g_{s_i}^{(a)}||R_i|\tau_i^{(a)}\})$ (shown in Table G10).

**Table G9**: Character table of projective irreps $M_A^l$ of $\tilde{L}_{MS}^{A}$.

| | $g_1^{(1)}$ | $g_2^{(1)}$ | $g_3^{(1)}$ | $g_4^{(1)}$ |
|---|---|---|---|---|
| $M_A^1(1)$ | 1 | $e^{i\frac{\pi}{4}}$ | $i$ | $e^{i\frac{3\pi}{4}}$ |
| $M_A^2(1)$ | 1 | $e^{-i\frac{3\pi}{4}}$ | $i$ | $e^{-i\frac{\pi}{4}}$ |
| $M_A^3(1)$ | 1 | $e^{-i\frac{\pi}{4}}$ | $-i$ | $e^{-i\frac{3\pi}{4}}$ |
| $M_A^4(1)$ | 1 | $e^{i\frac{3\pi}{4}}$ | $-i$ | $e^{i\frac{\pi}{4}}$ |

**Table G10**: Character table of irreps $d_A^l$ of $\tilde{L}_{MS}^A$.

| | $g_1^{(1)}$ | $g_2^{(1)}$ | $g_3^{(1)}$ | $g_4^{(1)}$ |
|---|---|---|---|---|
| $d_A^1(1)$ | $1$ | $1$ | $1$ | $1$ |
| $d_A^2(1)$ | $1$ | $-1$ | $1$ | $-1$ |
| $d_A^3(1)$ | $1$ | $-i$ | $-1$ | $i$ |
| $d_A^4(1)$ | $1$ | $i$ | $-1$ | $-i$ |

After considering the anti-unitary operator $g_5^{(1)} = TA = T\{1||1|0\ 0\ 1/2\}$, one can apply the modified Dimmock and Wheeler's character sum rule in Eq. (E17): (note that $T^2 = -1$)

$$\sum_{g_i^{(a)} \in \tilde{L}_{MS}^A} \chi^1\left(\left(TAg_i^{(a)}\right)^2\right) = 1 + 1 + 1 + 1 = 4 \cdots (a), \tag{G30}$$

$$\sum_{g_i^{(a)} \in \tilde{L}_{MS}^A} \chi^2\left(\left(TAg_i^{(a)}\right)^2\right) = 1 + 1 + 1 + 1 = 4 \cdots (a), \tag{G31}$$

$$\sum_{g_i^{(a)} \in \tilde{L}_{MS}^A} \chi^3\left(\left(TAg_i^{(a)}\right)^2\right) = 1 - 1 + 1 - 1 = 0 \cdots (c), \tag{G32}$$

$$\sum_{g_i^{(a)} \in \tilde{L}_{MS}^A} \chi^4\left(\left(TAg_i^{(a)}\right)^2\right) = 1 - 1 + 1 - 1 = 0 \cdots (c). \tag{G33}$$

Consequently, the co-rep matrices of $d_A^1$ and $d_A^2$ have identical dimensions with themselves while $d_A^3$ and $d_A^4$ could be paired by the anti-unitary operator $g_5^{(1)}$ and form a two-dimensional co-irreps. The result is given in Table G12. Besides, the co-irreps matrices of the MSG of $CeAuAl_3$ at some high-symmetry points are also provided in Tables G11 to G16 and these results are consistent with the matrices of COREPRESENTATIONS on Bilbao [9, 91].

**Table G11:** Co-irreps of $CeAuAl_3$'s MSG $P_c4_1$ (No. 76.10) at $\Gamma(0,0,0)$. The number in the parentheses in the first column denotes the dimension of the co-irrep.

| | $\{4_{001}^1\|\|4_{001}^1\|0\ 0\ 1/4\}$ | $\{T\|\|1\|0\ 0\ 1/2\}$ |
|---|---|---|
| $D_\Gamma^1(2)$ | $\begin{pmatrix} e^{i\frac{\pi}{4}} & 0 \\ 0 & e^{-i\frac{\pi}{4}} \end{pmatrix}$ | $\begin{pmatrix} 0 & -1 \\ 1 & 0 \end{pmatrix}$ |
| $D_\Gamma^2(2)$ | $\begin{pmatrix} e^{i\frac{3\pi}{4}} & 0 \\ 0 & e^{-i\frac{3\pi}{4}} \end{pmatrix}$ | $\begin{pmatrix} 0 & -1 \\ 1 & 0 \end{pmatrix}$ |

**Table G12:** Co-irreps of CeAuAl$_3$'s MSG $P_c4_1$ (No. 76.10) at $A(1/2,1/2,1/2)$. The number in the parentheses in the first column denotes the dimension of the co-irrep.

| | $\{4_{001}^1\|\|4_{001}^1\|0\ 0\ 1/4\}$ | $\{T\|\|1\|0\ 0\ 1/2\}$ |
|---|---|---|
| $D_A^1(1)$ | $1$ | $1$ |
| $D_A^2(1)$ | $-1$ | $1$ |
| $D_A^3(2)$ | $\begin{pmatrix} -i & 0 \\ 0 & i \end{pmatrix}$ | $\begin{pmatrix} 0 & 1 \\ 1 & 0 \end{pmatrix}$ |

**Table G13:** Co-irreps of CeAuAl$_3$'s MSG $P_c4_1$ (No. 76.10) at $M(1/2,1/2,0)$. The number in the parentheses in the first column denotes the dimension of the co-irrep.

| | $\{4_{001}^1\|\|4_{001}^1\|0\ 0\ 1/4\}$ | $\{T\|\|1\|0\ 0\ 1/2\}$ |
|---|---|---|
| $D_M^1(2)$ | $\begin{pmatrix} e^{i\frac{\pi}{4}} & 0 \\ 0 & e^{-i\frac{\pi}{4}} \end{pmatrix}$ | $\begin{pmatrix} 0 & -1 \\ 1 & 0 \end{pmatrix}$ |
| $D_M^2(2)$ | $\begin{pmatrix} e^{i\frac{3\pi}{4}} & 0 \\ 0 & e^{-i\frac{3\pi}{4}} \end{pmatrix}$ | $\begin{pmatrix} 0 & -1 \\ 1 & 0 \end{pmatrix}$ |

**Table G14:** Co-irreps of CeAuAl$_3$'s MSG $P_c4_1$ (No. 76.10) at $R(0,1/2,1/2)$. The number in the parentheses in the first column denotes the dimension of the co-irrep.

| | $\{2_{001}\|\|2_{001}\|0\ 0\ 1/2\}$ | $\{T\|\|1\|0\ 0\ 1/2\}$ |
|---|---|---|
| $D_R^1(1)$ | $1$ | $1$ |
| $D_R^2(1)$ | $-1$ | $1$ |

**Table G15:** Co-irreps of CeAuAl$_3$'s MSG $P_c4_1$ (No. 76.10) at $X(0,1/2,0)$. The number in the parentheses in the first column denotes the dimension of the co-irrep.

| | $\{2_{001}\|\|2_{001}\|0\ 0\ 1/2\}$ | $\{T\|\|1\|0\ 0\ 1/2\}$ |
|---|---|---|
| $D_X^1(2)$ | $\begin{pmatrix} -i & 0 \\ 0 & i \end{pmatrix}$ | $\begin{pmatrix} 0 & -1 \\ 1 & 0 \end{pmatrix}$ |

**Table G16:** Co-irreps of CeAuAl$_3$'s MSG $P_c4_1$ (No. 76.10) at $Z(0,0,1/2)$. The number in the parentheses in the first column denotes the dimension of the co-irrep.

| | $\{4_{001}^1\|\|4_{001}^1\|0\ 0\ 1/4\}$ | $\{T\|\|1\|0\ 0\ 1/2\}$ |
|---|---|---|
| $D_Z^1(2)$ | $\begin{pmatrix} i & 0 \\ 0 & -i \end{pmatrix}$ | $\begin{pmatrix} 0 & 1 \\ 1 & 0 \end{pmatrix}$ |
| $D_Z^2(1)$ | $1$ | $1$ |
| $D_Z^3(1)$ | $-1$ | $1$ |

## H. Band representations in noncoplanar SSGs

In this section, we provide the derivation of the co-irrep matrices and character tables of the projective co-irreps of the little groups at high-symmetry points and lines in noncoplanar SSG.

The nontrivial SSG of $CoNb_3S_6$ is $P^{3^2_{001}}6_3{}^{m_{100}}2^{m_{010}}2|(2_{24-3},2_{423},1)$ (4.182.4.2). Considering the spin-only group $G_{SO}^{np}=E$ for noncoplanar magnetic structure, it turns into $P^{3^2_{001}}6_3{}^{m_{100}}2^{m_{010}}2|(2_{24-3},2_{423},1)$. The $M$(1/2,0,0) point is taken as an example to show the process of calculating projective irreps of noncoplanar SSG using the CSCO method.

First, the spin Brillouin zone and the magnetic Brillouin zone with all high-symmetric points/lines for noncoplanar antiferromagnet $CoNb_3S_6$ are given in Figs. H1 and H2, respectively.

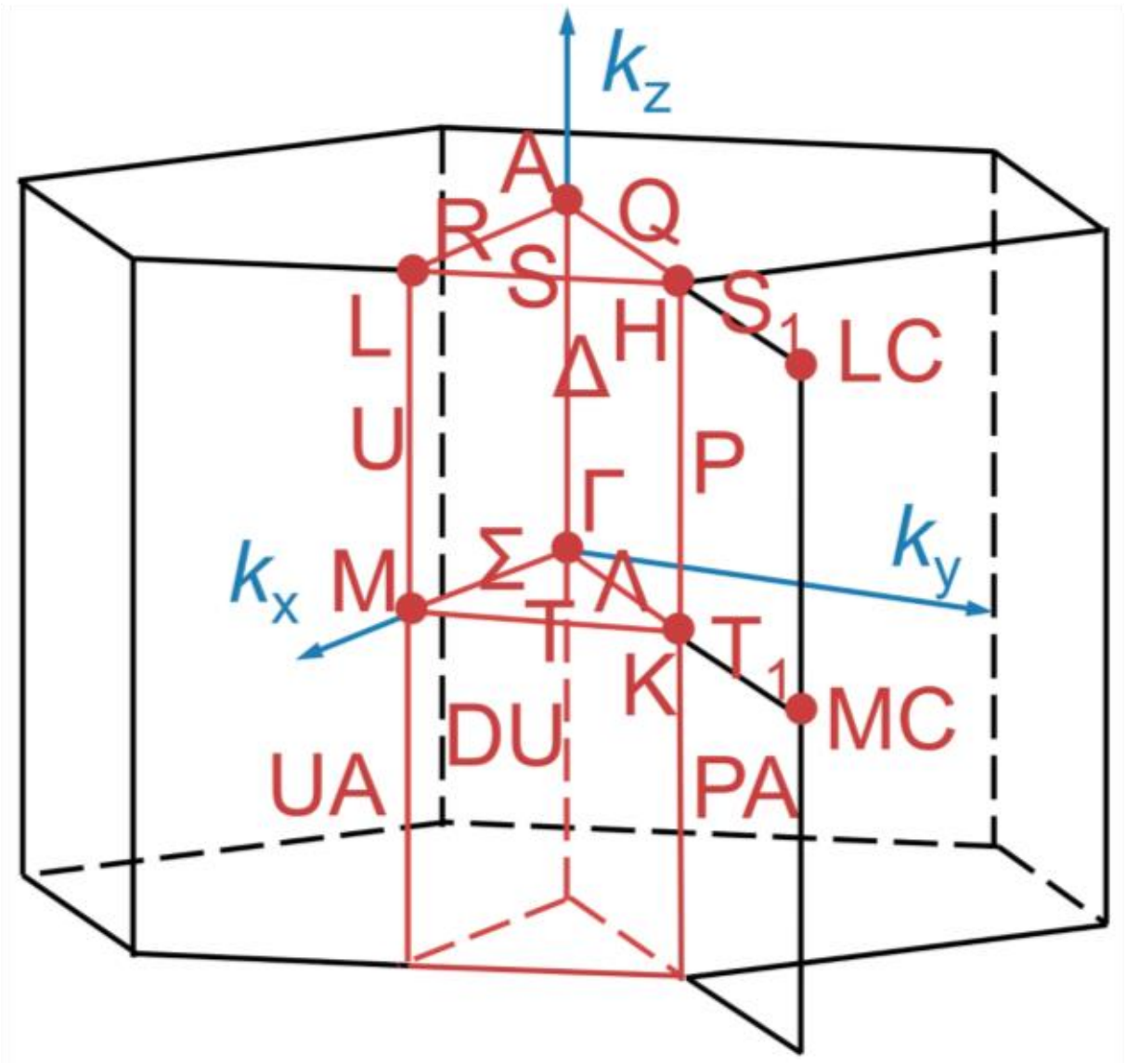

**Fig. H1:** Schematic of spin Brillouin zone for SSG $P^{3^2_{001}}6_3{}^{m_{100}}2^{m_{010}}2|(2_{24-3},2_{423},1)$.

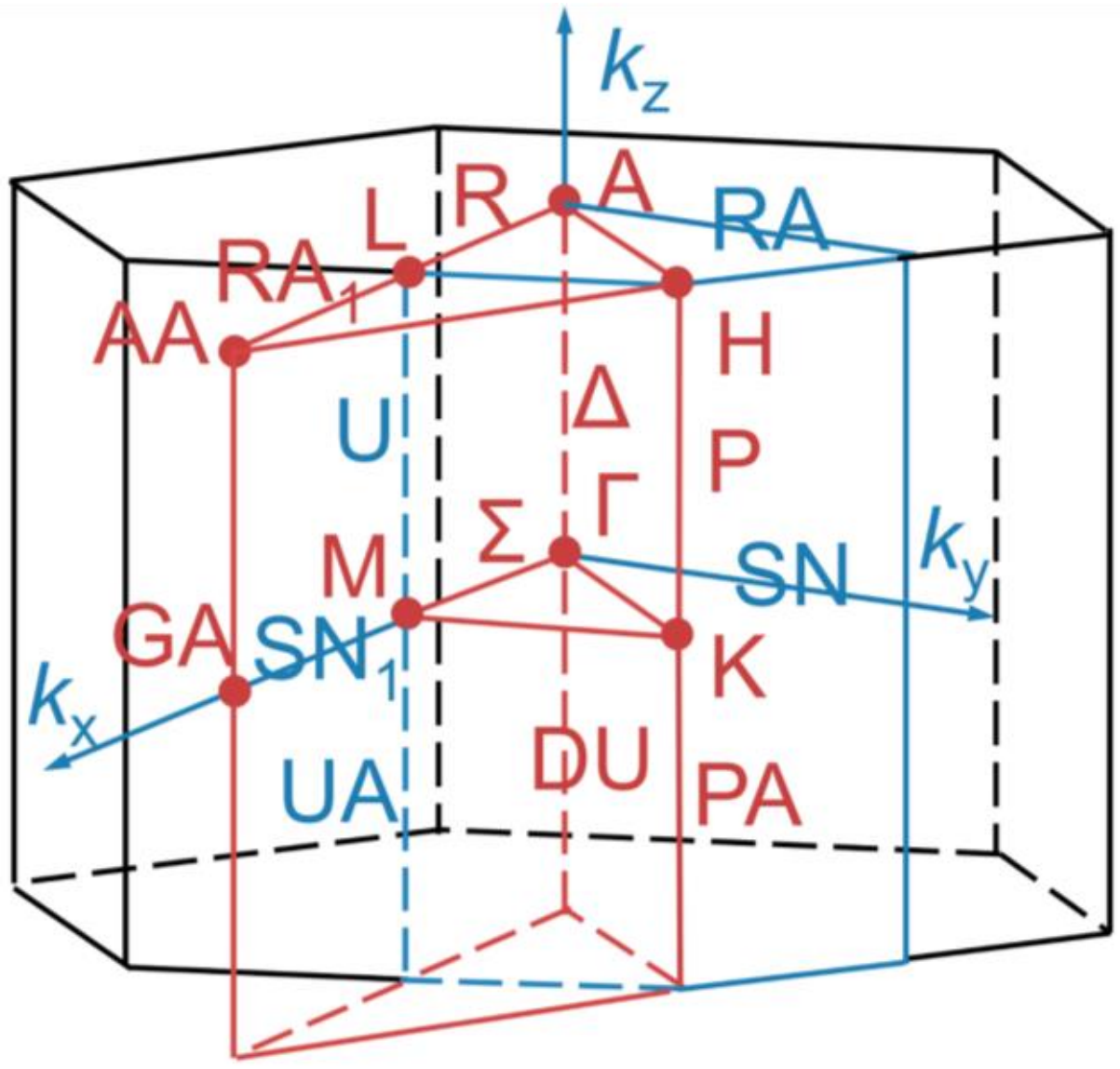

**Fig. H2:** Schematic of magnetic Brillouin zone for MSG $P32'1$.

## 1. Irreducible co-representations of $CoNb_3S_6$ at $M$(1/2,0,0)

Similar approach are employed to obtain co-irreps of the little group $G_{SS}^{M}$ of $CoNb_3S_6$ at $M$(1/2,0,0). The SSG of $CoNb_3S_6$ is $P^{3_{001}^{2}}6_3{}^{m_{100}}2^{m_{010}}2|(2_{24-3},2_{423},1)$ (4.182.4.2). The finite quotient group $\tilde{G}_{SS}^{M}=G_{SS}^{M}/T(L_0)$ is given by:

$$\begin{aligned} &g_1^{(1)}=\{1||1|0\}, && g_1^{(2)}=\{2_{24-3}||1|1\ 0\ 0\},\\ &g_1^{(3)}=\{2_{423}||1|0\ 1\ 0\}, && g_1^{(4)}=g_1^{(2)}g_1^{(3)},\\ &g_2^{(1)}=\{1||2_{001}|0\ 0\ 1/2\} && g_2^{(2)}=g_1^{(2)}g_2^{(1)},\\ &g_2^{(3)}=g_1^{(3)}g_2^{(1)} && g_2^{(4)}=g_1^{(4)}g_2^{(1)}, \end{aligned} \quad \text{(H1)}$$

and the anti-unitary operator:

$$g_3^{(1)}=TA=T\{2_{010}||2_{210}|0\}, \quad \text{(H2)}$$

First, we consider the maximal unitary subgroup $\tilde{L}_{SS}^{M}\cong mmm$ of $\tilde{G}_{SS}^{M}$. According to the matrix element of the left regular projective reps:

$$M_k\left(g_i^{(a)}\right)_{g,g_j^{(a)}}=\left\langle g\middle|g_i^{(a)}\middle|g_j^{(a)}\right\rangle=(-1)^{\xi\left(\phi\left(g_{s_l}^{(a)}\right)\right)}exp\left(-iK_i\cdot\tau_j^{(a)}\right)\delta_{g,g_i^{(a)}g_j^{(a)}}, \quad \text{(H3)}$$

and the matrix element of the intrinsic group:

$$M_k\left(\bar{g}_i^{(a)}\right)_{g,g_j^{(a)}}=\left\langle g\middle|\bar{g}_i^{(a)}\middle|g_j^{(a)}\right\rangle=(-1)^{\xi\left(\phi\left(g_{s_m}^{(a)}\right)\right)}\exp\left(-iK_j\cdot\tau_i^{(a)}\right)\delta_{g,g_j^{(a)}g_i^{(a)}}, \quad \text{(H4)}$$

the rep matrices of the generators of MUSG $\tilde{L}_{SS}^{M}$ can be obtained:

$$M_M\left(g_1^{(2)}\right) = \begin{pmatrix} 0 & 0 & -1 & 0 & 0 & 0 & 0 & 0 \\ 0 & 0 & 0 & -1 & 0 & 0 & 0 & 0 \\ 1 & 0 & 0 & 0 & 0 & 0 & 0 & 0 \\ 0 & 1 & 0 & 0 & 0 & 0 & 0 & 0 \\ 0 & 0 & 0 & 0 & 0 & 0 & -1 & 0 \\ 0 & 0 & 0 & 0 & 0 & 0 & 0 & -1 \\ 0 & 0 & 0 & 0 & 1 & 0 & 0 & 0 \\ 0 & 0 & 0 & 0 & 0 & 1 & 0 & 0 \end{pmatrix}, \tag{H5}$$

$$M_M\left(g_1^{(3)}\right) = \begin{pmatrix} 0 & 0 & 0 & 0 & -1 & 0 & 0 & 0 \\ 0 & 0 & 0 & 0 & 0 & -1 & 0 & 0 \\ 0 & 0 & 0 & 0 & 0 & 0 & 1 & 0 \\ 0 & 0 & 0 & 0 & 0 & 0 & 0 & 1 \\ 1 & 0 & 0 & 0 & 0 & 0 & 0 & 0 \\ 0 & 1 & 0 & 0 & 0 & 0 & 0 & 0 \\ 0 & 0 & -1 & 0 & 0 & 0 & 0 & 0 \\ 0 & 0 & 0 & -1 & 0 & 0 & 0 & 0 \end{pmatrix}, \tag{H6}$$

$$M_M\left(g_2^{(1)}\right) = \begin{pmatrix} 0 & 1 & 0 & 0 & 0 & 0 & 0 & 0 \\ 1 & 0 & 0 & 0 & 0 & 0 & 0 & 0 \\ 0 & 0 & 0 & -1 & 0 & 0 & 0 & 0 \\ 0 & 0 & -1 & 0 & 0 & 0 & 0 & 0 \\ 0 & 0 & 0 & 0 & 0 & 1 & 0 & 0 \\ 0 & 0 & 0 & 0 & 1 & 0 & 0 & 0 \\ 0 & 0 & 0 & 0 & 0 & 0 & 0 & -1 \\ 0 & 0 & 0 & 0 & 0 & 0 & -1 & 0 \end{pmatrix}. \tag{H7}$$

The corresponding intrinsic group $\overline{\tilde{L}_{SS}^{M}}$ can be written as follows:

$$M_M\left(\bar{g}_1^{(2)}\right) = \begin{pmatrix} 0 & 0 & -1 & 0 & 0 & 0 & 0 & 0 \\ 0 & 0 & 0 & 1 & 0 & 0 & 0 & 0 \\ 1 & 0 & 0 & 0 & 0 & 0 & 0 & 0 \\ 0 & -1 & 0 & 0 & 0 & 0 & 0 & 0 \\ 0 & 0 & 0 & 0 & 0 & 0 & 1 & 0 \\ 0 & 0 & 0 & 0 & 0 & 0 & 0 & -1 \\ 0 & 0 & 0 & 0 & -1 & 0 & 0 & 0 \\ 0 & 0 & 0 & 0 & 0 & 1 & 0 & 0 \end{pmatrix}, \tag{H8}$$

$$M_M\left(g_1^{(3)}\right) = \begin{pmatrix} 0 & 0 & 0 & 0 & -1 & 0 & 0 & 0 \\ 0 & 0 & 0 & 0 & 0 & -1 & 0 & 0 \\ 0 & 0 & 0 & 0 & 0 & 0 & -1 & 0 \\ 0 & 0 & 0 & 0 & 0 & 0 & 0 & -1 \\ 1 & 0 & 0 & 0 & 0 & 0 & 0 & 0 \\ 0 & 1 & 0 & 0 & 0 & 0 & 0 & 0 \\ 0 & 0 & 1 & 0 & 0 & 0 & 0 & 0 \\ 0 & 0 & 0 & 1 & 0 & 0 & 0 & 0 \end{pmatrix}, \tag{H9}$$

$$M_M\left(g_2^{(1)}\right) = \begin{pmatrix} 0 & 1 & 0 & 0 & 0 & 0 & 0 & 0 \\ 1 & 0 & 0 & 0 & 0 & 0 & 0 & 0 \\ 0 & 0 & 0 & 1 & 0 & 0 & 0 & 0 \\ 0 & 0 & 1 & 0 & 0 & 0 & 0 & 0 \\ 0 & 0 & 0 & 0 & 0 & 1 & 0 & 0 \\ 0 & 0 & 0 & 0 & 1 & 0 & 0 & 0 \\ 0 & 0 & 0 & 0 & 0 & 0 & 0 & 1 \\ 0 & 0 & 0 & 0 & 0 & 0 & 1 & 0 \end{pmatrix}, \quad \text{(H10)}$$

The canonical subgroup chain is $\tilde{L}_{SS}^{M} > \tilde{L}_{SS1}^{M}$, where $\tilde{L}_{SS1}^{M} \cong C_2$ consists of $g_1^{(1)}$ and $g_2^{(1)}$ and the corresponding intrinsic subgroup chain $\overline{\tilde{L}_{SS1}^{M}}$ consists of $\bar{g}_1^{(1)}$ and $\bar{g}_2^{(1)}$.

By constructing the linear combination of class operators $C = \sum_{i,a} k_i^{(a)} C_i^{(a)}$ ($C_i^{(a)}$ are class operators and $k_i^{(a)}$ are arbitrary real coefficients to avoid accidental elimination of the same class operators), we get the CSCO-I $C$ of $\tilde{L}_{SS}^{M}$, $C(s)$ of the canonical subgroup chain and the $\bar{C}(s)$ of the intrinsic subgroup chain with all arbitrary real constants $k_i^{(a)} = 1$ for all $i = 1,2$ and $a = 1,2,3,4$:

$$C = \sum_{\substack{i=1,2\\ a=1,2,3,4}} C_i^{(a)} = \begin{pmatrix} 8 & 0 & 0 & 0 & 0 & -8 & 0 & 0 \\ 0 & 8 & 0 & 0 & -8 & 0 & 0 & 0 \\ 0 & 0 & 8 & 0 & 0 & 0 & 0 & -8 \\ 0 & 0 & 0 & 8 & 0 & 0 & -8 & 0 \\ 0 & 8 & 0 & 0 & 8 & 0 & 0 & 0 \\ 8 & 0 & 0 & 0 & 0 & 8 & 0 & 0 \\ 0 & 0 & 0 & 8 & 0 & 0 & 8 & 0 \\ 0 & 0 & 8 & 0 & 0 & 0 & 0 & 8 \end{pmatrix}, \quad \text{(H11)}$$

$$C(s) = \sum_{\substack{i=1,2\\ a=1}} C_{i1}^{(a)} = \begin{pmatrix} 2 & 2 & 0 & 0 & 0 & 0 & 0 & 0 \\ 2 & 2 & 0 & 0 & 0 & 0 & 0 & 0 \\ 0 & 0 & 2 & -2 & 0 & 0 & 0 & 0 \\ 0 & 0 & -2 & 2 & 0 & 0 & 0 & 0 \\ 0 & 0 & 0 & 0 & 2 & 2 & 0 & 0 \\ 0 & 0 & 0 & 0 & 2 & 2 & 0 & 0 \\ 0 & 0 & 0 & 0 & 0 & 0 & 2 & -2 \\ 0 & 0 & 0 & 0 & 0 & 0 & -2 & 2 \end{pmatrix}, \quad \text{(H12)}$$

$$\bar{C}(s) = \sum_{\substack{i=1,2\\ a=1}} \bar{C}_{i1}^{(a)} = \begin{pmatrix} 2 & 2 & 0 & 0 & 0 & 0 & 0 & 0 \\ 2 & 2 & 0 & 0 & 0 & 0 & 0 & 0 \\ 0 & 0 & 2 & 2 & 0 & 0 & 0 & 0 \\ 0 & 0 & 2 & 2 & 0 & 0 & 0 & 0 \\ 0 & 0 & 0 & 0 & 2 & 2 & 0 & 0 \\ 0 & 0 & 0 & 0 & 2 & 2 & 0 & 0 \\ 0 & 0 & 0 & 0 & 0 & 0 & 2 & 2 \\ 0 & 0 & 0 & 0 & 0 & 0 & 2 & 2 \end{pmatrix}, \quad \text{(H13)}$$

The complete set $(C, C(s), \bar{C}(s))$ and the unitary transformation matrix $U$ constructed by the common eigenvectors are applied to lift the degeneracy.

Simultaneous block diagonalization of the left regular projective rep matrices of all group elements is achieved as:

$$U^{-1}M_M\left(g_1^{(2)}\right)U=\begin{pmatrix} 0 & 1 & 0 & 0 & 0 & 0 & 0 & 0 \\ -1 & 0 & 0 & 0 & 0 & 0 & 0 & 0 \\ 0 & 0 & 0 & 1 & 0 & 0 & 0 & 0 \\ 0 & 0 & -1 & 0 & 0 & 0 & 0 & 0 \\ 0 & 0 & 0 & 0 & 0 & -1 & 0 & 0 \\ 0 & 0 & 0 & 0 & 1 & 0 & 0 & 0 \\ 0 & 0 & 0 & 0 & 0 & 0 & 0 & -1 \\ 0 & 0 & 0 & 0 & 0 & 0 & 1 & 0 \end{pmatrix}, \quad \text{(H14)}$$

$$U^{-1}M_M\left(g_1^{(3)}\right)U=\begin{pmatrix} -i & 0 & 0 & 0 & 0 & 0 & 0 & 0 \\ 0 & i & 0 & 0 & 0 & 0 & 0 & 0 \\ 0 & 0 & i & 0 & 0 & 0 & 0 & 0 \\ 0 & 0 & 0 & -i & 0 & 0 & 0 & 0 \\ 0 & 0 & 0 & 0 & i & 0 & 0 & 0 \\ 0 & 0 & 0 & 0 & 0 & -i & 0 & 0 \\ 0 & 0 & 0 & 0 & 0 & 0 & -i & 0 \\ 0 & 0 & 0 & 0 & 0 & 0 & 0 & i \end{pmatrix}, \quad \text{(H15)}$$

$$U^{-1}M_M\left(g_2^{(1)}\right)U=\begin{pmatrix} -1 & 0 & 0 & 0 & 0 & 0 & 0 & 0 \\ 0 & 1 & 0 & 0 & 0 & 0 & 0 & 0 \\ 0 & 0 & -1 & 0 & 0 & 0 & 0 & 0 \\ 0 & 0 & 0 & 1 & 0 & 0 & 0 & 0 \\ 0 & 0 & 0 & 0 & 1 & 0 & 0 & 0 \\ 0 & 0 & 0 & 0 & 0 & -1 & 0 & 0 \\ 0 & 0 & 0 & 0 & 0 & 0 & 1 & 0 \\ 0 & 0 & 0 & 0 & 0 & 0 & 0 & -1 \end{pmatrix}, \quad \text{(H16)}$$

With the assistance of the character table of projective irreps in Table H1, we find that there are two different projective irreps $M_M^1$:

$$M_M^1\left(g_1^{(2)}\right)=\begin{pmatrix} 0 & 1 \\ -1 & 0 \end{pmatrix}\quad M_M^1\left(g_1^{(3)}\right)=\begin{pmatrix} -i & 0 \\ 0 & i \end{pmatrix}\quad M_M^1\left(g_2^{(1)}\right)=\begin{pmatrix} -1 & 0 \\ 0 & 1 \end{pmatrix}, \quad \text{(H17)}$$

and $M_M^2$:

$$M_M^2\left(g_1^{(2)}\right)=\begin{pmatrix} 0 & 1 \\ -1 & 0 \end{pmatrix}\quad M_M^2\left(g_1^{(3)}\right)=\begin{pmatrix} i & 0 \\ 0 & -i \end{pmatrix}\quad M_M^2\left(g_2^{(1)}\right)=\begin{pmatrix} -1 & 0 \\ 0 & 1 \end{pmatrix}, \quad \text{(H18)}$$

Furthermore, irreps $d_M^l$ can be obtained by $d_M^l(\{g_{s_i}^{(a)}||R_i|\tau_i^{(a)}\})=exp(-iM\cdot\tau_i^{(a)})M_M^l(\{g_{s_i}^{(a)}||R_i|\tau_i^{(a)}\})$:

$$d_M^1\left(g_1^{(2)}\right)=\begin{pmatrix} 0 & -1 \\ 1 & 0 \end{pmatrix}\quad d_M^1\left(g_1^{(3)}\right)=\begin{pmatrix} -i & 0 \\ 0 & i \end{pmatrix}\quad d_M^1\left(g_2^{(1)}\right)=\begin{pmatrix} -1 & 0 \\ 0 & 1 \end{pmatrix}, \quad \text{(H19)}$$

and:

$$d_M^2\left(g_1^{(2)}\right)=\begin{pmatrix} 0 & -1 \\ 1 & 0 \end{pmatrix}\quad d_M^2\left(g_1^{(3)}\right)=\begin{pmatrix} i & 0 \\ 0 & -i \end{pmatrix}\quad d_M^2\left(g_2^{(1)}\right)=\begin{pmatrix} -1 & 0 \\ 0 & 1 \end{pmatrix}, \quad \text{(H20)}$$

**Table H1:** Character table of projective irreps $M_M^1$ and $M_M^2$ of $\tilde{L}_{SS}^M$.

| | $g_1^{(1)}$ | $g_1^{(2)}$ | $g_1^{(3)}$ | $g_1^{(4)}$ | $g_2^{(1)}$ | $g_2^{(2)}$ | $g_2^{(3)}$ | $g_2^{(4)}$ |
|---|---|---|---|---|---|---|---|---|
| $M_M^1$ | 2 | 0 | 0 | 0 | 0 | 0 | $2i$ | 0 |
| $M_M^2$ | 2 | 0 | $2i$ | 0 | 0 | 0 | $-2i$ | 0 |

To identify the reality type of co-irrep associated with the irreps of MUSG, the modified Dimmock and Wheeler's character sum rule in Eq. (E17) are performed:

$$\sum_{g_i^{(a)}\in\tilde{L}_{SS}^M}\chi^1\left(\left(TAg_i^{(a)}\right)^2\right)=2+2+0+0+2+2+0+0=8\cdots(a),\qquad\text{(H21)}$$

$$\sum_{g_i^{(a)}\in\tilde{L}_{SS}^M}\chi^3\left(\left(TAg_i^{(a)}\right)^2\right)=2+2+0+0+2+2+0+0=8\cdots(a),\qquad\text{(H22)}$$

The co-rep matrices $D_M^l$ have identical dimensions with irrep $d_M^l$, and the result is shown in Table H2. Besides, the co-irreps matrices of the SSG of $CoNb_3S_6$ at some high-symmetry points are provided in Tables H2 to H7 and co-irreps of the MSG of $CoNb_3S_6$ calculated by CSCO method are also tabulated in Tables H8 to H13.

## 2. Irreducible co-representations of the SSG and MSG of $CoNb_3S_6$

In the following tables, constants $\alpha_1=0.5774i$, $\alpha_2=0.4082+0.7071i$, $\alpha_3=0.8161+0.0244i$, $\alpha_4=0.8165$, $\alpha_5=0.4143+0.7036i$, $\alpha_6=0.5000+0.2887i$, $\alpha_7=0.3870+0.7190i$, $\alpha_8=0.8165+0.0071i$, $\alpha_9=0.8161+0.0269i$, $\alpha_{10}=0.4313+0.6933i$, $\alpha_{11}=0.3847+0.7202i$, $\alpha_{12}=0.3827+0.7213i$, $\alpha_{13}=0.8160+0.0292i$, $\alpha_{14}=0.4333+0.6920i$, $\alpha_{15}=0.2406+0.9706i$ are employed for simplicity.

**Table H2:** Co-irreps of $CoNb_3S_6$'s SSG at Γ(0,0,0). The number in the parentheses in the first column denotes the dimension of the co-irrep.

| | $\{1\|\|2_{001}\|0\ 0\ 1/2\}$ | $\{3_{001}^1\|\|3_{001}^1\|0\}$ | $\{2_{24\bar{3}}\|\|1\|1\ 0\ 0\}$ | $\{2_{423}\|\|1\|0\ 1\ 0\}$ | $\{T2_{110}\|\|2_{110}\|0\}$ |
|---|---|---|---|---|---|
| $D_\Gamma^1(2)$ | $\begin{pmatrix}1&0\\0&1\end{pmatrix}$ | $\begin{pmatrix}e^{i\frac{\pi}{3}}&0\\0&e^{-i\frac{\pi}{3}}\end{pmatrix}$ | $\begin{pmatrix}\alpha_1^*&\alpha_2^*\\-\alpha_2&\alpha_1\end{pmatrix}$ | $\begin{pmatrix}\alpha_1&-\alpha_2^*\\\alpha_2&\alpha_1^*\end{pmatrix}$ | $\begin{pmatrix}1&0\\0&1\end{pmatrix}$ |
| $D_{\Gamma}^{2}(2)$ | $\begin{pmatrix}1 & 0\\0 & 1\end{pmatrix}$ | $\begin{pmatrix}e^{i\frac{\pi}{3}} & 0\\0 & -1\end{pmatrix}$ | $\begin{pmatrix}\alpha_1 & -\alpha_3\\\alpha_3^* & \alpha_1^*\end{pmatrix}$ | $\begin{pmatrix}\alpha_1^* & -\alpha_7\\\alpha_7^* & \alpha_1\end{pmatrix}$ | $\begin{pmatrix}1 & 0\\0 & 1\end{pmatrix}$ |
| $D_{\Gamma}^{3}(2)$ | $\begin{pmatrix}1 & 0\\0 & 1\end{pmatrix}$ | $\begin{pmatrix}e^{-i\frac{\pi}{3}} & 0\\0 & -1\end{pmatrix}$ | $\begin{pmatrix}\alpha_1^* & -\alpha_3^*\\\alpha_3 & \alpha_1\end{pmatrix}$ | $\begin{pmatrix}\alpha_1 & -\alpha_7^*\\\alpha_7 & \alpha_1^*\end{pmatrix}$ | $\begin{pmatrix}1 & 0\\0 & 1\end{pmatrix}$ |
| $D_{\Gamma}^{4}(2)$ | $\begin{pmatrix}-1 & 0\\0 & -1\end{pmatrix}$ | $\begin{pmatrix}e^{-i\frac{\pi}{3}} & 0\\0 & e^{i\frac{\pi}{3}}\end{pmatrix}$ | $\begin{pmatrix}\alpha_1 & \alpha_4\\-\alpha_4 & \alpha_1^*\end{pmatrix}$ | $\begin{pmatrix}\alpha_1^* & \alpha_2\\-\alpha_2^* & \alpha_1\end{pmatrix}$ | $\begin{pmatrix}1 & 0\\0 & 1\end{pmatrix}$ |
| $D_{\Gamma}^{5}(2)$ | $\begin{pmatrix}-1 & 0\\0 & -1\end{pmatrix}$ | $\begin{pmatrix}e^{-i\frac{\pi}{3}} & 0\\0 & -1\end{pmatrix}$ | $\begin{pmatrix}\alpha_1^* & -\alpha_5\\\alpha_5^* & \alpha_1\end{pmatrix}$ | $\begin{pmatrix}\alpha_1 & -\alpha_8^*\\\alpha_8 & \alpha_1^*\end{pmatrix}$ | $\begin{pmatrix}1 & 0\\0 & 1\end{pmatrix}$ |
| $D_{\Gamma}^{6}(2)$ | $\begin{pmatrix}-1 & 0\\0 & -1\end{pmatrix}$ | $\begin{pmatrix}e^{i\frac{\pi}{3}} & 0\\0 & -1\end{pmatrix}$ | $\begin{pmatrix}\alpha_1 & -\alpha_5^*\\\alpha_5 & \alpha_1^*\end{pmatrix}$ | $\begin{pmatrix}\alpha_1^* & -\alpha_8\\\alpha_8^* & \alpha_1\end{pmatrix}$ | $\begin{pmatrix}1 & 0\\0 & 1\end{pmatrix}$ |

**Table H3:** Co-irreps of $CoNb_3S_6$'s SSG at M(1/2,0,0). The number in the parentheses in the first column denotes the dimension of the co-irrep.

| | $\{1\|\|2_{001}\|0\ 0\ 1/2\}$ | $\{2_{24\bar{3}}\|\|1\|1\ 0\ 0\}$ | $\{2_{423}\|\|1\|0\ 1\ 0\}$ | $\{T2_{010}\|\|2_{210}\|0\}$ |
|---|---|---|---|---|
| $D_{M}^{1}(2)$ | $\begin{pmatrix}-1 & 0\\0 & 1\end{pmatrix}$ | $\begin{pmatrix}0 & -1\\1 & 0\end{pmatrix}$ | $\begin{pmatrix}i & 0\\0 & -i\end{pmatrix}$ | $\begin{pmatrix}1 & 0\\0 & 1\end{pmatrix}$ |
| $D_{M}^{2}(2)$ | $\begin{pmatrix}-1 & 0\\0 & 1\end{pmatrix}$ | $\begin{pmatrix}0 & -1\\1 & 0\end{pmatrix}$ | $\begin{pmatrix}-i & 0\\0 & i\end{pmatrix}$ | $\begin{pmatrix}1 & 0\\0 & 1\end{pmatrix}$ |

**Table H4:** Co-irreps of $CoNb_3S_6$'s SSG at K(1/3,1/3,0). The number in the parentheses in the first column denotes the dimension of the co-irrep.

| | $\{3_{001}^{1}\|\|3_{001}^{1}\|0\}$ | $\{2_{24\bar{3}}\|\|1\|1\ 0\ 0\}$ | $\{2_{423}\|\|1\|0\ 1\ 0\}$ | $\{T2_{010}\|\|2_{210}\|0\}$ |
|---|---|---|---|---|
| $D_{K}^{1}(2)$ | $\begin{pmatrix}-1 & 0\\0 & e^{i\frac{\pi}{3}}\end{pmatrix}$ | $\begin{pmatrix}\alpha_6 & \alpha_9\\\alpha_{10} & -\alpha_6\end{pmatrix}$ | $\begin{pmatrix}-\alpha_6 & \alpha_{10}^*\\-\alpha_{11}^* & \alpha_6\end{pmatrix}$ | $\begin{pmatrix}1 & 0\\0 & 1\end{pmatrix}$ |
| $D_{K}^{2}(2)$ | $\begin{pmatrix}-1 & 0\\0 & e^{-i\frac{\pi}{3}}\end{pmatrix}$ | $\begin{pmatrix}-\alpha_6 & \alpha_{10}^*\\-\alpha_{11}^* & \alpha_6\end{pmatrix}$ | $\begin{pmatrix}\alpha_6 & \alpha_9\\\alpha_{10} & -\alpha_6\end{pmatrix}$ | $\begin{pmatrix}1 & 0\\0 & 1\end{pmatrix}$ |
| $D_{K}^{3}(2)$ | $\begin{pmatrix}e^{i\frac{\pi}{3}} & 0\\0 & e^{-i\frac{\pi}{3}}\end{pmatrix}$ | $\begin{pmatrix}\alpha_6 & \alpha_{13}^*\\\alpha_{12} & -\alpha_6\end{pmatrix}$ | $\begin{pmatrix}-\alpha_6 & \alpha_{12}^*\\-\alpha_{14}^* & \alpha_6\end{pmatrix}$ | $\begin{pmatrix}1 & 0\\0 & 1\end{pmatrix}$ |

**Table H5:** Co-irrep of $CoNb_3S_6$'s SSG at A(0,0,1/2). The number in the parentheses in the first column denotes the dimension of the co-irrep.

| | $\{1\|\|2_{001}\|0\ 0\ 1/2\}$ | $\{3_{001}^{1}\|\|3_{001}^{1}\|0\}$ | $\{2_{24\bar{3}}\|\|1\|1\ 0\ 0\}$ | $\{2_{423}\|\|1\|0\ 1\ 0\}$ | $\{T2_{110}\|\|2_{110}\|0\}$ |
|---|---|---|---|---|---|
| $D_{A}^{1}(2)$ | $\begin{pmatrix}-i & 0\\0 & -i\end{pmatrix}$ | $\begin{pmatrix}e^{i\frac{\pi}{3}} & 0\\0 & e^{-i\frac{\pi}{3}}\end{pmatrix}$ | $\begin{pmatrix}\alpha_1^* & \alpha_2^*\\-\alpha_2 & \alpha_1\end{pmatrix}$ | $\begin{pmatrix}\alpha_1 & -\alpha_2\\\alpha_2^* & \alpha_1^*\end{pmatrix}$ | $\begin{pmatrix}1 & 0\\0 & 1\end{pmatrix}$ |
| $D_A^2(2)$ | $\begin{pmatrix}-i & 0\\ 0 & -i\end{pmatrix}$ | $\begin{pmatrix}e^{i\frac{\pi}{3}} & 0\\ 0 & -1\end{pmatrix}$ | $\begin{pmatrix}\alpha_1 & -\alpha_3\\ \alpha_3^* & \alpha_1^*\end{pmatrix}$ | $\begin{pmatrix}\alpha_1^* & -\alpha_7\\ \alpha_7^* & \alpha_1\end{pmatrix}$ | $\begin{pmatrix}1 & 0\\ 0 & 1\end{pmatrix}$ |
| $D_A^3(2)$ | $\begin{pmatrix}-i & 0\\ 0 & -i\end{pmatrix}$ | $\begin{pmatrix}e^{-i\frac{\pi}{3}} & 0\\ 0 & -1\end{pmatrix}$ | $\begin{pmatrix}\alpha_1^* & -\alpha_3^*\\ \alpha_3 & \alpha_1\end{pmatrix}$ | $\begin{pmatrix}\alpha_1 & -\alpha_7^*\\ \alpha_7 & \alpha_1^*\end{pmatrix}$ | $\begin{pmatrix}1 & 0\\ 0 & 1\end{pmatrix}$ |
| $D_A^4(2)$ | $\begin{pmatrix}i & 0\\ 0 & i\end{pmatrix}$ | $\begin{pmatrix}e^{-i\frac{\pi}{3}} & 0\\ 0 & e^{i\frac{\pi}{3}}\end{pmatrix}$ | $\begin{pmatrix}\alpha_1 & \alpha_4\\ -\alpha_4 & \alpha_1^*\end{pmatrix}$ | $\begin{pmatrix}\alpha_1^* & \alpha_2\\ -\alpha_2^* & \alpha_1\end{pmatrix}$ | $\begin{pmatrix}1 & 0\\ 0 & 1\end{pmatrix}$ |
| $D_A^5(2)$ | $\begin{pmatrix}i & 0\\ 0 & i\end{pmatrix}$ | $\begin{pmatrix}e^{-i\frac{\pi}{3}} & 0\\ 0 & -1\end{pmatrix}$ | $\begin{pmatrix}\alpha_1^* & -\alpha_5\\ \alpha_5^* & \alpha_1\end{pmatrix}$ | $\begin{pmatrix}\alpha_1 & -\alpha_8^*\\ \alpha_8 & \alpha_1^*\end{pmatrix}$ | $\begin{pmatrix}1 & 0\\ 0 & 1\end{pmatrix}$ |
| $D_A^6(2)$ | $\begin{pmatrix}i & 0\\ 0 & i\end{pmatrix}$ | $\begin{pmatrix}e^{i\frac{\pi}{3}} & 0\\ 0 & -1\end{pmatrix}$ | $\begin{pmatrix}\alpha_1 & -\alpha_5^*\\ \alpha_5 & \alpha_1^*\end{pmatrix}$ | $\begin{pmatrix}\alpha_1^* & -\alpha_8\\ \alpha_8^* & \alpha_1\end{pmatrix}$ | $\begin{pmatrix}1 & 0\\ 0 & 1\end{pmatrix}$ |

**Table H6:** Co-irreps of $CoNb_3S_6$'s SSG at L(1/2,0,1/2). The number in the parentheses in the first column denotes the dimension of the co-irrep.

| | $\{1\|\|2_{001}\|0\ 0\ 1/2\}$ | $\{2_{24\bar{3}}\|\|1\|1\ 0\ 0\}$ | $\{2_{423}\|\|1\|0\ 1\ 0\}$ | $\{T2_{010}\|\|2_{210}\|0\}$ |
|---|---|---|---|---|
| $D_L^1(2)$ | $\begin{pmatrix}\alpha_{15} & 0\\ 0 & -\alpha_{15}\end{pmatrix}$ | $\begin{pmatrix}0 & -1\\ -1 & 0\end{pmatrix}$ | $\begin{pmatrix}-i & 0\\ 0 & i\end{pmatrix}$ | $\begin{pmatrix}1 & 0\\ 0 & 1\end{pmatrix}$ |
| $D_L^2(2)$ | $\begin{pmatrix}\alpha_{15} & 0\\ 0 & -\alpha_{15}\end{pmatrix}$ | $\begin{pmatrix}0 & 1\\ 1 & 0\end{pmatrix}$ | $\begin{pmatrix}i & 0\\ 0 & -i\end{pmatrix}$ | $\begin{pmatrix}1 & 0\\ 0 & 1\end{pmatrix}$ |

**Table H7:** Co-irreps of $CoNb_3S_6$'s SSG at H (1/3,1/3,1/2). The number in the parentheses in the first column denotes the dimension of the co-irrep.

| | $\{3_{001}^1\|\|3_{001}^1\|0\}$ | $\{2_{24\bar{3}}\|\|1\|1\ 0\ 0\}$ | $\{2_{423}\|\|1\|0\ 1\ 0\}$ | $\{T2_{010}\|\|2_{210}\|0\}$ |
|---|---|---|---|---|
| $D_H^1(2)$ | $\begin{pmatrix}-1 & 0\\ 0 & e^{i\frac{\pi}{3}}\end{pmatrix}$ | $\begin{pmatrix}\alpha_6 & \alpha_{11}^*\\ -\alpha_{10}^* & -\alpha_6\end{pmatrix}$ | $\begin{pmatrix}-\alpha_6 & -\alpha_{10}\\ -\alpha_9 & \alpha_6\end{pmatrix}$ | $\begin{pmatrix}1 & 0\\ 0 & 1\end{pmatrix}$ |
| $D_H^2(2)$ | $\begin{pmatrix}-1 & 0\\ 0 & e^{-i\frac{\pi}{3}}\end{pmatrix}$ | $\begin{pmatrix}-\alpha_6 & \alpha_{10}^*\\ -\alpha_{11}^* & \alpha_6\end{pmatrix}$ | $\begin{pmatrix}\alpha_6 & \alpha_9\\ \alpha_{10} & -\alpha_6\end{pmatrix}$ | $\begin{pmatrix}1 & 0\\ 0 & 1\end{pmatrix}$ |
| $D_H^3(2)$ | $\begin{pmatrix}e^{i\frac{\pi}{3}} & 0\\ 0 & e^{-i\frac{\pi}{3}}\end{pmatrix}$ | $\begin{pmatrix}\alpha_6 & \alpha_{13}^*\\ \alpha_{12} & -\alpha_6\end{pmatrix}$ | $\begin{pmatrix}-\alpha_6 & \alpha_{12}^*\\ -\alpha_{14}^* & \alpha_6\end{pmatrix}$ | $\begin{pmatrix}1 & 0\\ 0 & 1\end{pmatrix}$ |

**Table H8:** Co-irreps of $CoNb_3S_6$'s MSG $P32'1$ (150.27) at $\Gamma(0,0,0)$. The number in the parentheses in the first column denotes the dimension of the co-irrep.

| | $\{3_{001}^1\|\|3_{001}^1\|0,0,0\}$ | $\{T2_{110}\|\|2_{110}\|0,0,0\}$ |
|---|---|---|
| $D_\Gamma^1(1)$ | $e^{i\frac{\pi}{3}}$ | 1 |
| $D_\Gamma^2(1)$ | $e^{-i\frac{\pi}{3}}$ | 1 |
| $D_{\Gamma}^{3}(1)$ | $-1$ | 1 |

**Table H9:** Co-irreps of $CoNb_3S_6$'s MSG $P32'1$ (150.27) at $A(0,0,1/2)$. The number in the parentheses in the first column denotes the dimension of the co-irrep.

| | $\{3_{001}^{1}\|\|3_{001}^{1}\|0,0,0\}$ | $\{T2_{110}\|\|2_{110}\|0,0,0\}$ |
|---|---|---|
| $D_{A}^{1}(1)$ | $e^{i\frac{\pi}{3}}$ | 1 |
| $D_{A}^{2}(1)$ | $e^{-i\frac{\pi}{3}}$ | 1 |
| $D_{A}^{3}(1)$ | $-1$ | 1 |

**Table H10:** Co-irreps of $CoNb_3S_6$'s MSG $P32'1$ (150.27) at $H(1/3,1/3,1/2)$. The number in the parentheses in the first column denotes the dimension of the co-irrep.

| | $\{3_{001}^{1}\|\|3_{001}^{1}\|0,0,0\}$ |
|---|---|
| $D_{H}^{1}(1)$ | $e^{i\frac{\pi}{3}}$ |
| $D_{H}^{2}(1)$ | $e^{-i\frac{\pi}{3}}$ |
| $D_{H}^{3}(1)$ | $-1$ |

**Table H11:** Co-irreps of $CoNb_3S_6$'s MSG $P32'1$ (150.27) at $K(1/3,1/3,0)$. The number in the parentheses in the first column denotes the dimension of the co-irrep.

| | $\{3_{001}^{1}\|\|3_{001}^{1}\|0,0,0\}$ |
|---|---|
| $D_{K}^{1}(1)$ | $e^{i\frac{\pi}{3}}$ |
| $D_{K}^{2}(1)$ | $e^{-i\frac{\pi}{3}}$ |
| $D_{K}^{3}(1)$ | $-1$ |

**Table H12:** Co-irreps matrices of $CoNb_3S_6$'s MSG $P32'1$ (150.27) at $HA(-1/3,-1/3,1/2)$. The number in the parentheses in the first column denotes the dimension of the co-irrep.

| | $\{3_{001}^{1}\|\|3_{001}^{1}\|0,0,0\}$ |
|---|---|
| $D_{HA}^{1}(1)$ | $e^{i\frac{\pi}{3}}$ |
| $D_{HA}^{2}(1)$ | $e^{-i\frac{\pi}{3}}$ |
| $D_{HA}^{3}(1)$ | $-1$ |

**Table H13:** Co-irreps matrices of $CoNb_3S_6$'s MSG $P32'1$ (No. 150.27) at $KA(-1/3,-1/3,0)$. The number in the parentheses in the first column denotes the dimension of the co-irrep.

| | $\{3_{001}^{1}\|\|3_{001}^{1}\|0,0,0\}$ |
|---|---|
| $D_{KA}^{1}(1)$ | $e^{i\frac{\pi}{3}}$ |
| $D_{KA}^{2}(1)$ | $e^{-i\frac{\pi}{3}}$ |
| $D_{KA}^{3}(1)$ | $-1$ |

# I. Online program in SSG identification

Here we will demonstrate how to use our online program (https://findspingroup.com/) to obtain the desired information from the $RuO_2$.mcif file. On the website, the sentence "All experimentally determined magnetic structures available in the MAGNDATA database have been identified and provided here" presents the complete list of SSGs of magnetic structures. This file includes the nontrivial SSG, spin-only group for all commensurate magnetic materials in MAGNDATA.

First, put the of $RuO_2$.mcif in our program.

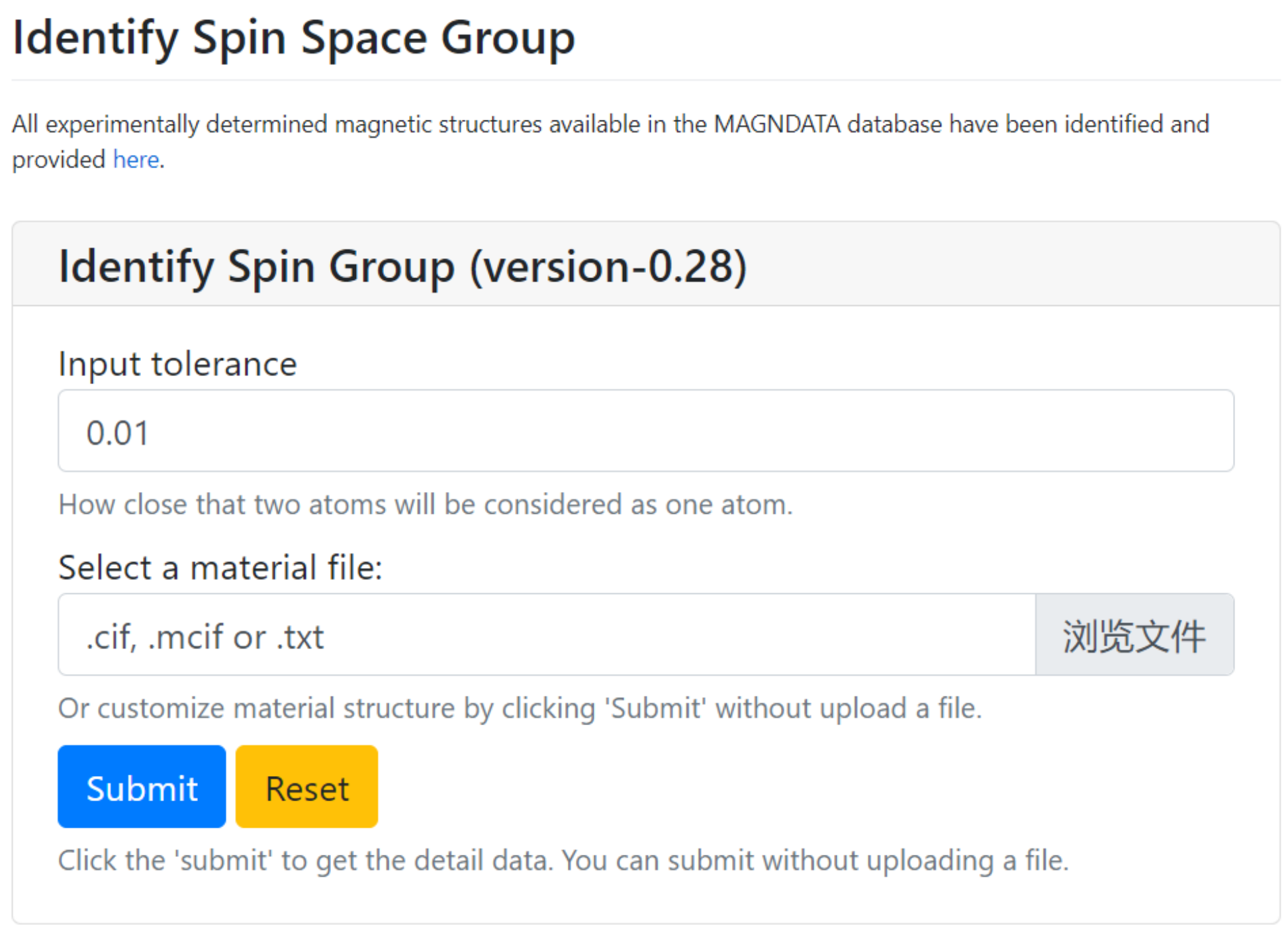

**Fig. I1:** The interface allows users to input the material file on the website.

Then you can refine the cell. For example, change the angles of moments by changing the coordinates $M_x$ $M_y$ $M_z$ which are written under the input lattice.

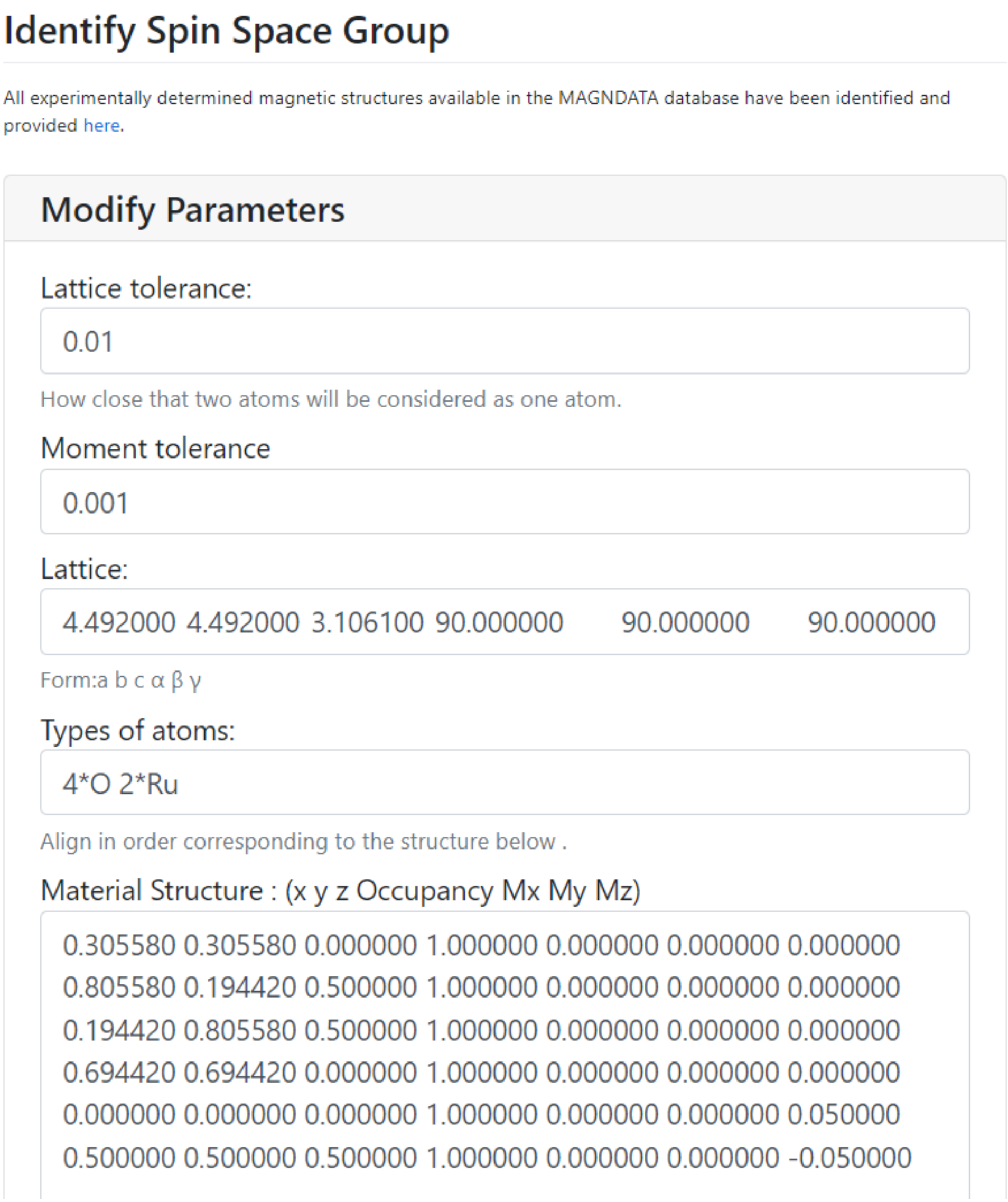

**Fig. I2:** After the submission of the material file, the second interface is displayed. In this interface, users can modify the lattice tolerance, the moment tolerance, the lattice constant, the types of atoms, atomic sites and magnetic structure.

Click submit then after a few seconds you will obtain the information normally. For some special cases which have a lot of atoms and high symmetry, it might take a few minutes. In Fig. I3, the nontrivial SSG, the magnetic configuration, $\boldsymbol{G_0}$, $\boldsymbol{L_0}$, $\boldsymbol{G^s}$, $\boldsymbol{i_t}$, $\boldsymbol{i_k}$, the SG and the MSG are provided. In addition, the standard POSCAR in the default basis and all standard spin space symmetries are also given in the online program.

## Identify Spin Space Group

All experimentally determined magnetic structures available in the MAGNDATA database have been identified and provided here.

| Filename | nSSG Symbol | Material Configuration | G0 Symbol | L0 Symbol | Spin part PG | it | ik |
|---|---|---|---|---|---|---|---|
| $0.607/RuO2$ | $P^{-1}4_2/^{1}m^{-1}n^{1}m$ | $Collinear$ | $P4_2/mnm(136)$ | $Cmmm(65)$ | $-1$ | $2$ | $1$ |

| G Symbol | MSG Symbol (BNS) | MSG type | PTcheck |
|---|---|---|---|
| $P4_2/mnm(136)$ | $P4_2'/mnm'(136.499)$ | $3$ | $without\ PT/without\ PTtau/without\ effectivePT$ |

**Fig. I3:** The results of SSG identification,

## J. First-principles methods

All DFT calculations herein are performed using projector augmented wave method, implemented in Vienna ab initio simulation package (VASP) [92, 93]. The generalized gradient approximation of the Perdew-Burke-Ernzerhof-type exchange-correlation potential [94] is adopted. To include the effect of electron correlation, the DFT+U approach within the rotationally invariant formalism [95] were performing with $U_{eff}$ = 2.0 eV for Ru 4d ($RuO_2$), $U_{eff}$ = 4.0 eV for Ce 4f ($CeAuAl_3$) and $U_{eff}$ = 1.0 eV for Co 3d ($CoNb_3S_6$). Tight-binding models are constructed from DFT bands using the WANNIER90 package [96, 97]; WannierTools package is used for the calculations of anomalous hall conductivity [98].